%% file: paper.tex
\documentclass[prd,twocolumn,showpacs,showkeys,preprintnumbers,floatfix,nofootinbib,superscriptaddress]{revtex4-1}
\usepackage{amsfonts} % AMS
\usepackage{amssymb} % AMS
\usepackage{amsmath} % AMS
\usepackage{graphicx} % Include figure files
\usepackage{subfigure} % Include figure files
\usepackage{array} % array
\usepackage{dcolumn} % Align table columns on decimal point
\usepackage{bm} % bold math
\usepackage{latexsym} % latex symbols
\usepackage{longtable} % long tables
\usepackage{hyperref} % hypertext links 
\usepackage{bm}
\usepackage{slashed}
\usepackage{bbold}
\usepackage{color}
\usepackage{multirow}
\usepackage{rotating}
\usepackage{comment}

\graphicspath{{./figsX/}}

\newcommand{\tsep}{\mathop{t_{\rm sep}}\nolimits}
\newcommand{\tsepi}{\mathop{\tau \to \infty}\nolimits}
\newcommand{\tskip}{\mathop{t_{\rm skip}}\nolimits}
\newcommand{\gpNN}{\mathop{g_{\pi {\rm NN}}}\nolimits}

\newcommand{\gsim}{\raisebox{-0.7ex}{$\stackrel{\textstyle >}{\sim}$ }}

\providecommand{\matrixe}[3]{\langle#1\lvert#2\rvert#3\rangle}

\definecolor{green}{rgb}{0.1, 0.8, 0.1}

\newcolumntype{.}[1]{D{.}{.}{#1}}

\begin{document}

%%%

\title{Axial Vector Form Factors of the Nucleon from Lattice QCD}
%
%

%% \author{Vincenzo Cirigliano}
%% \email{cirigliano@lanl.gov}
%% \affiliation{Los Alamos National Laboratory, Theoretical Division T-2, Los Alamos, NM 87545}

\author{Rajan Gupta}
\email{rajan@lanl.gov}
\affiliation{Los Alamos National Laboratory, Theoretical Division T-2, Los Alamos, NM 87545}

\author{Yong-Chull Jang}
\email{ypj@lanl.gov}
\affiliation{Los Alamos National Laboratory, Theoretical Division T-2, Los Alamos, NM 87545}

\author{Huey-Wen Lin}
\email{hwlin@pa.msu.gov}
\affiliation{Department of Physics and Astronomy, Michigan State University, MI, 48824, U.S.A}

\author{Boram Yoon}
\email{boram@lanl.gov}
\affiliation{Los Alamos National Laboratory, Computer Computational and Statistical Sciences, CCS-7, Los Alamos, NM 87545}

\author{Tanmoy Bhattacharya}
\email{tanmoy@lanl.gov}
\affiliation{Los Alamos National Laboratory, Theoretical Division T-2, Los Alamos, NM 87545}

\collaboration{Precision Neutron Decay Matrix Elements (PNDME) Collaboration}
\preprint{LA-UR-17-23678}
\date{\today}
\begin{abstract}
We present results for the form factors of the isovector axial vector
current in the nucleon state using large scale simulations of lattice
QCD.  The calculations were done using eight ensembles of gauge
configurations generated by the MILC collaboration using the HISQ
action with 2+1+1 dynamical flavors. These ensembles span three
lattice spacings $a \approx 0.06, 0.09$ and $0.12$~fm and light-quark
masses corresponding to the pion masses $M_\pi \approx 135, 225$ and
$310$ MeV. High-statistics estimates allow us to quantify systematic
uncertainties in the extraction of $G_A(Q^2)$ and the induced
pseudoscalar form factor $\tilde{G}_P(Q^2)$.  We perform a
simultaneous extrapolation in the lattice spacing, lattice volume and
light-quark masses of the axial charge radius $r_A$ data to obtain
physical estimates. Using the dipole ansatz to fit the $Q^2$ behavior
we obtain $r_A|_{\rm dipole} = 0.49(3)$~fm, which corresponds to
${\cal M}_A = 1.39(9)$~GeV, and is consistent with ${\cal M}_A =
1.35(17)$~GeV obtained by the miniBooNE collaboration. The estimate
obtained using the $z$-expansion is $r_A|_{z-{\rm expansion}} =
0.46(6)$~fm, and the combined result is $r_A|_{\rm combined} =
0.48(4)$~fm. Analysis of the induced pseudoscalar form factor
$\tilde{G}_P(Q^2)$ yields low estimates for $g_P^\ast$ and $g_{\pi
  {\rm NN}}$ compared to their phenomenological values.  To understand
these, we analyze the partially conserved axial current (PCAC)
relation by also calculating the pseudoscalar form factor. We find
that these low values are due to large deviations in the PCAC relation
between the three form factors and from the pion-pole dominance
hypothesis.
\end{abstract}
\maketitle
%
%
%
%\nopagebreak
%
%%%%%%%%%%%%%%%%%%%%%%%%%%%%%%%%%%%%%%%%%%%%%%%%%%%%%%%%%%%%%%%%%%%%%
%%%  SECTION                                                      %%%
%%%%%%%%%%%%%%%%%%%%%%%%%%%%%%%%%%%%%%%%%%%%%%%%%%%%%%%%%%%%%%%%%%%%%
\section{Introduction}
\label{sec:into}

Spurred by the demonstration of neutrino
oscillations~\cite{Fukuda:1998mi,Ahmad:2001an,Ahmad:2002jz,neutrinoOS}, 
a number of neutrino experiments are
underway worldwide~\cite{Olive:2016xmw,neutrino:expts} to probe
more detailed properties of neutrinos including CP violation in the
lepton sector, the mass hierarchy, the absolute mass scale and whether the
neutrino is its own antiparticle, i.e., a Majorana neutrino.  A major
challenge to many of these experiments is the precise determination of
the flux of neutrino beams and their cross-sections off nuclear
targets. The standard model provides the strength and nature (V$-$A) of
the interactions of the neutrinos with quarks through charged and
neutral current interactions. To describe the interactions of
neutrinos with nuclei, these elementary interactions have to be first
corrected for the interaction between quarks and gluons, described by
QCD, to account for the binding of quarks into nucleons and then by
nuclear effects such as the binding of the nucleons within the
nuclei. Since the energy scale of both neutrino oscillations and
neutrino-less double $\beta$-decay ($0\nu \beta \beta$) experiments is
less than a few GeV, non-perturbative analyses are needed for both QCD
and nuclear effects.

\begin{figure*}[tbp]%1
\centering
\includegraphics[width=0.32\linewidth]{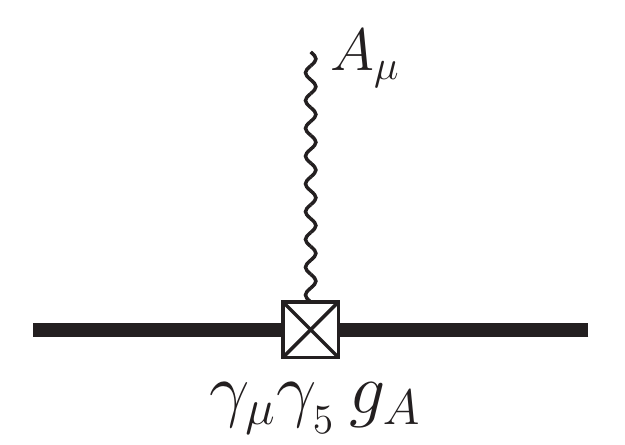}
\includegraphics[width=0.32\linewidth]{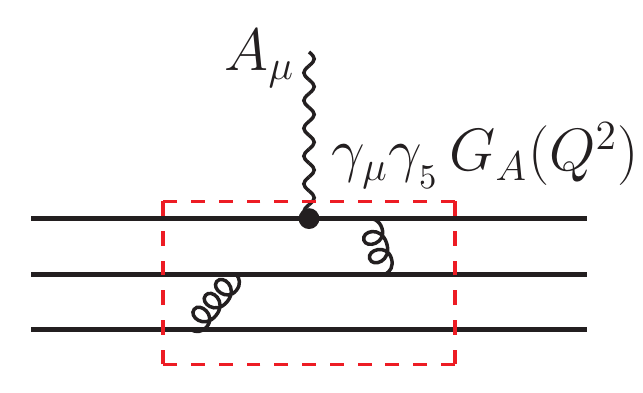}
\includegraphics[width=0.32\linewidth]{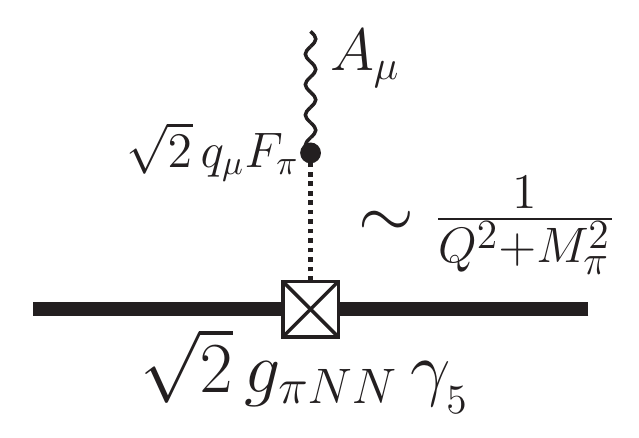}
\caption{The Feynman diagrams illustrating the decomposition of the
  matrix element of the axial current $A_\mu = \overline{u}
  \gamma_\mu \gamma_5 d$ within a nucleon state in terms of form
  factors.  The plot on the left represents the interation at $Q^2=0$
  in which case the axial current interacts with the nucleon with
  strength $g_A$.  The middle panel shows one of the lowest order
  two-gluon exchange Feynman diagrams that contributes to $G_A(Q^2)$, and provides the
  basis for the dipole ansatz. The diagram on the right is the leading
  contribution to the induced pseudoscalar form factor
  $\tilde{G}_P(Q^2)$ by a pion intermediate
  state. Its coupling to the nucleon at the pion pole defines
  $\gpNN$.}
\label{fig:feynman}
\end{figure*}

There is little experimental data, beyond old bubble chamber results, on neutrino
scattering off nucleons. A recent analysis of the data off
deuterium is given in Ref.~\cite{Meyer:2016oeg}.  The best data are
for heavier nuclei such as carbon, oxygen and iron. The current
approach used to extract the axial vector form factors of nucleons
from these data is a combination of phenomenology and modeling of
nuclear effects~\cite{Carlson:2014vla,AguilarArevalo:2010zc}.  As an alternate, first principle determinations of
nucleon form factors using lattice QCD can be convoluted with nuclear
effects to make predictions and determine the cross-sections of
neutrinos off nuclei needed to analyze experimental data.

The charged current interaction of the neutrino with the nucleon is
given by the matrix element of the isovector axial vector current, defined to be 
$A_\mu = \overline{u} \gamma_\mu \gamma_5 d$, 
within the nucleon state $N$. It is expressed in terms of two form factors
through the relativistically covariant decomposition
\begin{multline}
\label{eq:AFFdef}
\left\langle N(\vec{p}_f) | A_\mu (\vec{q}) | N(\vec{p}_i)\right\rangle  = \\
{\overline u}_N(\vec{p}_f)\left( G_A(q^2) \gamma_\mu
+ q_\mu \frac{\tilde{G}_P(q^2)}{2 M_N}\right) \gamma_5 u_N(\vec{p}_i),
\end{multline}
where $G_A(q^2)$ is the axial vector form factor, $\tilde{G}_P(q^2)$
is the induced pseudoscalar form factor and the momentum transfer
$\vec{q}=\vec{p}_f-\vec{p}_i$. In this paper, we will express the form factors in terms
of the space-like four-momentum transfer $Q^2 \equiv {\bf p}^2 - (E-m)^2 =
-q^2$.  Also, in the decomposition in Eq.~\eqref{eq:AFFdef}, we
neglect the induced tensor form factor $\tilde{G}_T$ since it vanishes
in the limit of isospin symmetry that is implicit in this
work~\cite{Bhattacharya:2011qm}, i.e., the up and down quarks are taken
to be degenerate.  

We also define the pseudoscalar form factor
$G_P$
\begin{equation}
\left\langle N(\vec{p}_f) | P (\vec{q}) | N(\vec{p}_i)\right\rangle  = 
{\overline u}_N(\vec{p}_f) G_P(q^2) \gamma_5 u_N(\vec{p}_i) \,, 
\label{eq:PSdef}
\end{equation}
where the operator $P = \overline{u} \gamma_5 d$.
Contracting Eq.~\eqref{eq:AFFdef} with $q^\mu$ and using the 
partially conserved axial current (PCAC) identity gives the following relation 
between the three form factors 
\begin{equation}
2 {\widehat m} G_P(Q^2) = 2 M_N G_A(Q^2) - \frac{Q^2}{2M_N} {\tilde G}_P(Q^2) \,,
\label{eq:PCAC}
\end{equation}
where we define ${\widehat m} \equiv Z_m Z_P (m_u +m_d)/2 Z_A$, the
common mass of the u and d quarks in our isospin symmetric theory
multiplied by the appropriate renormalization constants arising in
Eq.~\eqref{eq:PCAC}.  This mass parameter, ${\widehat m} $, can be
measured directly on the lattice again using PCAC from the
pseudoscalar two-point correlation function, i.e., by requiring that,
up to lattice artifacts, $\Gamma(t) = \langle \Omega|(\partial_\mu
A_\mu - 2 {\widehat m} P)_t P_0 | \Omega \rangle=0$ for all Euclidean
times $t$. Note that ${\tilde G}_P(Q^2)$ and $G_P(Q^2)$ cannot be 
extracted at $Q^2=0$. 

The three form factors can be extracted directly from the two matrix
elements defined in Eqs.~\eqref{eq:AFFdef} and~\eqref{eq:PSdef}. PCAC
relates them, and $G_A(Q^2)$ and ${\tilde G}_P(Q^2)$ are usually taken
to be the two independent form factors.  Since PCAC is an operator
relation, it should be satisfied at all values of $a$, $M_\pi$ and
$Q^2$ up to lattice discretization effects.  The first goal of large
scale simulations of lattice QCD is, therefore, to calculate these
three form factors with control over all systematics and show that
they satisfy the PCAC relation. Only then can one compare them with
phenomenological extractions to constrain/guide the modeling of
nuclear effects in the calculation of the cross-section of neutrinos
off nuclei.

A diagrammatic description of these form factors is as follows.  At
$Q^2=0$ the axial current interacts with the nucleon with strength
given by the axial charge $g_A$ as shown in Fig.~\ref{fig:feynman}
(left).  At high $Q^2$, the lowest order Feynman diagram contributing
to $G_A(Q^2)$ requires two gluons to be exchanged between the three quarks
in all possible combinations as illustrated in Fig.~\ref{fig:feynman}
(middle).  This two gluon exchange amplitude at large $Q^2$ behaves as
$1/Q^4$, and is the historical motivation for the dipole ansatz we
discuss below.  In Fig.~\ref{fig:feynman} (right), we show the
interaction via a pion intermediate state, i.e., the axial current
creates a pion intermediate state with coupling $\sqrt{2} q^\mu
F_\pi$. This pion state propagates with the factor $1/(Q^2 + M_\pi^2)$
before interacting with the nucleon with strength $\sqrt{2}
\gpNN$. This diagram constitutes the lowest order contribution to the
induced pseudoscalar form factor $\tilde{G}_P(Q^2)$ and provides the
motivation for analyzing it using the pion pole-dominance ansatz.

In this paper we present results for the isovector part of $G_A$ and
$\tilde{G}_P$ in the range $0.05 < Q^2 \lesssim 0.8$~GeV${}^2$ using
first principle simulations of lattice QCD on eight ensembles covering
the range of lattice spacings ($0.06 \, \lesssim a \, \lesssim
0.12$~fm), pion masses ($135 \, \lesssim M_\pi \, \lesssim 320$~MeV)
and lattice volumes ($3.3\, \lesssim M_\pi L\, \lesssim 5.5$). These
ensembles were generated using $2+1+1$-flavors of highly improved
staggered quarks (HISQ)~\cite{Follana:2006rc} by the MILC
collaboration~\cite{Bazavov:2012xda}.  On four of thesze ensembles we
have also calculated the pseudoscalar form factor ${G}_P$ that is
needed to check the PCAC relation.

The axial radius of the nucleon is determined from the slope of
$G_A(Q^2)$ in the $Q^2 \to 0$ limit:
\begin{equation}
\langle r_A^2\rangle = -6\frac{d}{dQ^2}\left.\left(\frac{G_A(Q^2)}{G_A(0)}\right)\right|_{Q^2=0}.
\label{eq:rdef}
\end{equation}
The challenge to the direct calculation of the slope using a discrete
derivative is that the value of the smallest momenta and the intervals
between the lowest few lattice momenta in typical lattice simulations
are large. In our calculations, the lowest non-zero momenta is $\gsim
220$~MeV. It is, therefore, customary to fit the data using a
physically motivated ansatz for $G_A(Q^2)$ and then use the result to
evaluate the derivative given in Eq.~\eqref{eq:rdef}. This modeling of
$G_A$ introduces a systmatic uncertainty in the value of $\langle
r_A^2\rangle$ that we estimate by comparing results using different
fit ansatz. 

An ansatz that is commonly used to fit the experimental
data is the dipole approximation
\begin{equation}
G_A(Q^2) = \frac{G_A(0)}{(1+Q^2/{\cal M}_A^2)^{2}} \quad \Longrightarrow \quad \langle r_A^2\rangle = \frac{12}{{\cal M}_A^2} \,, 
\label{eq:dipole}
\end{equation}
where ${\cal M}_A$ is the axial dipole mass. 
It is the simplest one parameter form that is normalized to $G_A(0)
\equiv g_A$ at $Q^2=0$ and goes as $Q^{-4}$ in the $Q^2 \to \infty$
limit in accord with the leading contribution in perturbation theory
as shown in the middle panel of Fig.~\ref{fig:feynman}. Estimates of
the RMS charge radius $ r_A  \equiv \sqrt{\langle r_A^2 \rangle} $ 
obtained from (i) a weighted world average of (quasi)elastic neutrino and 
anti-neutrino scattering data~\cite{Bernard:2001rs}, (ii) charged pion
electroproduction experiments~\cite{Bernard:2001rs}, and (iii) a
reanalysis of the deuterium target data~\cite{Meyer:2016oeg} are
\begin{eqnarray}
r_A&=& 0.666(17)~{\rm fm}   \qquad \nu,{\overline \nu}-{\rm scattering}    \,,  \nonumber \\
r_A&=& 0.639(10)~{\rm fm}   \qquad {\rm Electroproduction} \,,  \nonumber \\
r_A&=& 0.68(16)~{\rm fm}  \, \  \qquad {\rm Deuterium} \,,
\label{eq:rA_expt}
\end{eqnarray}
which correspond to the dipole masses 
\begin{eqnarray}
{\cal M}_A  &=& 1.026(21)~{\rm GeV}   \qquad \nu,{\overline \nu}-{\rm scattering}    \,,  \nonumber \\
{\cal M}_A  &=& 1.069(16)~{\rm GeV}   \qquad {\rm Electroproduction} \,,  \nonumber \\
{\cal M}_A  &=& 1.00(24)~{\rm GeV}    \ \qquad {\rm Deuterium} \,.
\label{eq:MA_expt}
\end{eqnarray}
On the other hand, the MiniBooNE Collaboration, using the dipole
ansatz and a relativistic Fermi gas model~\cite{Smith:1972xh}, find
that ${\cal M}_A=1.35(17)$~GeV reproduces their double differential
cross-section for charged current quasi elastic neutrino and
antineutrino scattering data off carbon~\cite{AguilarArevalo:2010zc}.
Lattice QCD, by providing first-principle estimates of $G_A(Q^2)$ for
nucleons, aims to resolve the difference in the phenomenological
estimates and to pin down the $Q^2$ behavior of the form factors. 

The analysis presented here shows that the dipole ansatz fits the
lattice data surprisingly well, however, our result, $r_A|_{\rm
  dipole} = 0.49(3)$, is smaller than the phenomenological estimates
given in Eq.~\eqref{eq:rA_expt}.

The second ansatz we use is a model-independent parameterization
called the $z$-expansion~\cite{Hill:2010yb,Bhattacharya:2011ah}:
\begin{equation}
 \frac{G_A(Q^2)}{G_A(0)} = \sum_{k=0}^{\infty} a_k z(Q^2)^k  \,,
\label{eq:Zexpansion}
\end{equation}
where the $a_k$ are fit parameters and $z$ is defined as 
\begin{equation}
  z = \frac{\sqrt{t_\text{cut}+Q^2}-\sqrt{t_\text{cut}+{\overline t}_0}}
           {\sqrt{t_\text{cut}+Q^2}+\sqrt{t_\text{cut}+{\overline t}_0}} \,, 
\label{eq:Zdef}
\end{equation}
with $t_\text{cut} \equiv Q^2_\text{cut} = 9M_\pi^2$.  The nearest
singularity in the form factor $G_A(Q^2)$ is the three-pion branch cut
at $Q^2=9M_\pi^2$. In terms of $z$, the domain of analyticity of
$G_A(Q^2)$ is mapped into the unit circle with the three-pion branch
cut at $t_\text{cut}=9M_\pi^2$ moved to
$z=1$~\cite{Bhattacharya:2011ah}. The value of the constant
${\overline t}_0$ is typically chosen to be in the middle of the range
of $Q^2$ of interest to minimize $z_{\rm max}$ and possibly improve
the convergence of the $z$-expansion. The choice of ${\overline t}_0$
could have been important in our calculation because we have data at
only the five lowest values of momenta on most ensembles and can,
therefore, perform an analysis keeping terms only up to $O(z^4)$. Our
analysis of the data with ${\overline t}_0=0$ and ${\overline t}_0 =
{\overline t}_0^{\rm mid} \equiv \{0.12,\ 0.20,\ 0.40\}$~GeV${}^2$,
corresponding to the approximate midpoint of the range of $Q^2$ on the
$M_\pi \approx \{130,\ 220,\ 310\}$~MeV ensembles, respectively,
however shows that the quality of the fits and the results are
insensitive to the choice of ${\overline t}_0$.  For presenting our
final results, we choose the midpoint values, ${\overline t}_0^{\rm
  mid}$.

The asymptotic requirement, that
$G_A(Q^2) \to Q^{-4}$ as $Q^2 \to \infty$, requires $Q^n G_A(Q^2) \to
0$ for $n=0, 1, 2, 3$~\cite{Lee:2015jqa}. These constraints can be incorporated 
into the $z$-expansion as four sum rules
\begin{equation}
\sum_{k=n}^{k_{\rm max}} k(k-1) \ldots (k-n+1) a_k = 0 \qquad n=0,1,2,3 \,, 
\label{eq:sumrule}
\end{equation}
where for $n=0$ it is $\sum_{k=0}^{k_{\rm max}} a_k = 0$.
Incorporating these sumrules ensures that the $a_k$ are not only
bounded but must also decrease at large $k$~\cite{Lee:2015jqa}.  We
have six data points (zero and five non-zero momentum cases) for all
but the two physical quark mass ensembles, $a09m130$ and $a06m135$.
The analysis was therefore done using $k_{\rm max} = 5,\ 6,\ 7$ and
$8$. Including the four sum rules, these values of $k_{\rm max}$
correspond to 4, 3, 2, and 1 degrees of freedom, respectively.  We use
the quality of the fits and the stability of the value of the axial
charge radius squared $\langle r_A^2\rangle$ obtained from them as
checks on the consistency of the analysis, ensemble by ensemble. Based
on these checks, we drop $k_{\rm max} = 5$ fits as the associated
$\chi^2/{\rm d.o.f.}$ are not good and the $k_{\rm max} = 8$ fits, as
they are unstable in many cases.

Our final result, $r_A|_{z-{\rm expansion}} =
0.46(6)$~fm, is obtained as an average of the $k_{\rm max} = 6$ and
$7$ analyses, which we label $k^{2+4}$ and $k^{3+4}$ to make explicit
that four powers of $z$ are constrained by the sumrules. This lattice
estimate is again smaller than the current phenomenological estimates
given in Eq.~\eqref{eq:rA_expt}. The uncertainty in the estimates,
ensemble by ensemble, is larger with the $z$-expansion versus
the dipole ansatz.

The induced pseudoscalar form factor ${\tilde G}_P(Q^2)$ is typically
analyzed assuming the pion pole-dominance ansatz: 
\begin{equation} 
{\tilde G}_P(Q^2) \propto  {G}_A(Q^2)  \left[\frac{1}{Q^2+M_{\pi}^2}\right] \,, 
\label{eq:GPpole}
\end{equation}
where the coefficient of proportionality is often taken to $4M_{N}^2$
as suggested by the Goldberger-Trieman
relation~\cite{Goldberger:1958vp}.  This behavior is consistent with
the PCAC relation, Eq.~\eqref{eq:PCAC}, only if $2 {\widehat m}
G_P(Q^2) = (M_\pi^2 / 2 M_N) {\tilde G}_P(Q^2)$.  If this ansatz is a
good approximation, then there is only one independent form factor,
which can be taken to be ${G}_A(Q^2)$ or ${\tilde G}_P(Q^2)$.

%%%%%%%%%%%%%%%%%%%%%%%%%%%%%%%%%%%%%%%%%%%%%%%%%%%%%%%%%%%%%%%%%%%%%
%
\begin{table*}[tbp]%1
\begin{center}
\renewcommand{\arraystretch}{1.2} % Change horizontal spacing
\begin{ruledtabular}
\begin{tabular}{l|ccc|cc|cccc}
Ensemble ID & $a$ (fm) & $M_\pi^{\rm sea}$ (MeV) & $M_\pi^{\rm val}$ (MeV) & $L^3\times T$    & $M_\pi^{\rm val} L$ & $t_\text{sep}/a$ & $N_\text{conf}$  & $N_{\rm meas}^{\rm HP}$  & $N_{\rm meas}^{\rm AMA}$  \\
\hline
a12m310    & 0.1207(11) & 305.3(4) & 310.2(2.8) & $24^3\times 64$ & 4.55 &  $\{8,10,12\}$& 1013 & 8104  &   64,832   \\
%% a12m220S   & 0.1202(12) & 218.1(4) & 225.0(2.3) & $24^3\times 64$ & 3.29 & $\{8, 10, 12\}$   & 1000 & 24000 &           \\
%% a12m220    & 0.1184(10) & 216.9(2) & 227.9(1.9) & $32^3\times 64$ & 4.38 & $\{8, 10, 12\}$   & 958  & 7664  &           \\
a12m220L   & 0.1189(09) & 217.0(2) & 227.6(1.7) & $40^3\times 64$ & 5.49 &  $\{8, 10, 12, 14\}$         & 1010 & 8080  &  68,680   \\
\hline                                                                                                      
a09m310    & 0.0888(08) & 312.7(6) & 313.0(2.8) & $32^3\times 96$ & 4.51 & $\{10,12,14\}$    & 881  & 7048  &           \\
a09m220    & 0.0872(07) & 220.3(2) & 225.9(1.8) & $48^3\times 96$ & 4.79 & $\{10,12,14\}$    & 890  & 7120  &           \\
a09m130    & 0.0871(06) & 128.2(1) & 138.1(1.0) & $64^3\times 96$ & 3.90 & $\{10,12,14\}$    & 883  & 7064  &  60,044   \\
\hline                                                                                                      
a06m310    & 0.0582(04) & 319.3(5) & 319.6(2.2) & $48^3\times 144$& 4.52 & $\{16,20,22,24\}$ & 1000 & 8000  &  64,000   \\
a06m220    & 0.0578(04) & 229.2(4) & 235.2(1.7) & $64^3\times 144$& 4.41 & $\{16,20,22,24\}$ & 650  & 2600  &  41,600   \\
a06m135    & 0.0570(01) & 135.5(2) & 135.6(1.4) & $96^3\times 192$& 3.7  & $\{16,18,20,22\}$ & 322  & 1610  &  51,520   \\
\end{tabular}
\end{ruledtabular}
\caption{Parameters, including the Goldstone pion mass $M_\pi^{\rm
    sea}$, of the eight 2+1+1- flavor HISQ lattices generated by the
  MILC collaboration and analyzed in this study are quoted from
  Ref.~\cite{Bazavov:2012xda}.  All fits are made versus $M_\pi^{\rm
    val}$ and finite-size effects are analyzed in terms of $M_\pi^{\rm
    val} L$.  Estimates of $M_\pi^{\rm val}$, the clover-on-HISQ pion
  mass, are the same as given in Ref.~\cite{Bhattacharya:2015wna} and
  the error is governed mainly by the uncertainty in the lattice
  scale. In the last four columns, we give, for each ensemble, the
  values of the source-sink separation $t_{\rm sep}$ used in the
  calculation of the three-point functions, the number of
  configurations analyzed, and the number of measurements made using
  the HP and AMA methods.  The HP calculation on the $a12m220L$
  ensemble has been done with a single $t_{\rm sep}=10$ while the LP
  analysis has been done with $t_{\rm sep}=\{8,10,12,14\}$.  }
\label{tab:ens}
\end{center}
\end{table*}

Experimentally, ${\tilde G}_P(Q^2)$ is probed in muon capture by a
proton, $\mu^{-} + p \to \nu_\mu +
n$~\cite{Andreev:2012fj,Andreev:2015evt}. From these measurements,
the induced pseudoscalar charge $g_P^\ast$ is defined as 
\begin{equation} 
g_P^\ast \equiv  \frac{m_\mu}{2M_N} {\tilde G}_P(Q^2 = Q^{\ast\,2} \equiv 0.88m_\mu^2)   \,.
\label{eq:GPstar}
\end{equation}
Current estimates from the MuCap
experiment~\cite{Andreev:2012fj,Andreev:2015evt}, and from chiral
perturbation theory~\cite{Schindler:2006it,Bernard:2001rs} are 
\begin{eqnarray}
g_P^\ast|_{\rm MuCap}     &=& 8.06(55) \,,  \nonumber \\
g_P^\ast|_{\chi{\rm PT}}  &=& 8.29^{+0.24}_{-0.13} \pm 0.52 \,.
\label{eq:GPstar_pheno}
\end{eqnarray} 

On the lattice, once the modeling of the $Q^2$ behavior of ${\tilde
  G}_P(Q^2)$ is under control, one can determine $g_P^\ast$ by
extrapolation to $Q^2 = Q^{\ast\,2} \equiv 0.88m_\mu^2$ and the
pion-nucleon coupling $\gpNN$ as the residue at $Q^2 = -M_\pi^2$.  To
compare our lattice QCD estimates with these phenomenological values,
we first extract $g_P^\ast$ from fits to ${\tilde G}_P(Q^2)$ versus
$Q^2$ for each ensemble, and then extrapolate these data to $a=0$ and
$M_\pi=135$~MeV. The result is a surprisingly low value, $g_P^\ast
=4.44(18)$, compared to the values given in
Eq.~\eqref{eq:GPstar_pheno}. This discrepency arises due to large
deviations from the PCAC relation involving the three form factors as
discussed further in Sec.~\ref{sec:GP}. We also show that using just
the pion-pole ansatz to extrapolate $g_P^\ast(Q^{\ast\, 2})$ obtained
from simulations at $M_\pi > 300 $~MeV to $M_\pi \to M_\pi^{\rm
  Physical} = 135$~MeV does not match our lattice data at $M_\pi =
220$ or $135$~MeV.

%\clearpage

Lastly, we evaluate the pion-nucleon coupling $\gpNN$ using the
Goldberger-Treiman (GT) relation $g_{\pi N N} = M_N g_A/F_\pi $, and
as the residue at the pion pole at $Q^2=-M_\pi^2$ of ${\tilde
  G}_P(Q^2)$. As discussed in Sec.~\ref{sec:GTR}, our estimate, $\gpNN
= M_N g_A/F_\pi = 12.87(34)$ using the lattice data is consistent with
that obtained using the experimental values. Our direct calculation of
$\gpNN$, as the residue of ${\tilde G}_P(Q^2)$ at the pion pole,
suffers from the same problem as the analysis of $g_P^\ast$ and
gives $\gpNN =5.78(57)$, much smaller than the phenomenological
estimate $13.69 \pm 0.12 \pm 0.15$ obtained from the $\pi N$
scattering length analysis~\cite{Baru:2011bw}.

This paper is organized as follows. In Sec.~\ref{sec:Methodology}, we
describe the parameters of the gauge ensembles analyzed and the
lattice methodology. The strategy used to isolate excited-state
contamination is described in Sec.~\ref{sec:excited}. In
Sec.~\ref{sec:2pt-results}, we present the analysis of the two-point
correlation functions. The extraction of the form factors from the
three-point functions is discussed in Sec.~\ref{sec:3pt-fits}, and of
the axial charge radius $r_A$ from these in
Sec.~\ref{sec:fits-rA}. Simultaneous fits in the lattice spacing $a$,
the pion mass $M_\pi$ and the lattice size $M_\pi L$ to obtain our
physical estimate of $r_A$ are presented in
Sec.~\ref{sec:rA_results}.  The analysis of the induced pseudoscalar
form factor is carried out in Sec.~\ref{sec:GP}, of $g_P^\ast$ in 
Sec.~\ref{sec:gPstar}, and of the
pion-nucleon coupling, $\gpNN$, in Sec.~\ref{sec:GTR}. In
Sec.~\ref{sec:heuristic}, we present a heuristic analysis to understand
violations of the PCAC relation between $G_A(Q^2)$, ${\tilde
  G}_P(Q^2)$, and ${G}_P(Q^2)$. We end with conclusions in
Sec.~\ref{sec:conclusions}.

%%%%%%%%%%%%%%%%%%%%%%%%%%%%%%%%%%%%%%%%%%%%%%%%%%%%%%%%%%%%%%%%%%%%%%%%

\begin{table}[htbp]%2
\centering
\begin{ruledtabular}
\begin{tabular}{l|lc|c}
\multicolumn1c{ID}       & \multicolumn1c{$m_l$} &  $c_{\text{SW}}$ & Smearing    \\
         &       &                  & Parameters  \\
\hline
a12m310  & $-0.0695$  & 1.05094 & \{5.5, 70\}  \\
a12m220L & $-0.075$   & 1.05091 & \{5.5, 70\}  \\ 
\hline
a09m310  & $-0.05138$ & 1.04243 & \{5.5, 70\}  \\
a09m220  & $-0.0554$  & 1.04239 & \{5.5, 70\}  \\
a09m130  & $-0.058$   & 1.04239 & \{5.5, 70\}  \\
\hline
a06m310  & $-0.0398$  & 1.03493 & \{6.5, 70\}  \\
a06m220  & $-0.04222$ & 1.03493 & \{5.5, 70\}  \\
a06m135  & $-0.044$   & 1.03493 & \{9.0, 150\}  \\
\end{tabular}
\end{ruledtabular}
\caption{The parameters used in the calculation of clover propagators.
  The hopping parameter $\kappa$ in the clover action is given by
  $2\kappa_{l} = 1/(m_{l}+4)$.  The Gaussian smearing parameters are
  defined by $\{\sigma, N_{\text{KG}}\}$ where $N_{\text{KG}}$ is the
  number of applications of the Klein-Gordon operator and the width of
  the smearing is controlled by the coefficient $\sigma$, both in
  Chroma convention~\cite{Edwards:2004sx}.  $m_l$ is tuned to achieve
  $M_\pi^{\rm val} \approx  M_\pi^\text{sea}$. }
  \label{tab:cloverparams}
\end{table}

%%%%%%%%%%%%%%%%%%%%%%%%%%%%%%%%%%%%%%%%%%%%%%%%%%%%%%%%%%%%%%%%%%%%%

%%%%%%%%%%%%%%%%%%%%%%%%%%%%%%%%%%%%%%%%%%%%%%%%%%%%%%%%%%%%%%%%%%%%%
%%%  SECTION                                                      %%%
%%%%%%%%%%%%%%%%%%%%%%%%%%%%%%%%%%%%%%%%%%%%%%%%%%%%%%%%%%%%%%%%%%%%%
\section{Lattice Methodology}
\label{sec:Methodology}

The eight ensembles used in the analysis cover a range of lattice
spacings ($0.06 \, \lesssim a \, \lesssim 0.12$~fm), pion masses ($135
\, \lesssim M_\pi \, \lesssim 320$~MeV) and lattice volumes ($3.3\,
\lesssim M_\pi L\, \lesssim 5.5$). These were generated using
$2+1+1$-flavors of highly improved staggered quarks
(HISQ)~\cite{Follana:2006rc} by the MILC
collaboration~\cite{Bazavov:2012xda} and their parameters are summarized
in Table~\ref{tab:ens}.  Results for the isovector charges,
$g_A^{u-d}$, $g_S^{u-d}$ and $g_T^{u-d}$ on these ensembles have
already been published in
Refs.~\cite{Bhattacharya:2015wna,Bhattacharya:2016zcn}. In this work
we follow the same computational strategy, so we only summarize 
the important issues and point the reader to the appropriate
references for details.

%% \begin{itemize}
% \item

The correlation functions used to calculate the matrix elements on
these HISQ ensembles are constructed using Wilson-clover fermions
after the lattices have been smoothed using hypercubic (HYP)
smearing~\cite{Hasenfratz:2001hp}. This mixed-action, clover-on-HISQ
approach, leads to a non-unitary lattice formulation that at small,
but {\it a priori} unknown, quark masses suffers from the problem of
exceptional configurations. As described in
Ref.~\cite{Bhattacharya:2015wna}, tests performed by us did not find
configurations exhibiting large deviations from the mean behavior on
these ensembles.
%5 \item

The parameters used to construct the quark propagators with the 
clover action are given in Table~\ref{tab:cloverparams}. The
Sheikholeslami-Wohlert coefficient~\cite{Sheikholeslami:1985ij} used
in the clover action is fixed to its tree-level value with tadpole
improvement, i.e., $c_\text{sw} = 1/u_0^3$, where $u_0$ is the fourth root of
the plaquette expectation value calculated on the HYP 
smeared HISQ lattices.
%% \item

The masses of light clover quarks were tuned so that the
clover-on-HISQ pion masses, $M^{\rm val}_\pi$, match the HISQ-on-HISQ
Goldstone ones, $M_\pi^{\rm sea}$. Both estimates are given in
Table~\ref{tab:ens}. All fits in $M_\pi^2$ to study the chiral
behavior are made using the clover-on-HISQ $M^{\rm val}_{\pi}$ since 
the correlation functions, and thus the observables, have a greater
sensitivity to it. Henceforth, we denote the clover-on-HISQ pion mass as $M_\pi$. 
%% \item

On six ensembles, we have used the truncated solver method with bias
correction (labeled the AMA method)~\cite{Bali:2009hu,Blum:2012uh} to
cost-effectively increase the statistics in the calculation of the
two- and three-point correlation functions. The details of our
implementation are given in
Refs.~\cite{Bhattacharya:2015wna,Bhattacharya:2016zcn,Yoon:2016jzj}.
%% \item

The two- and three-point correlation functions were constructed using
the nucleon interpolating operator 
\begin{align}
 \chi(x) = \epsilon^{abc} \left[ {q_1^a}^T(x) C \gamma_5
            \frac{(1 \pm \gamma_4)}{2} q_2^b(x) \right] q_1^c(x)
\label{eq:nucl_op}
\end{align}
with color indices $\{a, b, c\}$, charge conjugation matrix
$C=\gamma_0 \gamma_2$, and $q_1$ and $q_2$ denoting the two different
flavors of light Dirac quarks.
The non-relativistic projection $(1 \pm \gamma_4)/2$ is inserted to
improve the signal, with the plus (minus) sign applied to the forward
(backward) propagation in Euclidean time as described in
Refs.~\cite{Bhattacharya:2015wna,Bhattacharya:2016zcn,Yoon:2016jzj}. On
the other hand, the $\gamma_4$ part introduces mixing with spin $3/2$
states at non-zero momentum, with concomitant excited-state
contamination.
%% \item

All errors are determined using a single-elimination Jackknife
procedure. We first construct the configuration average, i.e., the
mean of the correlation functions over multiple measurements on each
configuration, and then implement the Jackknife process over these
configuration averages. In all the fits to the two- and three-point
correlation functions based on minimizing the $\chi^2/{\rm d.o.f.}$,
we used the full covariance matrix as described in
Ref.~\cite{Yoon:2016jzj}.
%% \item

The value of the axial radius from each ensemble was extracted from
the form factors using two fit ansatz: the model-independent
$z$-expansion, and the dipole fit.  ${\tilde G}_P$ was analyzed using the 
PCAC relation and the pion pole-dominance ansatz.
%% \item

All estimates, such as $\langle r_A^2 \rangle$ and $g_P^\ast$ obtained
on the eight ensembles, were simultaneously fit versus the three
variables, the lattice spacing $a$, the pion mass $M_\pi$, and the
lattice size parameterized by $M_\pi L$, keeping only the leading
order correction terms in each. From these fits, the final value was
obtained at the physical pion mass $M_\pi=135$~MeV with extrapolation
to the continuum and the infinite volume limits.

%% \item

The renormalization factor for the axial current cancels in the ratios
used in the extraction of the axial charge radius, defined in
Eq.~\eqref{eq:rdef}, and in the analysis of ${\tilde G}_P(Q^2)$ using
the pole-dominance hypothesis given in Eq.~\eqref{eq:GPpole}.  Thus,
all results presented in this work are the same as for
renormalized operators.
%% \end{itemize}

Further details of the analysis are given at appropriate places 
when discussing the results.

%%%%%%%%%%%%%%%%%%%%%%%%%%%%%%%%%%%%%%%%%%%%%%%%%%%%%%%%%%%%%%%
\section{Controlling Excited-State Contamination}
\label{sec:excited}
%%%%%%%%%%%%%%%%%%%%%%%%%%%%%%%%%%%%%%%%%%%%%%%%%%%%%%%%%%%%%%%

To extract the desired nucleon form factors we need to evaluate the
matrix elements of the axial current between ground-state
nucleons. The lattice nucleon interpolating operator given in
Eq.~\eqref{eq:nucl_op}, however, couples to the nucleon, all 
excitations and multiparticle states with the same quantum
numbers. Three strategies are used to reduce excited-state
contamination as described in
Refs.~\cite{Bhattacharya:2015wna,Bhattacharya:2016zcn,Yoon:2016jzj}.
\begin{itemize}
\item
The overlap between the nucleon operator and the excited states in the
construction of the two- and three-point functions is reduced by using
tuned smeared sources when calculating the quark propagators on
the HYP smeared HISQ lattices. We
construct gauge-invariant Gaussian smeared sources by applying the
three-dimensional Laplacian operator, $\nabla^2$, a fixed number,
$N_{\rm GS}$, of times, i.e., $(1 + \sigma^2\nabla^2/(4N_{\rm
  GS}))^{N_{\rm GS}}$.  The smearing parameters $\{\sigma, N_{\rm
  GS}\}$ for each ensemble are given in Table~\ref{tab:cloverparams}.
\item
The analysis of the nucleon two-point functions, $C^\text{2pt}$, was carried
out keeping four states in the spectral decomposition:
\begin{align}
C^\text{2pt}
  &(t,\bm{p}) = \nonumber \\
  &{|{\cal A}_0|}^2 e^{-E_0 t} + {|{\cal A}_1|}^2 e^{-E_1 t}\,+ \nonumber \\
  &{|{\cal A}_2|}^2 e^{-E_2 t} + {|{\cal A}_3|}^2 e^{-E_3 t}\,, 
\label{eq:2pt}
\end{align}
where the amplitudes and the energies with momentum $\bm{p}$ of the
four states are denoted by ${\cal A}_i$ and $E_i$, respectively.  The
strategy for the selection of non-trivial priors for the masses and
amplitudes used in the fits is the same as described in
Ref.~\cite{Yoon:2016jzj}. A comparison between 2- and 4-state fits is
shown in Figs.~\ref{fig:Meffa12m310}--\ref{fig:Meffa06m135} in
Appendix~\ref{sec:appendix1}. In the 4-state fits used in the final
analysis, the starting time slice in the fit, $t_{\rm min}$, is chosen
to be small to include as much data as possible while maintaining the
stability of the fit parameters. Since the excited-state contamination is 
observed to be similar, $t_{\rm min}$ is chosen to be the same for
all momenta for a given ensemble.

The analysis of the three-point functions, 
$C_\Gamma^{(3\text{pt})} (t;\tau;\bm{p}^\prime,\bm{p})$  
was carried out keeping two states in the spectral decomposition:  
\begin{align}
  C_\Gamma^{(3\text{pt})}&(t;\tau;\bm{p}^\prime,\bm{p}) = \nonumber \\
   &\; {\mathcal{A}_0^\prime} {\mathcal{A}_0}\matrixe{0^\prime}{\mathcal{O}_\Gamma}{0} e^{-E_0t - M_0(\tau-t)} + \nonumber \\
   &\; {\mathcal{A}_1^\prime} {\mathcal{A}_1}\matrixe{1^\prime}{\mathcal{O}_\Gamma}{1} e^{-E_1t - M_1(\tau-t)} + \nonumber \\
   &\; {\mathcal{A}_0^\prime} {\mathcal{A}_1}\matrixe{0^\prime}{\mathcal{O}_\Gamma}{1} e^{-E_0t - M_1(\tau-t)}  + \nonumber \\
   &\; {\mathcal{A}_1^\prime} {\mathcal{A}_0}\matrixe{1^\prime}{\mathcal{O}_\Gamma}{0} e^{-E_1t -M_0(\tau-t)}  \,,
  \label{eq:3pt}
\end{align}
%% Using absolute value or not? It is true that the amplitudes
%% can all be chosen real and positive (there are four states and four
%% amplitudes, so the phase can be absorbed into the matrix elements),
%% but the absolute value is wrong unless the amplitudes are chosen to be
%% real and positive.
where the source point is translated to $t=0$, the operator is
inserted at time $t$, and nucleon state is annihilated at the sink
time slice $\tau \equiv t_\text{sep}$.  The states $|0\rangle$ and
$|1\rangle$ represent the ground and all higher states that we
collectively label the ``first excited'' state, respectively. The
label ${\cal A}_i^\prime$ denotes the amplitude for the creation of
state $i$ with momentum $\bm{p}^\prime$ by the nucleon interpolating
operator $\chi$.  To extract the matrix elements, we need the four
amplitudes ${\cal A}_0$, ${\cal A}_1$, ${\cal A}_0^\prime$ and ${\cal
  A}_1^\prime$, which we obtain from the 4-state fits to the two-point
functions.  Note that the insertion of the nucleon at the sink
timeslice $t = \tau = t_\text{sep}$ is at $\bm{p}=0$ in all cases, and
the insertion of the current at time $t$ is at a definite momentum
$\bm{p}^\prime$. To ensure a good signal for all $\bm{p}^\prime$, the
nucleon state at the source timeslice, constructed from smeared
sources, should have a large overlap with all momentum states
analyzed. The data in in
Figs.~\ref{fig:Meffa12m310}--\ref{fig:Meffa06m135} show that with the
smeared sources used, a decent signal is achieved for $Q^2 \lesssim
1$~GeV${}^2$.
\item
We calculate the three-point correlation functions for a number of
values of the source-sink separation $t_{\rm sep}$ that are listed in
Table~\ref{tab:ens}. We fit the data at all $t_{\rm sep}$
simultaneously using the 2-state ansatz given in Eq.~\eqref{eq:3pt}.
In these fits, we skip $\tskip$ points adjacent to the source and sink
for each $\tsep$ as these points have the largest excited state
contamination. As a result, more points with larger $\tsep$ that have
less excited-state contamination and larger statistical errors are
included. The value of $\tskip$ for each ensemble is chosen to be same
for all momenta since the onset of the plateau in the effective-mass
plot is observed to start at roughly the same timeslice, independent
of the momenta, as shown in
Figs.~\ref{fig:Meffa12m310}--\ref{fig:Meffa06m135}.
\end{itemize}
From these fits we get $\matrixe{0^\prime}{\mathcal{O}_\Gamma}{0}$,
the desired $\tsepi$ estimate. The above procedure has been 
followed for all values of momentum insertion and on each ensemble.

%%%%%%%%%%%%%%%%%%%%%%%%%%%%%%%%%%%%%%%%%%%%%%%%%%%%%%%%%%%%%%%%%%%%%%%%%%%%%%%%
% Section results on 2-pt functions
%%%%%%%%%%%%%%%%%%%%%%%%%%%%%%%%%%%%%%%%%%%%%%%%%%%%%%%%%%%%%%%%%%%%%%%%%%%%%%%%
\section{Fits to the two-point functions}
\label{sec:2pt-results}

On each ensemble, we performed 2-, 3- and 4-state fits to the
two-point correlation function data to extract the amplitudes and the
masses. On all ensembles, we collected data for momenta $\textbf{p} =
2\pi \textbf{n} /aL$ with $\textbf{n} = \{(0,0,0), (1,0,0), (1,1,0),
(1,1,1), (2,0,0), (2,1,0)\}$. On the $a09m130$ and $a06m135$
ensembles, we also collected data for $\textbf{n}=\{(2,1,1), (2,2,0),
(2,2,1), (3,0,0), (3,1,0)\}$.

We illustrate the quality of the two-point data  by 
plotting the effective-energy defined as 
\begin{equation}
E_{\rm eff}(t) =  \log \frac{C^{\rm 2pt}(t)}{C^{\rm 2pt}(t+1)} \,, 
\label{eq:effmass}
\end{equation}
in
Figs.~\ref{fig:Meffa12m310},~\ref{fig:Meffa12m220L},~\ref{fig:Meffa09m310},~\ref{fig:Meffa09m220},~\ref{fig:Meffa09m130},~\ref{fig:Meffa06m310},~\ref{fig:Meffa06m220},
and~\ref{fig:Meffa06m135} given in Appendix~\ref{sec:appendix1}. In
each panel, we show the data for the various momentum channels
analyzed. The panels on the left (right) show results of the 2-state
(4-state) fits to the two-point function data for the different
momenta. The data with the largest errors and the least convincing
plateau at the larger momenta are from (i) the $a09m310$ and $a09m220$
ensembles that have lower statistics as they have not been analyzed
using the AMA method, and (ii) the $a06m220$ and $a06m135$ ensembles
at the weakest coupling that have the fewest gauge configurations
analyzed. Also, on a number of ensembles, we observe correlated
fluctuations in the data for $E_{\rm eff}$; both over $t$ for a given momenta
and at a given $t$ over the various momenta.  The former are 
taken into account by using the full covariance matrix in the fits to 
correlators at a given momenta. Since data at each momentum are  analyzed 
separately, the latter are ignored.

The results for the $M_i$ and the ${\cal A}_i$ are given 
in Tables~\ref{tab:multistates-twopt-mom-a12m310AMA-3},
\ref{tab:multistates-twopt-mom-a12m220LAMA-3},
\ref{tab:multistates-twopt-mom-a09m310-3},
\ref{tab:multistates-twopt-mom-a09m220-3},
\ref{tab:multistates-twopt-mom-a09m130LP1-3},
\ref{tab:multistates-twopt-mom-a06m310AMA-3},
\ref{tab:multistates-twopt-mom-a06m220AMA-3} and~\ref{tab:multistates-twopt-mom-a06m130AMA-3} in Appendix~\ref{sec:appendix1}.
The results from the 2-state fit shown in these tables are
slightly different from those presented in
Ref.~\cite{Bhattacharya:2016zcn} because, in this study, we use the full
covariance matrix when doing the fits, whereas in
Ref.~\cite{Bhattacharya:2016zcn} only the diagonal elements were used. 

As shown in Tables~\ref{tab:multistates-twopt-mom-a12m310AMA-3}--\ref{tab:multistates-twopt-mom-a06m130AMA-3}, the ground
state parameters, $E_0$ and ${\cal A}_0$ are consistent between the
2-, 3- and 4-state fits. The parameters for the first excited state,
$E_1$ and ${\cal A}_1$, also needed in 2-state fits to three-point
functions show stability only between the 3- and 4-state fits. When
analyzing the three-point correlation functions, we, therefore, used
estimates obtained from the 4-state fits for all four parameters,
$M_0$, ${\cal A}_0$, $M_1$ and ${\cal A}_1$.  It is worth noting the
change in the ratio $\Delta M_1/M_0$ for the two ensembles $a06m220$ and
$a06m135$ to about $0.85$ compared to $\lesssim 0.6$ for the other six
ensembles. With the current data, we cannot ascertain whether this
change is a statistical fluctuation or implies that the combination
and/or the nature of excited-states contributing have changed.

\begin{figure*}[tbp]%6
\centering
\includegraphics[width=0.97\linewidth]{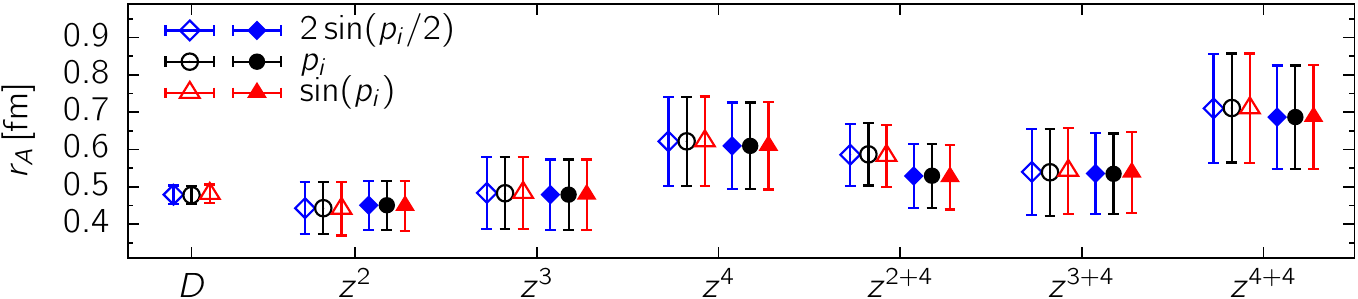}
\caption{Comparison of results for $r_A$ obtained
  using three possible definitions of lattice momenta. For the six
  $z$-expansion fits we also show variation of estimates between the 
  two values of ${\overline t}_0$: ${\overline t}_0=0$ shown using
  open symbols and ${\overline t}_0= {\overline t}_0^{\rm mid} = 0.12$
  GeV${}^2$ with filled symbols.  The label D stands for the dipole
  ansatz.  The data are from the $a06m135$ ensemble.}
\label{fig:6fits-a06m135}
\end{figure*}

When analyzing the three-point data to extract the form factors, we
need to decide what definition of momenta to use, i.e., whether one
should use $ap_i$ or $\sin(ap_i)$ or $2\sin(ap_i/2)$ for the lattice
momenta in the expression $Q^2 = \bm{p}^2 - (E-m)^2$.  Since
the three versions differ at $O(a^2)$ and our calculation has errors
starting at $O(a)$, there is no theoretical reason to prefer one over
the other. For guidance, we examined the dispersion relation for the
nucleon, $(aE)^2 - \sum_i f_i^2 = (aM)^2$, for the three cases $f_i =
ap_i$, $\sin(ap_i)$ and $2\sin(ap_i/2)$ in Fig.~\ref{fig:dispersion}
(Appendix~\ref{sec:appendix1}), for four
ensembles, two with the largest values of $\bf{p}$ and the two
physical mass ensembles. We find that, with our statistics, the
difference between the three forms is insignificant in all cases for
$(a\bm{p})^2 < 0.1$. Only the data at the highest momenta on the
$a12m310$, $a12m220L$ and $a09m310$ ensembles, that have results at
$(a\bm{p})^2 \gsim 0.1$, do we see some variation.  In short, no one
form is uniformly preferred by the data on all the
ensembles.\footnote{We did not investigate using alterate forms for energy such as 
$\sinh{aE}$.}  Nevertheless, we carried through the analysis to extract
the axial charge radius $r_A$ from fits to $G_A(Q^2)$
using all three forms, and found no sensitivity to the choice of the
form. As illustrated in Fig.~\ref{fig:6fits-a06m135}, the difference
between the three forms is not significant enough to even estimate an
associated systematic uncertainty.  We, therefore, present our final
estimates using the simplest version, $f_i = ap_i$.

\begin{figure*}[tbp]%2
\centering
\subfigure{
\includegraphics[width=0.47\linewidth]{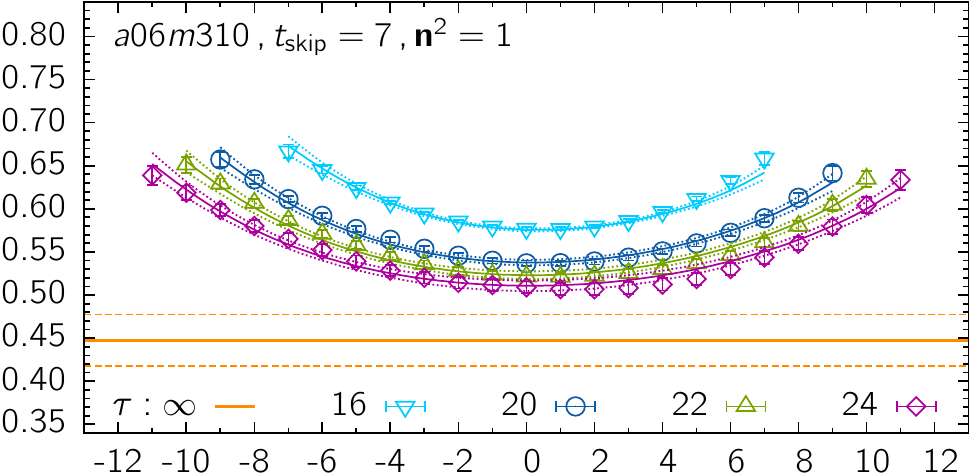}
\includegraphics[width=0.47\linewidth]{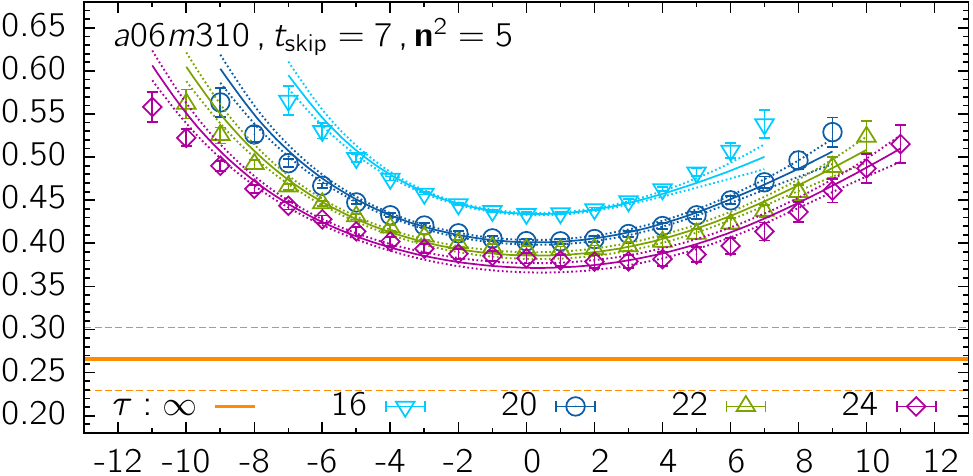}
}
\subfigure{
\includegraphics[width=0.47\linewidth]{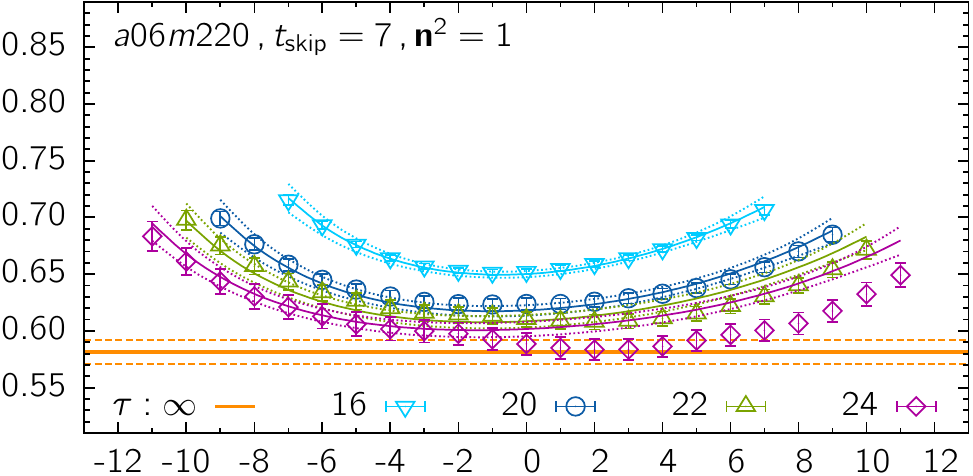}
\includegraphics[width=0.47\linewidth]{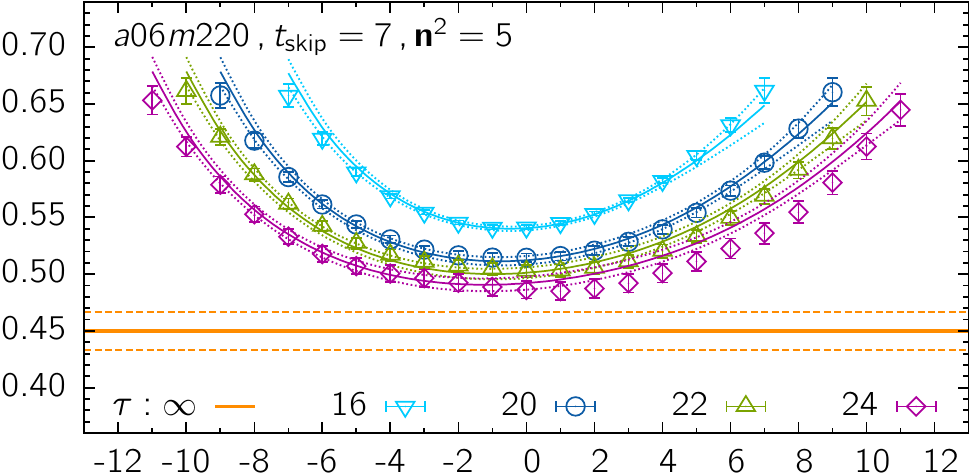}
}
\subfigure{
\includegraphics[width=0.47\linewidth]{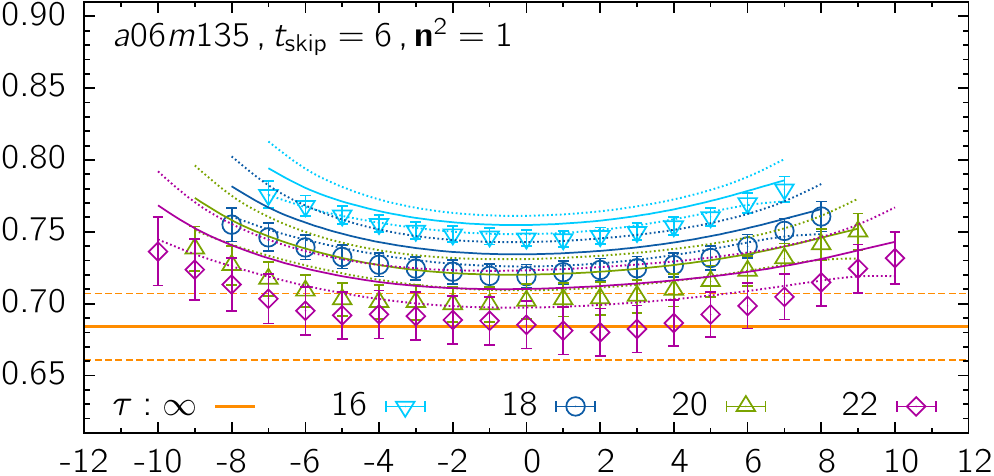}
\includegraphics[width=0.47\linewidth]{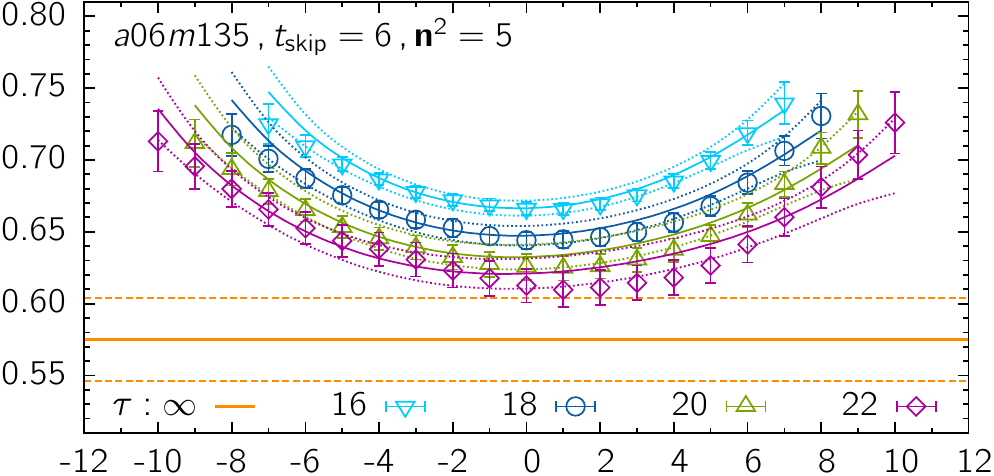}
}
\caption{The three-point data for ${\cal R}_{53}$ defined in
  Eq.~\protect\eqref{eq:ratio} versus the operator insertion time $t$,
  shifted by $\tau/2$. The labels give the ensemble ID, the number of
  points, $\tskip$, skipped on either end in the fits, the
  momentum label ${\bf n}^2$ and the values of $\tau$ simulated.
  Prediction of the 2-state fit for various values of the source-sink
  separation $\tau$ is shown in the same color as the data. The result
  for the matrix elements in the $\tsepi$ limit is shown by the
  horizontal band.  The plots on the top row are for the $a06m310$
  ensemble, middle row for the $a06m220$, and those on the bottom row
  for the $a06m135$ ensemble.  The plots on the left are for momenta
  $\bm{p}^2= {\bf n}^2 (2\pi/La)^2$ with ${\bf n}^2 = 1$, while those
  on the right are with ${\bf n}^2 =5$.  }
\label{fig:me-a06}
\end{figure*}

In Fig.~\ref{fig:dispersion} (bottom panels), we show the two points
with momentum components $n_i = (2,2,1)$ and $n_i = (3,0,0)$,
corresponding to ${\bf n^2} = 9$, in the $a09m130$ and $a06m135$
data. The difference between these two estimates is a measure of
the effect of the breaking of the rotational symmetry on the lattice to
the cubic group. Throughout this work, we keep these two data points
separate when analyzing the $a09m130$ and the $a06m135$ ensembles data.

%%%%%%%%%%%%%%%%%%%%%%%%%%%%%%%%%%%%%%%%%%%%%%%%%%%%%%%%%%%%%%%%%%%%%%%%%%%%%%%%
% Section: Isovector Axial Form Factors of the Nucleon
%%%%%%%%%%%%%%%%%%%%%%%%%%%%%%%%%%%%%%%%%%%%%%%%%%%%%%%%%%%%%%%%%%%%%%%%%%%%%%%%
\section{Extracting form factors from fits to the three-point Functions}
\label{sec:3pt-fits}

To display the data for the three-point correlation functions with the 
insertion of the axial current, we 
construct the following ratio, ${\cal R}_{5\Gamma}$, of the three-point
to the two-point correlation functions, 
\begin{align}
{\cal R}_{\gamma_5\Gamma}&(t, \tau, \bm{p}^\prime, \bm{p}) =   \frac{C^{(3\text{pt})}_\Gamma(t,\tau;\bm{p}^\prime,\bm{p})}{C^{(2\text{pt})}(\tau,\bm{p}^\prime)} \, \times \, \nonumber \\
&
  \left[ \frac{C^{(2\text{pt})}(t,\bm{p}^\prime) C^{(2\text{pt})}(\tau,\bm{p}^\prime) C^{(2\text{pt})}(\tau-t,\bm{p})}{C^{(2\text{pt})}(t,\bm{p}) C^{(2\text{pt})}(\tau,\bm{p}) C^{(2\text{pt})}(\tau-t,\bm{p}^\prime)}
  \right]^{1/2} \,.
\label{eq:ratio}
\end{align}
This ratio gives the desired ground state matrix element in the limit
$\tau \to \infty$, $t \to \infty$ and $(\tau-t) \to \infty$.  For all
the two-point correlation functions, we used the results of the
4-state fit. When calculating the matrix elements of the axial vector
current, defined in Eq.~\eqref{eq:AFFdef}, we use the spin projection
operator ${\cal P} = (1 + \gamma_4)(1 + i\gamma_5 \gamma_3)/2$.  As a
result, the imaginary part of the following three ratios of
correlators have a signal and give the desired form factors in the
limit $t$, $\tau-t$ and $\tau\to\infty$ :
\begin{align}
  {\cal R}_{51} \rightarrow&\; \frac{1}{\sqrt{(2 E_p (E_p+M))}} \left[ -\frac{q_1 q_3}{2M} {\tilde G}_P \right] \,,
  \label{eq:r2ff-GPGA1} \\
  {\cal R}_{52} \rightarrow&\; \frac{1}{\sqrt{(2 E_p (E_p+M))}} \left[ -\frac{q_2 q_3}{2M} {\tilde G}_P \right] \,,
  \label{eq:r2ff-GPGA2} \\
  {\cal R}_{53} \rightarrow&\; \frac{1}{\sqrt{(2 E_p (E_p+M))}} \left[ -\frac{q_3^2}{2M} {\tilde G}_P + (M+E) G_A \right]  \,.
  \label{eq:r2ff-GPGA3} 
%  \Re {\cal R}_{54} =&\; 4M q_3 \left[ \frac{M-E}{2M} G_P + G_A \right]
%  \label{eq:r2ff-GPGA4}
\end{align}
where ${\cal R}_{5i}$ implies the tensor structure ${\cal R}_{\gamma_5
  \gamma_i}$. We do not consider the ${\cal R}_{54}$ channel as the
signal in it is poor.  The pseudoscalar form factor $G_P(Q^2)$ is
given by the real part of ${\cal R}_{5} \equiv {\cal R}_{\gamma_5}$:
\begin{align}
  {\cal R}_{5} \rightarrow&\; \frac{1}{\sqrt{(2 E_p (E_p+M))}} \left[ q_3 {G}_P \right] \,. 
  \label{eq:r2ff-GP} 
\end{align}
In Fig.~\ref{fig:me-a06} (and in Figs.~\ref{fig:me-a12}
and~\ref{fig:me-a09} in Appendix~\ref{sec:appendix2}), we give plots
of the ratio ${\cal R}_{53}$, i.e., the ratio with tensor structure
$\gamma_5 \gamma_3$ for the axial current, defined in
Eq.~\eqref{eq:r2ff-GPGA3}.  The data are shown for all values of
$\tsep$ and for two values of momenta, $\bm{p}=(1,0,0) 2\pi/La$ and
$\bm{p}=(2,1,0) 2\pi/La$. Note that both $G_A$ and ${\tilde G}_P$
contribute to this ratio.  It is clear from the plots that the
excited-state contamination is significant in the data with $\tsep
\approx 1$~fm for our choice of the nucleon interpolating operator,
Eq.~\eqref{eq:nucl_op}, and the smearing parameters given in
Table~\ref{tab:cloverparams}.

From these data, the matrix element within the ground state is
obtained using Eq.~\eqref{eq:3pt}, i.e., keeping two intermediate
states in the fit to the three-point correlation function. The values
of $\tau \equiv \tsep$ and $\tskip$ used in the fit are given in the
figure's legend.  All values of $\tsep$ are fit simultaneously and the
resulting $\tsepi$ estimates are shown by the horizontal
band. Prediction of the fit for various values of $\tsep$ are also
shown as lines with error bands using the same color as the data points.
%% The data for the $a\approx 0.12$ and
%% $a\approx 0.09$~fm ensembles are given in Figs.~\ref{fig:me-a12}
%% and~\ref{fig:me-a09} in Appendix~\ref{sec:appendix2}.
We note that the $\tsepi$ estimate for some cases, such as on the
$a09m220$, $a09m130$ and $a06m310$ ensembles with ${\bf n}^2 = 5$, is
significantly below the data. Fits using only the diagonal elements of
the covariance matrix give $\tsepi$ results closer to the data.  This
could reflect that the statistical precision of the covariance matrix
is inadequate. However, for consistency, we keep fits using the full
covariance matrix in all cases.

\begin{figure*}[tbp]%3
\centering
\subfigure{
\includegraphics[width=0.32\linewidth]{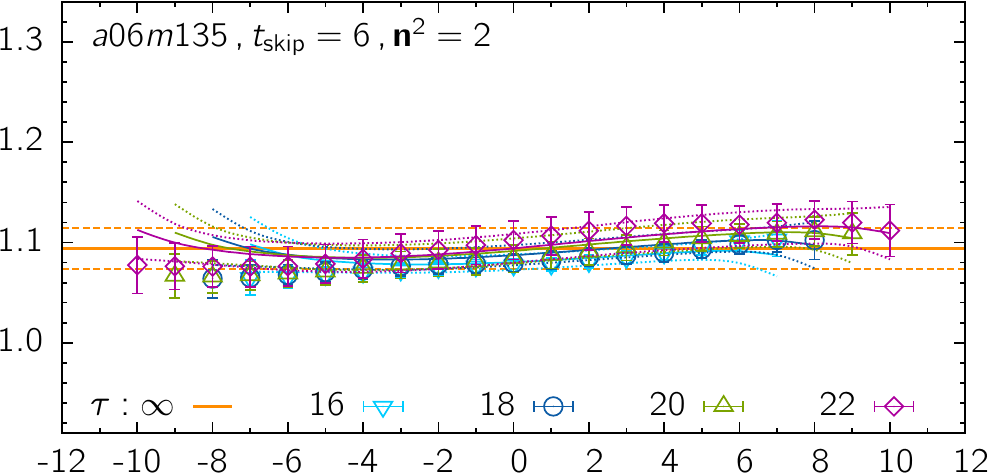}
\includegraphics[width=0.32\linewidth]{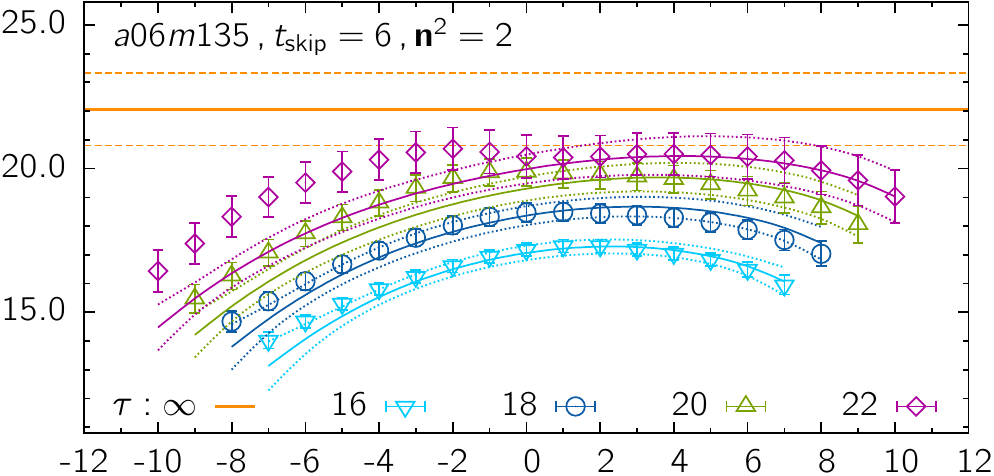}
\includegraphics[width=0.32\linewidth]{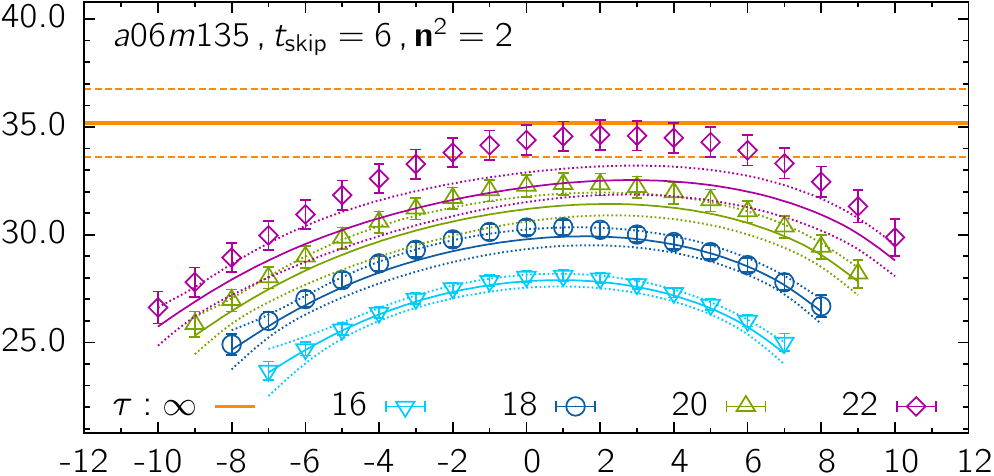}
}
\subfigure{
\includegraphics[width=0.32\linewidth]{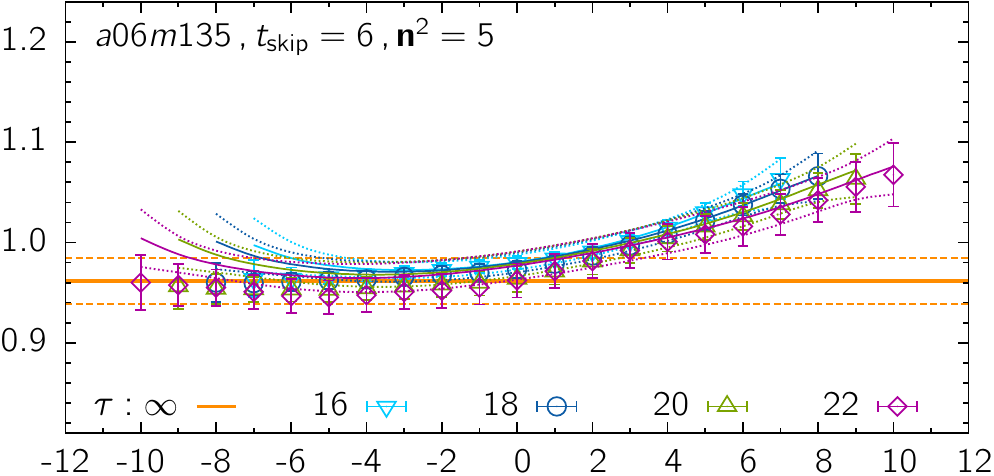}
\includegraphics[width=0.32\linewidth]{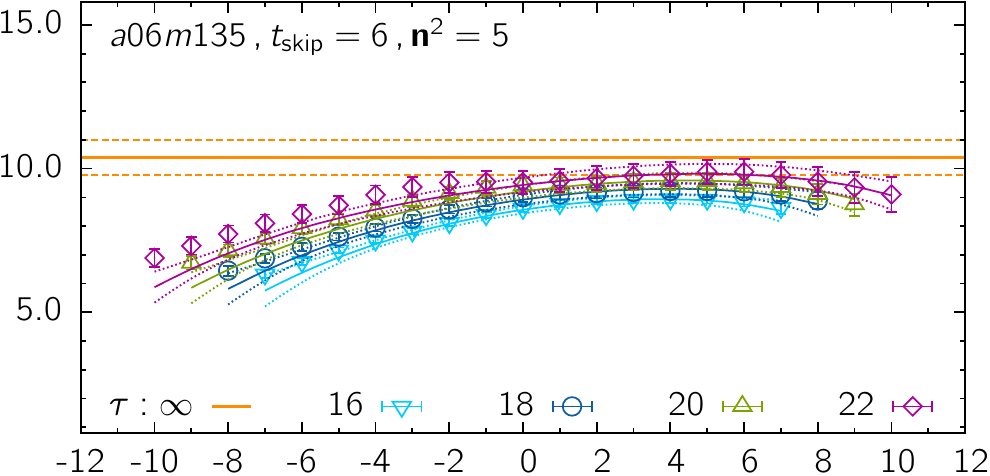}
\includegraphics[width=0.32\linewidth]{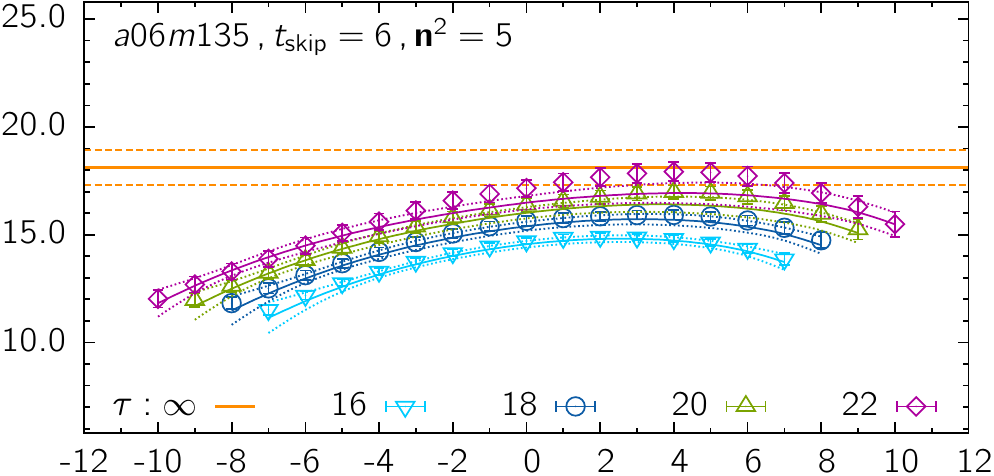}
}
\subfigure{
\includegraphics[width=0.32\linewidth]{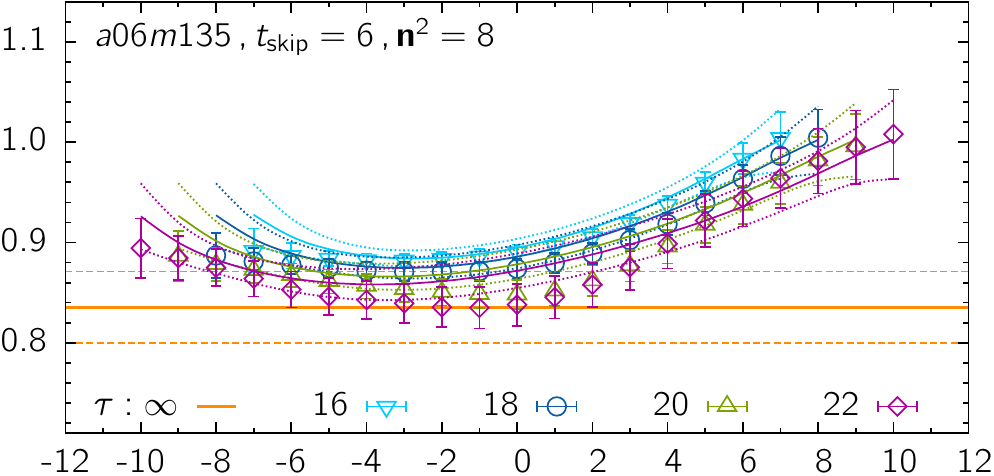}
\includegraphics[width=0.32\linewidth]{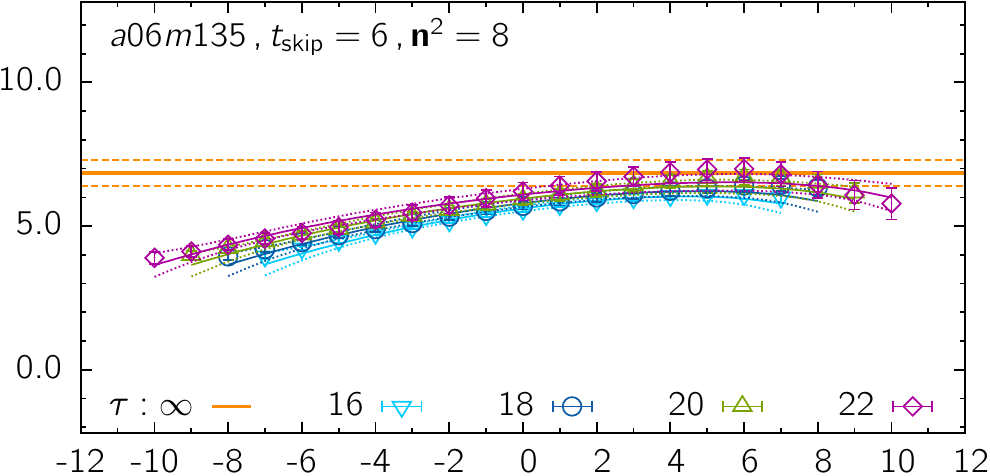}
\includegraphics[width=0.32\linewidth]{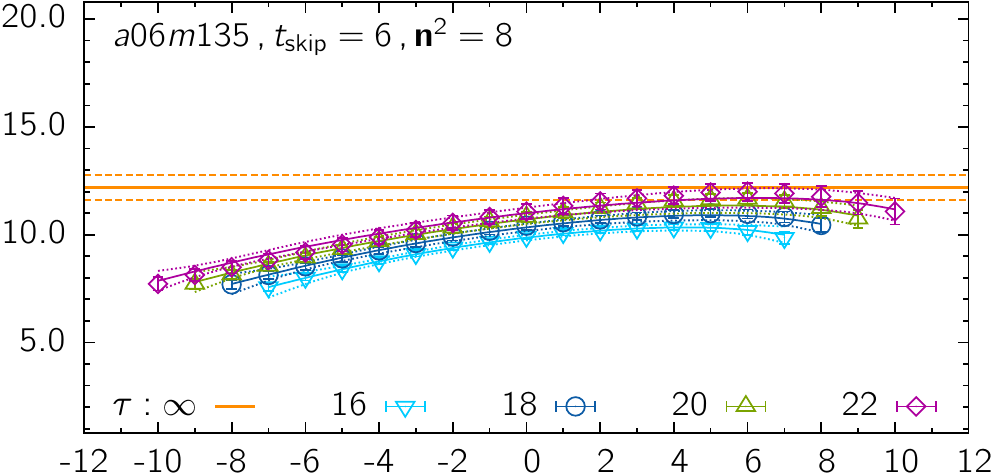}
}
\subfigure{
\includegraphics[width=0.32\linewidth]{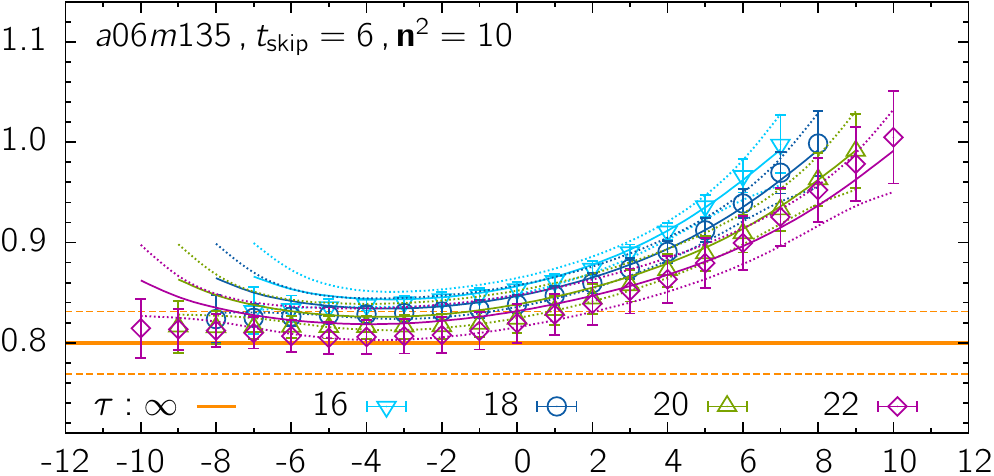}
\includegraphics[width=0.32\linewidth]{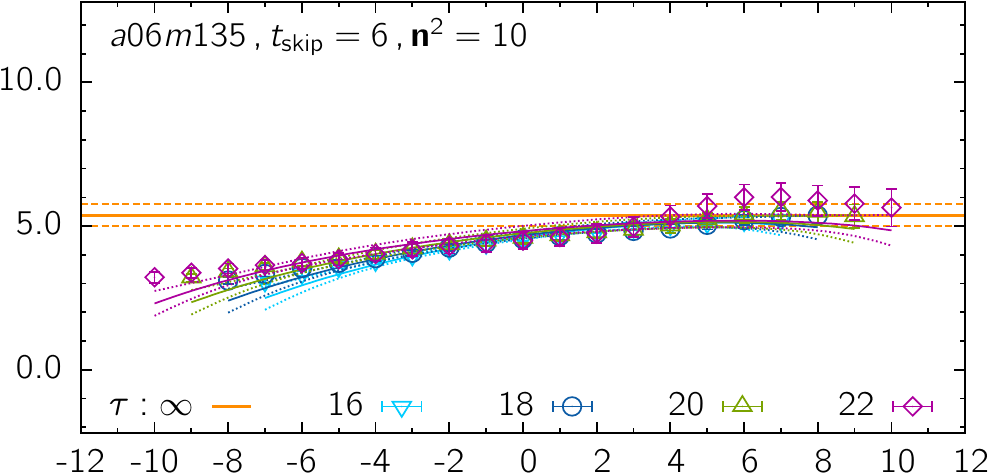}
\includegraphics[width=0.32\linewidth]{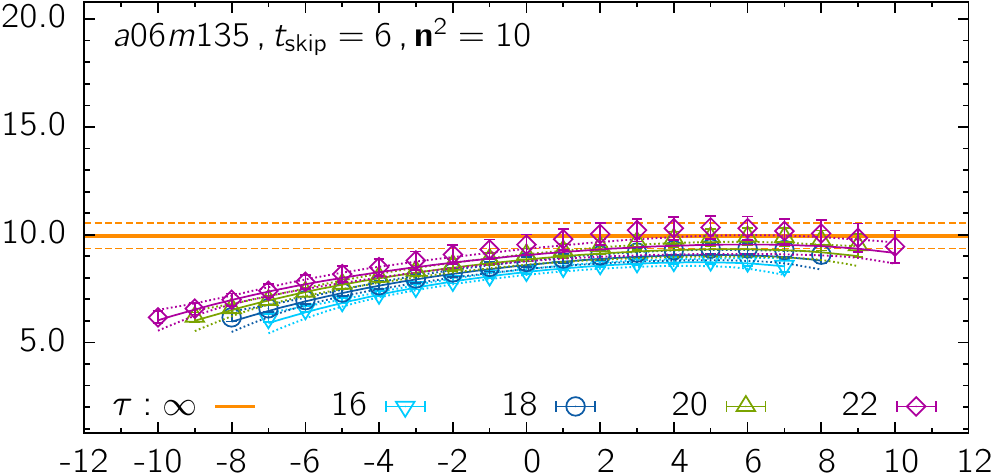}
}
\caption{Plots of the ratios ${\cal R}_{i}$ that give the three form
  factors: $G_A$ from ${\cal R}_{53}$ with $q_3=0$ but $q_{1,2} \neq
  0$ (left), ${\tilde G}_P$ from ${\cal R}_{51}$ (middle), and the
  pseudoscalar $G_P$ from ${\cal R}_{5}$ (right) versus the operator
  insertion time $t$ shifted by $\tau/2$ for the $a06m135$
  ensemble. The figures in the four rows are for data with ${\bf p}^2
  = {\bf n}^2 (2\pi/La)^2$ where ${\bf n}^2 = 2,\ 5,\ 8,$ and $10$,
  respectively.}
\label{fig:GP-GA-a06m135}
\end{figure*}

We also illustrate how the excited-state contamination impacts the
extraction of individual form factors $G_A$ and ${\tilde G}_P$ by
choosing two channels, ${\cal R}_{51}$ and ${\cal R}_{53}$ with
$q_3=0$ but non-zero $q_1$ or $q_2$, that give these directly. The data from the $a06m135$
ensemble are shown in Fig.~\ref{fig:GP-GA-a06m135}, while the data
from ensembles $a12m310$, $a06m310$ and $a06m220$ are given in
Figs.~\ref{fig:GP-GA-a12m310},~\ref{fig:GP-GA-a06m310},~\ref{fig:GP-GA-a06m220}
in Appendix~\ref{sec:appendix2}.  In these figures, fits to the
pseudoscalar form factor, defined in Eq.~\eqref{eq:r2ff-GP}, are also
shown where available. For the small ${\bf p}^2$ values, the
convergence of the three form factors with respect to $\tsep$ is from
below, i.e., excited state contamination leads to an underestimate.
The pattern of convergence changes for higher ${\bf p}^2$: $G_A(Q^2)$
starts to converge from above for ${\bf n}^2 \gtrsim 3$, and ${\tilde
  G}_P(Q^2)$ and $G_P(Q^2)$ for ${\bf n}^2 \gtrsim 10$ as shown in
Fig.~\ref{fig:GP-GA-a06m135}. Also illustrated in
Figs.~\ref{fig:GP-GA-a12m310},~\ref{fig:GP-GA-a06m310},
and~\ref{fig:GP-GA-a06m220} in Appendix~\ref{sec:appendix2}, the
transition ${\bf p}^2$ depends on the pion mass and the value of $Q^2$
in physical units. Note that these differences in trends in
convergence at low and high momenta act cohesively to increase the
slope of $G_A$ and ${\tilde G}_P$ with respect to $Q^2$, and thus the
values of $r_A$ and $g_P^\ast$ are larger compared to an analysis
neglecting excited state contamination.

\begin{figure}[tpb]%4
\centering
\subfigure{
\includegraphics[width=0.97\linewidth]{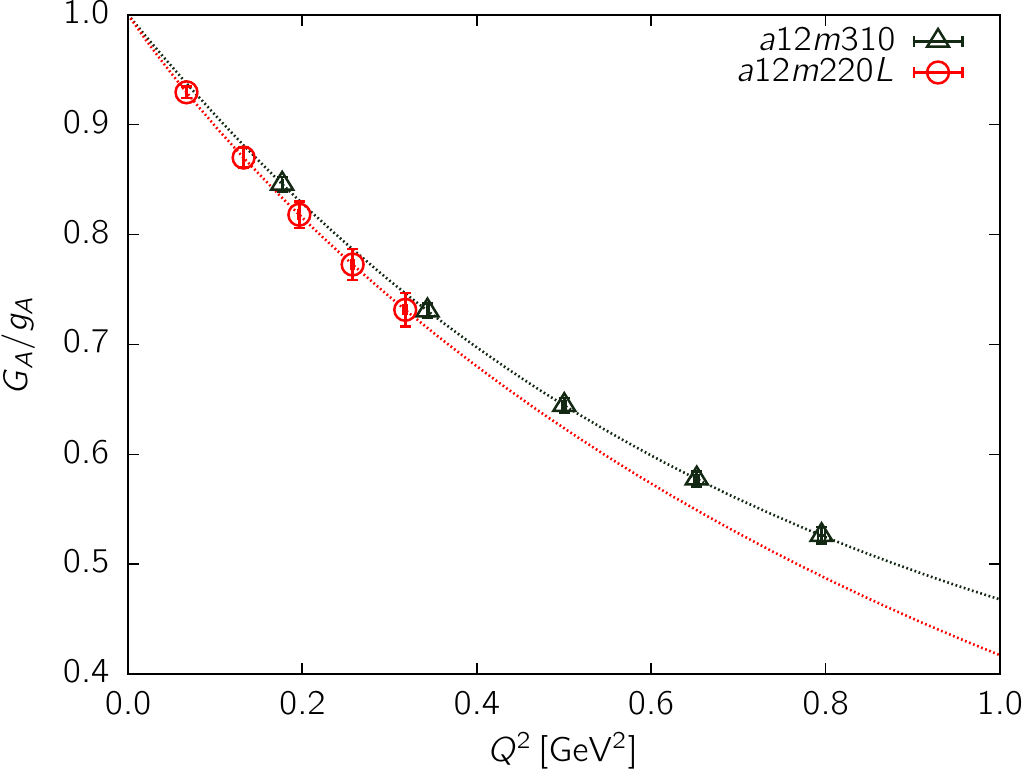} 
}
\subfigure{
\includegraphics[width=0.97\linewidth]{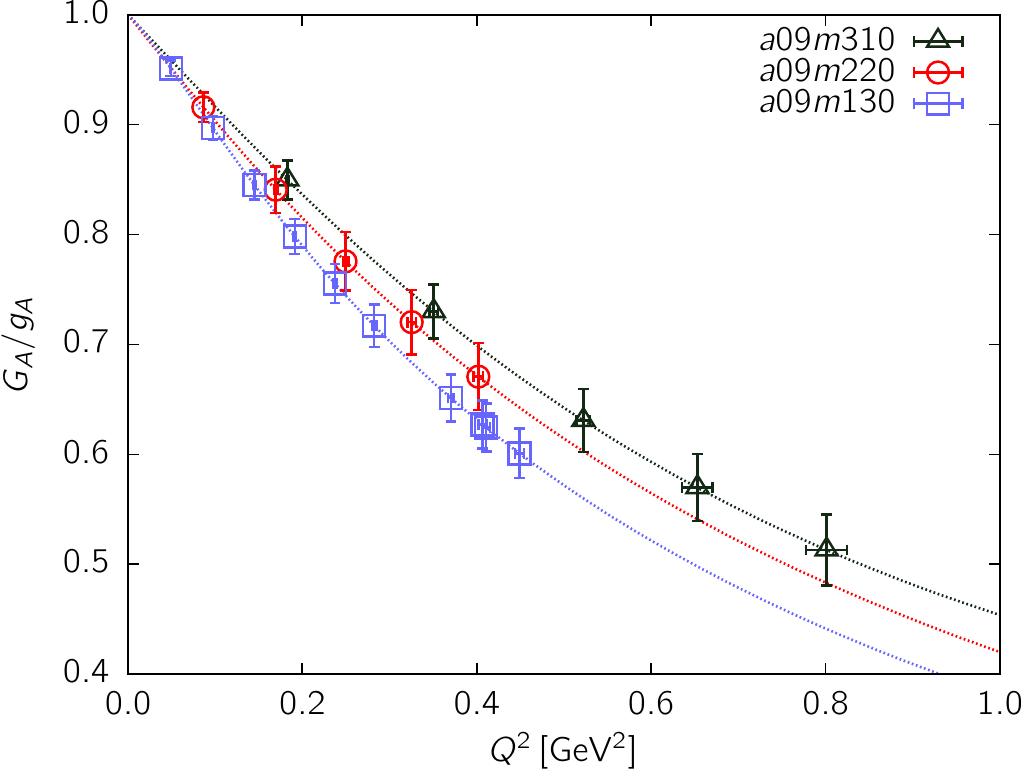} 
}
\subfigure{
\includegraphics[width=0.97\linewidth]{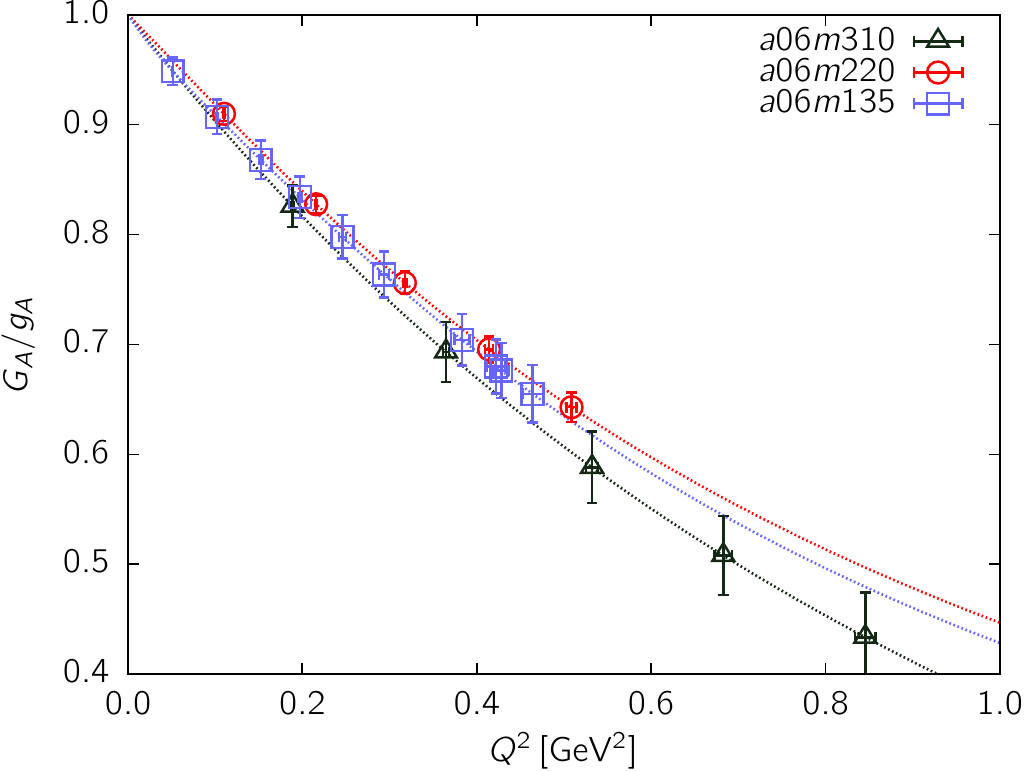} 
}
\caption{The data for the normalized axial form factor $G_A(Q^2)/g_A$
  versus $Q^2$ plotted to highlight the dependence on $M_\pi^2$ for
  fixed $a$. The top figure is for the $a\approx 0.12$~fm ensembles,
  the middle for the $a\approx 0.09$~fm ensembles, and the bottom for
  the $a\approx 0.06$~fm ensembles.  We also show the $z^{3+4}$ fit to
  the data for each ensemble; the corresponding value of $r_A$ 
  obtained from the slope at $Q^2=0$ is given in
  Table~\protect\ref{tab:rA_results}. The color scheme used is black
  for the $M_\pi \approx 310$, red for $M_\pi \approx 220$, and purple
  for the $M_\pi \approx 130$~MeV ensembles.}
\label{fig:ff-vsM}
\end{figure}

\begin{figure}[tpb]%5
\centering
\subfigure{
\includegraphics[width=0.97\linewidth]{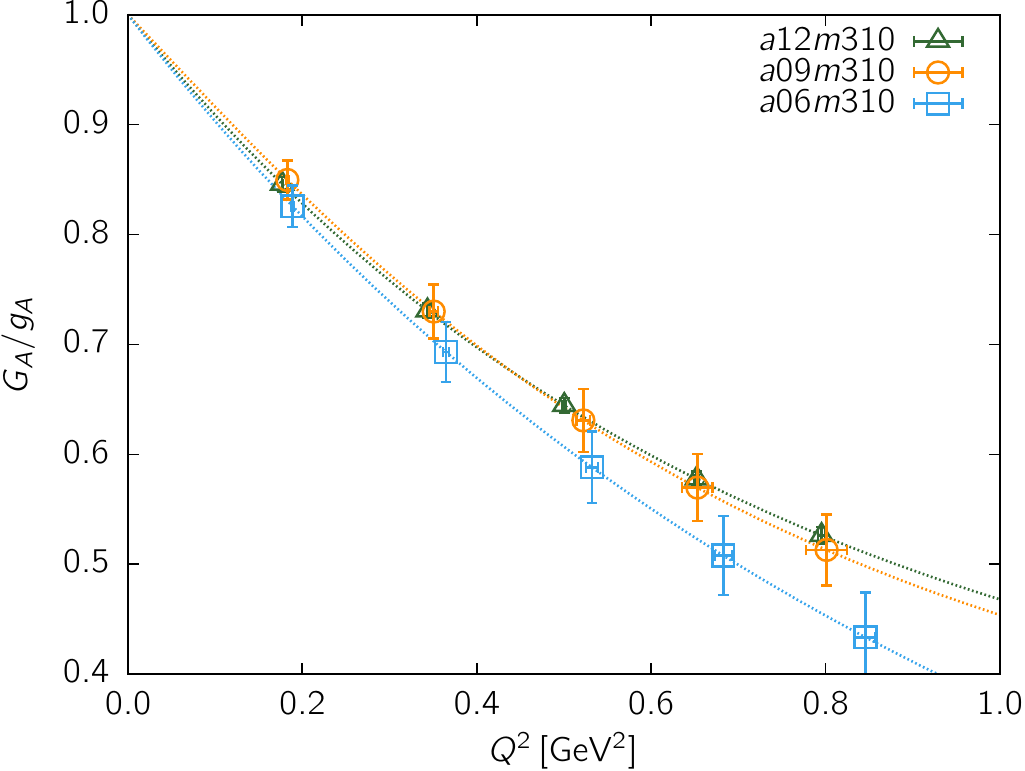}
}
\subfigure{
\includegraphics[width=0.97\linewidth]{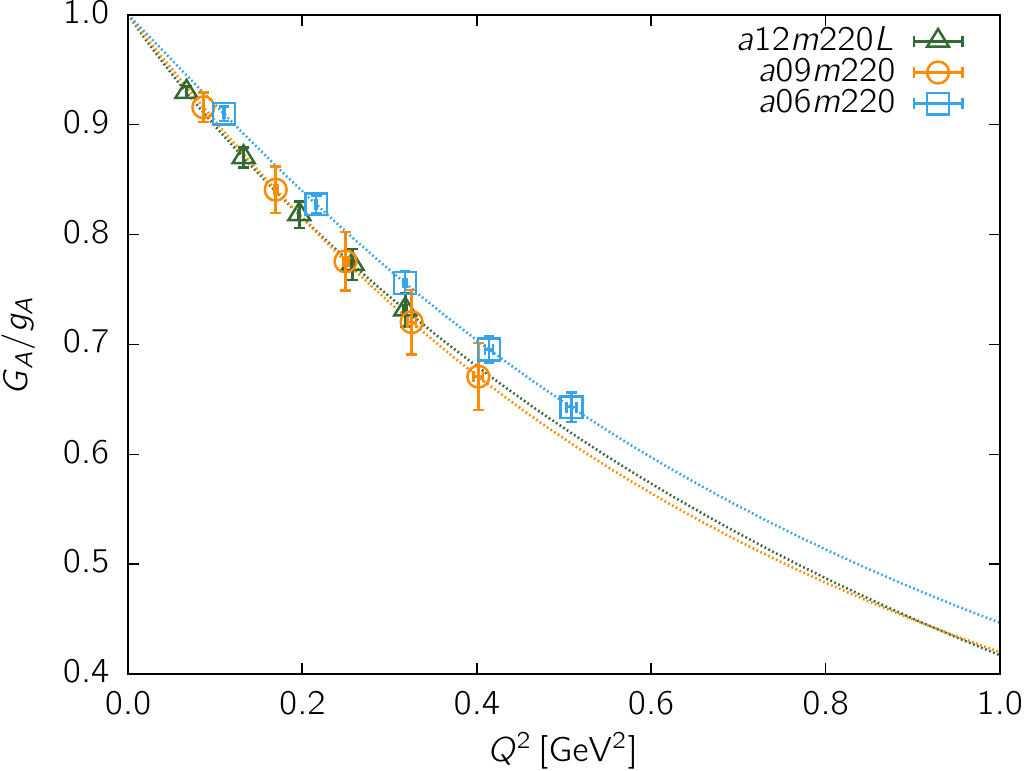}
}
\subfigure{
\includegraphics[width=0.97\linewidth]{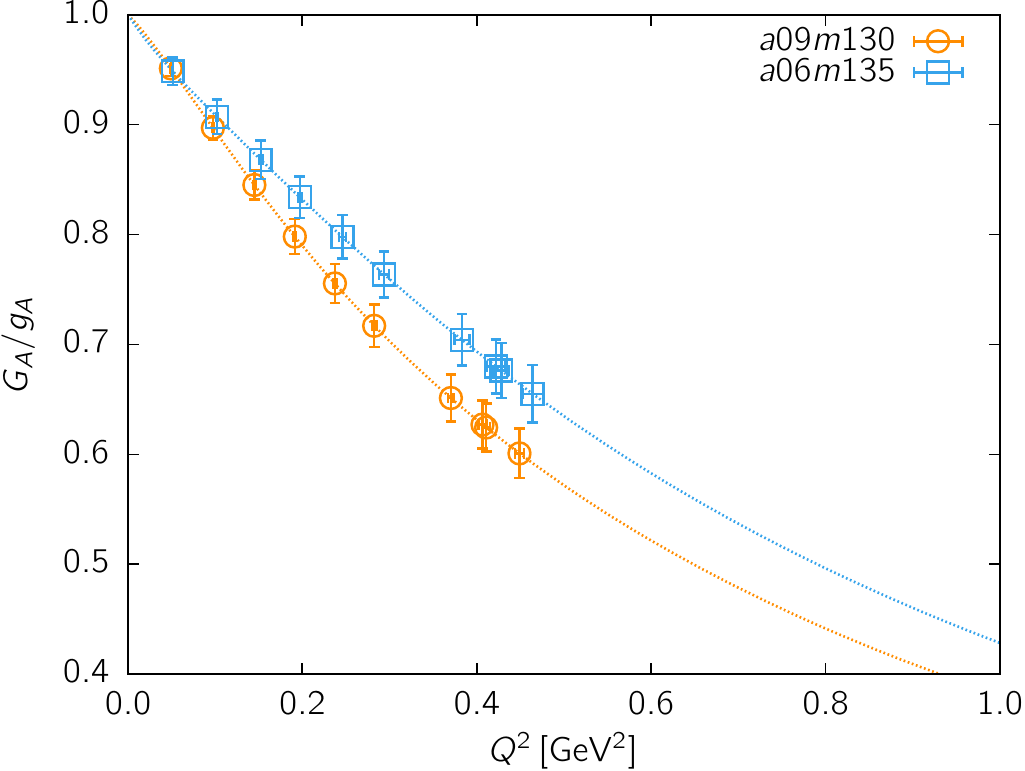}
}
\caption{The same data and fits for the normalized axial form factor
  $G_A(Q^2)/g_A$ versus $Q^2$ as shown in Fig.~\protect\ref{fig:ff-vsM}
  but plotted to highlight the dependence on $a$ for fixed $M_\pi$. The
  top figure is for the $M_\pi \approx 310$~MeV ensembles, the middle
  for the $M_\pi \approx 220$~MeV ensembles, and the bottom for the
  $M_\pi \approx 130$~MeV ensembles. The color scheme used is green
  for the $a \approx 0.12$, orange for $a \approx 0.09$ and blue 
  for the $a \approx 0.06$~fm ensembles. }
\label{fig:ff-vsa}
\end{figure}

\begin{table*}[tbp]%3
\centering
\vspace{0.5cm}
\renewcommand*{\arraystretch}{1.1}
\begin{ruledtabular}
  \begin{tabular}{c|c|cc|cc|cc|cc|cc}
                  & $\bm{n}^2=0$  & \multicolumn{2}{c|}{$\bm{n}^2=1$} & \multicolumn{2}{c|}{$\bm{n}^2=2$} & \multicolumn{2}{c|}{$\bm{n}^2=3$} & \multicolumn{2}{c|}{$\bm{n}^2=4$}  & \multicolumn{2}{c}{$\bm{n}^2=5$}  \\
                  & $G_A$         & $Q^2$ & $G_A(Q^2)$    & $Q^2$ & $G_A(Q^2)$    & $Q^2$ & $G_A(Q^2)$    & $Q^2$ & $G_A(Q^2)$    & $Q^2$ & $G_A(Q^2)$    \\
    \hline                                                                                                                                
    $a12m310$     &   1.270(12)   & 0.177 &   1.073(5)    & 0.344 &   0.929(8)    & 0.500 &   0.823(9)    & 0.652 &  0.723(12)    & 0.796 &   0.668(10)   \\
    \hline                                                                                                                                
    $a12m220L$    &   1.304(20)   & 0.067 &   1.211(14)   & 0.133 &   1.132(10)   & 0.197 &   1.058(9)    & 0.258 &  1.007(10)    & 0.318 &   0.950(11)   \\
    \hline                                                                                                                                
    $a09m310$     &   1.257(38)   & 0.183 &   1.073(18)   & 0.351 &   0.930(17)   & 0.522 &   0.793(23)   & 0.653 &   0.730(31)   & 0.801 &   0.660(31)   \\
    \hline                                                                                                                                
    $a09m220$     &   1.291(44)   & 0.086 &   1.178(29)   & 0.170 &   1.081(21)   & 0.250 &   0.984(21)   & 0.325 &   0.918(24)   & 0.402 &   0.856(24)   \\
    \hline                                                                                                                                
    $a09m130$     &   1.252(21)   & 0.049 &   1.193(17)   & 0.097 &   1.121(12)   & 0.145 &   1.052(11)   & 0.191 &   1.004(13)   & 0.237 &   0.945(14)   \\
                  &               & 0.282 &   0.897(16)   &       &               & 0.370 &   0.806(19)   & 0.411 &   0.783(19)   & 0.449 &   0.748(22)   \\
                  &               &       &               &       &               &       &               & 0.407 &   0.781(22)   &       &               \\
    \hline                                                                                                                                
    $a06m310$     &   1.231(25)   & 0.189 &   1.018(10)   & 0.365 &   0.853(19)   & 0.532 &   0.721(30)   & 0.683 &   0.635(36)   & 0.846 &   0.529(42)   \\
    \hline                                                                                                                                
    $a06m220$     &   1.206(14)   & 0.110 &   1.098(11)   & 0.216 &   0.997(10)   & 0.318 &   0.906(11)   & 0.414 &   0.845(14)   & 0.509 &   0.775(14)   \\
    \hline                                                                                                                                
    $a06m135$     &   1.204(24)   & 0.051 &   1.136(20)   & 0.102 &   1.094(20)   & 0.152 &   1.031(26)   & 0.197 &   1.005(20)   & 0.246 &   0.953(23)   \\
                  &               & 0.294 &   0.900(31)   &       &               & 0.383 &   0.837(35)   & 0.428 &   0.793(36)   & 0.464 &   0.793(32)   \\
                  &               &       &               &       &               &       &               & 0.422 &   0.824(37)   &       &               \\
  \end{tabular}
\end{ruledtabular}
  \caption{Results for the unrenormalized axial form factor $G_A(Q^2)$
    obtained from solving the overdetermined 
    set of equations,
    Eqs.~\protect\eqref{eq:r2ff-GPGA1}-\protect\eqref{eq:r2ff-GPGA3},
    relating the form factors to the matrix elements as
    described in the text. We also give the associated momentum
    transfer $Q^2$ in units of GeV${}^2$.  The label ${\bf n}^2 =
    \sum_i n_i^2$ gives the squared three-momentum in units of
    $(2\pi/La)^2$.  The second row for the ensembles $a09m130$ and
    $a06m135$ gives $G_A(Q^2)$ for momenta ${\bf n}^2 + 5$. The third
    row gives $G_A(Q^2)$ for momentum ${\bf n}^2=9$ with $n_i =
    (3,0,0)$, while the $n_i = (2,2,1)$ case is given in the second
    row.}
  \label{tab:ff_GAdata}
\end{table*}

\begin{table*}[tbp]%4
\centering
\renewcommand*{\arraystretch}{1.1}
\begin{ruledtabular}
  \begin{tabular}{c|ccccc}
    \multicolumn{1}{c|}{} & 
    \multicolumn{1}{c}{$\bm{n}^2=1$} & 
    \multicolumn{1}{c}{$\bm{n}^2=2$} &
    \multicolumn{1}{c}{$\bm{n}^2=3$} & 
    \multicolumn{1}{c}{$\bm{n}^2=4$} &
    \multicolumn{1}{c}{$\bm{n}^2=5$} 
    \\\hline
    $a12m310$     &   15.67(31)      &    9.08(20)     &    6.10(14)     &    4.18(10)     &    3.35(8)      \\
    \hline
    $a12m220L$    &   31.21(2.32)    &   20.79(1.56)   &   15.51(1.10)   &   12.27(83)     &    9.79(57)     \\
    \hline
    $a09m310$     &   15.21(82)      &    8.53(32)     &    5.79(31)     &    4.00(41)     &    2.99(31)     \\
    \hline
    $a09m220$     &   25.43(2.12)    &   16.57(1.46)   &    12.57(1.16)  &    9.50(75)     &    7.73(53)     \\
    \hline
    $a09m130$     &   37.93(1.84)    &   23.66(99)     &   17.57(70)     &   14.19(52)     &   11.21(35)     \\
                  &    9.62(28)      &                 &    7.06(18)     &    6.35(18)     &    5.45(16)     \\
                  &                  &                 &                 &    6.25(19)     &                 \\
    \hline
    $a06m310$     &   14.41(53)      &    8.16(27)     &    5.25(20)     &    3.78(19)     &    2.73(16)     \\
    \hline
    $a06m220$     &   19.94(45)      &   12.30(26)     &    8.60(20)     &    6.68(18)     &    5.25(13)     \\
    \hline
    $a06m135$     &   31.88(1.19)    &   22.00(82)     &   16.24(69)     &   13.05(43)     &    10.44(41)    \\
                  &    8.82(36)      &                 &    6.88(34)     &    5.79(26)     &     5.42(27)    \\
                  &                  &                 &                 &    6.19(31)     &                 \\
  \end{tabular}
\end{ruledtabular}
  \caption{Results for the unrenormalized induced pseudoscalar form
    factor $\tilde{G}_P(Q^2)$. The values of $Q^2$ for the various
    values of ${\bf n}^2 = \sum_i n_i^2$ and other details are the
    same as given in Table~\ref{tab:ff_GAdata}. }
  \label{tab:ff_GPdata}
\end{table*}

\begin{table*}[tbp]%5
\centering
\renewcommand*{\arraystretch}{1.1}
\begin{ruledtabular}
  \begin{tabular}{c|ccccc}
    \multicolumn{1}{c|}{} & 
    \multicolumn{1}{c}{$\bm{n}^2=1$} & 
    \multicolumn{1}{c}{$\bm{n}^2=2$} &
    \multicolumn{1}{c}{$\bm{n}^2=3$} & 
    \multicolumn{1}{c}{$\bm{n}^2=4$} &
    \multicolumn{1}{c}{$\bm{n}^2=5$} 
    \\\hline
    $a12m310$     &   19.24(41)      &   11.65(25)     &    8.06(18)     &    6.15(17)     &    4.84(11)     \\
    \hline
    $a09m130$     &   54.12(3.00)    &   36.58(1.92)   &   27.81(1.41)   &   22.05(1.00)   &    18.33(76)    \\
                  &   15.78(63)      &                 &   11.98(41)     &   10.75(38)     &     9.34(32)    \\
                  &                  &                 &                 &   10.67(38)     &                 \\
    \hline
    $a06m220$     &   28.00(74)      &   18.06(44)     &   13.13(35)     &   10.11(30)     &    8.13(21)     \\
    \hline
    $a06m135$     &   51.44(1.72)    &   35.21(1.23)   &   26.77(95)     &   21.18(77)     &    18.11(66)    \\
                  &   15.55(58)      &                 &   12.19(50)     &   10.77(48)     &     9.94(53)    \\
                  &                  &                 &                 &   10.89(58)     &                 \\
  \end{tabular}
\end{ruledtabular}
  \caption{Results for the unrenormalized pseudoscalar form factor
    ${G}_P(Q^2)$ obtained from the matrix element of the pseudoscalar
    operator $\overline{\psi} \gamma_5 \psi$ between nucleon states.
    The values of $Q^2$ for the various values of ${\bf n}^2 = \sum_i
    n_i^2$ and other details are the same as given in
    Table~\ref{tab:ff_GAdata}. }
  \label{tab:ff_GP5data}
\end{table*}

The final values of the two form factors, $G_A(Q^2)$ and
$\tilde{G}_P(Q^2)$, are extracted by solving the overdetermined set of
Eqs.~\eqref{eq:r2ff-GPGA1}--\eqref{eq:r2ff-GPGA3} for each momentum
$Q^2$. These results for $G_A(Q^2)$ are given in
Table~\ref{tab:ff_GAdata}, and those for ${\tilde G}_P(Q^2)$ in
Table~\ref{tab:ff_GPdata}. The data for the pseudoscalar form factor,
calculated from the matrix element of the operator $\overline{u}
\gamma_5 d$ using Eq.~\eqref{eq:r2ff-GP}, are given in
Table~\ref{tab:ff_GP5data} for the four ensembles analyzed, $a12m310$,
$a09m130$, $a06m220$, and $a06m135$.  The values for the PCAC mass,
determined from the pion two-point correlation functions, are
${\widehat m} a = a (m_u +m_d)Z_m Z_P /2 Z_A$ = 0.012119(18),
0.0015383(39), 0.0027984(23) and 0.0008840(18), respectively. The
difference from the corresponding values for the HISQ light quarks,
$(m_u +m_d)/2 = 0.0102$, 0.0012, 0.0024 and
0.00084~\cite{Bazavov:2012xda}, used in the generation of the
ensembles, can be attributed to the factor $Z_m Z_P / Z_A$ in the
clover formalism, which is unity for HISQ.

Results for $G_A(Q^2)$ are plotted as a function of $Q^2$ in
Figs.~\ref{fig:ff-vsM} and~\ref{fig:ff-vsa}.  The data in
Fig.~\ref{fig:ff-vsM} are organized to exhibit the dependence on the
light quark mass (equivalently, $M_\pi^2$) for fixed lattice spacing,
while Fig.~\ref{fig:ff-vsa} highlights the variation versus the
lattice spacing $a$ for fixed pion mass $M_\pi$. We also show the
$z$-expansion fit $z^{3+4}$, discussed in Sec.~\ref{sec:fits-rA},
which is used in obtaining the final estimate of $r_A$.  The data in
Fig.~\ref{fig:ff-vsM} show weak dependence on the light quark mass for
fixed $a$ on all ensembles but the $a09m130$ ensemble, for which they
are a little lower, and give a slightly larger $r_A$.  The trend in
the data versus the lattice spacing $a$ in Fig.~\ref{fig:ff-vsa} is a
small decrease with $a$ for the $M_\pi=310$ ensembles, but is reversed
in the $M_\pi \approx 220$ and 130~MeV data, suggesting that higher
precision data are needed to establish a possible trend.

%%%%%%%%%%%%%%%%%%%%%%%%%%%%%%%%%%%%%%%%%%%%%%%%%%%%%%%%%%%%%%%%%%%%%%%%%%%%%%%%
% Section: Fits to extract the axial charge radius
%%%%%%%%%%%%%%%%%%%%%%%%%%%%%%%%%%%%%%%%%%%%%%%%%%%%%%%%%%%%%%%%%%%%%%%%%%%%%%%%
\section{Fits to extract the axial charge radius}
\label{sec:fits-rA}

The data for $G_A(Q^2)$, given in Table~\ref{tab:ff_GAdata}, are fit
using seven ansatz to parameterize the $Q^2$ behavior: the dipole approximation given in
Eq.~\eqref{eq:dipole}; the $z^2$, $z^3$ and $z^4$ truncation of the
$z$-expansion given in Eq.~\eqref{eq:Zexpansion}; and these three
truncations of the $z$-expansion supplemented with the four sum rule
constraints given in Eq.~\eqref{eq:sumrule} and labeled $z^{2+4}$, $z^{3+4}$
and $z^{4+4}$.  From these fits we extract the axial charge radius squared,  
$r_A^2$, using Eq.~\eqref{eq:rdef}.

In the analyses using the $z$-expansion, we first investigated the
sensitivity of the fits on the choice of ${\overline t}_0$ in the definition of
$z$ and on the three choices for momenta, $f_i = ap_i$, $\sin(ap_i)$
and $2\sin(ap_i/2)$, in evaluating $Q^2$. The quality of the fits and
the results for $r_A$ are indistinguishable between
the three choices of $f_i$ and between ${\overline t}_0=0$ and the approximate
mid-point of $Q^2$ range, which we call ${\overline t}_0^{\rm mid}$. We illustrate
this insenstivity using the data from the $a06m135$ ensemble, that has
the largest number of $Q^2$ values, in
Fig.~\ref{fig:6fits-a06m135}. The same pattern is seen in all eight
ensembles.  Also, the fits in $z$ with and without using the sum
rules, for example, $z^2$ versus $z^{2+4}$, give consistent results
for $r_A$, however, as expected, the large $Q^2$
behavior is much more reasonable with fits including the sum rules. 

\begin{figure*}[tbp]%7
\centering
\subfigure{
\includegraphics[width=0.47\linewidth]{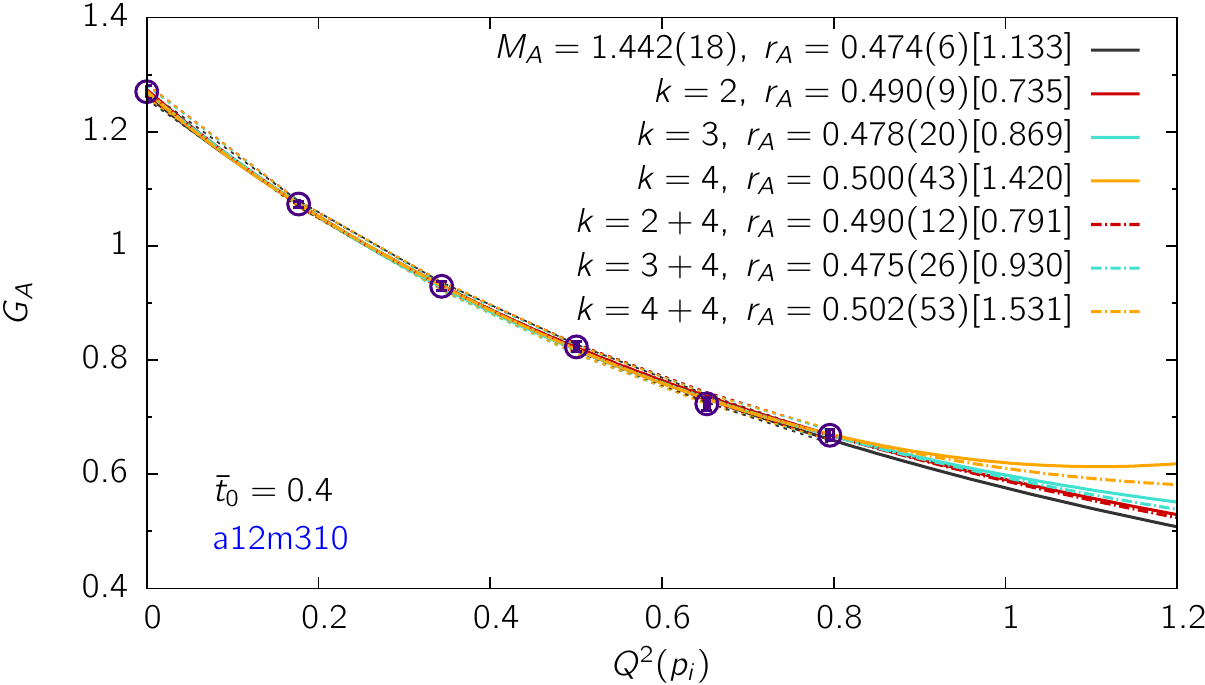}
\includegraphics[width=0.47\linewidth]{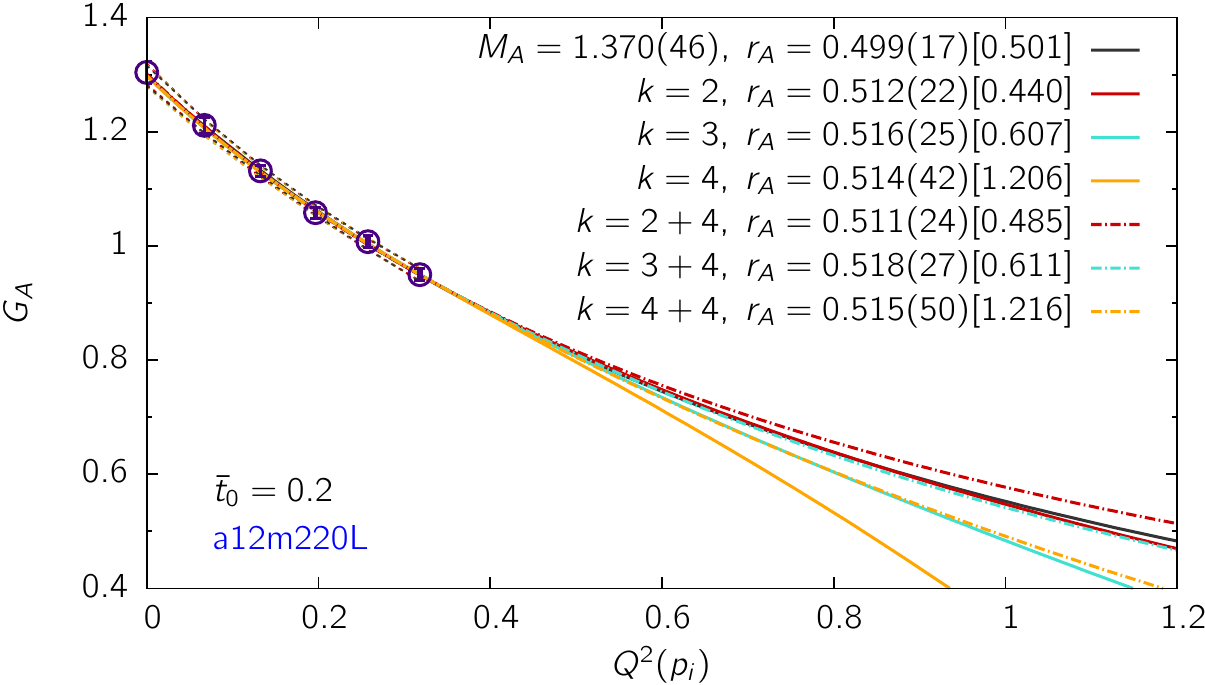}
}
\subfigure{
\includegraphics[width=0.47\linewidth]{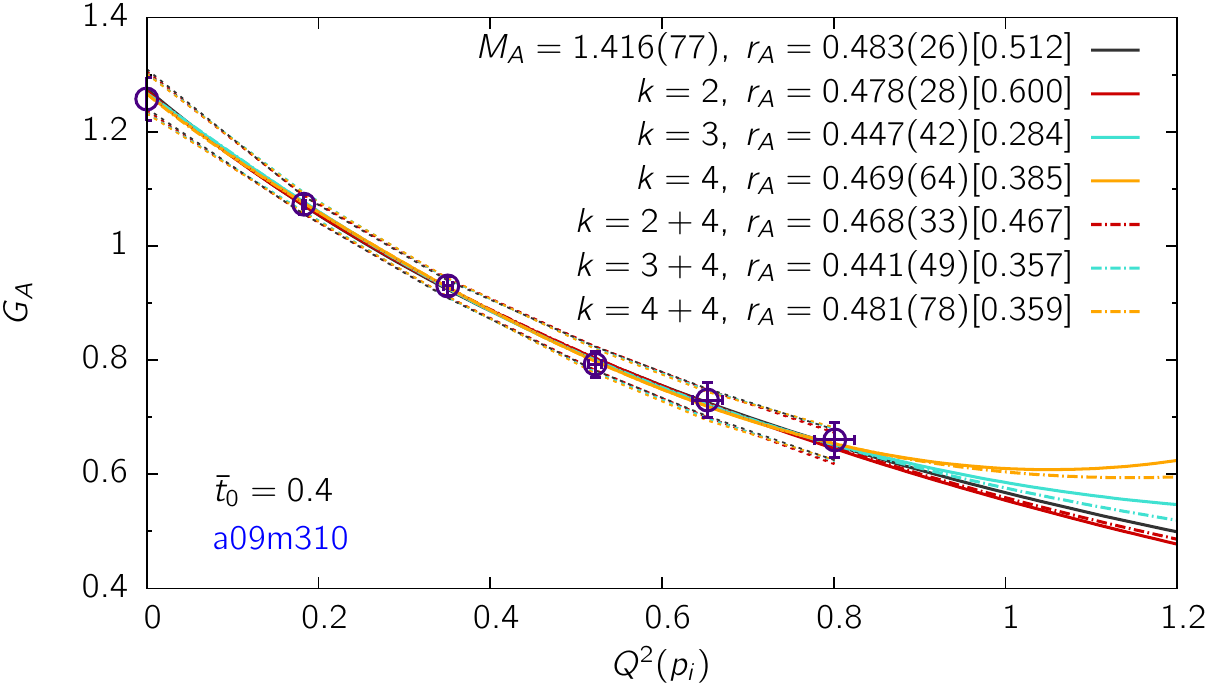}
\includegraphics[width=0.47\linewidth]{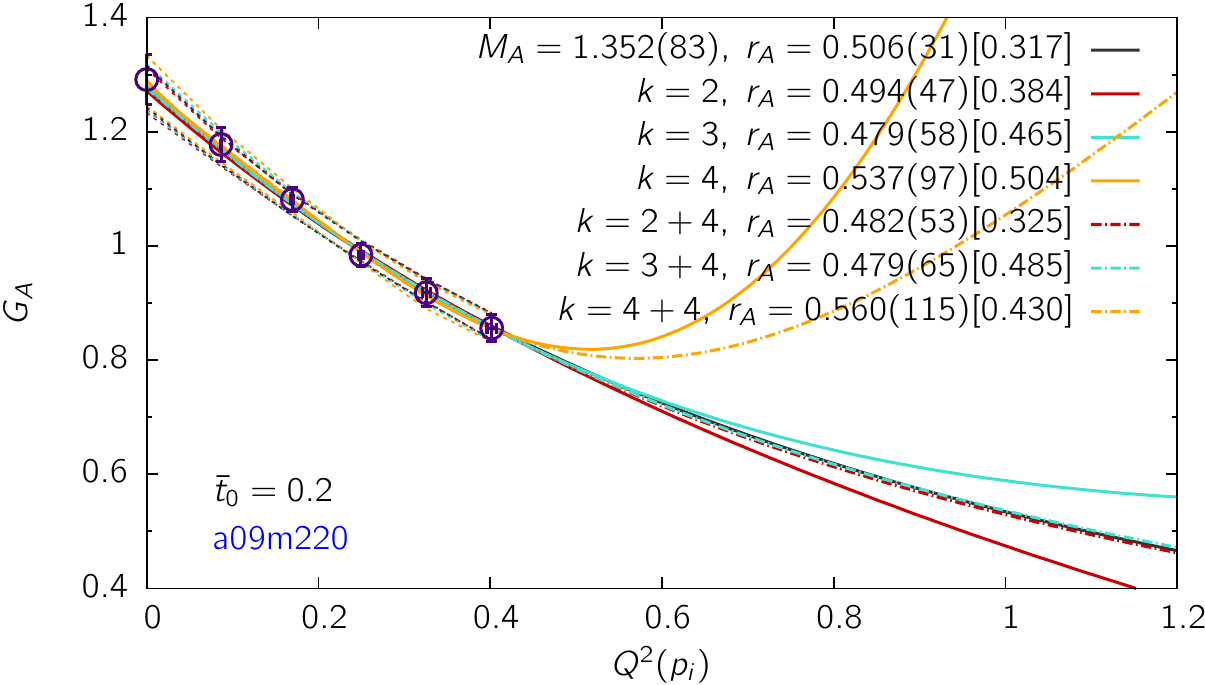}
}
\subfigure{
\includegraphics[width=0.47\linewidth]{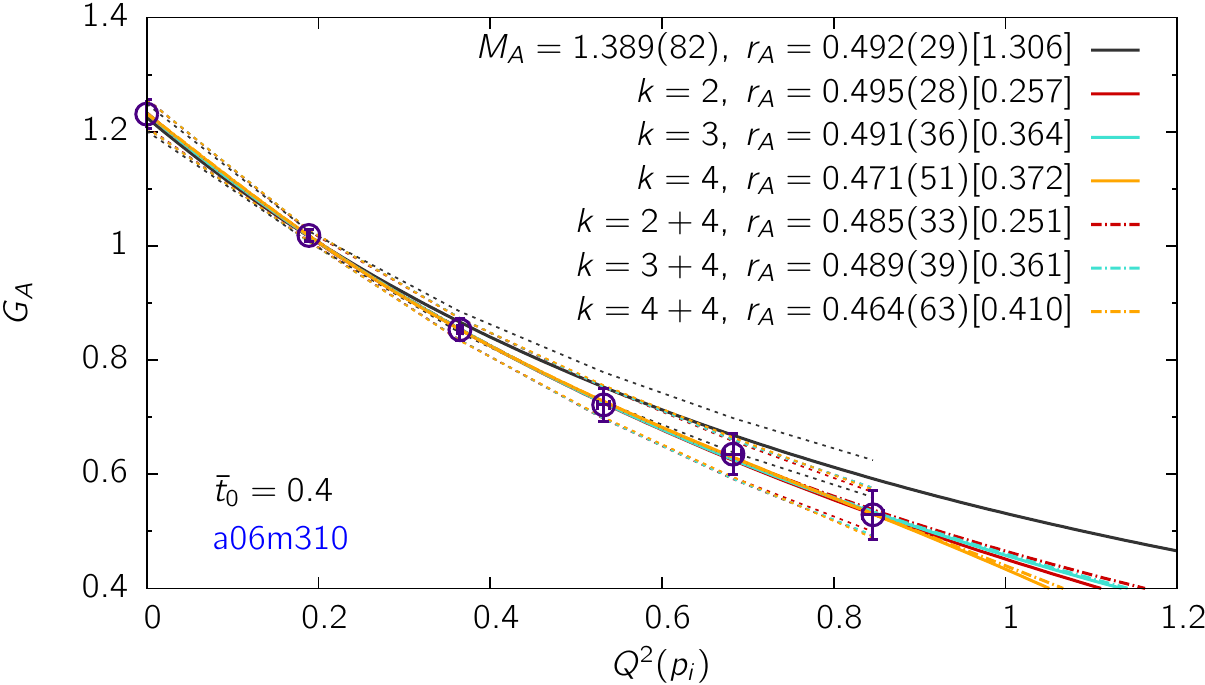}
\includegraphics[width=0.47\linewidth]{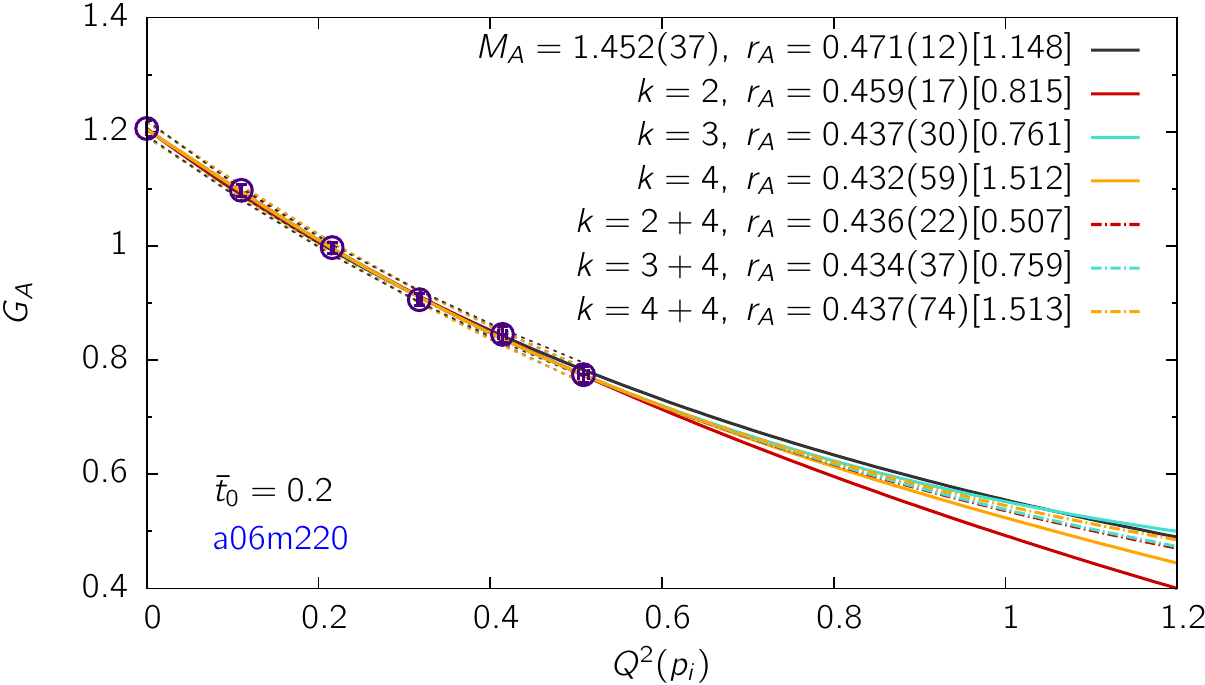}
}
\subfigure{
\includegraphics[width=0.47\linewidth]{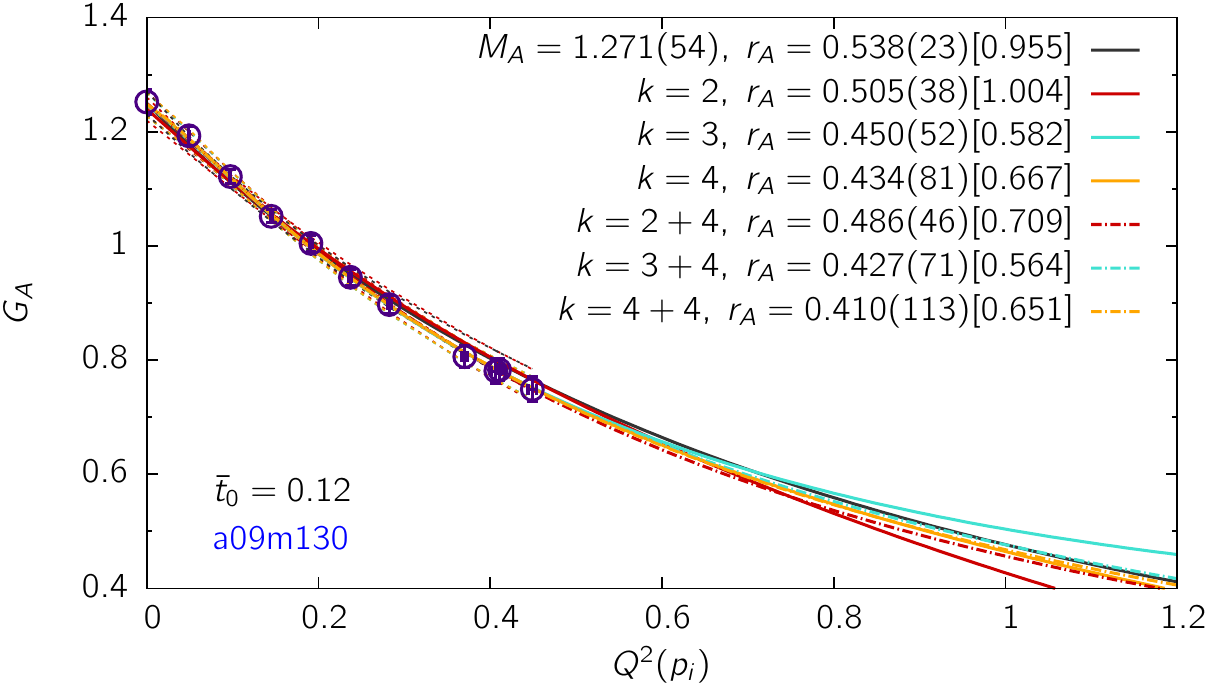}
\includegraphics[width=0.47\linewidth]{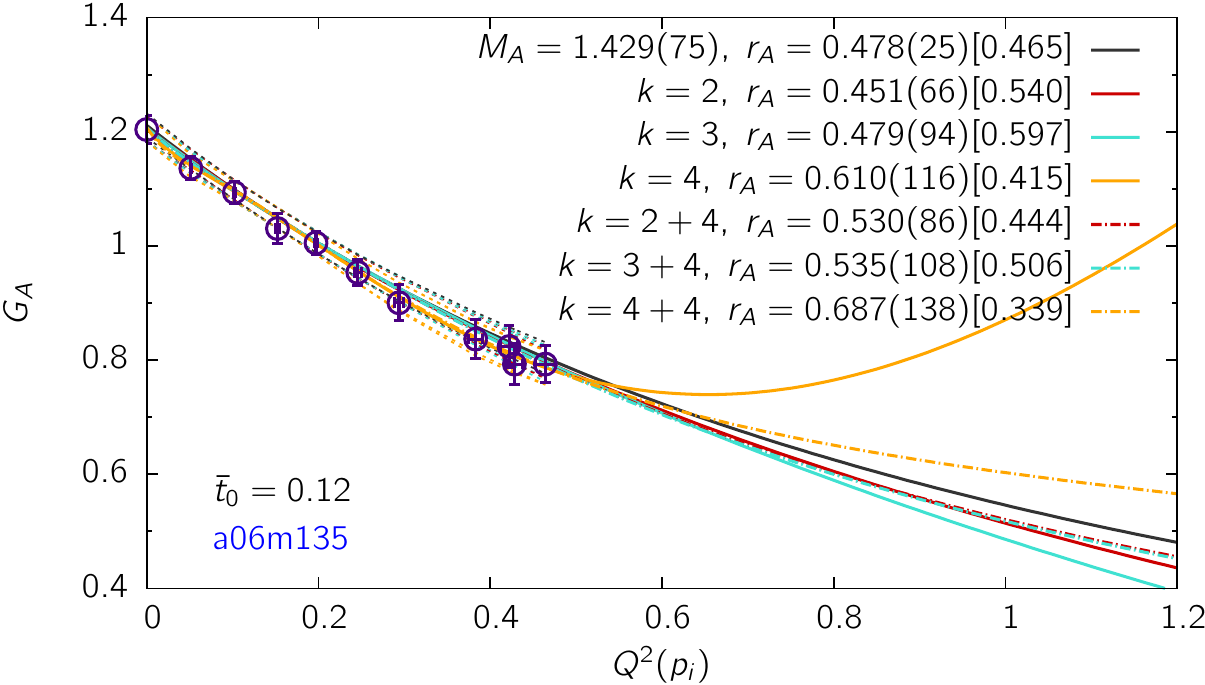}
}
\caption{Fits to the unrenormalized $G_A(Q^2)$ data (circles) versus
  $Q^2$ (GeV${}^2$) for the eight ensembles.  The top two panels show
  data and fits for the $a12m310$ and $a12m220L$ ensembles; the second
  row for $a09m310$ and $a09m220$; the third row for $a06m310$ and
  $a06m220$; and the final row for the two physical mass ensembles
  $a09m130$ and $a06m135$.  The axial radius $r_A$ is extracted from
  these fits using Eq.~\protect\eqref{eq:rdef}. Estimates of the mass
  ${\cal M}_A$ from the dipole fit and the axial radius $r_A$ from the
  various fits are given in the labels. The number within the square
  brackets is the $\chi^2/{\rm d.o.f.}$ of the fit. }
\label{fig:8fits-rA}
\end{figure*}

For our final results we use fits with $f_i= ap_i$, the mid-point
value, ${\overline t}_0^{\rm mid}$ as it minimizes $z_{\rm max}$, and
include the sum rules in the $z$-expansion.  These fits to $G_A(Q^2)$
versus $Q^2$ for the eight ensembles are shown in
Fig.~\ref{fig:8fits-rA}. The labels give the estimates of $r_A$ 
from the seven fit ansatz along with the $\chi^2/{\rm
  d.o.f.}$ within square brackets.\footnote{Fits to the ratio
  $G_A(Q^2)/G_A(Q^2=0)$ give essentially identical results for
  $r_A$ in all cases.}  The resulting values of
$r_A$ from the seven fits are collected together in
Table~\ref{tab:rA_results}. Overall, the dipole ansatz does a
remarkably good job of fitting the data as shown in
Fig.~\ref{fig:8fits-rA}.

We find that these estimates of $r_A$ from the seven
ansatz are, in most cases, consistent within the $ 1\sigma$ combined
statistical and fit uncertainty and show little dependence on the
lattice spacing or the pion mass.  The solid and dashed orange lines
in Fig.~\ref{fig:8fits-rA} show that the $k^4$ and the $k^{4+4}$ fits,
which have only one degree of freedom, and in many cases have a large curvature that becomes manifest outside 
the range of the data. For this reason, we do not include these
ansatz in our final estimates.

\begin{comment}
\begin{figure*}[tbp]%7
\centering
\subfigure{
\includegraphics[width=0.47\linewidth]{figsX/GA_incl0_a06m135AMA.pdf}
\includegraphics[width=0.47\linewidth]{figsX/GA_incl0_t0eq0p12_a06m135AMA}
}
\subfigure{
\includegraphics[width=0.47\linewidth]{figsX/GA_incl0_sinphalf_a06m135AMA}
\includegraphics[width=0.47\linewidth]{figsX/GA_incl0_t0eq0p12_sinphalf_a06m135AMA}
}
\subfigure{
\includegraphics[width=0.47\linewidth]{figsX/GA_incl0_sinp_a06m135AMA}
\includegraphics[width=0.47\linewidth]{figsX/GA_incl0_t0eq0p12_sinp_a06m135AMA}
}
\caption{Fits versus $Q^2$ (GeV${}^2$) to the $G_A(Q^2)$ data (circles) from the
  $a06m135$ ensemble. The three plots in each column illustrate the
  sensitivity of the data to the choice of the definition of lattice
  momenta. The two plots in each row compare fits with ${\overline t}_0=0$ versus
  ${\overline t}_0^{\rm mid} = 0.12$ GeV${}^2$.  Estimates of the dipole mass
  ${\cal M}_A$ and the axial radius $r_A$ from thbe various fits are
  given in the labels. The number with the square brackets is the
  $\chi^2/{\rm d.o.f.}$ of the fit. The resulting estimates are insensitive
  to the choice of the definition of lattice momenta and of ${\overline t}_0$.}
\label{fig:6fits-a06m135}
\end{figure*}
\end{comment}

\begin{table*}[tbp]%6
%  \vspace{-7mm}
  \centering
  \renewcommand{\arraystretch}{1.4}
  \setlength{\tabcolsep}{10pt}
%    \resizebox{0.95\textwidth}{!}{
   \begin{ruledtabular}
      \begin{tabular}{c|c|cc|cc|cc}
Ensemble   & Dipole         & $z^2$     & $z^{2+4}$      & $z^{3}$       & $z^{3+4}$  &  $z^{4}$       & $z^{4+4}$       \\
\hline
$a12m310$  &  0.225(05) & 0.240(09) & 0.240(12) & 0.228(19) & 0.225(24)  & 0.250(43)  & 0.252(53) \\
$a12m220L$ &  0.249(17) & 0.262(23) & 0.261(24) & 0.267(26) & 0.268(28)  & 0.264(43)  & 0.265(51) \\
$a09m310$  &  0.233(25) & 0.229(27) & 0.219(31) & 0.200(38) & 0.195(43)  & 0.220(60)  & 0.231(75) \\
$a09m220$  &  0.256(31) & 0.244(47) & 0.232(52) & 0.230(56) & 0.230(63)  & 0.288(10)  & 0.314(129) \\
$a09m130$  &  0.289(24) & 0.255(38) & 0.236(45) & 0.202(47) & 0.183(61)  & 0.188(70)  & 0.168(93) \\
$a06m310$  &  0.242(29) & 0.245(28) & 0.235(32) & 0.241(35) & 0.239(38)  & 0.222(48)  & 0.215(58) \\
$a06m220$  &  0.222(11) & 0.211(15) & 0.190(19) & 0.191(26) & 0.188(32)  & 0.187(51)  & 0.191(65) \\
$a06m135$  &  0.229(24) & 0.204(59) & 0.281(91) & 0.229(90) & 0.287(116) & 0.373(141) & 0.473(190) \\
\hline
$a12m310$  & 0.474(06)      & 0.490(09) & 0.490(12)      & 0.478(20)       & 0.475(26)     & 0.500(43)     & 0.502(53)       \\
$a12m220L$ & 0.499(17)      & 0.512(22) & 0.511(24)      & 0.516(25)       & 0.518(27)     & 0.514(42)     & 0.515(50)       \\
$a09m310$  & 0.483(26)      & 0.478(28) & 0.468(33)      & 0.447(42)       & 0.441(49)     & 0.469(64)     & 0.481(78)       \\
$a09m220$  & 0.506(31)      & 0.494(47) & 0.482(53)      & 0.479(58)       & 0.479(65)     & 0.537(97)     & 0.560(115)      \\
$a09m130$  & 0.538(23)      & 0.505(38) & 0.486(46)      & 0.450(52)       & 0.427(71)     & 0.434(81)     & 0.410(113)       \\
$a06m310$  & 0.492(29)      & 0.495(28) & 0.485(33)      & 0.491(36)       & 0.489(39)     & 0.471(51)     & 0.464(63)       \\
$a06m220$  & 0.471(12)      & 0.459(17) & 0.435(23)      & 0.436(31)       & 0.434(37)     & 0.432(60)     & 0.437(76)       \\
$a06m135$  & 0.478(25)      & 0.451(66) & 0.530(86)      & 0.479(94)       & 0.535(108)    & 0.610(116)    & 0.687(138)      \\
\end{tabular}
\end{ruledtabular}
%    }
\caption{The upper half of the table lists the isovector axial radius
  squared, $\langle r_A^2 \rangle$ in units of fm${}^2$, obtained from
  the dipole and six different $z$-expansion fits ($z^2$, $z^{2+4}$,
  $z^3$, $z^{3+4}$, $z^4$, and $z^{4+4}$) to the form factor
  $G_A(Q^2)$. The fits $z^{2+4}$, $z^{3+4}$, and $z^{4+4}$ include the
  four sumrule constraints given in
  Eq.~\protect\eqref{eq:sumrule}. For convinience, the bottom half of
  the table gives the radius, $r_A$, in units of fm.}
\label{tab:rA_results}
\end{table*}

\begin{figure*}[tbp]%8
{
    \includegraphics[width=0.32\linewidth]{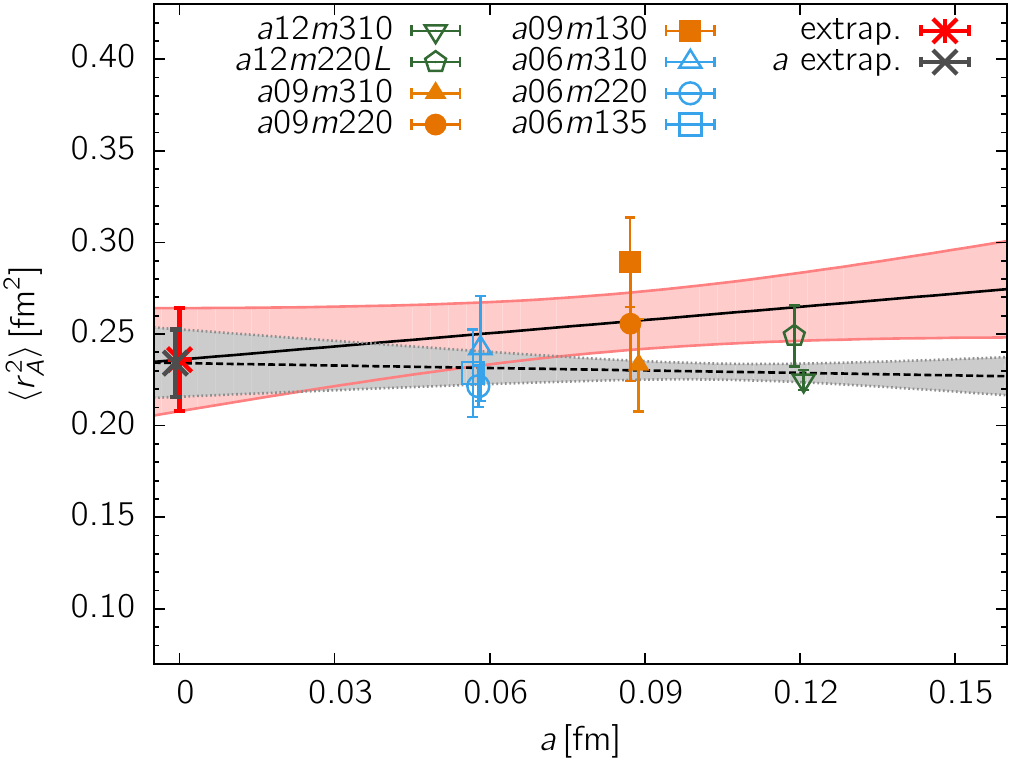}
    \includegraphics[width=0.32\linewidth]{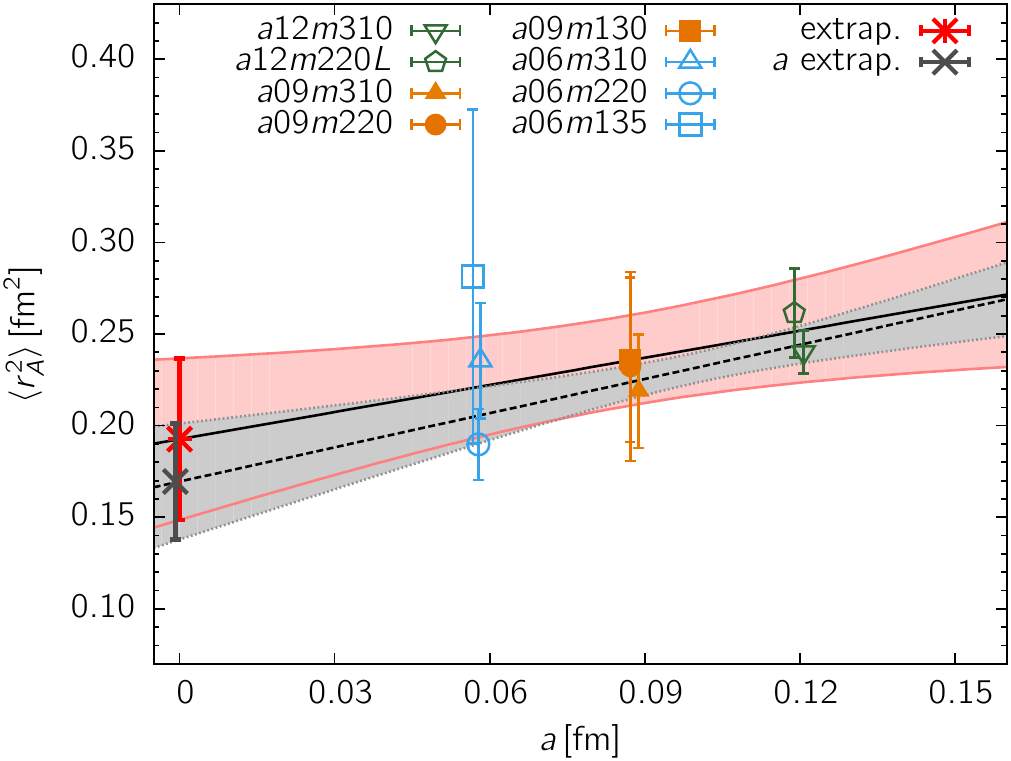}
    \includegraphics[width=0.32\linewidth]{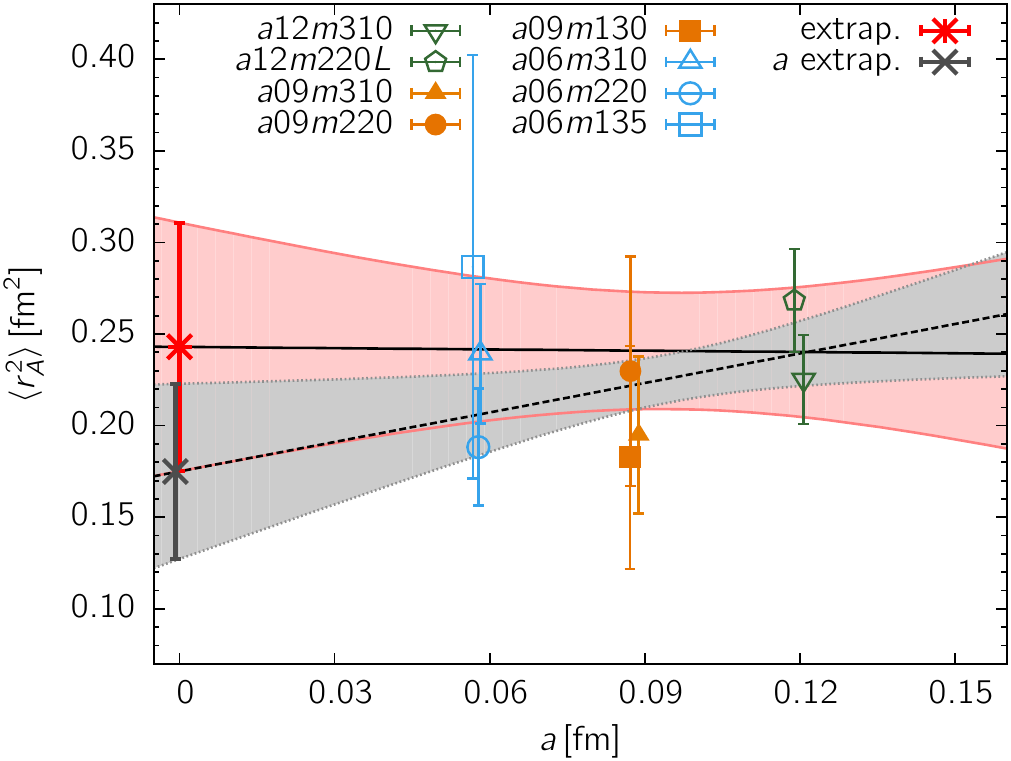}
}
{
    \includegraphics[width=0.32\linewidth]{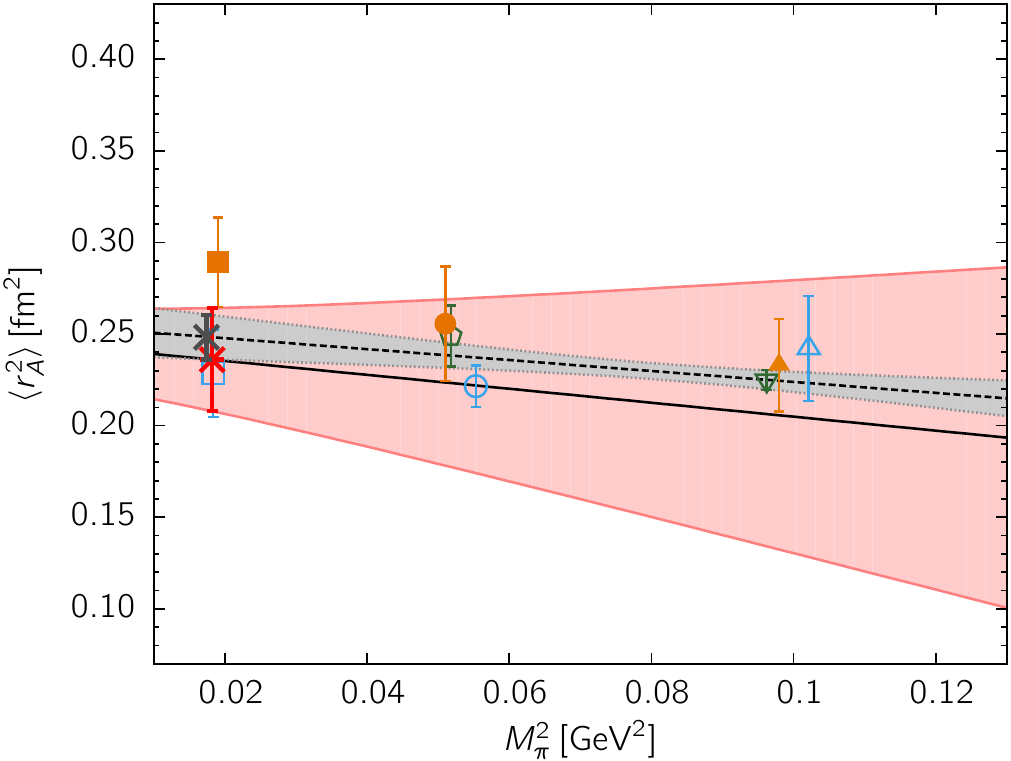}
    \includegraphics[width=0.32\linewidth]{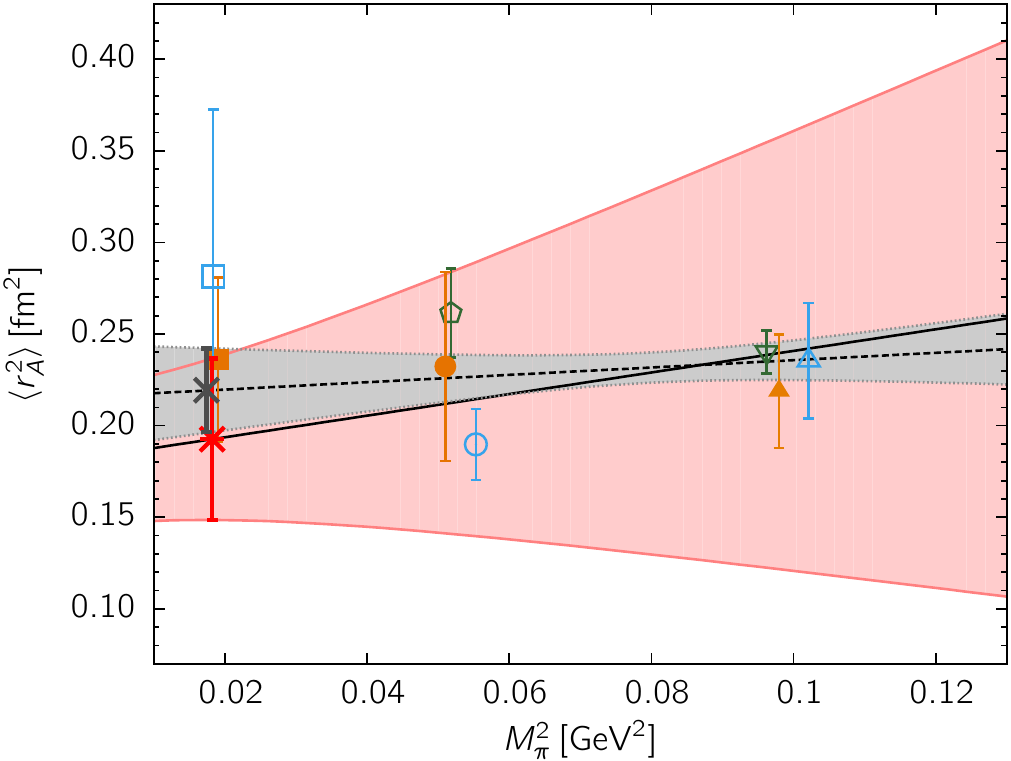}
    \includegraphics[width=0.32\linewidth]{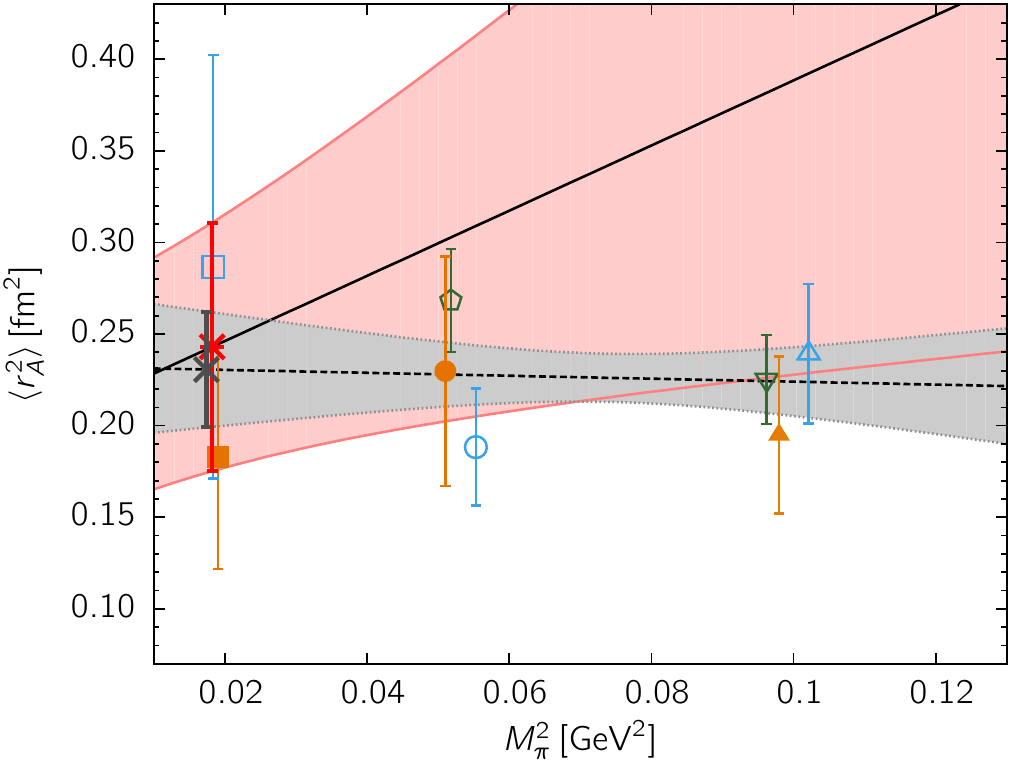}
}
{
    \includegraphics[width=0.32\linewidth]{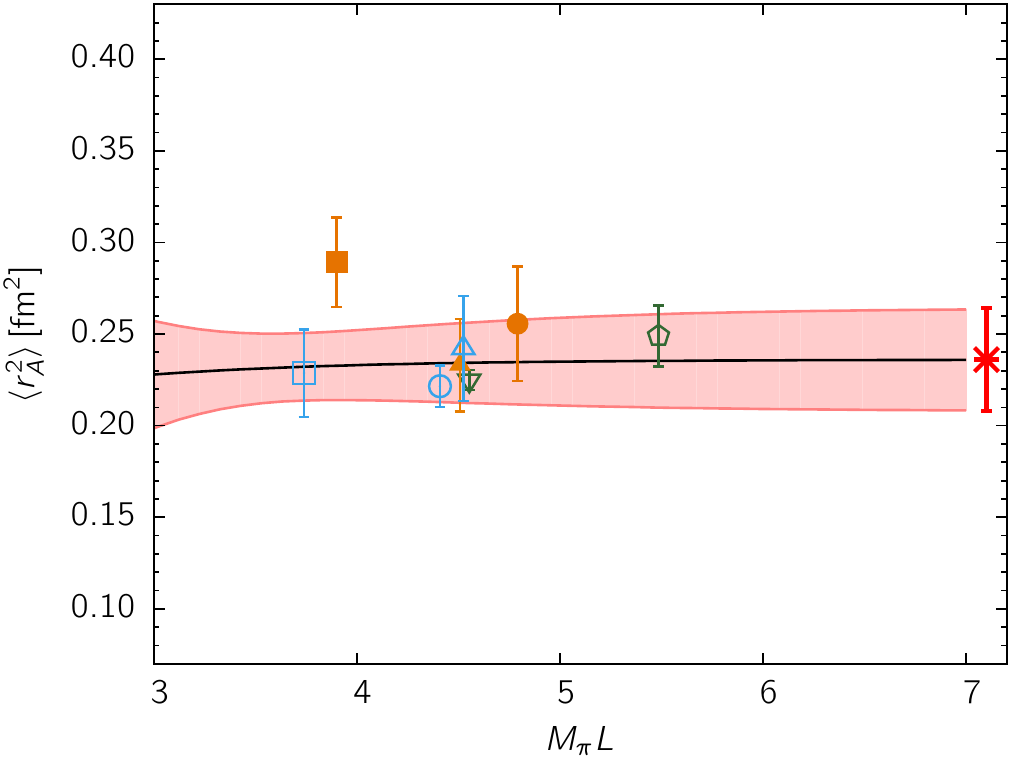}
    \includegraphics[width=0.32\linewidth]{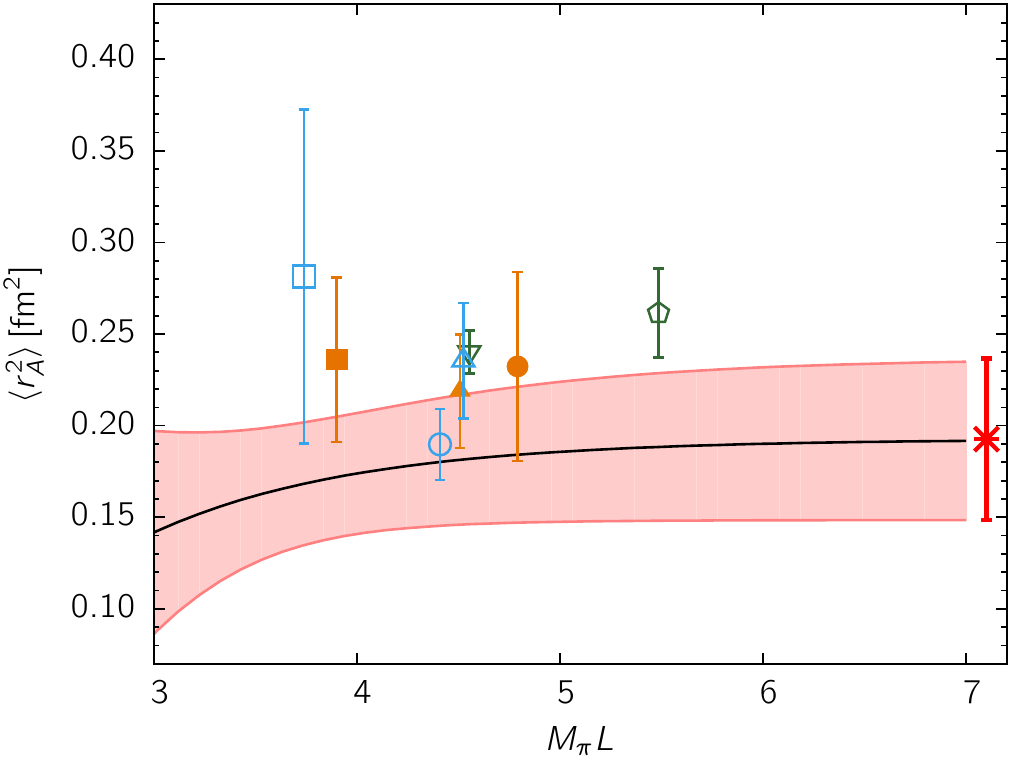}
    \includegraphics[width=0.32\linewidth]{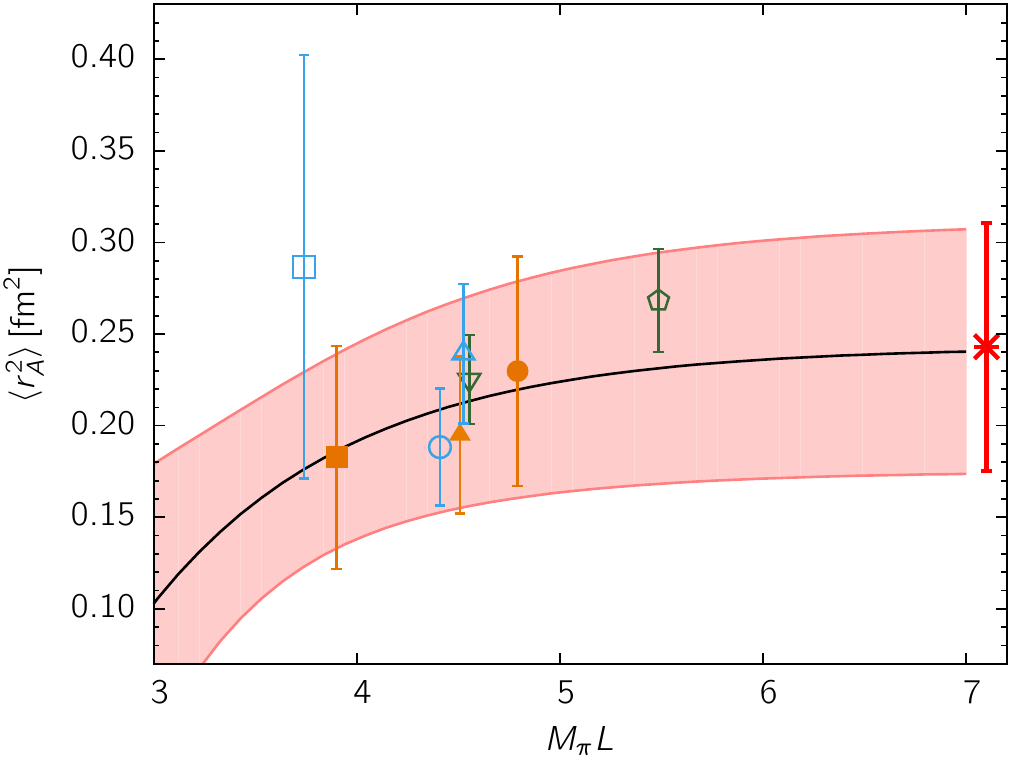}
}
\caption{The 8-point fit using the extrapolation ansatz
  Eq.~\protect\eqref{eq:extrap-rA} to the data for the axial radius
  squared $\langle r_A^2 \rangle$. Each panel shows the fit versus a
  single variable after the data have been extrapolated to the
  physical point in the other two variables.  The top row shows plots
  versus $a$, the middle versus $M_\pi^2$, and the bottom row versus
  $M_\pi L$. Each row shows $r_A$ extrapolated using the dipole ansatz
  (left); the $z^{2+4}$ ansatz (middle); and the $z^{3+4}$ ansatz
  (right). The extrapolated values are shown using the symbol red
  star.  The overlaid grey bands in the upper (middle) row are fits to
  the single variable $a$ ($M_\pi^2$), i.e., ignoring possible
  dependence on the other two variables.
  \label{fig:rA_extrap8}}
\end{figure*}
%
%%%%%%%%%%%%%%%%%%%%%%%%%%%%%%%%%%%%%%%%%%%%%%%%%%%%%%%%%%%%%%%%%%%%%
%%%  SECTION                                                      %%%
%%%%%%%%%%%%%%%%%%%%%%%%%%%%%%%%%%%%%%%%%%%%%%%%%%%%%%%%%%%%%%%%%%%%%
\section{Continuum, Chiral and Finite Volume Extrapolation of $\langle r_A^2 \rangle$}
\label{sec:rA_results}
To obtain results for the axial charge radius squared, $\langle r_A^2
\rangle$, in the limits $a \to 0$, $M_\pi \to 135$~MeV and $M_\pi L
\to \infty$, we extrapolate the data for $\langle r_A^2 \rangle$ given
in Table~\ref{tab:rA_results} and not the form factors themselves.
Since the $Q^2$ are different for each ensemble a more comprehensive
fit including dependence on $Q^2$ requires higher precision data.
Using the eight data points, including the two physical mass points,
we make a simultaneous fit in the three variables $a $, $M_\pi^2$ and
the lattice size $M_\pi L$ keeping only the lowest order correction
term in each~\cite{Bhattacharya:2016zcn}
\begin{align}
  r_{A}^2 (a,M_\pi,L) &=& c_1 + c_2a + c_3 M_\pi^2 +  c_4 M_\pi^2 {e^{-M_\pi L}} \,.
\label{eq:extrap-rA} 
\end{align}
A comparison of these ``8-point'' extrapolation fits using the
$z$-expansion and dipole ansatz data are shown in
Fig.~\ref{fig:rA_extrap8}.  We do not show the two
free parameter $z^{1+4}$ fits as the $\chi^2/{\rm d.o.f.}$ are not good. The
$z^{4+4}$ fits, with only one degree of freedom, are questionable 
outside the range of $Q^2$ values simulated, nevertheless, the data in
Table~\ref{tab:rA_results} show that they give values for $\langle r_A
\rangle$ that are consistent with the other fits. The error estimates,
on the other hand, grow steadily between the $z^{2+4}$ and the
$z^{4+4}$ cases.

\begin{table}[tbp]%7
\centering
\begin{ruledtabular}
\begin{tabular}{l|ccc|ccc}
  & \multicolumn{3}{c|}{Eq.~\protect\eqref{eq:extrap-rA}}  & \multicolumn{3}{c}{Eq.~\protect\eqref{eq:extrap-rA} with $c_4 =0$} \\
          & $\langle r_A^2 \rangle$ & $r_A$ & $\mathcal{M}_A$
          & $\langle r_A^2 \rangle$ & $r_A$ & $\mathcal{M}_A$ \\
\hline
dipole    & 0.24(3) & 0.49(3) & 1.41(08)  & 0.23(2) & 0.48(2) & 1.42(06) \\
$z^{2+4}$ & 0.19(4) & 0.44(5) & 1.56(18)  & 0.17(3) & 0.42(4) & 1.65(16) \\
$z^{3+4}$ & 0.24(7) & 0.49(7) & 1.39(19)  & 0.18(5) & 0.43(6) & 1.60(23) \\
\end{tabular}
\end{ruledtabular}
\caption{Results for $\langle r_A^2 \rangle$ in units of fm${}^2$
  after extrapolation to $a \to 0$, $M_\pi=135$~MeV, and $M_\pi L \to
  \infty$ using Eq.~\protect\eqref{eq:extrap-rA}. We also give the
  corresponding $r_A$ in units of fm and $M_A$ in
  units of GeV. The last three columns show results obtained by neglecting 
the finite volume correction term, i.e.,  $c_4=0$. }
  \label{tab:final_rA}
\end{table}

%%%%%%%%%%%%%%%%%%%%%%%%%%%%%%%%%%%%%%%%%%%%%%%%%%%%%%%%%%%%%%%%%%%%%%%%
%%%% dependence on momentum form and fit ansatz
%%%%%%%%%%%%%%%%%%%%%%%%%%%%%%%%%%%%%%%%%%%%%%%%%%%%%%%%%%%%%%%%%%%%%%%%

%
\begin{figure*}[tbp]%9
{
    \includegraphics[width=0.32\linewidth]{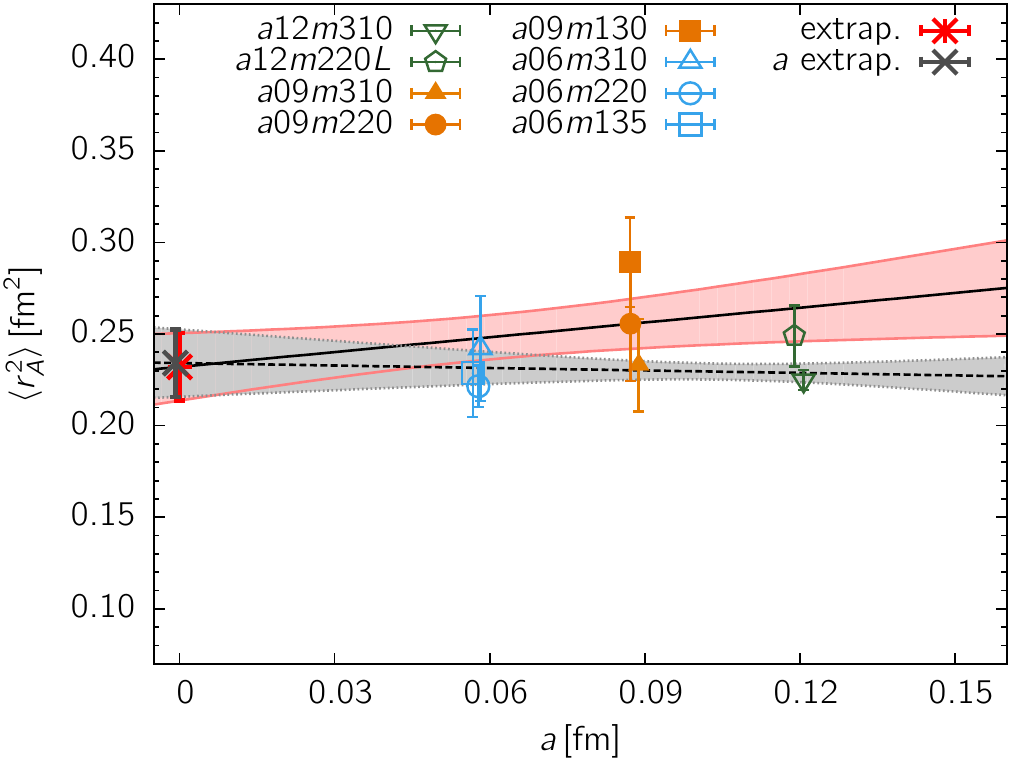}
    \includegraphics[width=0.32\linewidth]{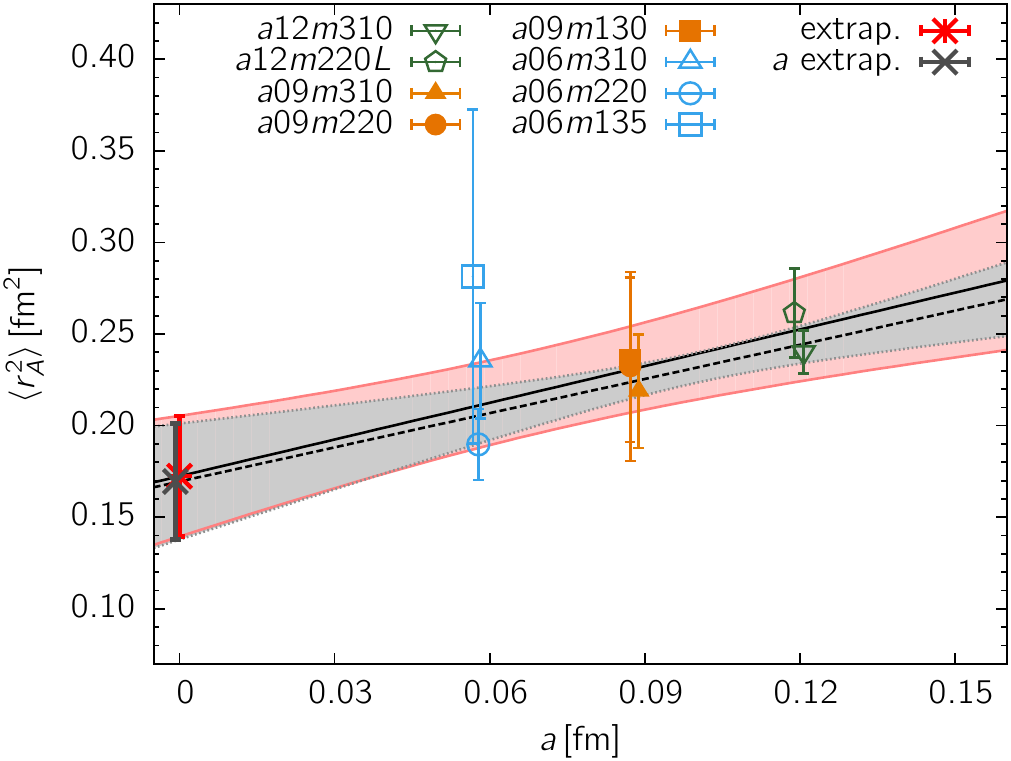}
    \includegraphics[width=0.32\linewidth]{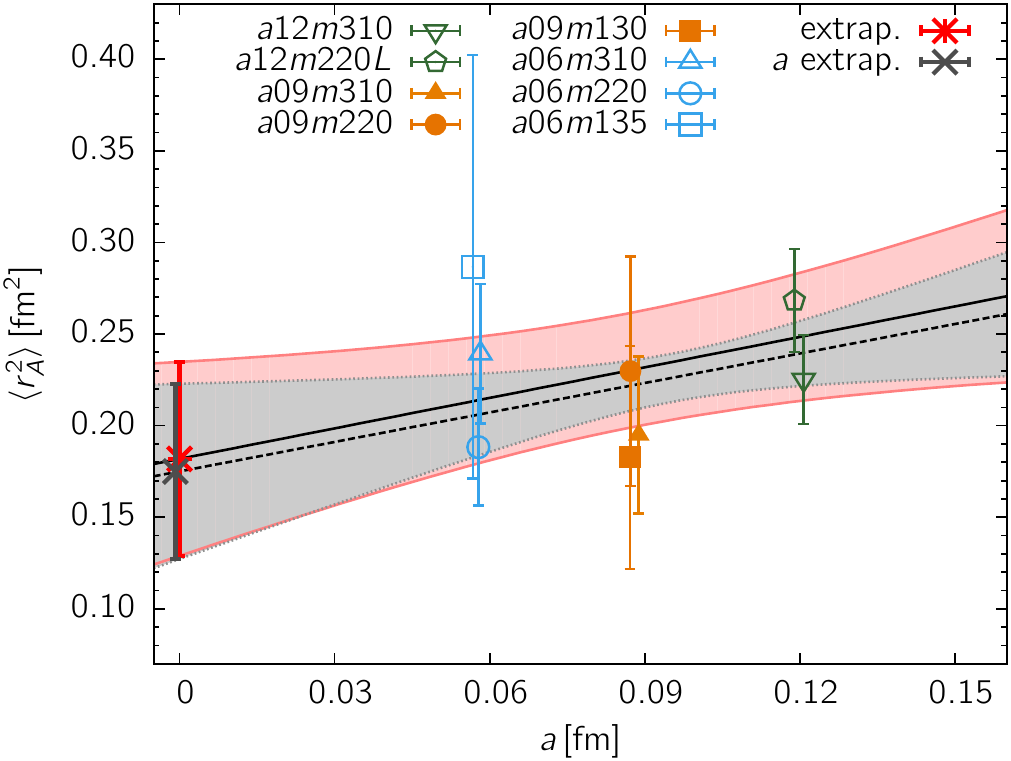}
}
{
    \includegraphics[width=0.32\linewidth]{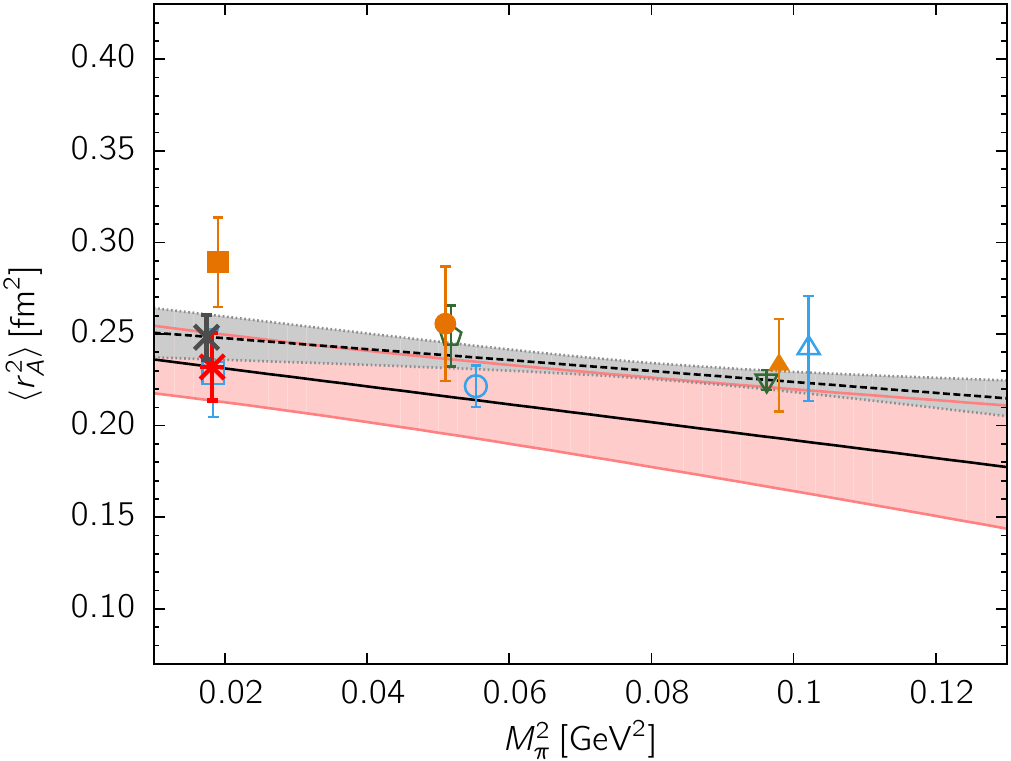}
    \includegraphics[width=0.32\linewidth]{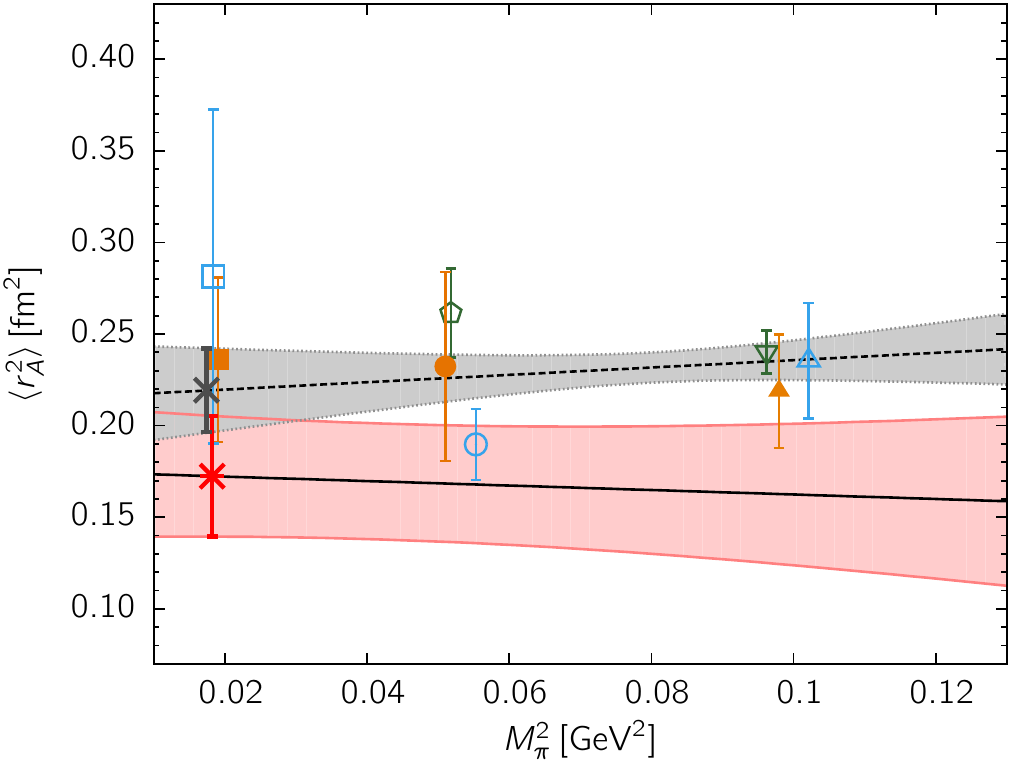}
    \includegraphics[width=0.32\linewidth]{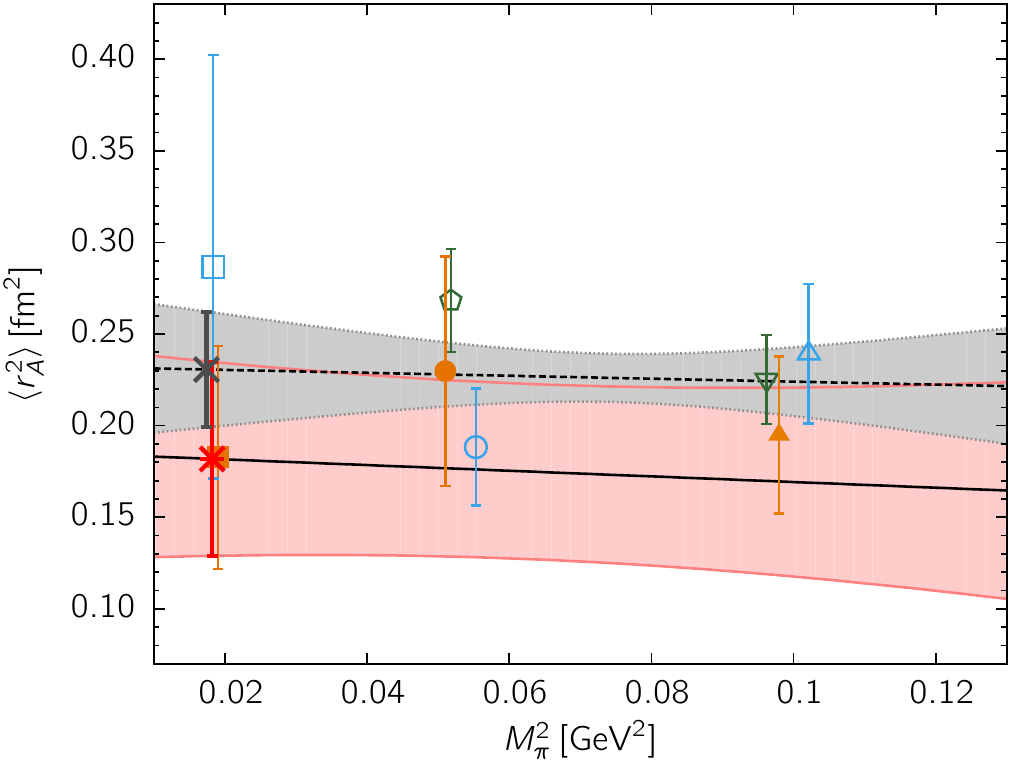}
}
\caption{The 8-point fit using Eq.~\protect\eqref{eq:extrap-rA}
  without the finite volume correction ($c_4=0$) to the data for the
  axial radius squared $\langle r_A^2 \rangle$. The overlaid grey
  bands in the upper (bottom) row are fits to the single variable $a$
  ($M_\pi^2$), i.e., ignoring possible dependence on the other 
  variable.  The rest is the same as in
  Fig~\protect\ref{fig:rA_extrap8}.
  \label{fig:rA_extrap8_nofv}}
\end{figure*}

\begin{figure*}[tbp]%10
{
    \includegraphics[width=0.485\linewidth]{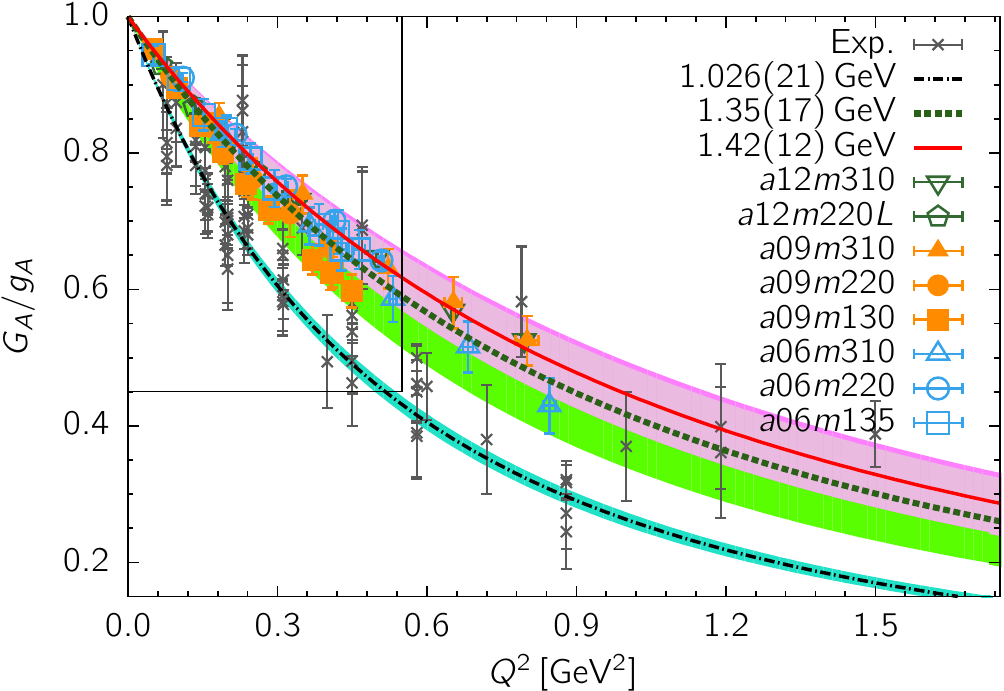}
    \includegraphics[width=0.485\linewidth]{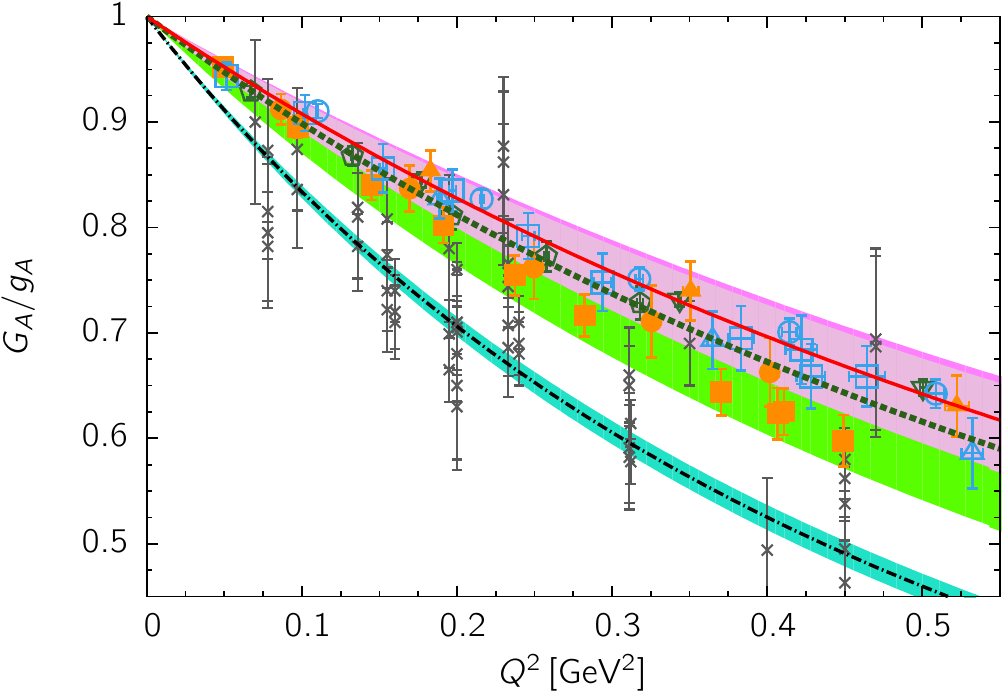}
}
\caption{(Left) The data for $G_A(Q^2)/g_A$ from the eight ensembles
  is plotted versus $Q^2$ (GeV${}^2$). We also show the dipole fit
  with the phenomenological estimates of the axial mass, ${\cal
    M}_A=1.026(21)$~GeV~\protect\cite{Bernard:2001rs} (turquoise
  band), the miniBooNE value ${\cal M}_A=1.35(17)$~GeV (green band),
  and our combined estimate ${\cal M}_A=1.42(12)$~GeV (magenta band)
  corresponding to $r_A|_{\rm dipole} = 0.49(3)$ given in
  Eq.~\protect\eqref{eq:rA_final}. The experimental data, reproduced
  from Ref.~\protect\cite{Bernard:2001rs}, were provided by Ulf
  Meissner. (Right) A magnified view of the data and the three dipole
  fits in the region $Q^2 < 0.5$~GeV${}^2$.
  \label{fig:dipole}}
\end{figure*}

The variation versus $a$, $M_\pi$ or $M_\pi L$ for the results from
the dipole, $z^{2+4}$ and $z^{3+4}$ fits are shown in
Fig.~\ref{fig:rA_extrap8}. The least well-determined coefficient is
the finite volume correction term, $c_4$ in Eq.~\eqref{eq:extrap-rA},
which is consistent with zero.  We, therefore, show the extrapolation
with $c_4=0$ in Fig.~\ref{fig:rA_extrap8_nofv}.  The results of fits,
with and without the $c_4$, are summarized in
Table~\ref{tab:final_rA}, and show that neglecting the finite volume
correction term $c_4$ does not significantly change the results, but
on comparing Figs.~\ref{fig:rA_extrap8} and~\ref{fig:rA_extrap8_nofv}
we find that the uncertainty versus $M_\pi^2$ is reduced on neglecting
$c_4$.  Overall, the results of the simultaneous fits to data obtained
using the three ansatz are consistent.  In Figs.~\ref{fig:rA_extrap8}
and \ref{fig:rA_extrap8_nofv}, we also show fits versus a single
variable ($a$ or $M_\pi^2$) as a grey band. Given the weak dependence
on $a$, $M_\pi$ or $M_\pi L$, they give estimates that are consistent
with results of the simultaneous fits but with smaller uncertainty.

Our final estimates, using the data summarized in
Table~\ref{tab:final_rA} for the case $c_4 \neq 0$, are
\begin{eqnarray}
r_A|_{\rm dipole}         &=&  0.49(3)\ {\rm fm}     \,, \nonumber \\
r_A|_{z-{\rm expansion}}  &=&  0.46(6)\ {\rm fm}     \,, \nonumber \\
r_A|_{\rm combined}       &=&  0.48(4)\ {\rm fm}     \,, \nonumber \\
{\cal M}_A|_{\rm dipole}                  &=&  1.39(9)\ {\rm GeV}    \,, \nonumber \\
{\cal M}_A|_{z-{\rm expansion}}           &=&  1.48(19)\ {\rm GeV}   \,,  \nonumber \\
{\cal M}_A|_{\rm combined}                &=&  1.42(12)\ {\rm GeV}   \,.
\label{eq:rA_final}
\end{eqnarray} 
The second two estimates are obtained by performing an average 
using the prescription for optimal correlation 
given in Ref.~\cite{Schmelling:1994pz}. For the $z$-expansion data, we have averaged the $z^{2+4}$
and the $z^{3+4}$ estimates with the lattice size correction term,
$c_4$, included.  The $r_A|_{\rm combined}$ result is
then obtained by averaging this $z$-expansion estimate with the dipole
result.  As remarked previously, the dipole ansatz fits our data
remarkably well and the final result is close to it.

In Fig.~\ref{fig:dipole}, we plot the data for $G_A(Q^2)$ from all
eight ensembles and compare them against a dipole fit using two
different estimates for the axial mass: the phenomenological value
${\cal M}_A = 1.026(17)$~GeV obtained from the combined neutrino
scattering and electroproduction data~\cite{Bernard:2001rs}, and the
value $1.35(17) $ used by the miniBooNE Collaboration to fit their
[anti-]neutrino cross-section data~\cite{AguilarArevalo:2010zc}.  We
also reproduce the data in Ref.~\cite{Bernard:2001rs} (provided by
Ulf Meissner) that was used to obtain the estimate ${\cal M}_A =
1.026(17)$~GeV. It is clear that the lattice data for $G_A(Q^2)$ show
little variation with the lattice spacing or the pion mass, and prefer
the larger values of ${\cal M}_A$ as shown in
Table~\ref{tab:rA_results}. The MiniBooNE value ${\cal M}_A=1.35(17)$
covers the spread in the lattice data, and our result ${\cal M}_A =
1.42(12)$~GeV is consistent with it. However, the bands showing our
and MiniBooNE results lie above most of the earlier experimental data
for the form factor, and the corresponding values of ${\cal M}_A$ are
larger than the phenomenological value, given in
Eq.~\eqref{eq:rA_expt}, extracted from the experimental data.

Two recent lattice QCD calculations give 
\begin{eqnarray}
{\cal M}_A|_{\rm dipole}                  &=&  1.32(7)\ {\rm GeV}    \ \qquad ({\rm ETMC}) \,, \nonumber \\
{\cal M}_A|_{z-{\rm expansion}}           &=&  1.14(15)\ {\rm GeV}   \qquad ({\rm Mainz})\,,  
\label{eq:rA_others}
\end{eqnarray}
where the first number is from the ETMC
collaboration~\cite{Alexandrou:2017hac} who use a dipole fit and
analyze a single $N_f=2$ twisted mass ensemble with $M_\pi \approx
130$~MeV and $a=0.093$~fm. The second number is from the Mainz
collaboration~\cite{Capitani:2017qpc} who use the $z$-{\rm expansion}
method on $N_f=2$ ensembles generated by the CLS
collaboration. Ensemble by ensemble, their data, which in all but one
case have been obtained with $M_\pi > 260$~MeV, are consistent with
what we find. The difference in the final results is a consequence of
their final extrapolation in $M_\pi \to 135$~MeV.  Our data at $M_\pi
\approx 220$ and $\approx 135$~MeV, and that by the ETMC collaboration
at $M_\pi \approx 130$~MeV, do not support the large increase in $r_A$
from their data to their value after extrapolation in $M_\pi$.

% edit [ypj] r_A^2, M_A columns are added. r_A and M_A are obtained from the extrapolated value of r_A^2.
%------------------------------------------------------------
% take z-expansion 2+4 and 3+4 results of r^2, then
% weighted avg. r_A^2             0.21(5)
% weighted avg. r_A               0.46(6)
% weighted avg. M_A               1.50(19)
%------------------------------------------------------------
% take dipole, z-expansion 2+4 and 3+4 results of r^2, then
% Use Schmelling for f=1 and optimal
%weighted avg. r_A^2             0.23(4)
%weighted avg. r_A               0.48(4)
%weighted avg. M_A               1.42(12)
%------------------------------------------------------------

\begin{figure}[tbp]%11
{
    \includegraphics[width=0.97\linewidth]{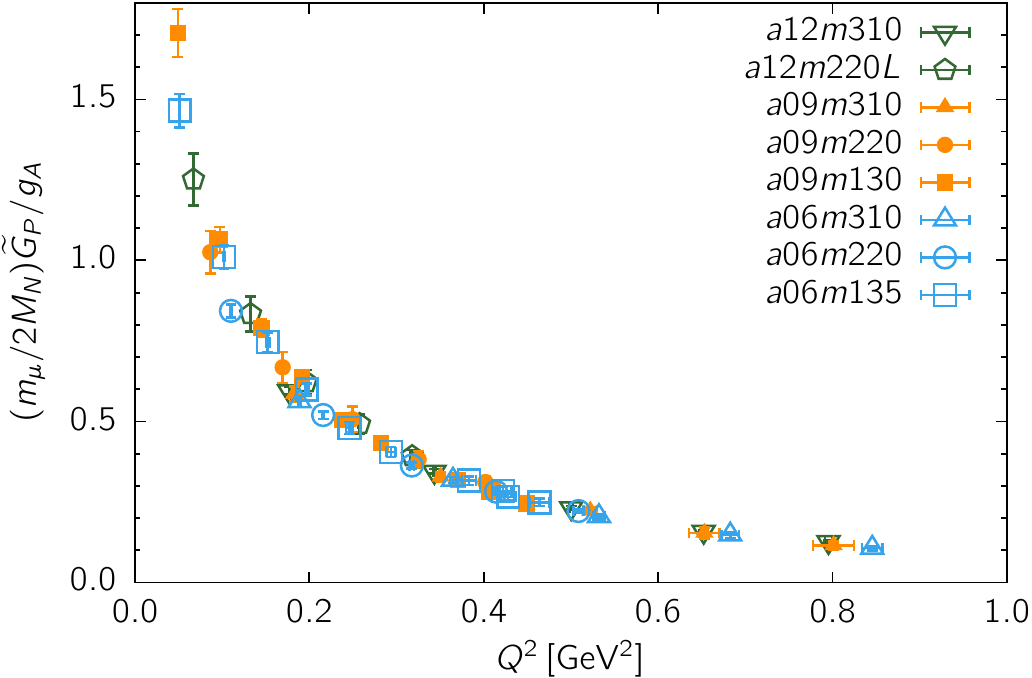}
}
\caption{The data for $(m_\mu/2 M_N) G_P (Q^2)/g_A$ from the eight
  ensembles is plotted versus $Q^2$ in units of GeV${}^2$. They show 
  little dependence on the lattice spacing $a$ or the pion mass $M_\pi$. 
  \label{fig:gPdata}}
\end{figure}
%

%%%%%%%%%%%%%%%%%%%%%%%%%%%%%%%%%%%%
\section{Analysis of the induced pseudoscalar form factor ${\tilde G}_P(Q^2)$}
\label{sec:GP}
%%%%%%%%%%%%%%%%%%%%%%%%%%%%%%%%%%%%

\begin{figure*}[tbp]%12
\centering
\subfigure{
\includegraphics[width=0.47\linewidth]{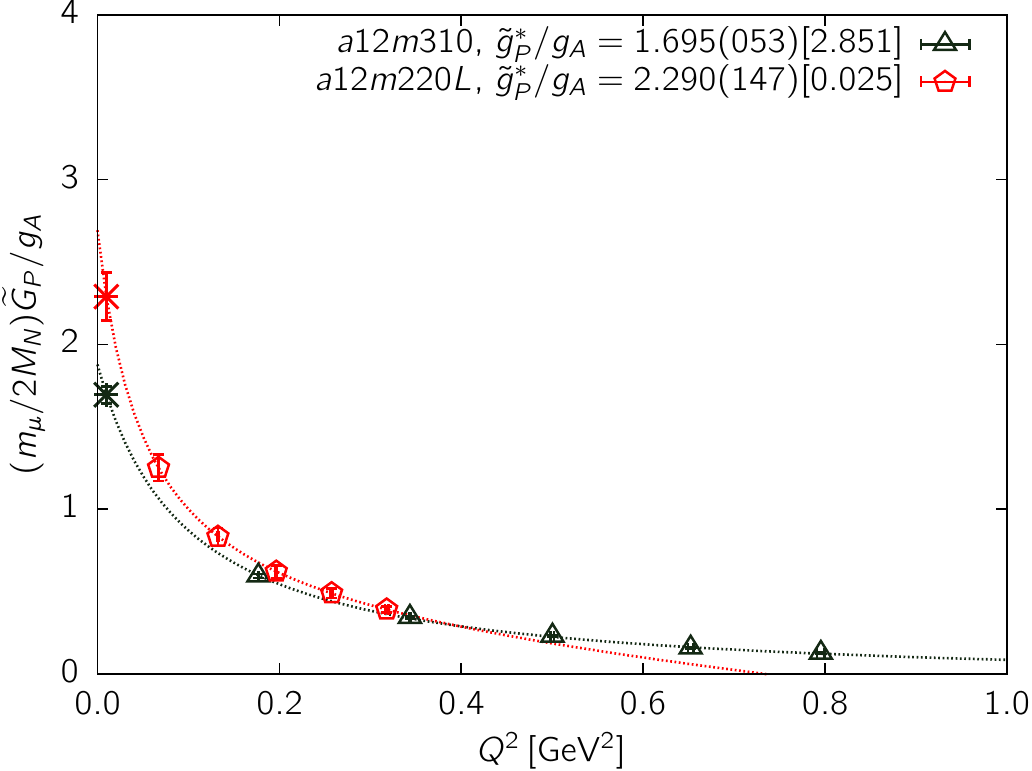} 
\includegraphics[width=0.47\linewidth]{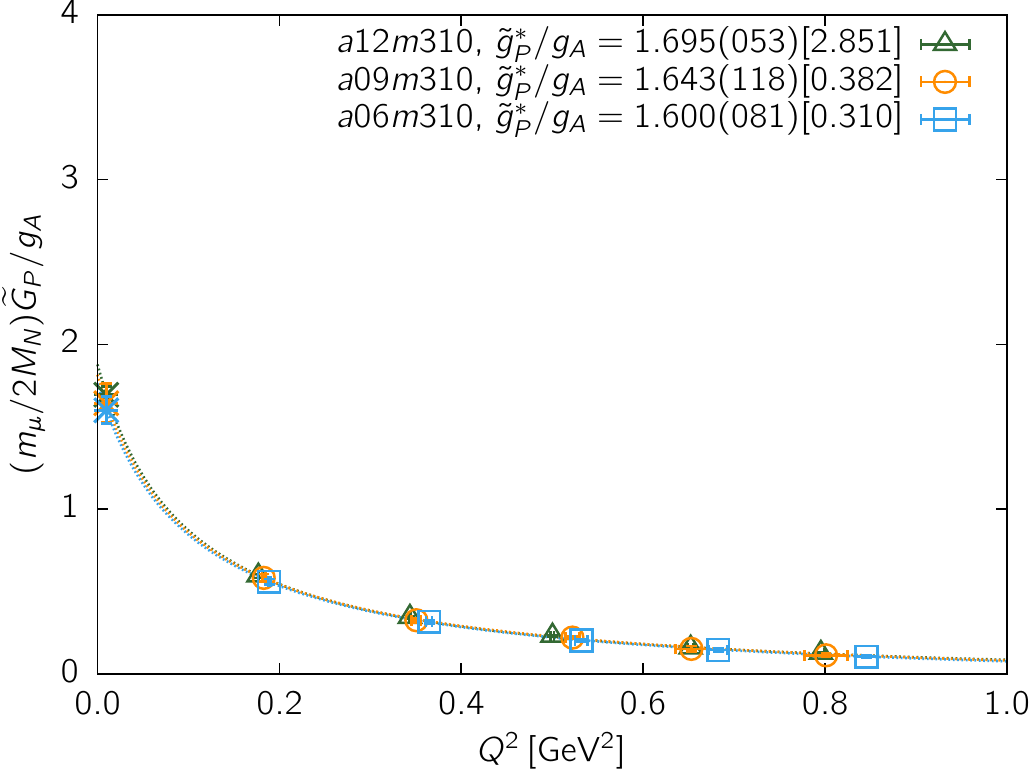} 
}
\subfigure{
\includegraphics[width=0.47\linewidth]{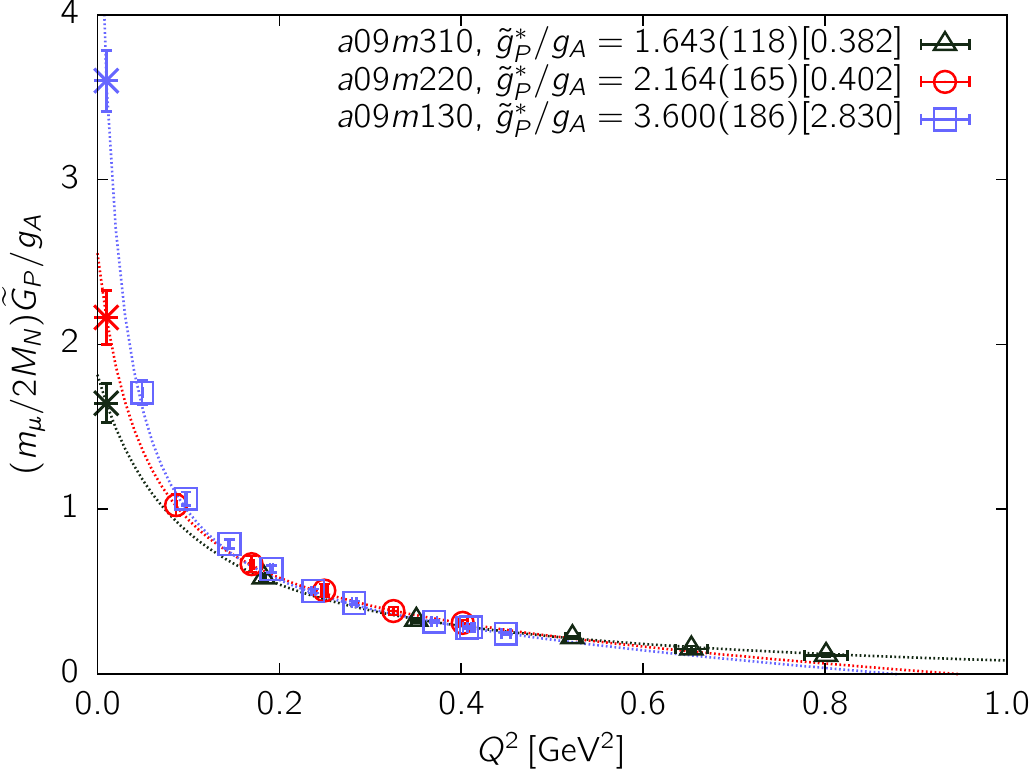} 
\includegraphics[width=0.47\linewidth]{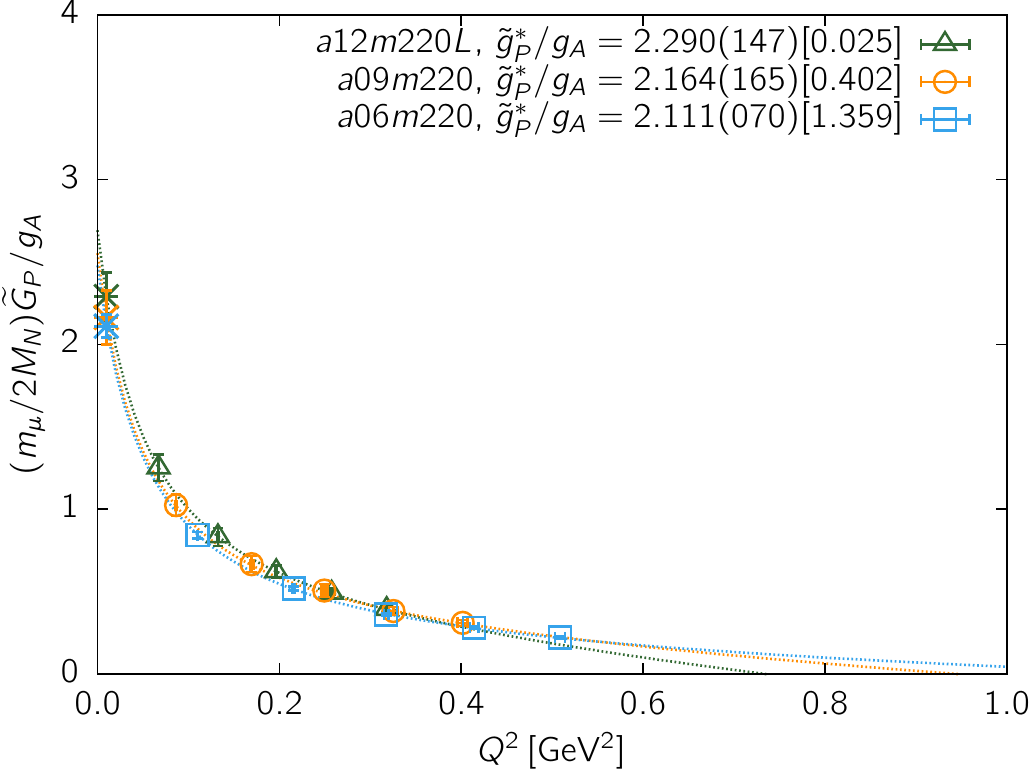} 
}
\subfigure{
\includegraphics[width=0.47\linewidth]{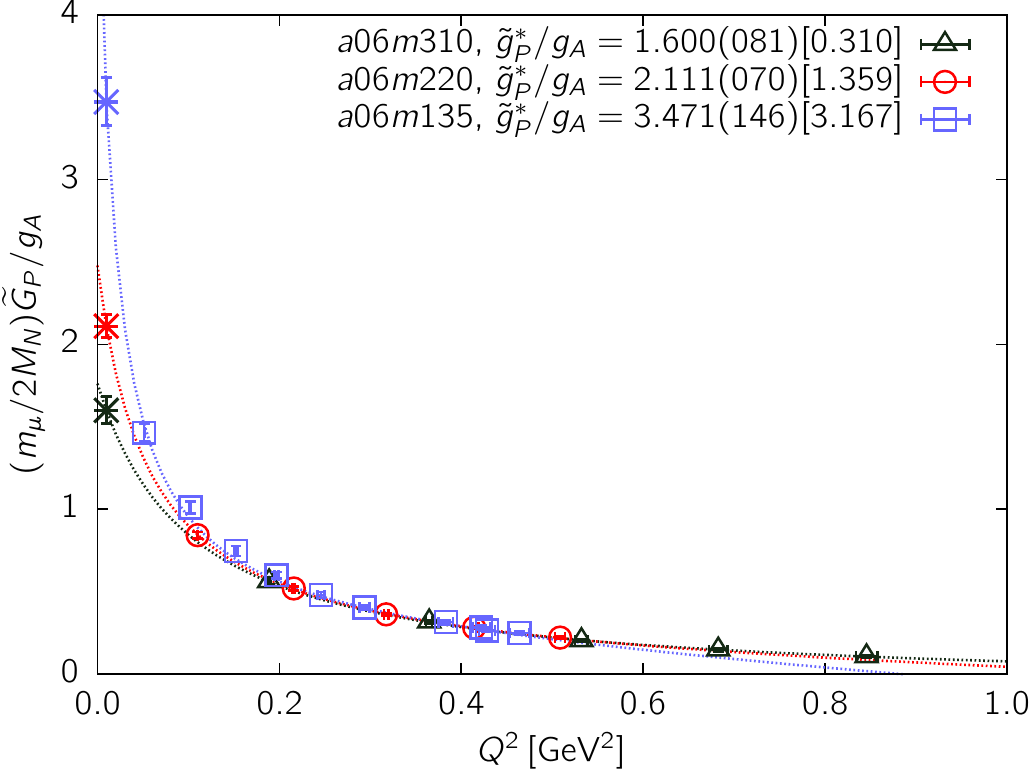} 
\includegraphics[width=0.47\linewidth]{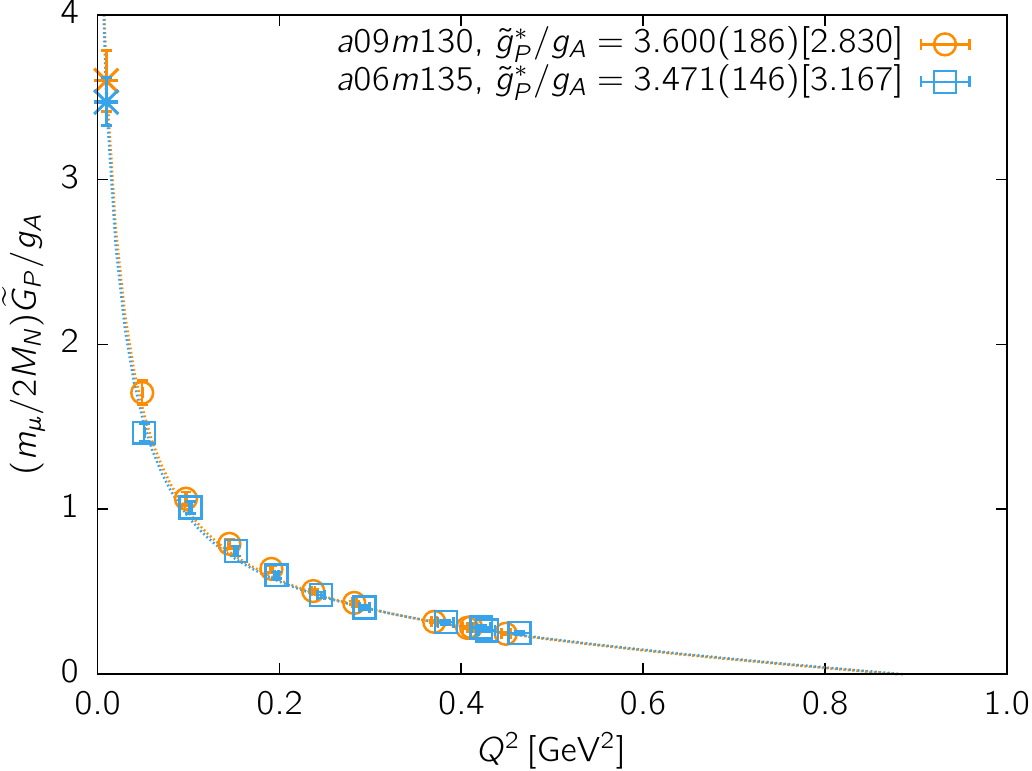} 
}
\caption{The data for the induced pseudoscalar form factor
  $(m_\mu/2M_N) {\tilde G}_P(Q^2)/g_A$ versus $Q^2$ in units of GeV${}^2$.  The left column
  highlights the dependence on $M_\pi^2$ for fixed $a$. The top panel
  is for the $a\approx 0.12$~fm, the middle for the $a\approx 0.09$~fm,
  and the bottom for the $a\approx 0.06$~fm ensembles.  The right
  column highlights the dependence on $a$ for fixed $M_\pi$. The top
  panel is for the $M_\pi \approx 310$~MeV, the middle for the $M_\pi
  \approx 220$~MeV, and the bottom for the $M_\pi \approx 130$~MeV
  ensembles.  The fits are made using Eq.~\protect\eqref{eq:PPDfit}
  with lattice estimates for the axial charge $g_A$ and nucleon mass
  $M_N$. The muon mass is $m_\mu =0.10566$~GeV. The number within the square
  brackets in the labels is the $\chi^2/{\rm d.o.f.}$ of the fit.}
\label{fig:gP-vsM-vsa}
\end{figure*}

The data for the normalized induced pseudoscalar form factor
  $(m_\mu/2M_N){\tilde G}_P(Q^2)/g_A$ versus $Q^2$ from the eight
  ensembles is summarized in Figs.~\ref{fig:gPdata}
  and~\ref{fig:gP-vsM-vsa}. Overall, the data show remarkably little
  dependence on the pion mass or the lattice spacing.

\begin{figure}[tbp]%13
{
    \includegraphics[width=0.97\linewidth]{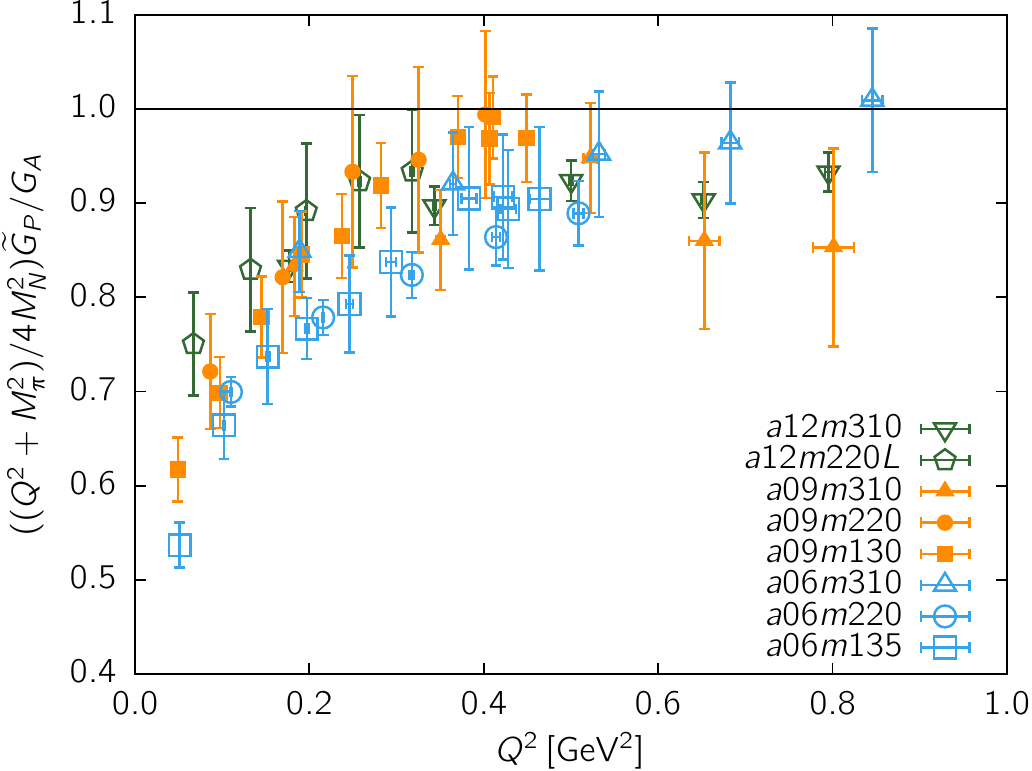}
}
\caption{Plot of the ratio $(Q^2+M_{\pi}^2) {\tilde G}_P(Q^2) /
  (4M_{p}^2 {G}_A(Q^2)) $ versus $Q^2$ for the eight
  ensembles. Validity of the pion pole-dominance hypothesis, given in
  Eq.~\protect\eqref{eq:GPpole}, requires that this ratio is unity for
  all $Q^2$. Our data show significant deviations, especially for $Q^2
  \lesssim 0.2$ GeV${}^2$. }
\label{fig:PPDH_test}
\end{figure}

The traditional starting point of the analysis of the $Q^2$ behavior
of ${\tilde G}_P(Q^2)$ data given in Table~\ref{tab:ff_GPdata} is the
pion pole-dominance ansatz given in Eq.~\eqref{eq:GPpole}. In
Fig.~\ref{fig:PPDH_test}, we show the data for $(Q^2+M_{\pi}^2)
{\tilde G}_P(Q^2) / (4M_{p}^2 {G}_A(Q^2)) $, which should be unity,
versus $Q^2$ from all eight ensembles.  We find that it tends to unity
for $Q^2 \gtrsim 0.5$ GeV${}^2$.  At low $Q^2$, however, there are
significant deviations suggesting that corrections to the pion
pole-dominance ansatz are large for $Q^2 2
\lesssim 0.2$ GeV${}^2$, precisely in the region in which it is
expected to work best.  Very similar behavior was reported in
Ref.~\cite{Bali:2014nma}.

\begin{figure*}[tbp]%14
\centering
\subfigure{
\includegraphics[width=0.47\linewidth]{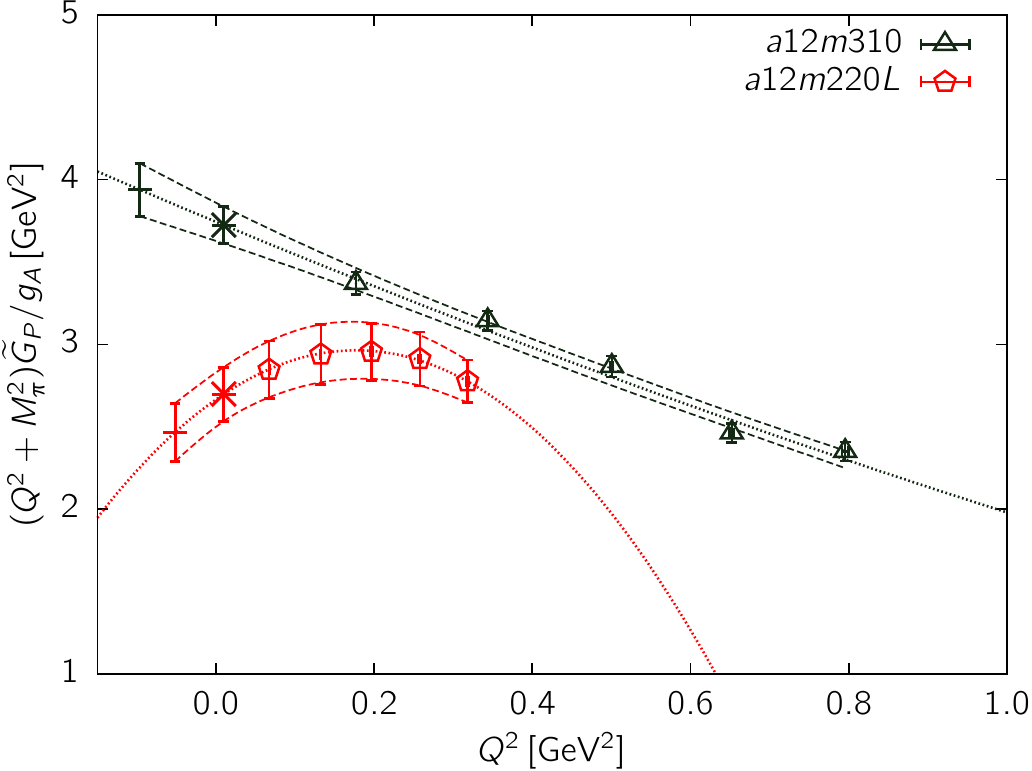}
\includegraphics[width=0.47\linewidth]{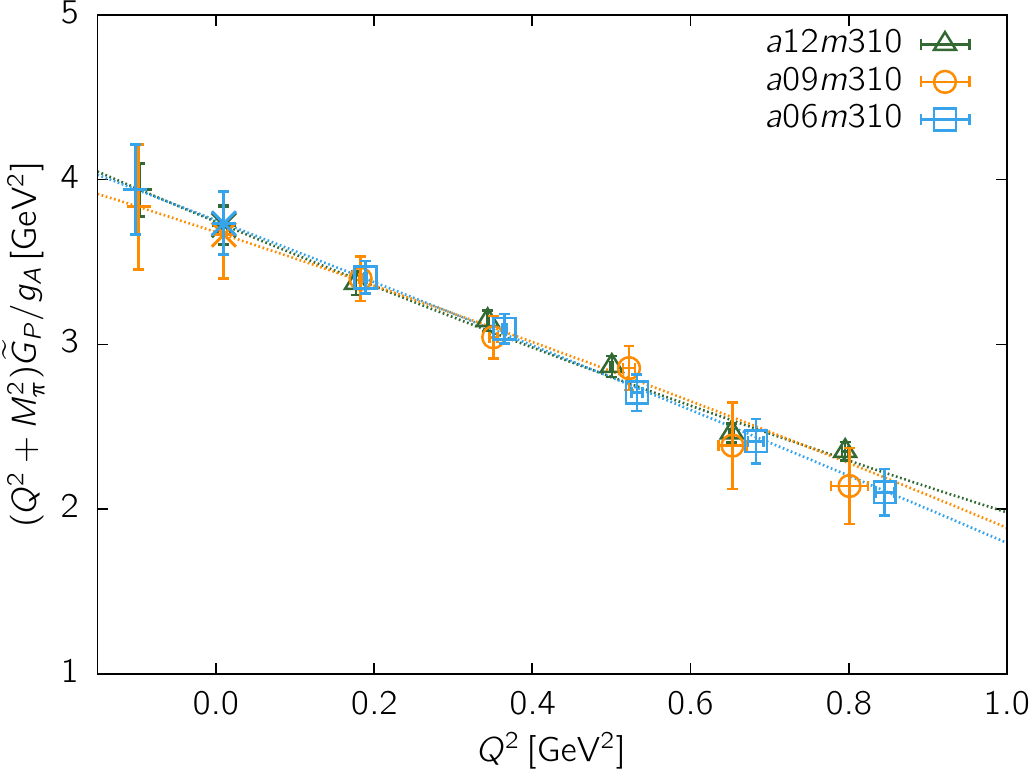}
}
\subfigure{
\includegraphics[width=0.47\linewidth]{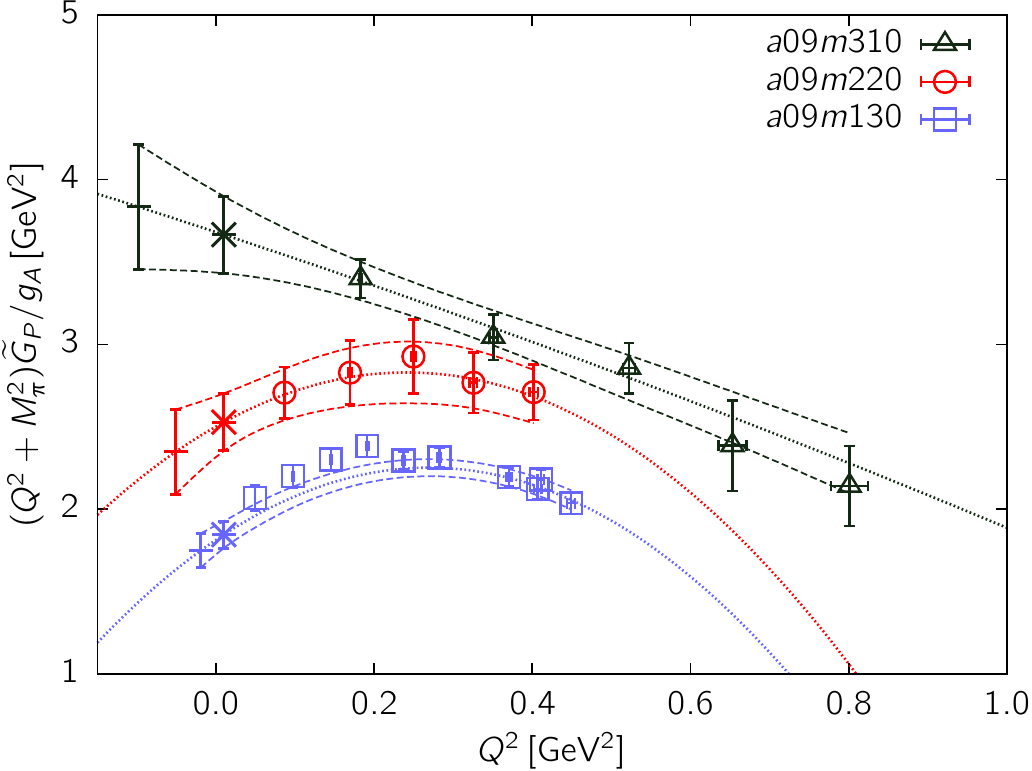}
\includegraphics[width=0.47\linewidth]{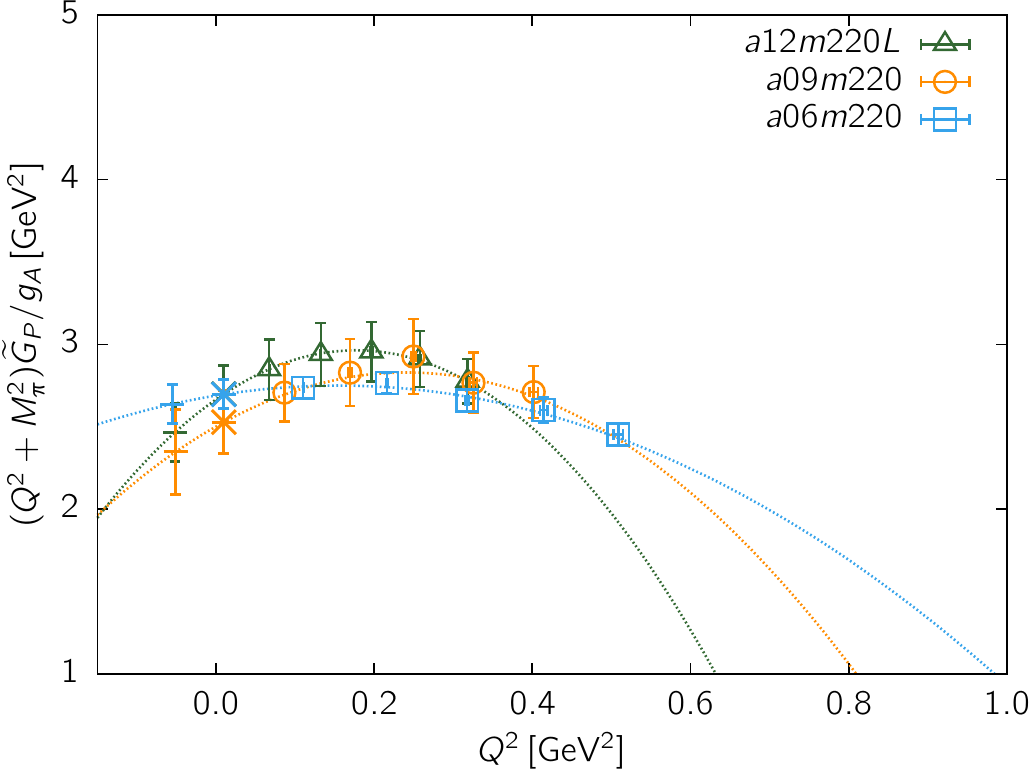}
}
\subfigure{
\includegraphics[width=0.47\linewidth]{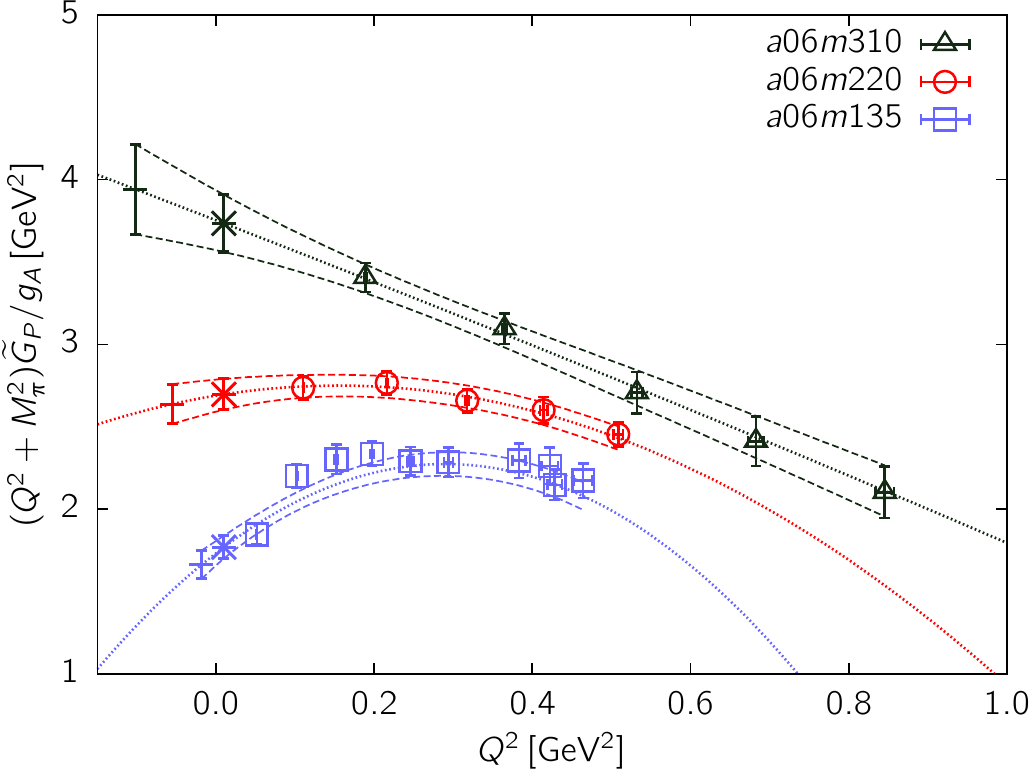}
\includegraphics[width=0.47\linewidth]{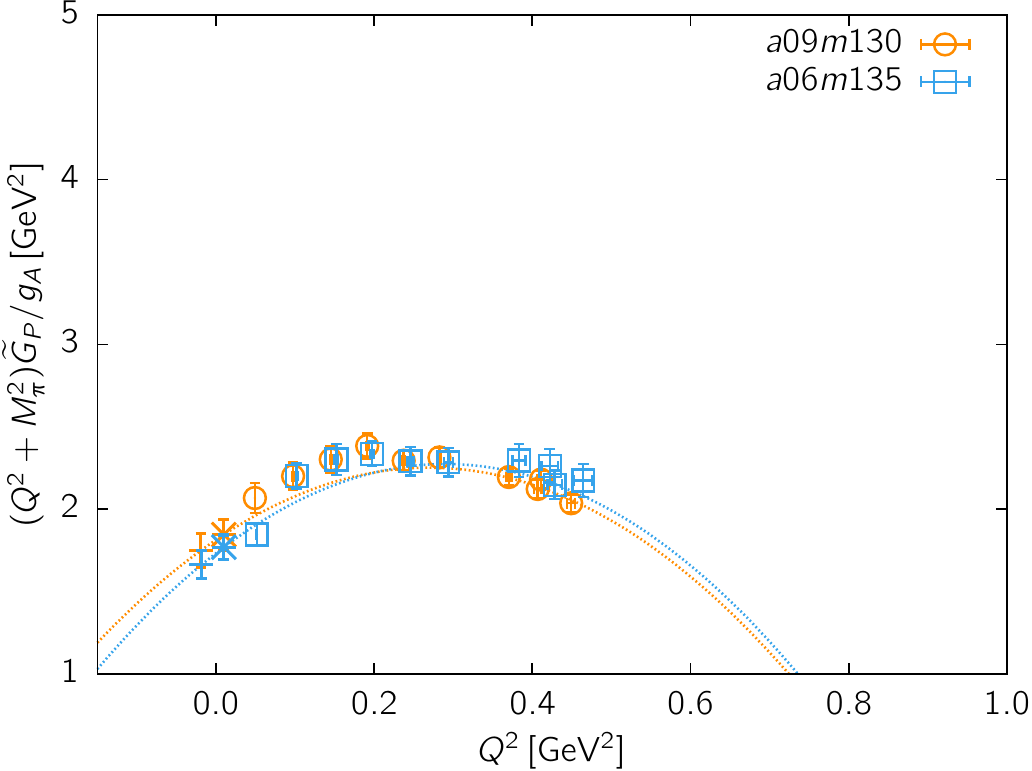}
}
\caption{The data for and the fits to the quantity $(Q^2+M_{\pi}^2) {\tilde G}_P(Q^2)/g_A$.
  The figures in the left column highlight the dependence on $M_\pi^2$
  for fixed $a$, and those in the right column highlight the
  dependence on $a$ for fixed $M_\pi$.  The fits versus $Q^2$
  (GeV${}^2$) are performed using Eq.~\protect\eqref{eq:PPDfit} with
  lattice estimates for the axial charge $g_A$ and the nucleon mass,
  $M_N$.  The muon mass is $m_\mu=0.10566$~GeV. The point
  with symbol star (plus) gives the value at $Q^2 = Q^{\ast\, 2} \equiv 0.88
  m_\mu^2$ ($Q^2 = -M_\pi^2$). We show the $1\sigma$ error band of the fits in 
  the panels on the left.}
\label{fig:gP-PD-detail}
\end{figure*}

To further evaluate the pion pole-dominance ansatz, we exhibit the
dependence of $(m_\mu/2 M_N) G_P (Q^2)/g_A$ on $M_\pi^2$ for fixed $a$
in Fig.~\ref{fig:gP-vsM-vsa} (left column), and on $a$ for fixed
$M_\pi^2$ (right column). These plots also show a fit using the
simplest small $Q^2$ expansion of
Eq.~\eqref{eq:PPDfit}~\cite{Bali:2014nma},
\begin{equation}
\frac{m_\mu}{2 M_N}  \frac{{\tilde G}_P(Q^2)}{g_A} = \frac{c_1}{M_\pi^2 + Q^2} + c_2 +c_3 Q^2 \,,
\label{eq:PPDfit}
\end{equation}
where the leading term is the pion-pole term and the polynomial
approximates the small $Q^2$ expansion of the dipole or the
$z$-expansion ansatz for $G_A$.  It is also the behavior predicted for
small $Q^2$ and $ M_\pi^2$ by the leading order chiral perturbation
theory~\cite{Bernard:2001rs}.\footnote{In our calculations, the $Q^2$
values are large, roughly 2--10 $M_\pi^2$ as can be
inferred from Table~\protect\ref{tab:ff_GAdata}.} We use lattice
estimates for the axial charge $g_A$ given in the
Table~\ref{tab:ff_GAdata}, the nucleon mass $M_N$ from
Tables~\ref{tab:multistates-twopt-mom-a12m310AMA-3}--\ref{tab:multistates-twopt-mom-a06m130AMA-3},
and the muon mass is $m_\mu=0.10566$~GeV.  The values of the fit
parameters $c_1,\ c_2$ and $c_3$, defined in Eq.~\eqref{eq:PPDfit},
are given in Table~\ref{tab:GPfit}.  Pion pole-dominance implies that
the contribution of terms proportional to $c_2$ and $c_3$ is
relatively small. The data in Table~\ref{tab:GPfit} show that both
$c_2$ and $c_3$ grow as $M_\pi$ is decreased, signaling that the pion
pole-dominance ansatz has large and growing corrections. This change
in behavior is exhibited in the Fig.~\ref{fig:gP-PD-detail}; as
$M_\pi$ decreases and contribution of the quadratic term becomes
larger.

\begin{table*}[tbp]%8
\centering
\begin{ruledtabular}
\begin{tabular}{l|cc|cc|cc}
ensemble   & $(Q^{\ast\, 2}+M_\pi^2){\tilde G}_P(Q^{\ast\, 2})/g_A$ & $g_P^\ast/g_A$   & $c_1*(2M_N/m_\mu)$ & $g_{\pi {\rm NN}}/g_A$ &$c_2*(2M_N/m_\mu)$   & $c_3*(2M_N/m_\mu)$   \\
           &                                                    & $(Q^2=Q^{\ast\, 2})$ & $[{\rm{GeV^2}}]$   & $(Q^2=-M_\pi^2)$       &                   & $[{\mathrm{GeV^{-2}}}]$ \\
\hline                                           
$a12m310 $ &    3.72(12)    &    1.70(05)      &    3.94(16)     &    8.35(38)     &   -2.02(48)        &    0.24(42)         \\
$a12m220L$ &    2.70(17)    &    2.29(15)      &    2.47(18)     &    5.86(43)     &    3.85(1.1)       &   -9.5(2.5)         \\
$a09m310 $ &    3.67(26)    &    1.64(12)      &    3.83(38)     &    8.20(83)     &   -1.56(1.2)       &   -0.21(94)         \\
$a09m220 $ &    2.53(19)    &    2.16(17)      &    2.35(26)     &    5.63(62)     &    3.01(1.9)       &   -5.7(3.3)         \\
$a09m130 $ &    1.85(10)    &    3.60(19)      &    1.75(10)     &    4.83(30)     &    3.37(43)        &   -6.04(63)         \\
$a06m310 $ &    3.74(19)    &    1.60(08)      &    3.94(27)     &    8.31(59)     &   -1.85(82)        &   -0.10(62)         \\
$a06m220 $ &    2.70(09)    &    2.11(07)      &    2.64(12)     &    6.28(30)     &    0.93(61)        &   -2.54(84)         \\
$a06m135 $ &    1.77(07)    &    3.47(15)      &    1.66(09)     &    4.56(26)     &    3.84(63)        &   -6.4(1.1)         \\
\end{tabular}
\end{ruledtabular}
\caption{Results obtained from fits to $(m_\mu/2M_N) {\tilde G}_P(Q^2) / g_A$
  using Eq.~\protect\eqref{eq:PPDfit}.  The second column gives $(Q^2
  + M_\pi^2) {\tilde G}_P(Q^{\ast\, 2})/g_A$ at
  $Q^2=Q^{\ast\, 2} =0.88m_\mu^2$~GeV${}^2$. These data are shown by the symbol star
  in Fig.~\ref{fig:gP-vsM-vsa}.  The third column gives $
  g_P^\ast/g_A$ using Eq.~\protect\eqref{eq:GPstar}.  The fit
  parameters $c_i$ are rescaled by $2M_N/m_\mu$ so that the fourth
  column gives the residue of the pole at $Q^2=-M_\pi^2$ from which
  $g_{\pi {\rm NN}}/g_A$, given in column five, is obtained by
  dividing by $4 M_N F_\pi$. Corrections to the pion pole-dominance ansatz 
  are proportional to the parameters $c_2$ and $c_3$. }
  \label{tab:GPfit}
\end{table*}

For each ensemble, the result for $g_P^\ast$, defined in
Eq.~\eqref{eq:GPstar} and obtained from the fit, is given in
Fig.~\ref{fig:gP-vsM-vsa} and in the third column of Table~\ref{tab:GPfit}. 
We find that the estimates from the
physical pion mass ensembles are about half the values obtained from
the muon capture experiment or the $\chi$PT analysis given in
Eq.~\eqref{eq:GPstar_pheno}.  It is, therefore, important to
understand how and where the analysis based on the pion pole-dominance
ansatz, Eq.~\eqref{eq:GPpole}, breaks down.

To do this, we start with the axial Ward
identity Eq.~\eqref{eq:PCAC} rewritten as
\begin{equation}
\frac{Q^2}{4 M_N^2} \frac{{\tilde G}_P(Q^2)}{G_A(Q^2)} + \frac{2 {\widehat m}}{2M_N} \frac{G_P(Q^2)}{G_A(Q^2)} = 1 \,.
\label{eq:testPCAC}
\end{equation}
This PCAC relation has to hold for each $Q^2$ and $M_\pi$ up to corrections
starting at $O(a)$ for the lattice action and operators used by us. If
the $O(a)$ improved axial current $A_\mu^{I} = Z_A(1+b_A m a)(A_\mu +
c_A a \partial_\mu P)$ is used, then Eq.~\eqref{eq:testPCAC} is
modified to
\begin{equation}
\frac{Q^2}{4 M_N^2} \frac{{\tilde G}_P^I(Q^2)}{G_A(Q^2)} + \frac{2 {\widehat m}}{2M_N} \frac{G_P(Q^2)}{G_A(Q^2)} = 1 \,.
\label{eq:testPCACI}
\end{equation}
where ${\tilde G}_P^I(Q^2) = {\tilde G}_P(Q^2) + 2 M_N a c_A G_P$.
Note that the extraction of $G_A(Q^2)$ is unchanged because the
improvement term contributes only to ${\tilde G}_P$.  Also, there is
no $O(a)$ correction to the pseudoscalar
density~\cite{Bhattacharya:2005rb}.  Typical estimates of the
improvement coefficient are $c_A \lesssim
-0.05$~\cite{Bulava:2015bxa}, and based on the values given there, we
take these to be $c_A=-0.05$, $-0.04$ and $-0.03$ for the $a=0.12$,
$0.09$ and 0.06~fm ensembles, respectively, for the purpose of the
test.  In the following discussion of tests of the PCAC relation, we
also ignore the differences in the mass dependent corrections $(1 +
b_i ma)$ to the renormalization constants $Z_i$ ($i \in m, A, P$) as
these are small ($ma < 0.01$) compared to the effects under
consideration.

The PCAC relation reduces to the pion pole-dominance ansatz given in
Eq.~\eqref{eq:GPpole} provided the relation
\begin{equation}
2 {\widehat m}\, G_P(Q^2) = (M_\pi^2 / 2 M_N) \, {\tilde G}_P^{[I]}(Q^2)
\label{eq:PPDcondition}
\end{equation}
also holds up to corrections starting at $O(a)$. Validation of both
the PCAC relation and the pion pole-dominance ansatz implies that only
one of the three form factors is independent.

\begin{figure*}[tbp]%15
\centering
\subfigure{
    \includegraphics[height=2.0in,trim={0.0cm 0.03cm 0 0},clip]{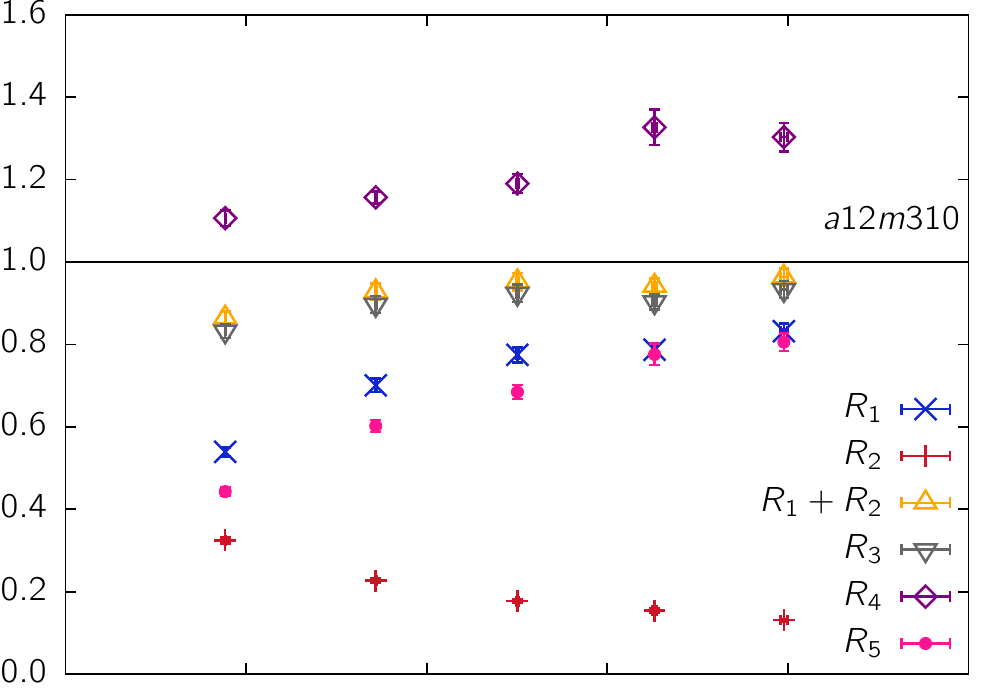} 
    \includegraphics[height=2.0in,trim={0.0cm 0.03cm 0 0},clip]{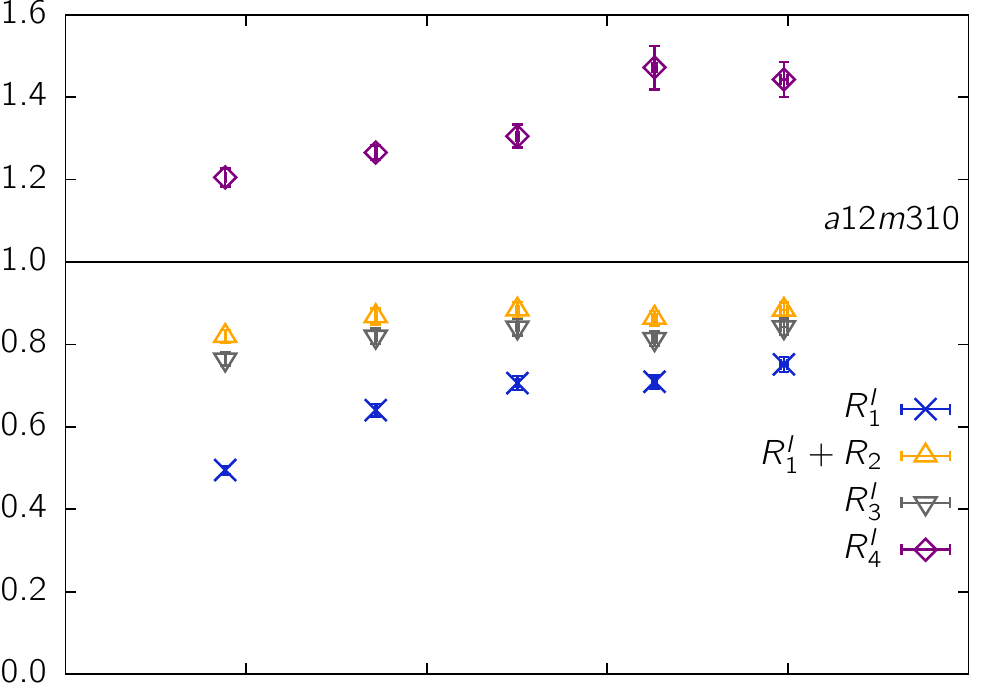}
}
\subfigure{
    \includegraphics[height=2.0in,trim={0.0cm 0.03cm 0 0},clip]{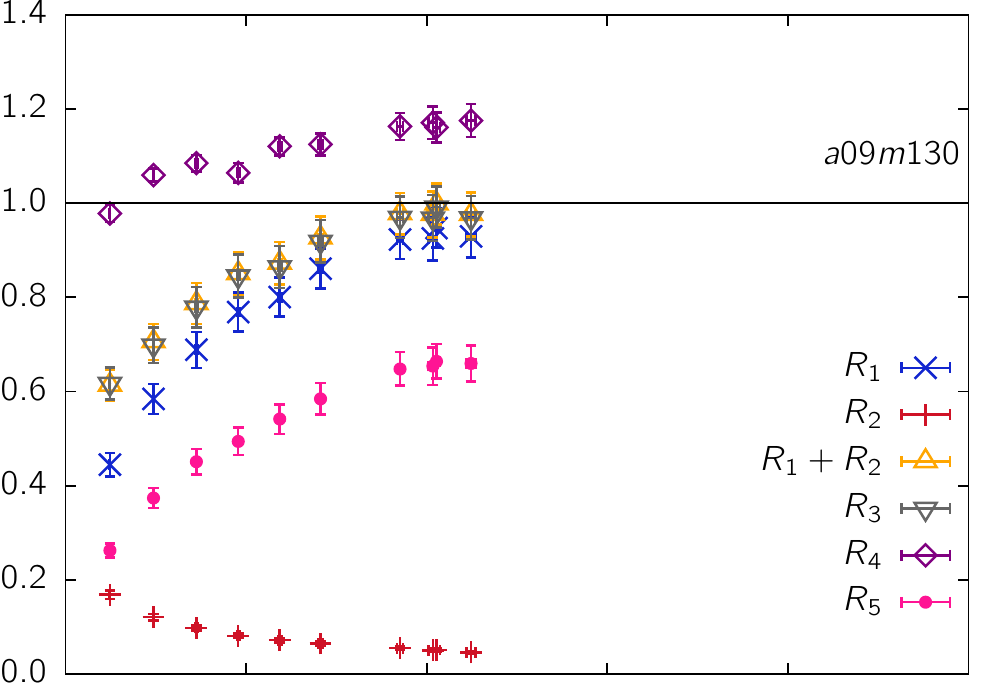} 
    \includegraphics[height=2.0in,trim={0.0cm 0.03cm 0 0},clip]{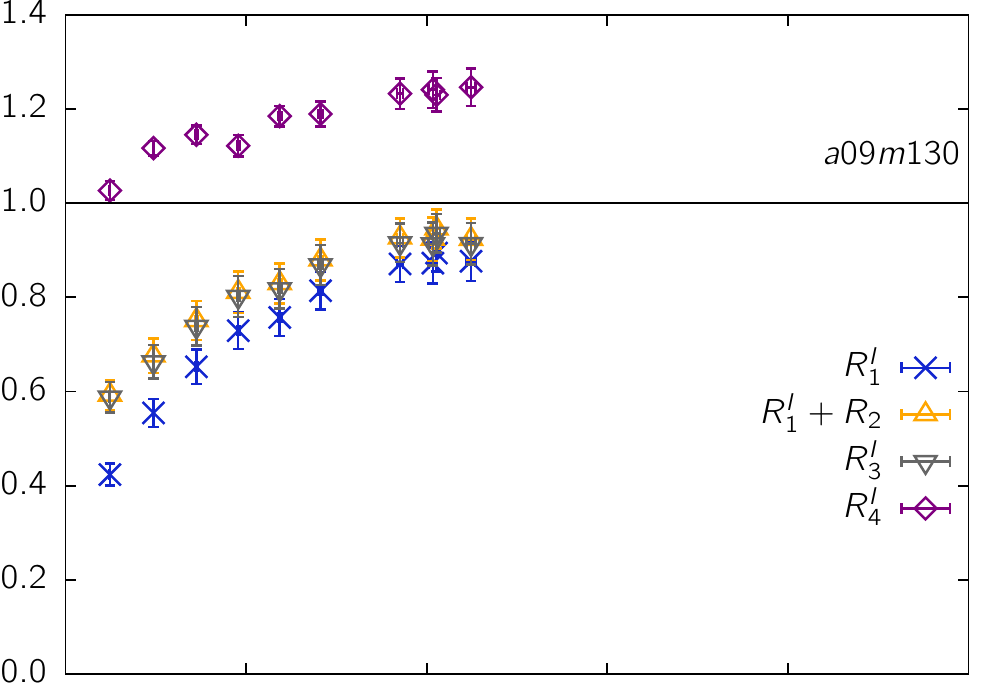}
}
\subfigure{
    \includegraphics[height=2.0in,trim={0.0cm 0.03cm 0 0},clip]{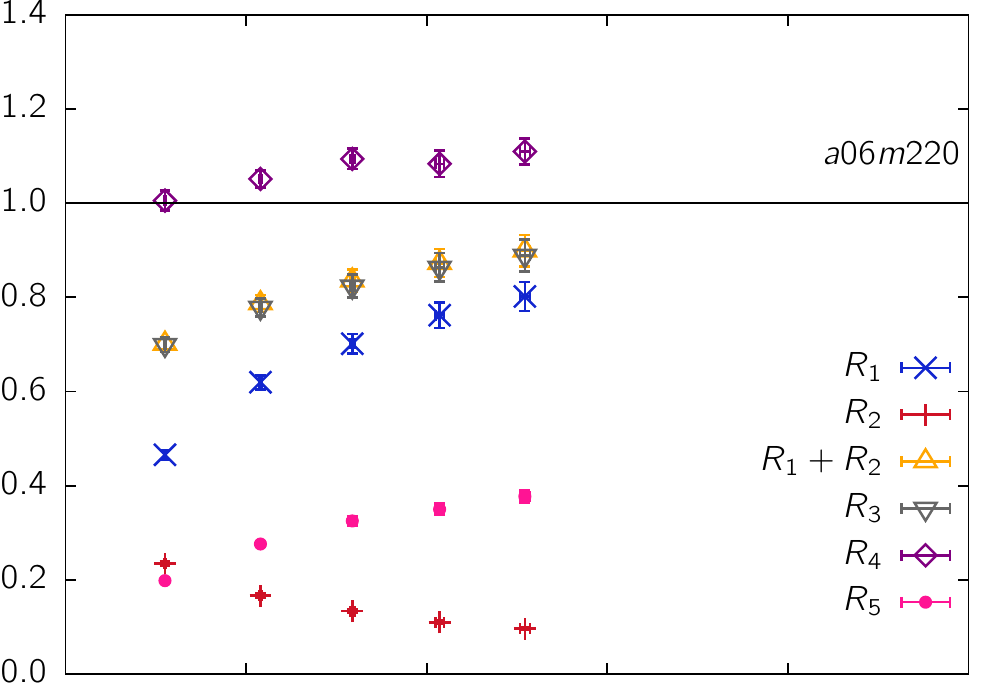}
    \includegraphics[height=2.0in,trim={0.0cm 0.03cm 0 0},clip]{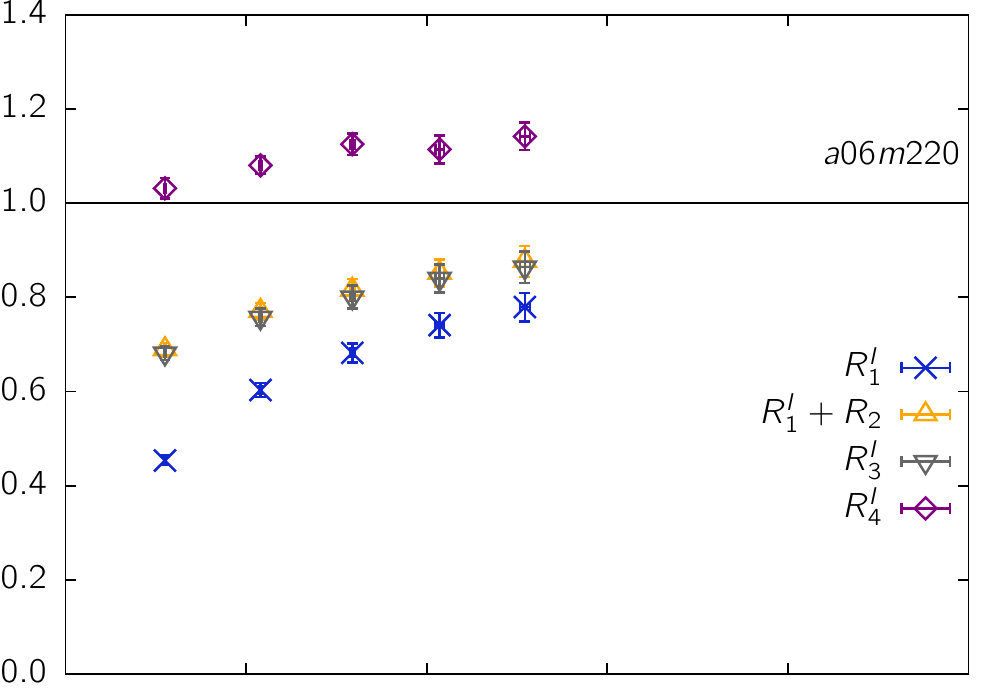}
}
\subfigure{
    \includegraphics[height=2.245in,trim={0.0cm 0.0cm 0 0},clip]{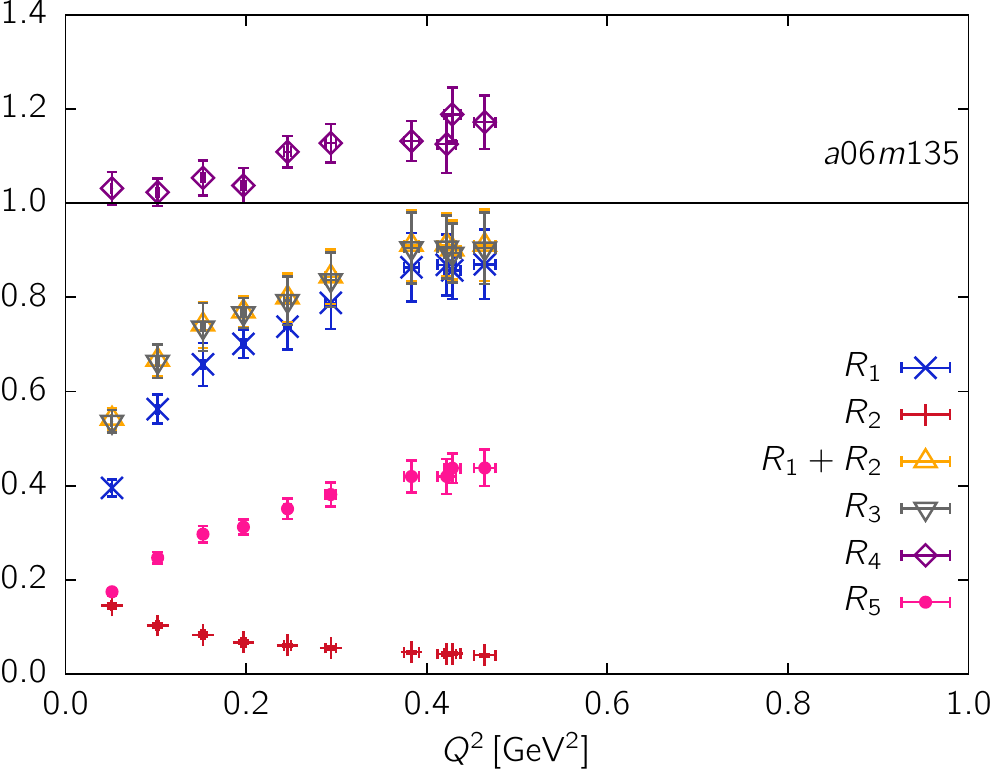}
    \includegraphics[height=2.245in,trim={0.0cm 0.0cm 0 0},clip]{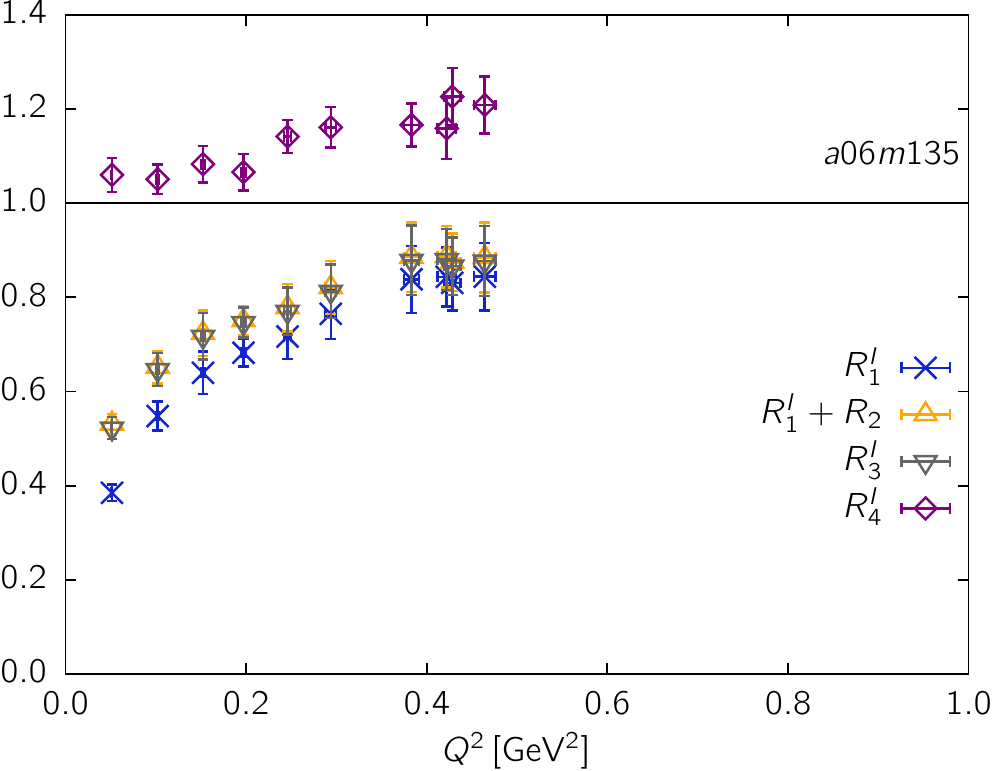}
}
\caption{(Left) The data for the five ratios $R_i$, defined in
  Eqs.~\protect\eqref{eq:defR1}--\protect\eqref{eq:defR5}. The four
  rows show data from the four ensembles $a12m310$, $a09m130$,
  $a06m220$ and $a06m135$.  Test of the PCAC relation,
  Eq.~\protect\eqref{eq:testPCAC}, is $R_1^{[I]} +R_2=1$; of the pion
  pole-dominance ansatz, Eq.~\protect\eqref{eq:GPpole}, is $R_3^{[I]}=1$;
  and of the relation given in Eq.~\protect\eqref{eq:PPDcondition} is
  $R_4^{[I]}=1$. (Right) Results for the four ratios defined in
  Eqs.~\protect\eqref{eq:defR1I}--\protect\eqref{eq:defR4I} using the
  $O(a)$ improved axial current.
  \label{fig:PCAC_test}}
\end{figure*}

We first test that the three form factors satisfy the PCAC relation,
Eq.~\eqref{eq:testPCAC}, by confirming that the quark mass ${\widehat
  m}$ obtained from the pion two-point correlation functions, $\langle
\Omega|(\partial_\mu A_\mu - 2 {\widehat m} P)_t P_0 | \Omega
\rangle=0$, is consistent with that from the three-point function
$\langle \Omega|\chi_\tau (\partial_\mu A_\mu - 2 {\widehat m} P)_t
\overline{\chi}_0 |\Omega \rangle=0$ for ${\bf p} =0$.\footnote{The
  full set of correlation functions needed to analyze the PCAC relation for the ${\bf p}
  \neq 0$ cases were, unfortunately, not calculated.} Using the more
accurate value of ${\widehat m}$ determined from the two-point
functions, we plot in Fig.~\ref{fig:PCAC_test} (left) the following
five quantities motivated by the PCAC relation given in
Eq.~\eqref{eq:testPCAC}:
\begin{align}
  R_1 =&\;  \frac{Q^2}{4 M_N^2} \frac{{\tilde G}_P(Q^2)}{G_A(Q^2)} \,,
\label{eq:defR1} \\
  R_2 =&\;  \frac{2 {\widehat m}}{2M_N} \frac{G_P(Q^2)}{G_A(Q^2)} \,,
\label{eq:defR2} \\
  R_3 =&\;  \frac{Q^2+M_\pi^2}{4 M_N^2} \frac{{\tilde G}_P(Q^2)}{G_A(Q^2)}   \,,
\label{eq:defR3} \\
  R_4 =&\;  \frac{4 {\widehat m} M_N}{M_\pi^2} \frac{{ G}_P(Q^2)}{{\tilde G}_P(Q^2)}   \,,
\label{eq:defR4} \\
  R_5 =&\;  \frac{aQ^2}{4 M_N} \frac{{ G}_P(Q^2)}{G_A(Q^2)}   \,,
\label{eq:defR5} 
\end{align}
for the four ensembles $a12m310$, $a09m130$, $a06m220$ and $a06m135$.
Including the $O(a)$ improvement of the axial current, the ratios in
Eqs~\eqref{eq:defR1},~\eqref{eq:defR3} and-\eqref{eq:defR4} become
\begin{align}
  R_1^I =&\;  \frac{Q^2}{4 M_N^2} \frac{{\tilde G}_P^I(Q^2)}{G_A(Q^2)} \,,
\label{eq:defR1I} \\
%%   R_2^I =&\;  \frac{2 {\widehat m} }{2M_N} \frac{G_P(Q^2)}{G_A(Q^2)} \,,
%% \label{eq:defR2I} \\
  R_3^I =&\;  \frac{Q^2+M_\pi^2}{4 M_N^2} \frac{{\tilde G}_P^I(Q^2)}{G_A(Q^2)}   \,,
\label{eq:defR3I} \\
  R_4^I =&\;  \frac{{2\widehat m } 2M_N}{M_\pi^2} \frac{{ G}_P(Q^2)}{{\tilde G}_P^I(Q^2)}   \,.
\label{eq:defR4I} 
\end{align}
The three improved ratios $R_{1,3,4}^I$ are shown in
Fig.~\ref{fig:PCAC_test} (right).  Note that $R_1^{[I]} +R_2=1$
checks the PCAC relation given in Eq.~\eqref{eq:testPCAC} (or Eq.~\eqref{eq:testPCACI});
$R_3^{[I]}=1$ tests the pion pole-dominance ansatz
Eq.~\eqref{eq:GPpole}; and $R_4^{[I]}=1$ tests the relation
Eq.~\eqref{eq:PPDcondition}. Comparing the two sets of panels in Fig.~\ref{fig:PCAC_test} shows
that improving the axial current has a very small effect. This is
because the value of the improvement coefficient $c_A$, that
multiplies the correction term $R_5^{[I]}$, is small.  Thus, improving
the axial current to $O(a)$ does not explain the large deviation of
$R_1^{[I]} +R_2$ from unity illustrated in Fig.~\ref{fig:PCAC_test}.

For all four ensembles, data in Fig.~\ref{fig:PCAC_test} show that
$R_1^{[I]} + R_2 \approx R_3^{[I]}$ for small $Q^2$, however, both
$R_1^{[I]} + R_2$ and $R_3^{[I]}$ are much smaller than unity.  The
deviation of $R_4^{[I]}$ from unity grows with $Q^2$, but decreases as
$a \to 0$ and $M_\pi \to M_\pi^{\rm Physical}$. This pattern is, in
general, consistent with these being discretization effects.  Note
that the corrections to $2 {\widehat m} G_P(Q^2) = (M_\pi^2 / 2 M_N)
{\tilde G}_P(Q^2)$, or to $R_4^{[I]} =1$, do not significantly impact
$R_1^{[I]} + R_2 \approx R_3^{[I]}$ because the dominant contribution
to both sides of this approximate equality comes from $R_1^{[I]}$.

The data for $R_3$ from all eight ensembles is plotted in
Fig.~\ref{fig:PPDH_test} and show that the deviations from unity
increase with decreasing $Q^2$, $a$ and $M_\pi^2$. For the physical
pion mass ensembles, the $O(50\%)$ deviation for $Q^2 < 0.2$~GeV${}^2$
is surprisingly large. Such $Q^2$ dependent deviations from the PCAC
relation are, generically, indicators of discretization artifacts. The
increase in the deviations with decreasing $a$ does not support this expectation,
and as shown in Fig.~\ref{fig:PCAC_test}, the $O(a)$ improvement of
the axial current does not reduce the deviations. Therefore, the
observed large deviation remains unexplained and requires further 
investigation. 

%

\begin{comment}
\begin{table}[htbp]
\centering
%\renewcommand{\arraystretch}{1.2} % Change vertical spacing
\begin{ruledtabular}
\begin{tabular}{l|l|l|l}
$g_P^\ast (a,M_\pi)/g_A$ & $M_\pi\approx 310$~MeV  & $M_\pi\approx 220$~MeV  & $M_\pi\approx 135$~MeV  \\
\hline
$a=0.12$~fm              & 1.70(05)                &  2.29(15)               &                    \\
$a=0.09$~fm              & 1.64(12)                &  2.16(17)               &  3.17(17)          \\
$a=0.06$~fm              & 1.60(08)                &  2.11(07)               &  3.47(15)          \\
\end{tabular}
\end{ruledtabular}
\caption{Results for $ g_P^\ast (a,M_\pi)/g_A$ obtained from fits 
  using Eq.~\protect\eqref{eq:PPDfit} that are shown in Fig.~\protect\ref{fig:gP-vsM-vsa}. 
  \label{tab:GPstarX}}
\end{table}
\end{comment}

%%%%%%%%%%%%%%%%%%%%%%%%%%%%%%%%%%%%
\section{Analysis of $g_P^\ast$}
\label{sec:gPstar}
%%%%%%%%%%%%%%%%%%%%%%%%%%%%%%%%%%%%

To determine $g_P^\ast/g_A$ and $g_{\pi {\rm NN}}/g_A$ we need to
evaluate ${\tilde G}_P(Q^2)$ at $Q^2 \equiv Q^{\ast\, 2} = 0.88
m_\mu^2$ and at $Q^2 = - M_\pi^2$. This is done using the ansatz given
in Eq.~\eqref{eq:PPDfit}.  In Fig.~\ref{fig:gP-PD-detail}, the data
for $(Q^2+M_{\pi}^2) {\tilde G}_P(Q^2)/g_A$ and the result of the fit
using Eq.~\eqref{eq:PPDfit}.  The extrapolated values are shown using
the symbol star at $Q^{\ast\, 2} = 0.88 m_\mu^2$ and by the symbol
plus at $Q^2 = - M_\pi^2$.  It is clear from
Fig.~\ref{fig:gP-PD-detail}, that there are enough free parameters in
Eq.~\eqref{eq:PPDfit} to fit the data and the values obtained at
$Q^{\ast\, 2}$ and $Q^2 = - M_\pi^2$ by extrapolation are
reasonable. However, the contributions of terms proportional to $c_2$
and $c_3$ (see Table~\ref{tab:GPfit}) increase as the lattice spacing
$a \to 0$ and $M_\pi \to 135$~MeV. The quantitative change in behavior
is already clear in all three $M_\pi \approx 220$~MeV ensembles. Thus,
it is unlikely that the change in behavior between the $M_\pi \approx
310$~MeV ensembles and those at lighter $M_\pi$ is a statistical
fluctuation.  Because of this change in behavior, we get low estimates
of $g_P^\ast/g_A$ and $g_{\pi {\rm NN}}/g_A$.

\begin{figure*}[tbp]%16
{
    \includegraphics[width=0.47\linewidth]{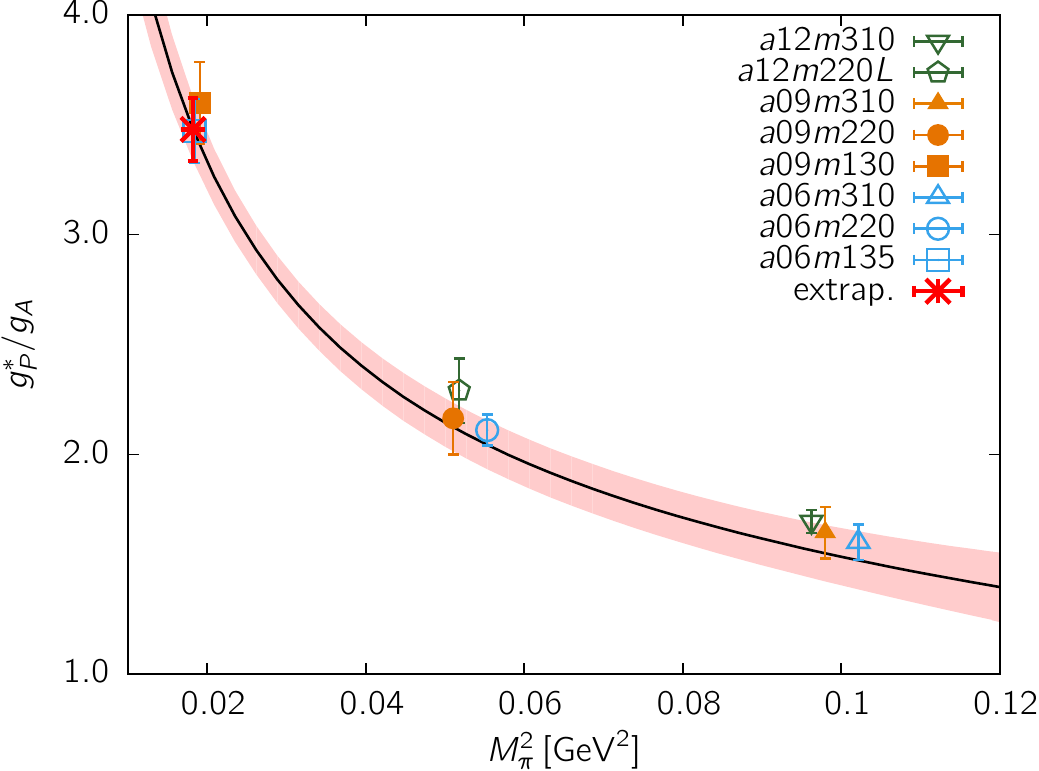}
    \includegraphics[width=0.47\linewidth]{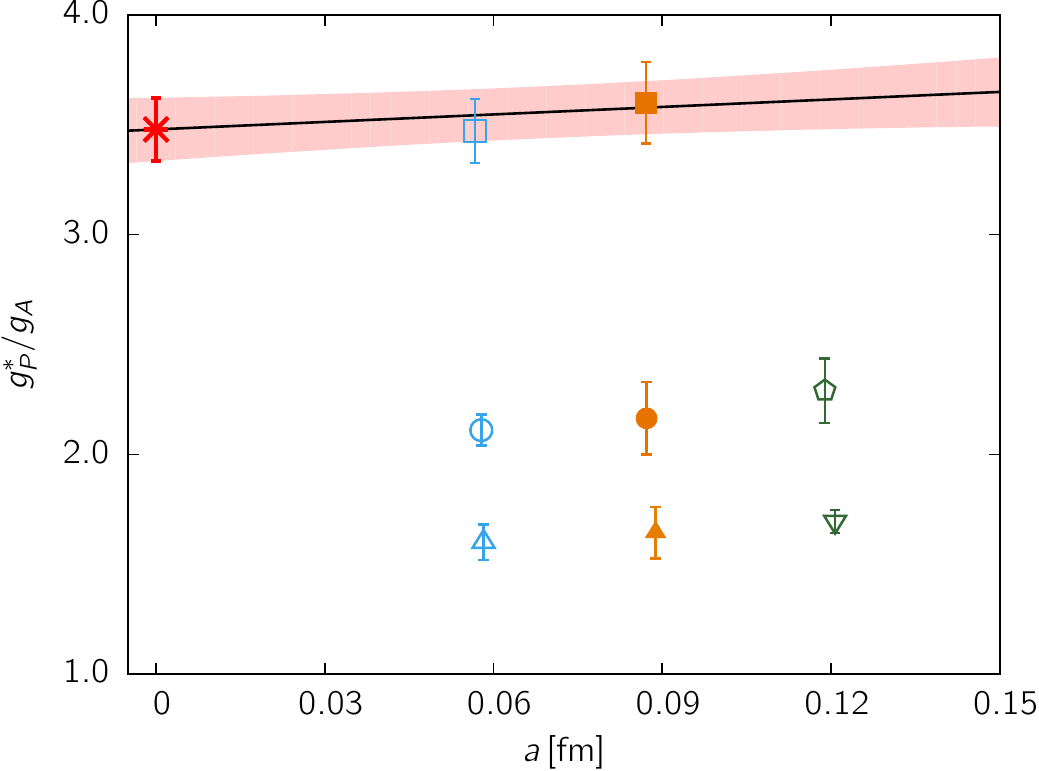}
}
\caption{The 8-point fit using Eq.~\protect\eqref{eq:extrap-gP} to the
  lattice data for $g_P^\ast /g_A$.  In the left (right) panel, the
  data are shown versus the single variable $M_\pi^2$ ($a$), whereas
  the fits include dependence on both variables simultaneously. The
  same symbols are used for data in both figures and defined in the
  left panel.
  \label{fig:gP_extrap8_nofv}}
\end{figure*}
\begin{figure}[tbp]%17
{
    \includegraphics[width=0.97\linewidth]{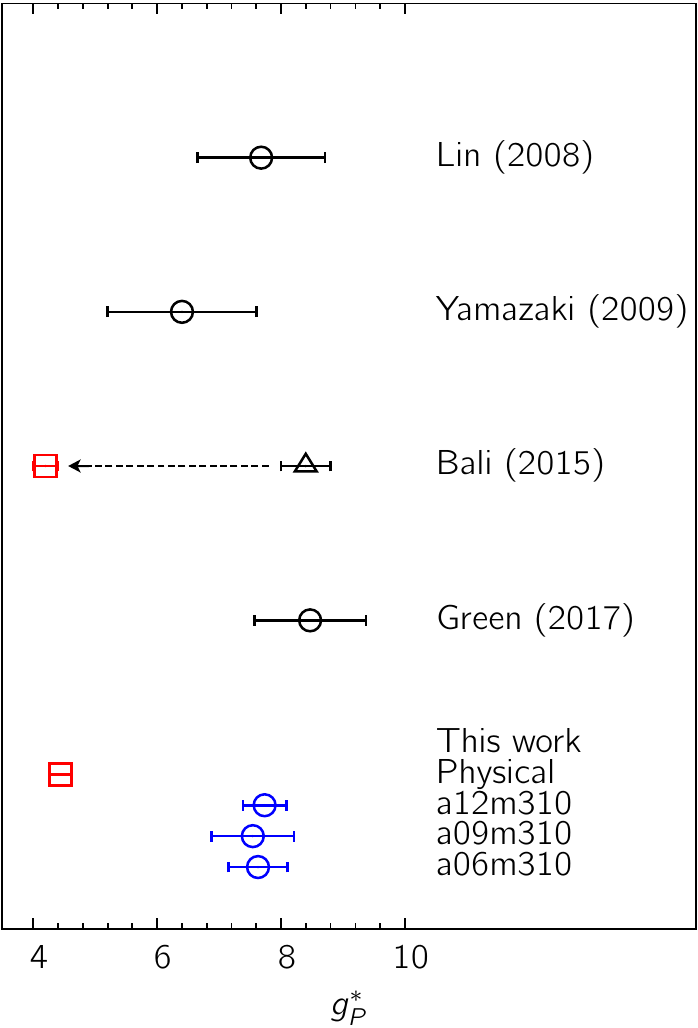}
}
\caption{Summary of lattice QCD results for the renormalized
  $g_P^\ast$. Previous results are from
  Lin(2008)~\protect\cite{Lin:2008uz},
  Yamazaki(2009)~\protect\cite{Yamazaki:2009zq},
  Bali(2015)~\protect\cite{Bali:2014nma}, and
  Green(2017)~\protect\cite{Green:2017keo}.  The data with open
  circles were obtained from simulations with $M_\pi > 300$~MeV and
  scaled to the physical pion mass $M_\pi=135$~MeV using just the
  pion-pole term as discussed in the text.
  \label{fig:gPstar}}
\end{figure}

Given the data in Table~\ref{tab:GPfit}, to estimate $g_P^\ast$ in the
limit $a\to 0$ and $M_\pi\to 135$~MeV, we make a fit using the ansatz
\begin{equation}
  g_P^\ast (a,M_\pi)/g_A =  d_1 + d_2 a + \frac{d_3}{ M_\pi^2 + 0.88 m_\mu^2} +  d_4 M_\pi^2  \,,
\label{eq:extrap-gP} 
\end{equation}
where the leading behavior in $M_\pi^2$ is taken to be the pion-pole
term evaluated at the experimental momentum scale of muon capture. We
neglect possible finite volume corrections in the data in obtaining
the estimates since the data do not show an obvious dependence
on $M_\pi L$.  The simultaneous fits in $a$ and $M_\pi$ are shown in
Fig.~\ref{fig:gP_extrap8_nofv}.  They give
\begin{eqnarray}
g_P^\ast/g_A &=& 3.48(14) \,,  \nonumber \\
g_P^\ast     &=& 4.44(18) \,,
\label{eq:GPstar_final}
\end{eqnarray} 
where the final value of $g_P^\ast$ is obtained by multiplying the
ratio obtained from the fit by the experimental value $g_A=1.276$.

We summarize lattice QCD results for $g_P^\ast$ in
Fig.~\ref{fig:gPstar}.  The results $g_P^\ast = 7.68 \pm 1.03$
(Lin(2008)~\protect\cite{Lin:2008uz}), $g_P^\ast = 6.4 \pm 1.2$
(Yamazaki(2009)~\protect\cite{Yamazaki:2009zq}), and $g_P^\ast =
8.47(21)(87)(2)(7)$ (Green(2017)~\protect\cite{Green:2017keo}) have
all been obtained on ensembles with $M_\pi > 300$~MeV and extrapolated
to $M_\pi^{\rm Physical}$ using just the pion-pole term, $(Q^{\ast\,
  2} + M_\pi^2) {\tilde G}_P(Q^{\ast\, 2})/((Q^{\ast\, 2} +
M_\pi^{2,{\rm Physical}})$.  Thus all estimates from $M_\pi > 300$~MeV
ensembles, including our three $M_\pi \approx 310$~MeV ensembles,
yield $g_P^\ast \approx 8$ after scaling in $M_\pi$ using the
pion-pole ansatz.  As we have discussed above, the $Q^2$ corrections
to the pion-pole ansatz become large for $M_\pi < 300$~MeV and our
direct simulations at $M_\pi \approx 220$ and 135~MeV show that using
just the pion-pole ansatz for scaling in $M_\pi^2$ is not justified.

Our estimate, $g_P^\ast = 4.44(18)$, is consistent with the value
$g_P^\ast = 4.20(20)$ extracted from Ref.~\cite{Bali:2014nma}, once their
result is corrected for by the factor 0.5 that was missed in their
definition of $g_P^\ast$.  Note that their analysis also shows the
change in the scaling behavior for $M_\pi < 300$, and they report results
analogous to our Fig.~\ref{fig:PCAC_test}.

To summarize, our low value, $g_P^\ast = 4.44(18)$, is about half of
the values obtained from the muon capture experiment and $\chi$PT as
summarized in Eq.~\eqref{eq:GPstar_pheno}.  Our data are well-fit by
the ansatz given in Eq.~\eqref{eq:PPDfit}, however, the corrections
proportional to the parameters $c_2$ and $c_3$ become large as $M_\pi
\to 135$~MeV. Thus, one cannot extrapolate to $M_\pi^{\rm Physical}$
using just the pion-pole term.  The underlying reason for a low value
of $g_P^\ast$ is the large deviation from unity of the ratios
$R_1^{[I]} + R_2$ and $R_3^{[I]}$, defined in
Eqs.~\eqref{eq:defR1}--~\eqref{eq:defR3}, at low $Q^2$.  The size of
the deviations are shown in Fig.~\ref{fig:PPDH_test}. Considering that
the $O(a)$ improvement of the axial current does not reduce the
deviation, the observed violation of the PCAC relation remains
unexplained.

%%%%%%%%%%%%%%%%%%%%%%%%%%%%%%%%%%%%
\section{Analysis of the pion-nucleon coupling, $\gpNN$}
\label{sec:GTR}
%%%%%%%%%%%%%%%%%%%%%%%%%%%%%%%%%%%%

\begin{figure*}[tbp]
{
    \includegraphics[width=0.47\linewidth]{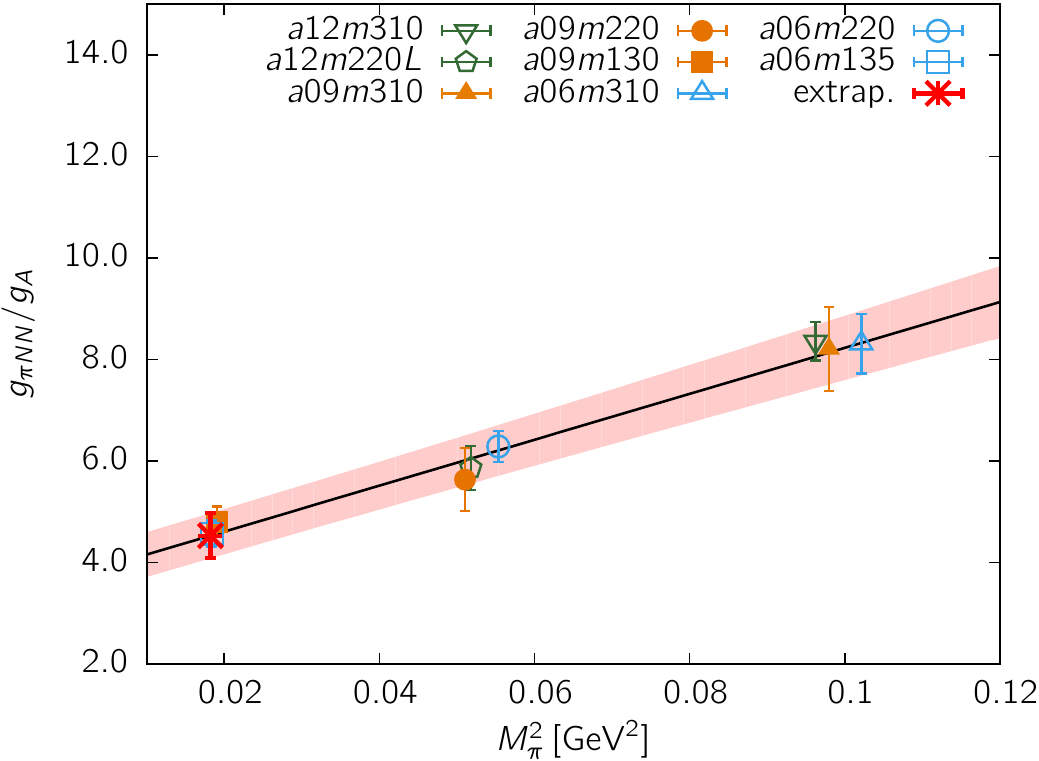}
    \includegraphics[width=0.47\linewidth]{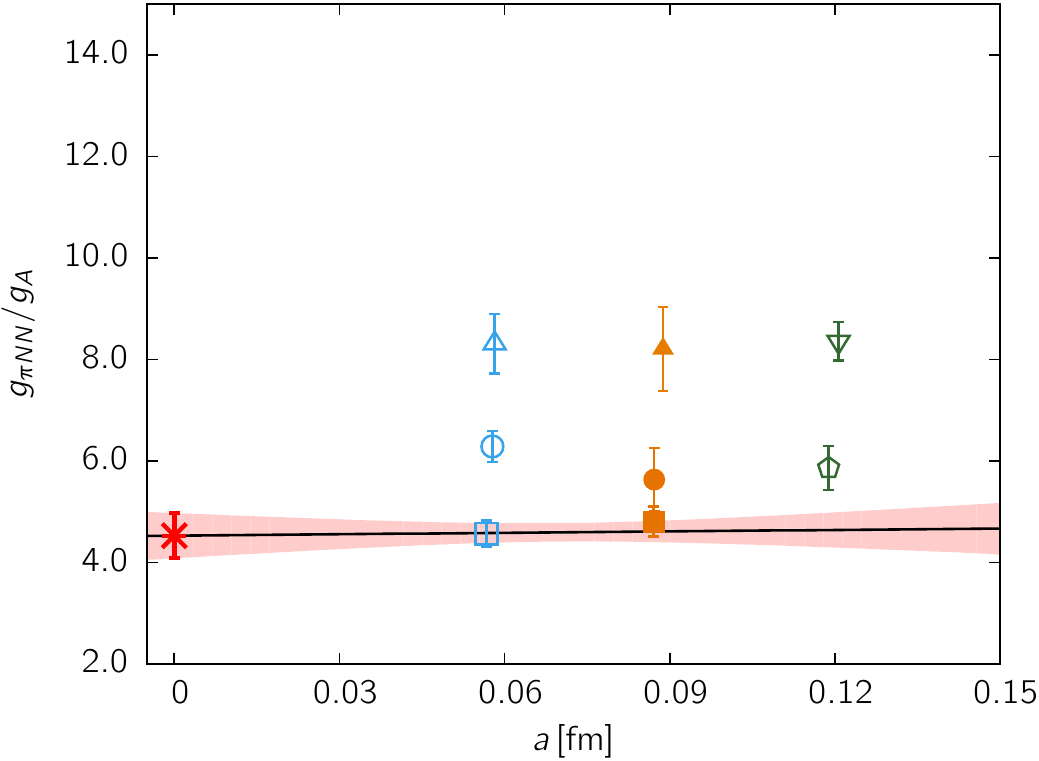}
}
\caption{The 8-point fit to the lattice data for $\gpNN /g_A$ using
  Eq.~\protect\eqref{eq:extrap-rA}. In this fit, We neglect the
  possible finite volume term.  In the left (right) panel, the data
  are shown versus the single variable $M_\pi^2$ ($a$), whereas the
  fit is to estimates extrapolated to the physical value in the other
  variable. The same symbols are used for data in both figures and
  defined in the left panel.
  \label{fig:gpNN_extrap8_nofv}}
\end{figure*}

The pion-nucleon coupling, $\gpNN$, is defined as the residue at
the pion pole of ${\tilde G}_P(Q^2)$, i.e., at $Q^2 = -q^2 = -M_\pi^2$.
Since all our data are obtained at positive values of $Q^2$, we
first fit ${\tilde G}_P(Q^2)$ using 
the ansatz in Eq.~\eqref{eq:PPDfit} and then calculate
\begin{equation}
\gpNN = \lim_{Q^2 \to -M_\pi^2} \frac{M_\pi^2 + Q^2}{4M_N F_\pi} {\tilde G}_P(Q^2) \,, 
\label{eq:gpiNN-fit}
\end{equation}
where $F_\pi$ is the pion decay constant. 
These estimates are given in the fifth column of Table~\ref{tab:GPfit}. To extrapolate 
to $a\to 0$ and $M_\pi \to 135$~MeV, we use the leading order ansatz
given in Eq.~\eqref{eq:extrap-rA}. The fit, shown in
Fig.~\ref{fig:gpNN_extrap8_nofv}, gives
\begin{eqnarray}
\gpNN /g_A &=& 4.53(45) \,,  \nonumber \\
\gpNN      &=& 5.78(57) \,,
\label{eq:gpNN_final}
\end{eqnarray} 
with $g_A=1.276$.  This lattice value has to be compared with $g_{\pi
  NN} = 13.69 \pm 0.12 \pm 0.15$ obtained from the $\pi N$ scattering
length analysis~\cite{Baru:2011bw}.  As discussed above in the
analysis of $g_P^\ast$, our low value is a consequence of the
unexplained deviation of the ratios $R_1^{[I]}+R_2$ and $R_3^{[I]}$
from unity at small $Q^2$ on the $M_\pi = 220$ and $135$~MeV
ensembles.

\begin{figure*}[tbp]
{
    \includegraphics[width=0.47\linewidth]{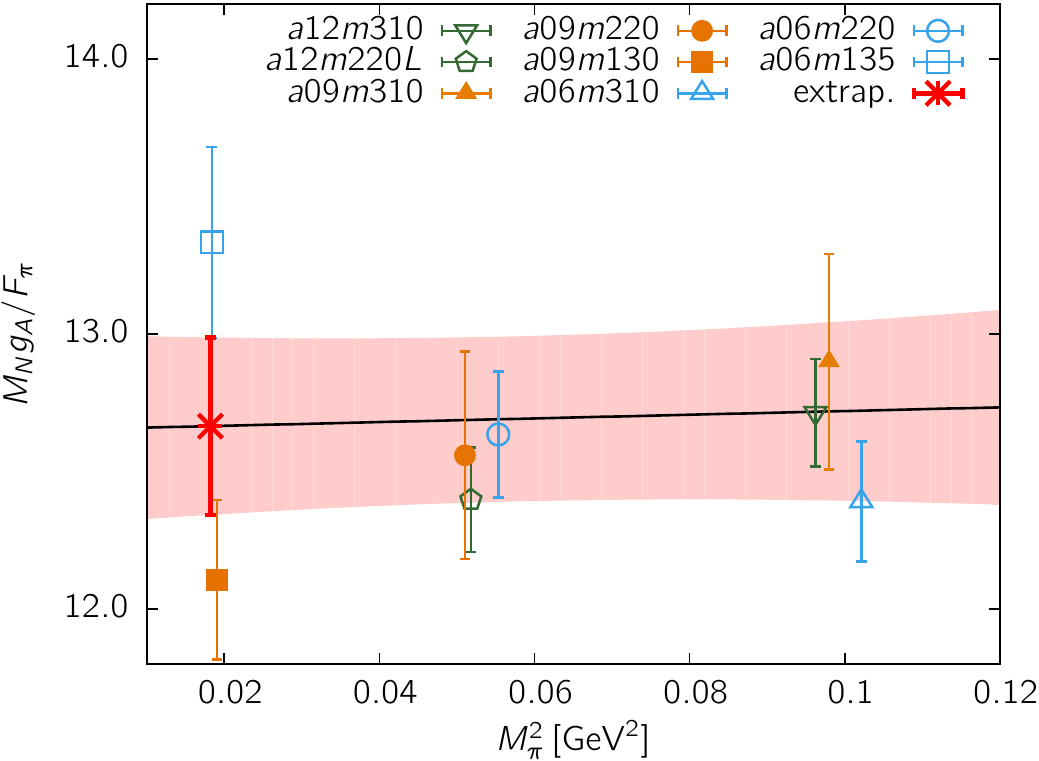}
    \includegraphics[width=0.47\linewidth]{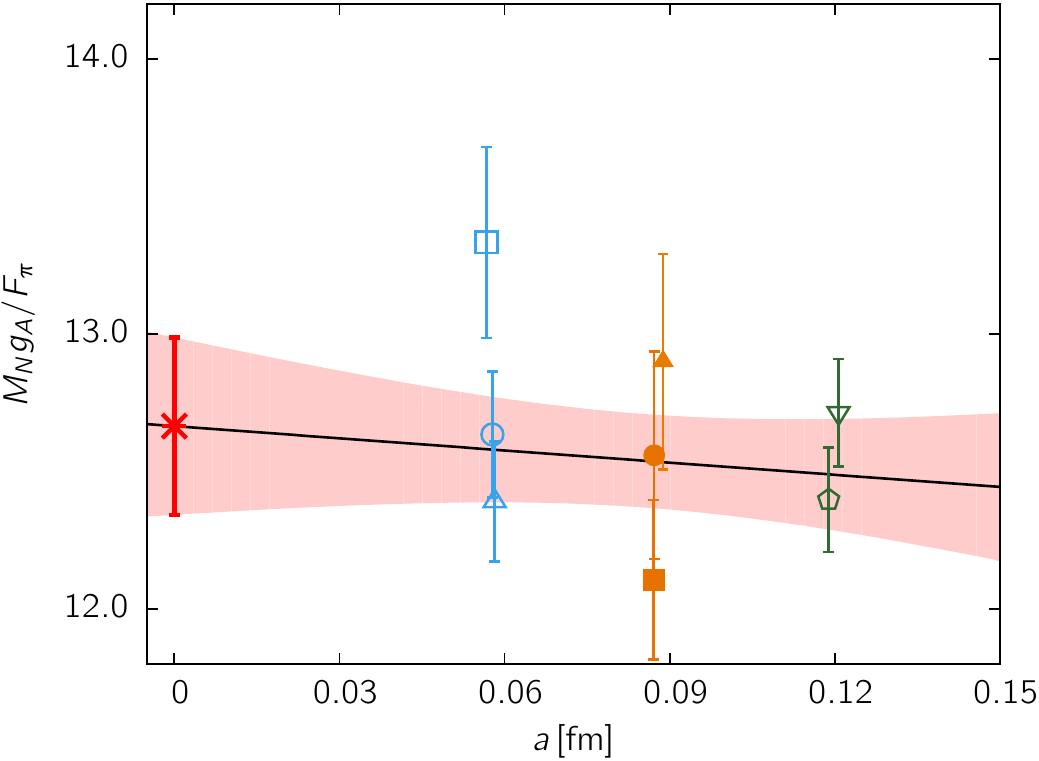}
}
\caption{The data for $\gpNN = M_N g_A/ F_\pi $ from the eight
  ensembles and the simultaneous fit in $a$ and $M_\pi^2$ to obtain
  the result in the limit $a\to 0$ and $M_\pi = 135$~MeV using the
  ansatz in Eq.~\protect\eqref{eq:extrap-rA}. The finite volume
  correction term is neglected in the fit. The rest is the same as in
  Fig.~\protect\ref{fig:gpNN_extrap8_nofv}.
  \label{fig:gpiNN}}
\end{figure*}

We can also estimate $\gpNN$ using the Goldberger-Treiman relation, 
\begin{equation}
\gpNN = \frac{M_N g_A}{F_\pi}  \,.
\label{eq:GTR}
\end{equation}
The resulting values of $\gpNN $ given in Table~\ref{tab:gpiNN} for
each ensemble are obtained using estimates of $g_A/F_\pi$ from
Ref.~\cite{Bhattacharya:2016zcn}.\footnote{The values for the
  $a06m135$ ensemble are $g_A/F_\pi=12.62(29)$ and $F_\pi=95.4(1.0)$.}
The extrapolation to $a\to 0$ and $M_\pi \to 135$~MeV using the ansatz
given in Eq.~\eqref{eq:extrap-rA} with just the leading order
corrections is shown in Fig.~\ref{fig:gpiNN}.  The result, $\gpNN =
12.87(34)$, is consistent with the value, $\gpNN = 13$, one gets by
using the experimental values, $g_A=1.276$, $M_N=939$~MeV and
$F_\pi=92.2$~MeV.  Note that this test of the Goldberger-Treiman
relation relies on our calculation of $g_A$ right to within $5\%$,
whereas direct calculations of $g_P^\ast$ and $\gpNN$ depend on
$\tilde{G}_P(Q^2)$, which we find shows large deviations from the PCAC
relation.

\begin{table}[htbp]%9
\centering
\begin{ruledtabular}
\begin{tabular}{l|l|l|l}
                & $M_\pi \approx 310$~MeV  & $M_\pi \approx 220$~MeV  & $M_\pi \approx 135$~MeV  \\
\hline
$a=0.12$~fm     & 12.7(2)            &  12.4(2)            &                    \\
$a=0.09$~fm     & 12.9(4)            &  12.6(4)            &  12.1(3)           \\
$a=0.06$~fm     & 12.4(2)            &  12.6(2)            &  13.3(3)           \\
\end{tabular}
\end{ruledtabular}
\caption{Results for $\gpNN = M_N g_A/ F_\pi$ determined using values
  of $M_N$, $g_A$ and $F_\pi$ obtained on the eight ensembles. }
\label{tab:gpiNN}
\end{table}

%

%%%%%%%%%%%%%%%%%%%%%%%%%%%%%%%%%%%%%%%%%%%%%%%%%%%%%%%%%%%%%%%%%%%%%
%%%  SECTION                                                      %%%
%%%%%%%%%%%%%%%%%%%%%%%%%%%%%%%%%%%%%%%%%%%%%%%%%%%%%%%%%%%%%%%%%%%%%
\section{A heuristic analysis}
\label{sec:heuristic}

\begin{figure}[tbp]
\centering
\subfigure{
\includegraphics[width=0.97\linewidth]{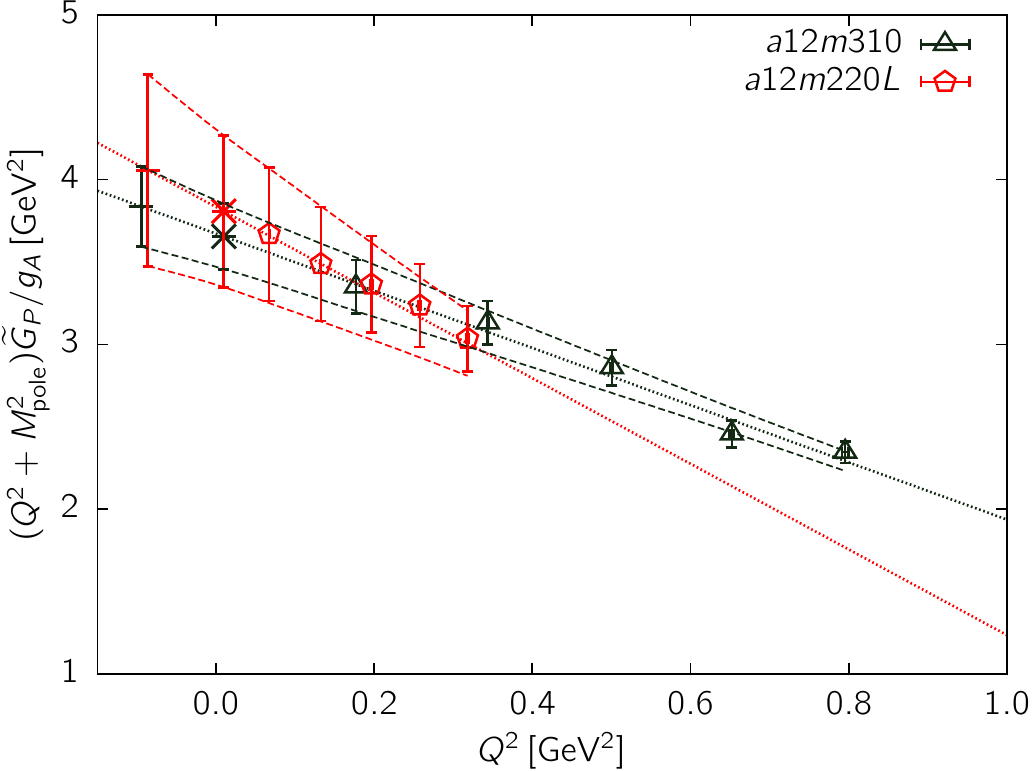}
}
\subfigure{
\includegraphics[width=0.97\linewidth]{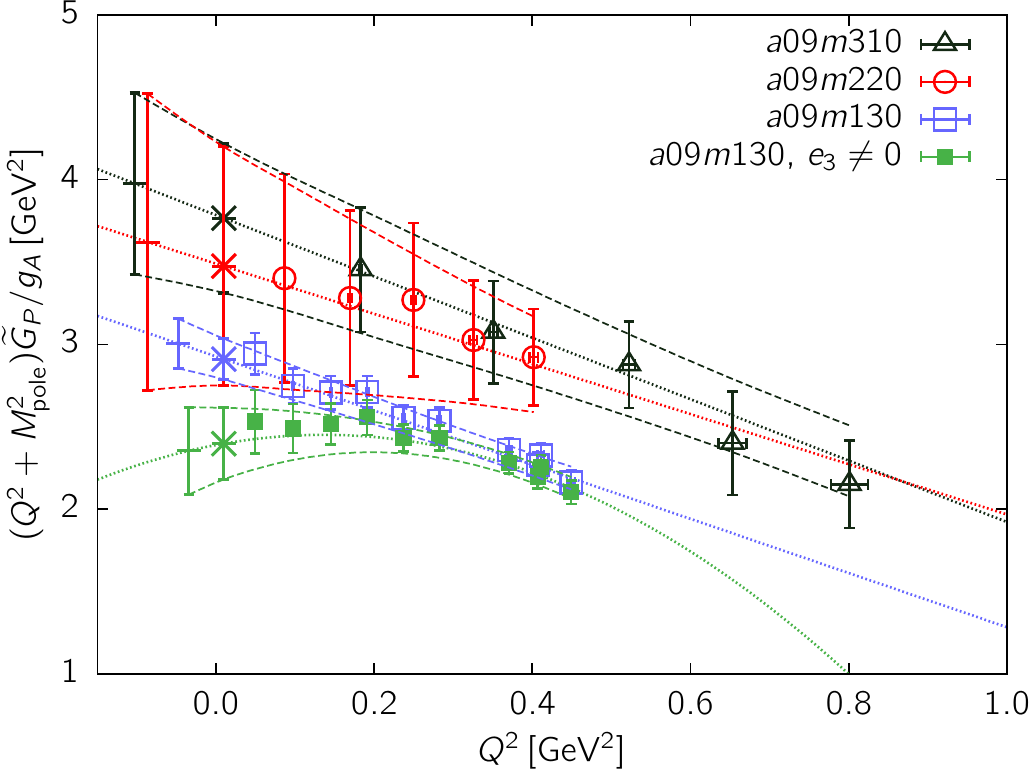}
}
\subfigure{
\includegraphics[width=0.97\linewidth]{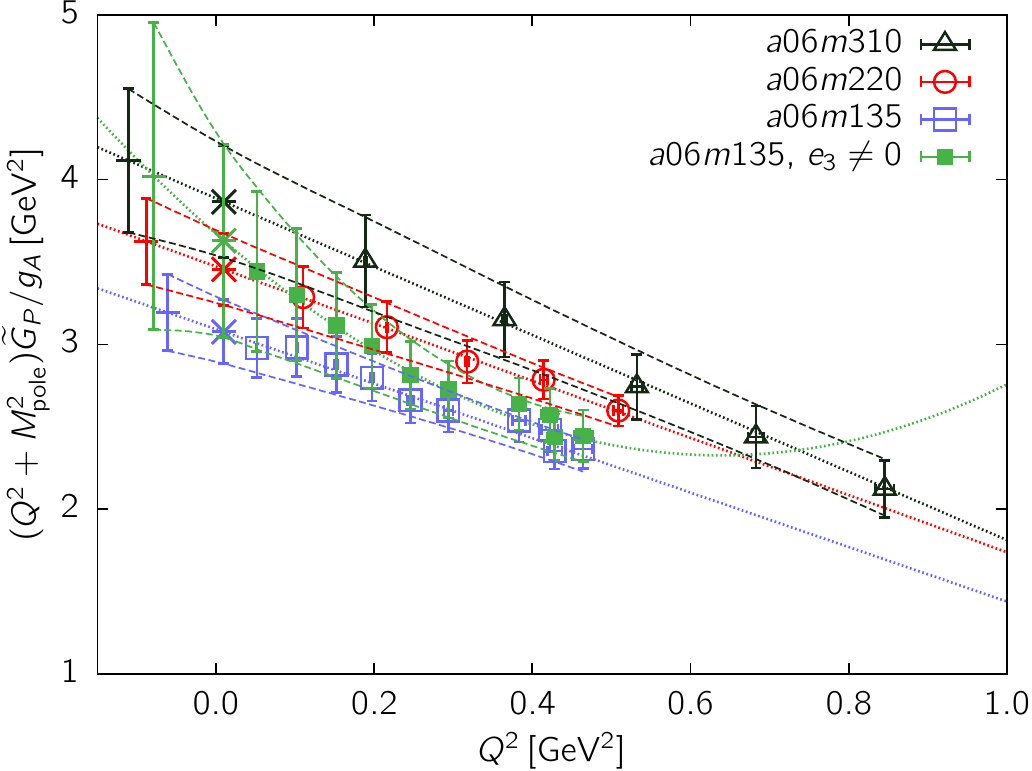}
}
\caption{The data for the quantity $(Q^2+M_{\rm pole}^2) {\tilde
    G}_P(Q^2)/g_A$.  The three panels highlight the dependence on
  $M_\pi^2$ for fixed $a$.  The fits versus $Q^2$ in units of
  GeV${}^2$ are performed using Eq.~\protect\eqref{eq:PPDfitX} with
  $e_3=0$. The point with symbol star (plus) gives the value at $Q^2 =
  Q^{\ast\, 2} \equiv 0.88 m_\mu^2$ ($Q^2 = -M_\pi^2$). We show the
  $1\sigma$ error band of the fits.  For the two physical mass
  ensembles, $a09m130$ and $a06m135$, we also show the data (solid
  green squares) and the fits (green lines) including the $e_3$ term
  defined in Eq.~\protect\eqref{eq:PPDfitX}. }
\label{fig:gP-PD-detailX}
\end{figure}

\begin{figure}[tbp]
\centering
\subfigure{
\includegraphics[width=0.97\linewidth]{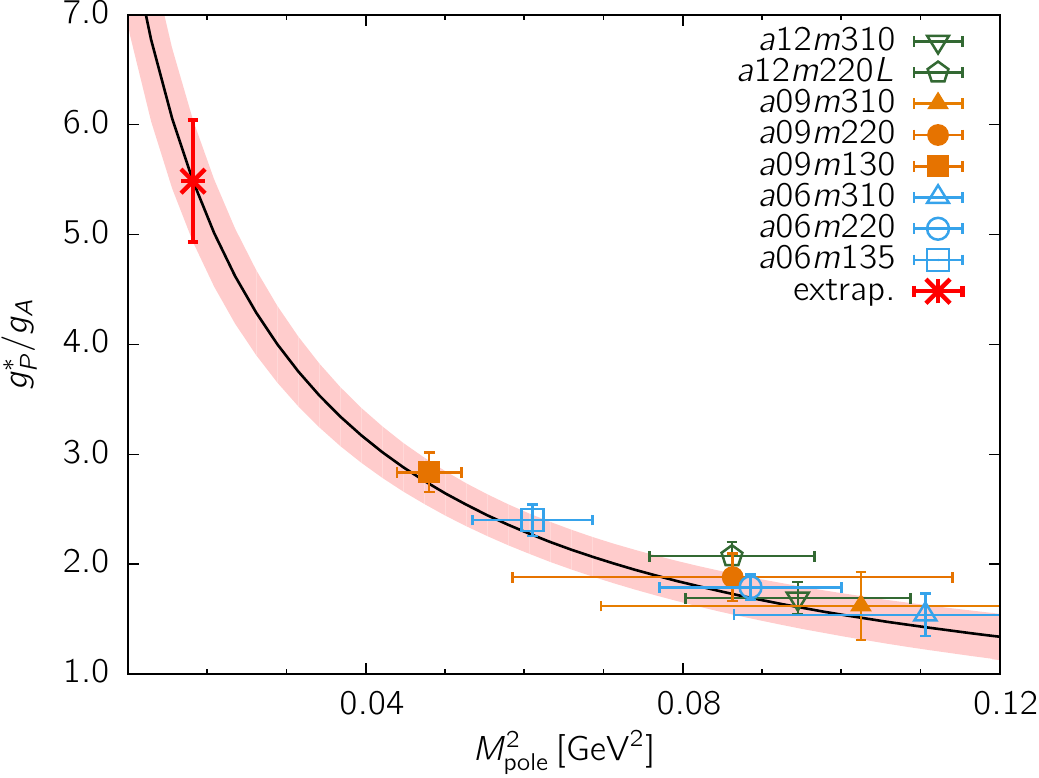}
}
\caption{The extrapolation of $g_P^\ast/g_A$ to the physical pion mass and
  the continuum limit using $h_0/(Q^{\ast\, 2} + M_{\rm pole}^2) + h_1 + h_2 a$, where 
  $M_{\rm pole}^2$ is a free parameter introduced in the heuristic analysis 
  to fit the data.  }
\label{fig:gPstarX}
\end{figure}

Testing the PCAC relation, Eq.~\eqref{eq:PCAC}, requires no input
outside of our lattice calculations: the three form factors,
$G_A(Q^2)$, ${\tilde G}_P(Q^2)$, and $G_P(Q^2)$, are obtained from our
lattice calculations of three-point functions, and $\widehat m$ is
obtained from the pion two-point correlations functions.  Thus, the
large deviations from the PCAC relation, as discussed in
Sections~\ref{sec:GP},~\ref{sec:gPstar} and~\ref{sec:GTR}, are troubling.  They
motivated us to examine alternatives to the single pion pole-dominance
ansatz.  The data in Fig.~\ref{fig:PCAC_test} suggest that the
deviation from the PCAC relation can be reduced by enhancing the
contribution of $R_2$, i.e., the relative size of the $M_\pi^2$ versus
the $Q^2$ term in pion pole-dominance ansatz.  We, therefore, fit the
data using
\begin{equation}
\frac{m_\mu}{2 M_N}  \frac{{\tilde G}_P(Q^2)}{g_A} = \frac{e_1}{M_{\rm pole}^2 + Q^2} + e_2 +e_3 Q^2 \,,
\label{eq:PPDfitX}
\end{equation}
where $M_{\rm pole}^2$ and $e_i$ are free parameters. The fits for
$e_3=0$ are shown in Fig.~\ref{fig:gP-PD-detailX} and the resulting
value of $M_{\rm pole}$ is given in Table~\ref{tab:heuristic}.  As
expected, allowing $M_{\rm pole}$ to be a free parameter changes the
fits very significantly and the results mimic the pion pole-dominance
behavior seen for the $M_\pi > 300$~MeV ensembles.  This can be seen
by comparing the fits in Fig.~\ref{fig:gP-PD-detailX} with those in
Fig.~\ref{fig:gP-PD-detail} which were obtained using the fit ansatz
given in Eq.~\eqref{eq:PPDfit}.  The surprise is the size of the
difference, $M_{\rm pole} - M_\pi$, that can be inferred from
Table~\ref{tab:heuristic}. While we expect some shift in $M_\pi$ to
correct for all the intermediate states that couple to the axial
current rather than just the groundstate pion, it is difficult to
explain the observed large shift. Nevertheless, continuing with this
heuristic analysis, we show in Fig.~\ref{fig:gPstarX} the
extrapolation of the estimates of $g_P^\ast$, given in
Table~\ref{tab:heuristic} and obtained with $e_3 = 0$, to the physical
pion mass and the continuum limit using the ansatz $h_0/(Q^{\ast\, 2}
+ M_{\rm pole}^2) + h_1 + h_2 a$.  This analysis gives $g_P^\ast =
7.0(7)$ and similarly $g_{\pi {\rm NN}} = 11.2(1.3)$. The large change is
mainly because the extrapolation is now being done from the larger values
of $M_{\rm pole}$.

For the physical pion mass ensembles, $a09m130$ and $a06m135$, the fits can be 
performed with $e_3$ a free parameter since we have data at ten
values of $Q^2$.  Adding $e_3$ to the fit ansatz give a significantly different value for
$M_{\rm pole}$ and, as a result, the violet open squares move to the filled green
squares in Fig.~\ref{fig:gP-PD-detailX}.  Even the curvature of the
fit has opposite sign in the two cases. Not surprisingly, the values of $g_P^\ast$ and
$\gpNN$ for the two physical mass ensembles change significantly and
in opposite direction on including the $e_3$ term in the fit.  In short, 
this heuristic analysis becomes unstable as $M_\pi \to 135$~MeV. 

Note that introducing $M_{\rm pole}^2$ as a free parameter is
analogous to tuning $\widehat m$ in the PCAC relation, 
Eq.~\ref{eq:PCAC}, by requiring $R_1 + R_2$, shown in
Fig.~\ref{fig:PCAC_test}, is unity independent of $Q^2$, rather than
using the value from the PCAC relation applied to the pion two-point
correlation function. The bottom line of such a heuristic analysis is
that the change, $M_\pi^2 \to M_{\rm pole}^2$ or in $\widehat m$, to
accomodate the data is much larger than what is expected from
discretization effects. Therefore, understanding why the three form
factors do not satisfy the PCAC relation remains our highest priority
for future work.

\begin{table}[htbp]
\centering
\begin{ruledtabular}
\begin{tabular}{l|l|l|l|l}
                & $M_\pi$   & $M_{\rm pole}$  & $g_P^*/g_A$ &    $g_{\pi {\rm NN}}/g_A$ \\
                & [MeV]     & [MeV]           &             &                           \\
\hline
$a12m310 $ &    310(3)      &    307(23)      &    1.69(15)      &    8.1(0.5)       \\
$a12m220L$ &    228(2)      &    294(18)      &    2.07(13)      &    9.6(1.4)       \\
$a09m310 $ &    313(3)      &    320(51)      &    1.62(31)      &    8.5(1.2)       \\
$a09m220 $ &    226(2)      &    294(47)      &    1.88(22)      &    8.7(2.2)       \\
$a09m130 $ &    138(1)      &    219(9)       &    2.84(18)      &    8.2(0.4)       \\
$a06m310 $ &    319(2)      &    332(36)      &    1.54(19)      &    8.7(0.9)       \\
$a06m220 $ &    235(2)      &    298(19)      &    1.79(11)      &    8.6(0.6)       \\
$a06m135 $ &    136(2)      &    247(15)      &    2.40(14)      &    8.8(0.7)       \\
\end{tabular}
\end{ruledtabular}
\caption{Results for $M_{\rm pole}$, $g_P^*$ and $\gpNN$ using the heuristic fit ansatz given in 
  Eq.~\protect\eqref{eq:PPDfitX} with $e_3=0$. }
\label{tab:heuristic}
\end{table}

%

%%%%%%%%%%%%%%%%%%%%%%%%%%%%%%%%%%%%%%%%%%%%%%%%%%%%%%%%%%%%%%%%%%%%%
%%%  SECTION                                                      %%%
%%%%%%%%%%%%%%%%%%%%%%%%%%%%%%%%%%%%%%%%%%%%%%%%%%%%%%%%%%%%%%%%%%%%%
\section{Conclusions}
\label{sec:conclusions}

We have presented high statistics results of the axial and the induced
pesudoscalar form factors on eight ensembles described in
Table~\ref{tab:ens} using a clover-on-HISQ approach. The pseudoscalar
form factor was calculated on four ensembles to test the PCAC
relation.

To fit the $Q^2$ dependence of the axial form factor, $G_A(Q^2)$, we
use the $z$-expansion and the dipole ansatz. Estimates from the
$z^{2+4}$ versus $z^{3+4}$ truncation of the $z$-expansion are
consistent within $1\sigma$ uncertainty. The dipole ansatz does a
remarkable job of fitting the data. The estimates of $r_A$ from these
three fit ansatz agree for all eight ensembles. The results, after
extrapolation in $a$ to the continuum limit and $M_\pi L \to \infty$,
and evaluated at $M_\pi = 135$~MeV, are $r_A|_{z-{\rm expansion}} =
0.46(6)$ and $r_A|_{\rm dipole} = 0.49(3)$. While
these results are consistent, they are smaller than the
phenomenological estimates given in Eq.~\eqref{eq:rA_expt}. Our
estimate $r_A|_{\rm dipole} = 0.49(3)$ corresponds to an axial mass
${\cal M}_A = 1.39(9)$ that is in good agreement with the value
obtained by the MiniBooNE
collaboration~\cite{AguilarArevalo:2010zc}. Our final estimate from
the combined dipole and the $z$-expansion analyses is $r_A|_{\rm
  combined} = 0.48(4)$.

The data for the induced pseudoscalar form factor ${\tilde G}_P(Q^2)$
versus $Q^2$ show little dependence on the lattice spacing $a$, the
pion mass $M_\pi$ or the lattice size $M_\pi L $. Our test of the PCAC
relation, including the contribution of the pseudoscalar form factor
$G_P(Q^2)$, show significant deviations for $Q^2 \lesssim
0.2$~GeV${}^2$, in particular for the physical mass
ensembles. Extrapolation in $Q^2$ using an ansatz based on the
pion pole-dominance hypothesis, Eq.~\eqref{eq:extrap-rA}, fits the
lattice data well but leads to very low estimates of the induced
pseudoscalar charge, $g_P^\ast = 4.44(18)$, and of the pion-nucleon
coupling $\gpNN = 5.78(57)$ estimated as the residue at the pole in
${\tilde G}_P(Q^2)$ at $Q^2 = -M_\pi^2$. These low estimates are a
consequence of the large deviations from the PCAC relation
for $Q^2 \lesssim 0.2$~GeV${}^2$. All previous estimates from $M_\pi >
300$~MeV ensembles that gave $g_P^\ast \approx 8$ were not sensitive
to this problem as discussed in Sec.~\ref{sec:GP}.

Work is under progress to improve the statistical and systematic
precision of the three form factors $G_A(Q^2)$, ${\tilde G}_P(Q^2)$
and ${G}_P(Q^2)$ and to understand the reason for the failure of these 
three form factors to satisfy the PCAC relation for $Q^2 \lesssim
0.2$~GeV${}^2$.

\begin{acknowledgments}
We thank the MILC Collaboration for providing the 2+1+1-flavor HISQ
lattices used in our calculations.  We acknowledge Gunnar Bali, Joseph
Carlson, Vincenzo Cirigliano, Sara Collins, Jeremy Green, Richard
Hill, Emanuele Mereghetti, Aaron Meyer and Saori Pastore for
discussions.  Simulations were carried out on computer facilities of
(i) the USQCD Collaboration, which are funded by the Office of Science
of the U.S. Department of Energy, (ii) the National Energy Research
Scientific Computing Center, a DOE Office of Science User Facility
supported by the Office of Science of the U.S. Department of Energy
under Contract No. DE-AC02-05CH11231; (iii) Institutional Computing at
Los Alamos National Laboratory; and (iv) the High Performance
Computing Center at Michigan State University.  The calculations used
the Chroma software suite~\cite{Edwards:2004sx}. This work is
supported by the U.S. Department of Energy, Office of Science of High
Energy Physics under contract number~DE-KA-1401020 and the LANL LDRD
program. The work of H-W. Lin was supported in part by the M. Hildred
Blewett Fellowship of the American Physical Society.
\end{acknowledgments}

%%%%%%%%%%%%%%%%%%%%%%%%%%%%%%%%%%%%%%%%%%%%%%%%%%%%%%%%%%%%%%%%%%

%\clearpage
%\newpage

\appendix

\section{Fits to two-point functions}
\label{sec:appendix1}

\begin{figure*}[tbp]
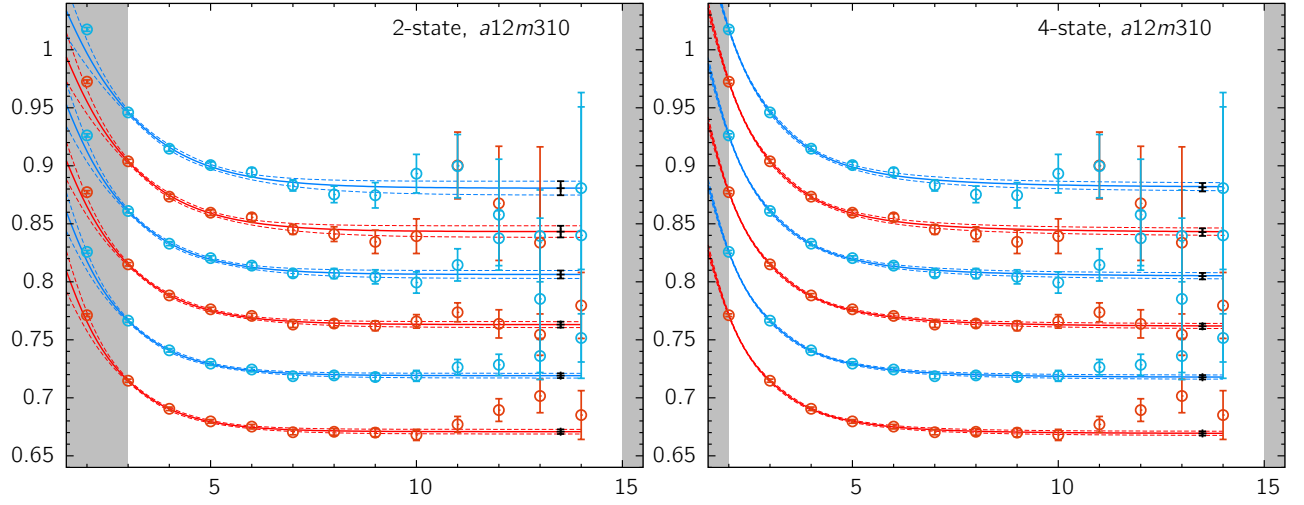

\centering
\includegraphics[width=0.47\linewidth]{{{figsX/meff_a12m310AMA.s2.fr03-15.full}}}
\includegraphics[width=0.47\linewidth]{{{figsX/meff_a12m310AMA.s4.fr02-15.full}}}
\caption{Plot of the effective-energy, $aE_0$, versus the Eucledian
  time $t$ for the $a12m310$ ensemble data.  The left
  panel shows the 2-state fits while the right panel shows the 4-state
  fits.  The lines with error bands show the result for $M_{\rm eff}$
  obtained from the 2-state (4-state) fit for the various momenta
  analyzed.  The unshaded region specifies the range of timeslices
  used in the fits. To help distinguish between the estimates for the
  various momenta, the data and fits for $M_{\rm eff}$ are shown using
  alternating red and blue colors. For each momenta, the point with
  error bars in black on the right of the $t-$interval used in the
  fits is the estimate of the ground-state energy $E_0$.  }
\label{fig:Meffa12m310}
\end{figure*}

\begin{figure*}[tbp]
\centering
  \subfigure{
\includegraphics[width=0.47\linewidth]{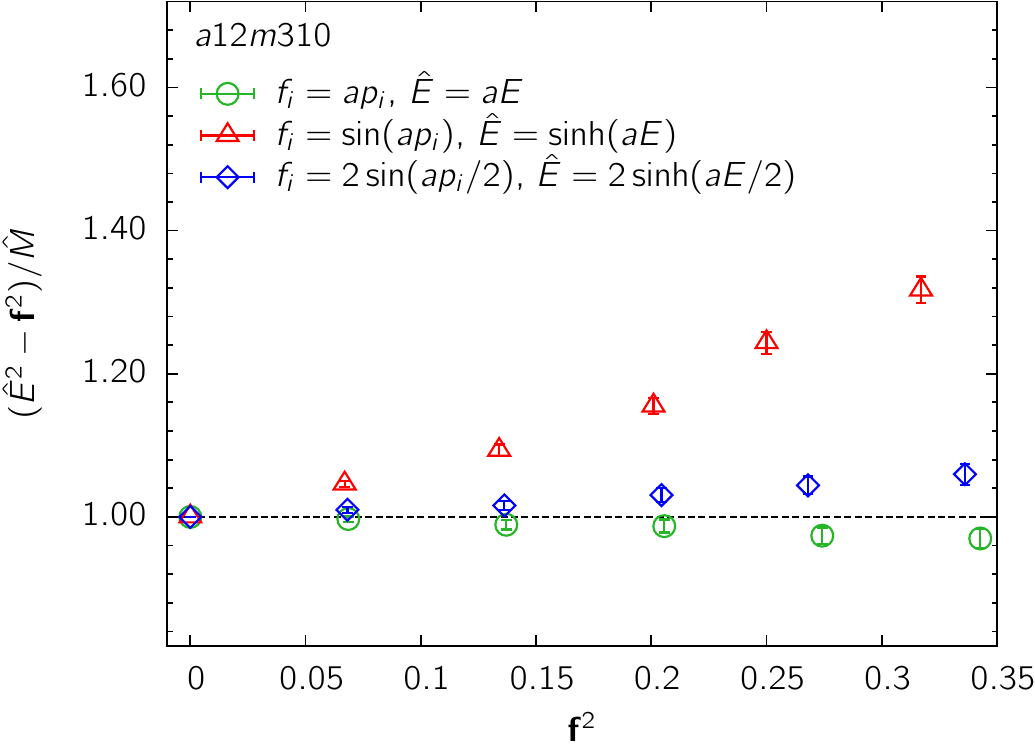}
\includegraphics[width=0.47\linewidth]{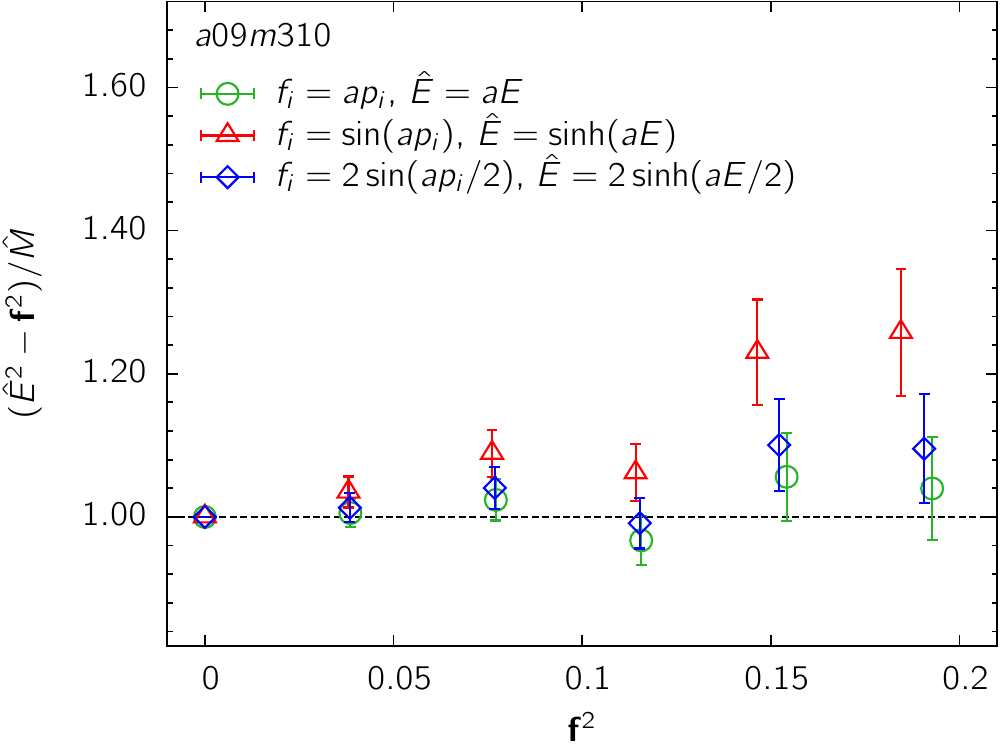}
}
  \subfigure{
\includegraphics[width=0.47\linewidth]{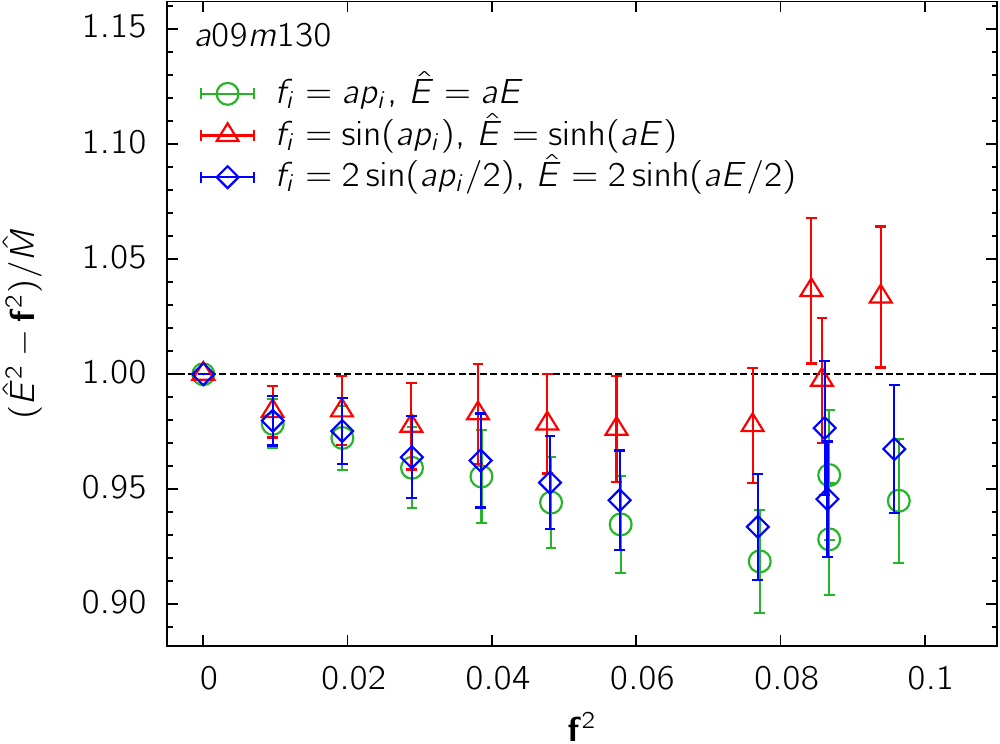}
\includegraphics[width=0.47\linewidth]{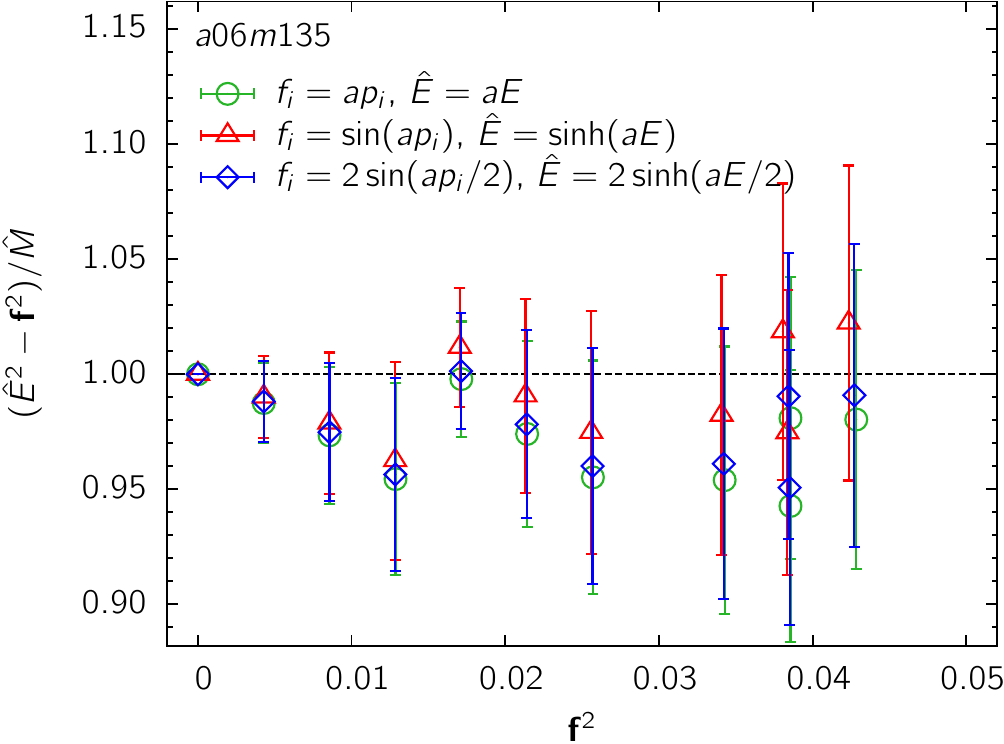}
}
\caption{Tests of the dispersion relation for the nucleon, $(aE)^2 -
  \sum_i f_i^2 = (aM)^2$, for the three cases $f_i = ap_i$,
  $\sin(ap_i)$ and $2\sin(ap_i/2)$. Data for the $a12m310$ and
  $a09m310$ ensembles are shown in the top panels and for the two
  physical mass ensembles, $a09m130$ and $a06m135$, in the bottom
  panels. The ideal behavior is a constant value given by the data at 
  ${\bf p}  = 0$. }
\label{fig:dispersion}
\end{figure*}

This appendix shows the 2- and 4-state fits to the
nucleon two-point correlation function on the eight ensembles 
Figs.~\ref{fig:Meffa12m310},~\ref{fig:Meffa12m220L},~\ref{fig:Meffa09m310},~\ref{fig:Meffa09m220},~\ref{fig:Meffa09m130},~\ref{fig:Meffa06m310},~\ref{fig:Meffa06m220},
and~\ref{fig:Meffa06m135}.  The
estimates of the nucleon energies and the amplitudes extracted from
these fits are collected together in Tables~\ref{tab:multistates-twopt-mom-a12m310AMA-3},
\ref{tab:multistates-twopt-mom-a12m220LAMA-3},
\ref{tab:multistates-twopt-mom-a09m310-3},
\ref{tab:multistates-twopt-mom-a09m220-3},
\ref{tab:multistates-twopt-mom-a09m130LP1-3},
\ref{tab:multistates-twopt-mom-a06m310AMA-3},
\ref{tab:multistates-twopt-mom-a06m220AMA-3} and~\ref{tab:multistates-twopt-mom-a06m130AMA-3}. 
Tests of the dispersion relation for the nucleon, $(aE)^2 - \sum_i f_i^2 = (aM)^2$,
for the three cases $f_i = ap_i$, $\sin(ap_i)$ and $2\sin(ap_i/2)$ are 
shown in Fig.~\ref{fig:dispersion} for four ensembles listed in the labels.

\begin{figure*}[tbp]
\centering
    \includegraphics[width=0.47\linewidth]{{{figsX/meff_a12m220LAMA.s2.fr04-15.full}}}
    \includegraphics[width=0.47\linewidth]{{{figsX/meff_a12m220LAMA.s4.fr02-15.full}}}
\caption{Plot of the effective-energy for the $a12m220L$ ensemble
  data.  The rest is the same as in
  Fig.~\protect\ref{fig:Meffa12m310}. }
\label{fig:Meffa12m220L}
\end{figure*}
%

%\clearpage

\begin{figure*}[tbp]
\centering
    \includegraphics[width=0.47\linewidth]{{{figsX/meff_a09m310.s2.fr05-20.full}}}
    \includegraphics[width=0.47\linewidth]{{{figsX/meff_a09m310.s4.fr03-20.full}}}
\caption{Plot of the effective-energy for the $a09m310$ ensemble data.
  The rest is the same as in Fig.~\protect\ref{fig:Meffa12m310}. }
\label{fig:Meffa09m310}
\end{figure*}

\begin{figure*}[tbp]
\centering
    \includegraphics[width=0.47\linewidth]{{{figsX/meff_a09m220.s2.fr05-20.full}}}
    \includegraphics[width=0.47\linewidth]{{{figsX/meff_a09m220.s4.fr03-20.full}}}
\caption{Plot of the effective-energy for the $a09m220$
  ensemble data.  The rest is the same as in Fig.~\protect\ref{fig:Meffa12m310}. }
\label{fig:Meffa09m220}
\end{figure*}

\begin{figure*}[tbp]
\centering
    \includegraphics[width=0.47\linewidth]{{{figsX/meff_a09m130LP1.s2.fr06-20.full}}}
    \includegraphics[width=0.47\linewidth]{{{figsX/meff_a09m130LP1.s4.fr04-20.full}}}
\caption{Plot of the effective-energy for the $a09m130$ ensemble 
  data.  The rest is the same as in
  Fig.~\protect\ref{fig:Meffa12m310}. }
\label{fig:Meffa09m130}
\end{figure*}
%

%\clearpage

\begin{figure*}[tbp]
\centering
    \includegraphics[width=0.47\linewidth]{{{figsX/meff_a06m310AMA.s2.fr10-30.full}}}
    \includegraphics[width=0.47\linewidth]{{{figsX/meff_a06m310AMA.s4.fr07-30.full}}}
\caption{Plot of the effective-energy for the $a06m310$ ensemble
  data.  The rest is the same as in
  Fig.~\protect\ref{fig:Meffa12m310}. }
\label{fig:Meffa06m310}
\end{figure*}

\begin{figure*}[tbp]
\centering
    \includegraphics[width=0.47\linewidth]{{{figsX/meff_a06m220AMA.s2.fr10-30.full}}}
    \includegraphics[width=0.47\linewidth]{{{figsX/meff_a06m220AMA.s4.fr07-30.full}}}
\caption{Plot of the effective-energy plots for the $a06m220$ ensemble
  data.  The rest is the same as in
  Fig.~\protect\ref{fig:Meffa12m310}. }
\label{fig:Meffa06m220}
\end{figure*}

\begin{figure*}[tbp]
\centering
    \includegraphics[width=0.47\linewidth]{{{figsX/meff_a06m135AMA.s2.fr08-30.full}}}
    \includegraphics[width=0.47\linewidth]{{{figsX/meff_a06m135AMA.s4.fr06-30.full}}}
    \caption{Plot of the effective-energy for the $a06m135$ ensemble 
  data.  The rest is the same as in
  Fig.~\protect\ref{fig:Meffa12m310}. }
\label{fig:Meffa06m135}
\end{figure*}

%

% \clearpage
\input{2pt_Tables/twopt-mom-a12m310AMA-3_All.tex}
\input{2pt_Tables/twopt-mom-a12m220LAMA-3_All.tex}
\input{2pt_Tables/twopt-mom-a09m310-3_All.tex}
%\clearpage
\input{2pt_Tables/twopt-mom-a09m220-3_All.tex}
\input{2pt_Tables/twopt-mom-a09m130LP1-3_All.tex}
\input{2pt_Tables/twopt-mom-a06m310AMA-3_All.tex}
%\clearpage
\input{2pt_Tables/twopt-mom-a06m220AMA-3_All.tex}
\input{2pt_Tables/twopt-mom-a06m130AMA-3_All.tex}

\cleardoublepage
\section{Fits to three-point functions}
\label{sec:appendix2}

\begin{figure*}[tbp]
\centering
\subfigure{
\includegraphics[width=0.47\linewidth]{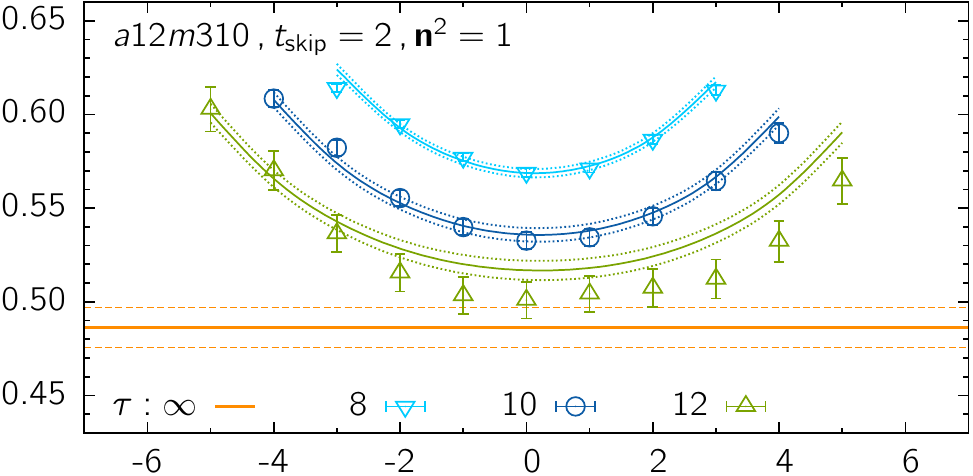}
\includegraphics[width=0.47\linewidth]{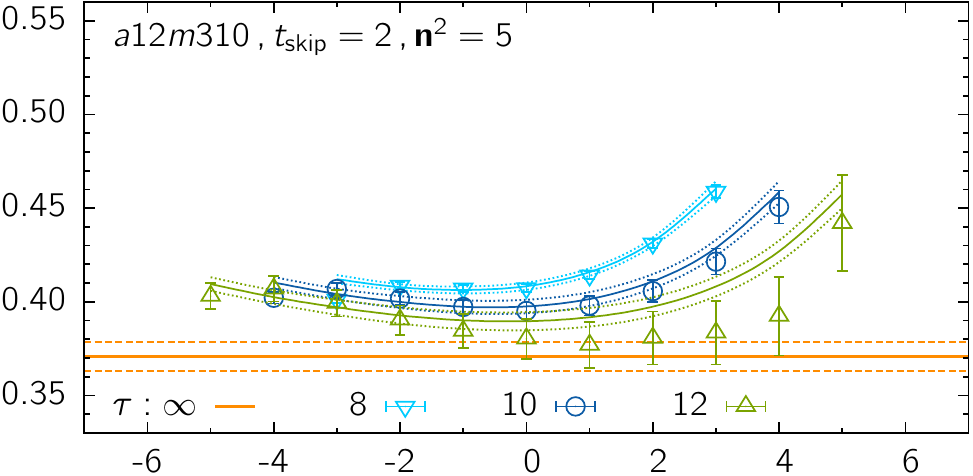}
}
\subfigure{
\includegraphics[width=0.47\linewidth]{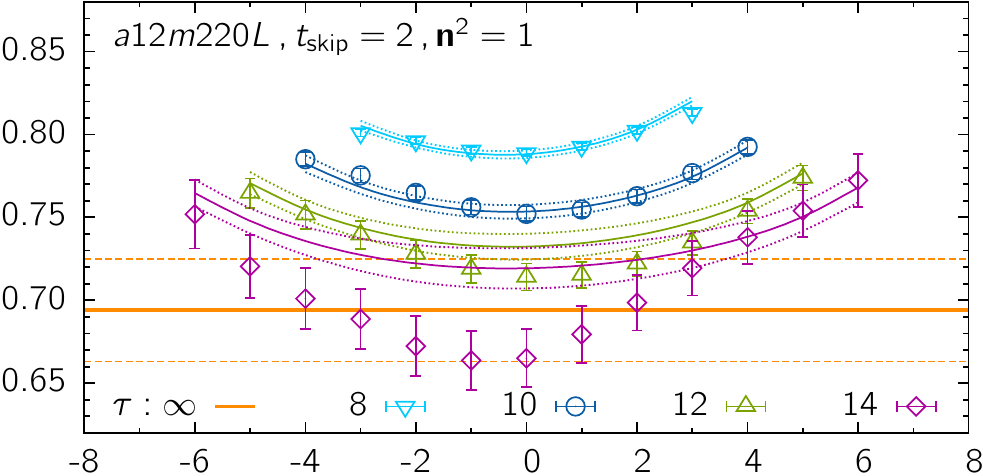}
\includegraphics[width=0.47\linewidth]{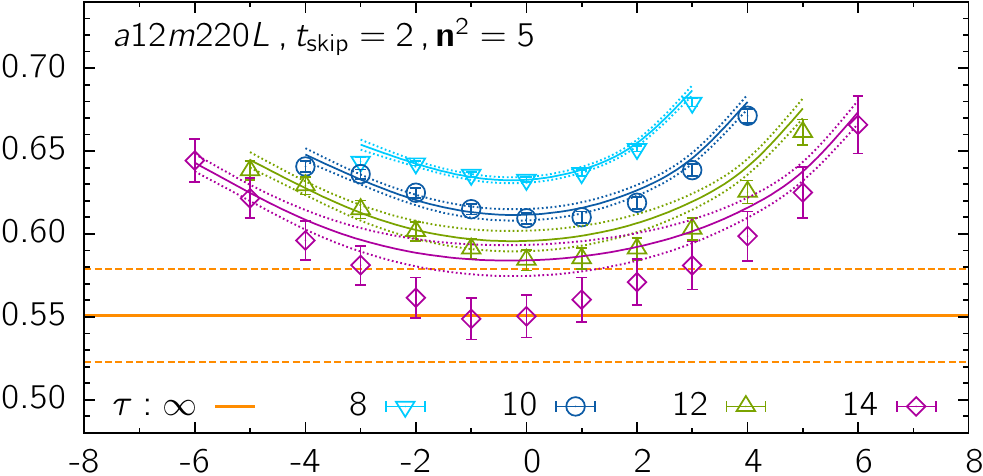}
}
\caption{The three-point data for ${\cal R}_{53}$ defined in 
  Eq.~\protect\eqref{eq:ratio} versus the operator insertion time
  $t$ shifted by $\tau/2$.  The label gives the values of $\tskip$ and $\tau$ used in the
  fits. Prediction of the 2-state fit for various values of $\tau$ is
  shown in the same color as the data. The result for the matrix
  elements in the $\tsepi$ limit is shown by the horizontal band.  The
  figures on top are for the $a12m310$ ensemble, and those on the
  bottom for the $a12m220L$ ensemble.  The plots on the left are for
  the lowest momenta $\bm{p}=(1,0,0) 2\pi/La$, while those on the
  right are for $\bm{p}=(2,1,0) 2\pi/La$.  }
\label{fig:me-a12}
\end{figure*}

\begin{figure*}[tbp]
\centering
\subfigure{
\includegraphics[width=0.47\linewidth]{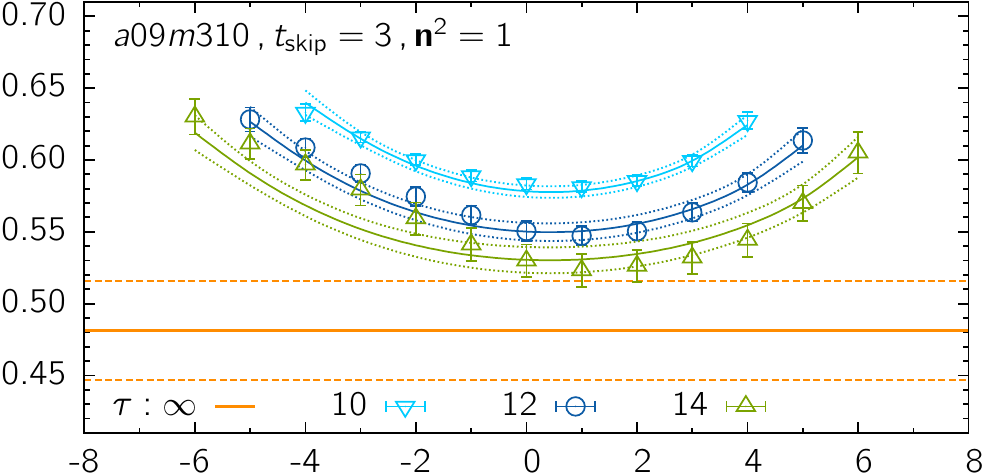}
\includegraphics[width=0.47\linewidth]{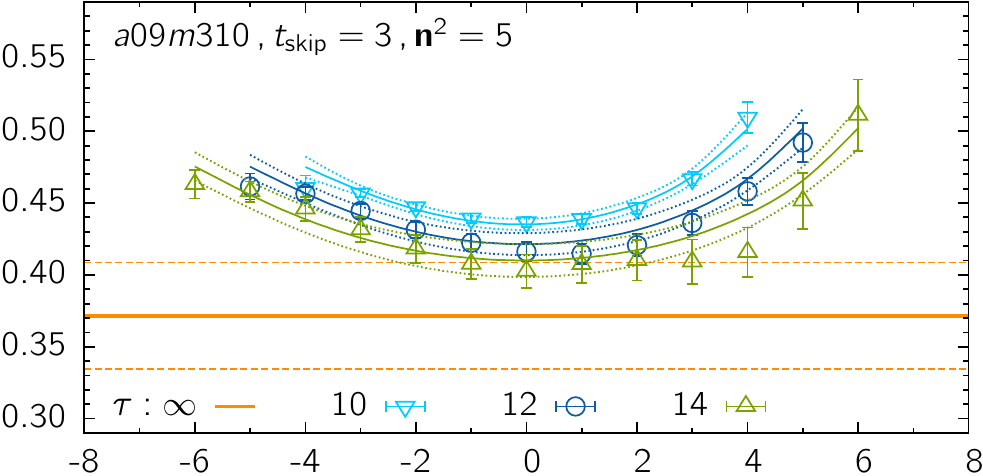}
}
\subfigure{
\includegraphics[width=0.47\linewidth]{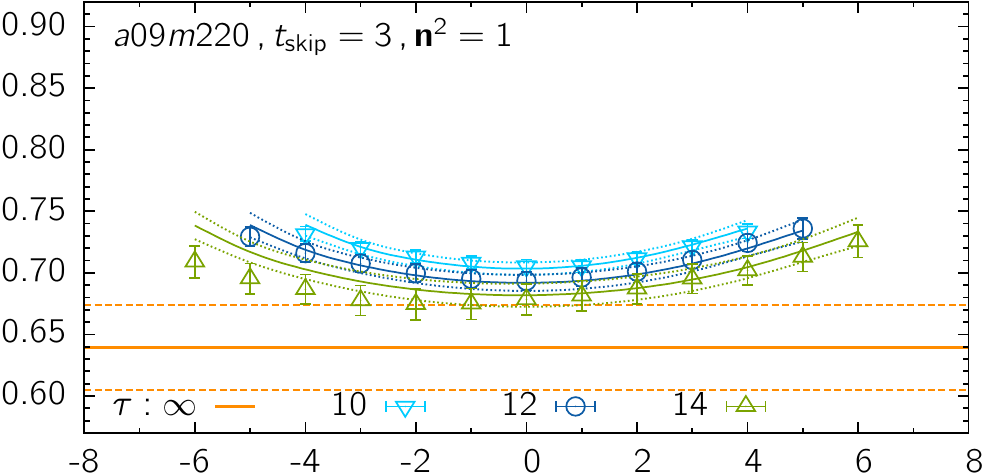}
\includegraphics[width=0.47\linewidth]{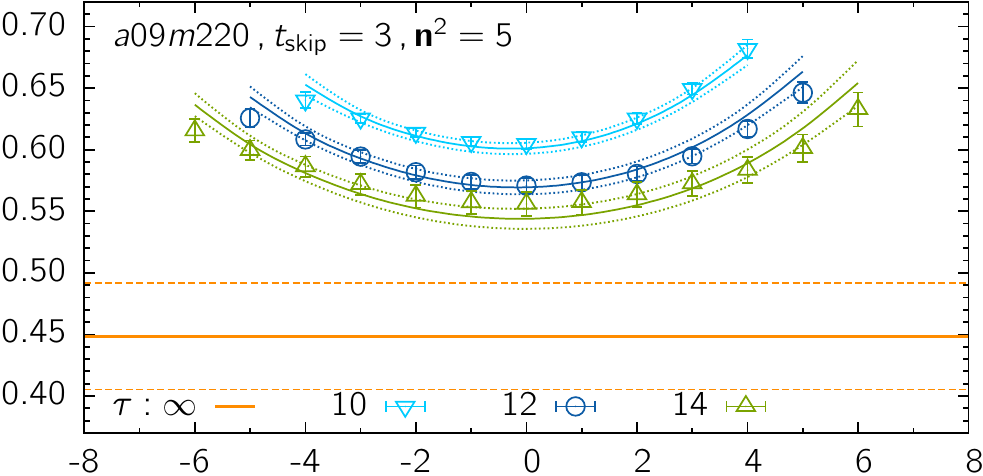}
}
\subfigure{
\includegraphics[width=0.47\linewidth]{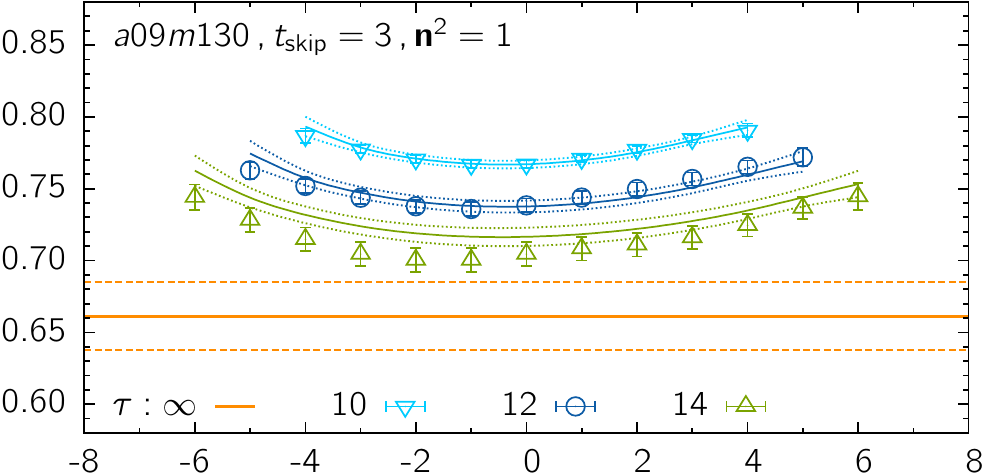}
\includegraphics[width=0.47\linewidth]{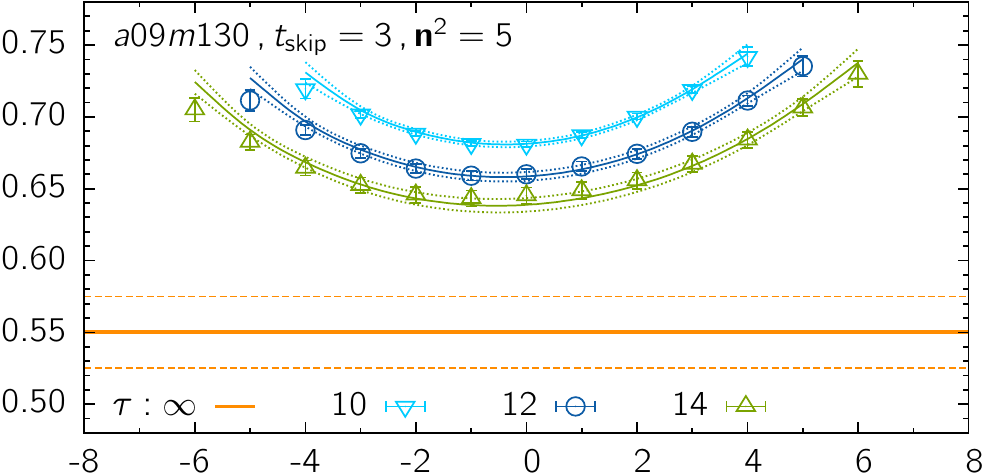}
}
\caption{The three-point data for ${\cal R}_{53}$ defined in
  \protect\eqref{eq:ratio} versus the operator insertion time
  $t$ shifted by $\tau/2$.  The label gives the value of $\tskip$ and $\tau$ used in the
  fits. Prediction of the 2-state fit for various values of $\tau$ is
  shown in the same color as the data. The result for the matrix
  elements in the $\tsepi$ limit is shown by the horizontal band.  The
  plots in the top row are for the $a09m310$ ensemble, middle are for
  the $a09m220$, and those on the bottom row for the $a09m130$
  ensemble.  The plots on the left are for the lowest momenta
  $\bm{p}=(1,0,0) 2\pi/La$, while those on the right are for the
  highest, $\bm{p}=(2,1,0) 2\pi/La$.  }
\label{fig:me-a09}
\end{figure*}

This appendix contains plots of fits to the data for three-point
functions from which the form factors are extracted. The data for the
ratio ${\cal R}_{53}$, defined in Eq.~\eqref{eq:ratio}, is plotted in
Figs.~\ref{fig:me-a12} and~\ref{fig:me-a09} for the $a=0.12$ and
$0.09$~fm ensembles. The horizontal band in these figures gives the
$\tsepi$ value defined in Eq.~\eqref{eq:r2ff-GPGA3}.

\begin{figure*}[tbp]
\centering
\subfigure{
\includegraphics[width=0.32\linewidth]{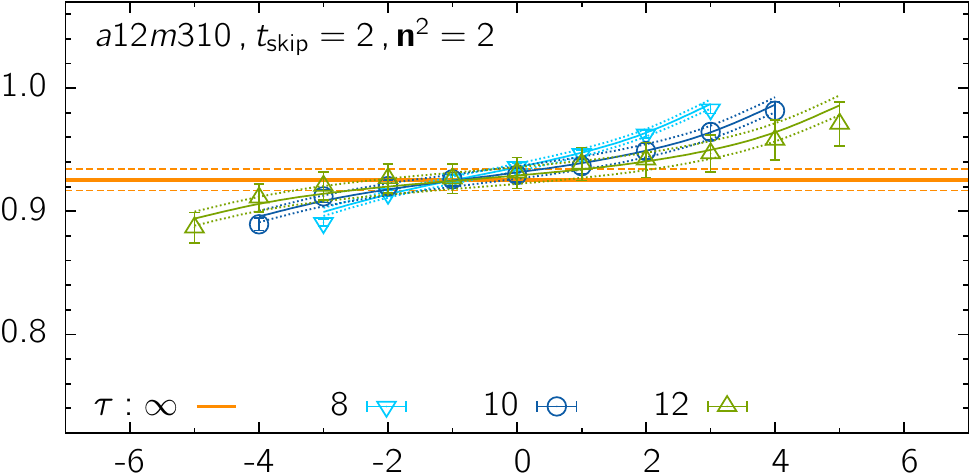}
\includegraphics[width=0.32\linewidth]{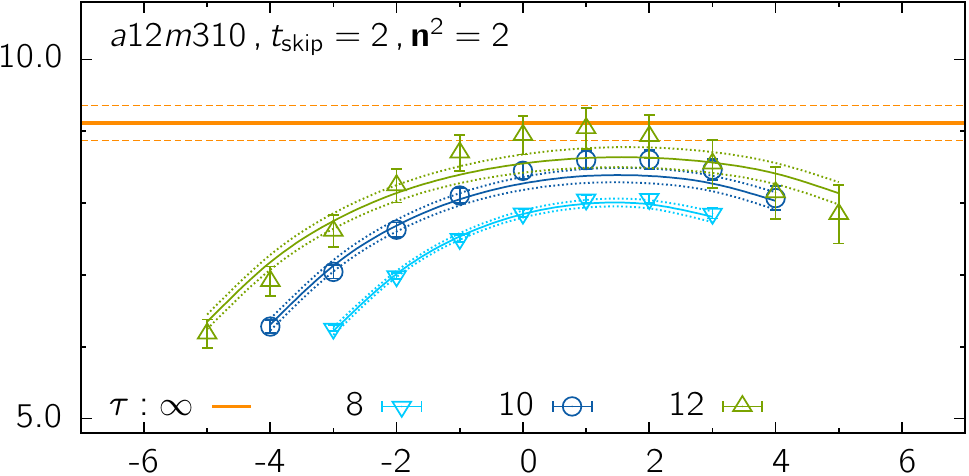}
\includegraphics[width=0.32\linewidth]{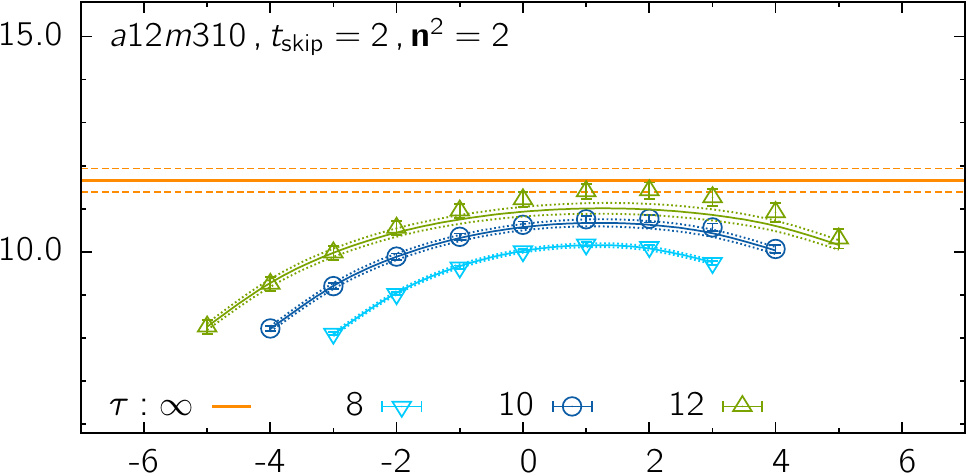}
}
\subfigure{
\includegraphics[width=0.32\linewidth]{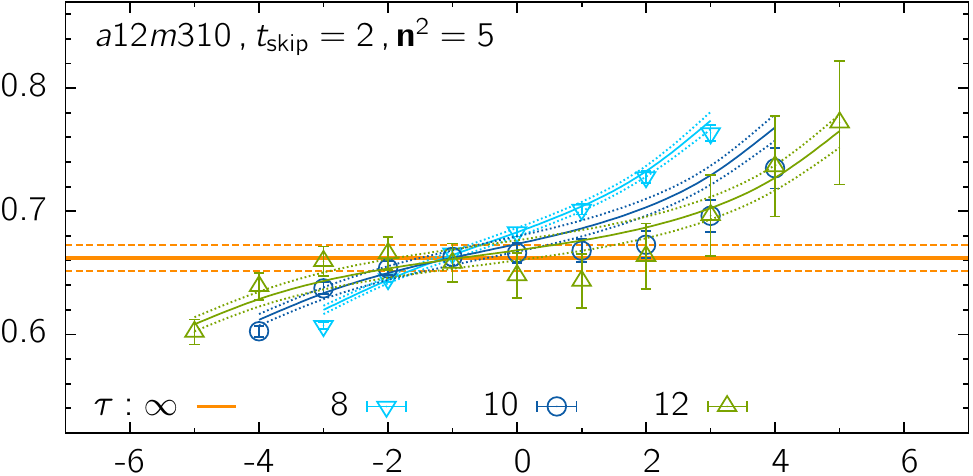}
\includegraphics[width=0.32\linewidth]{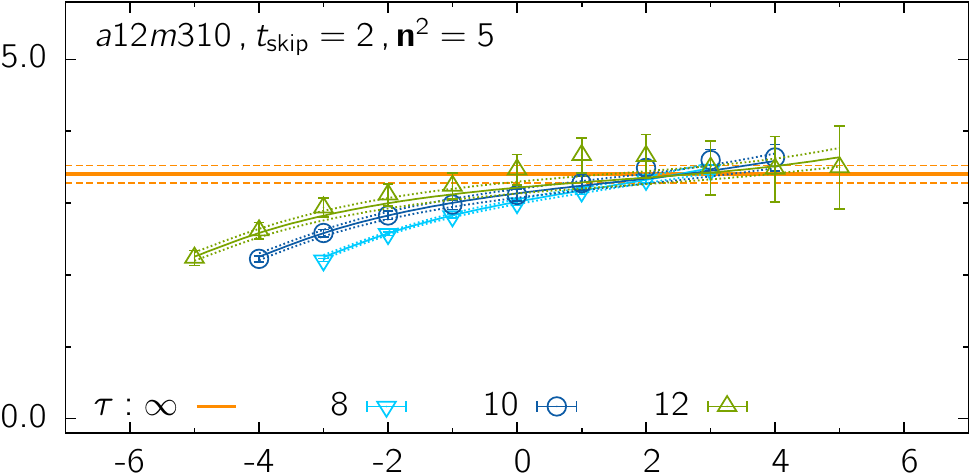}
\includegraphics[width=0.32\linewidth]{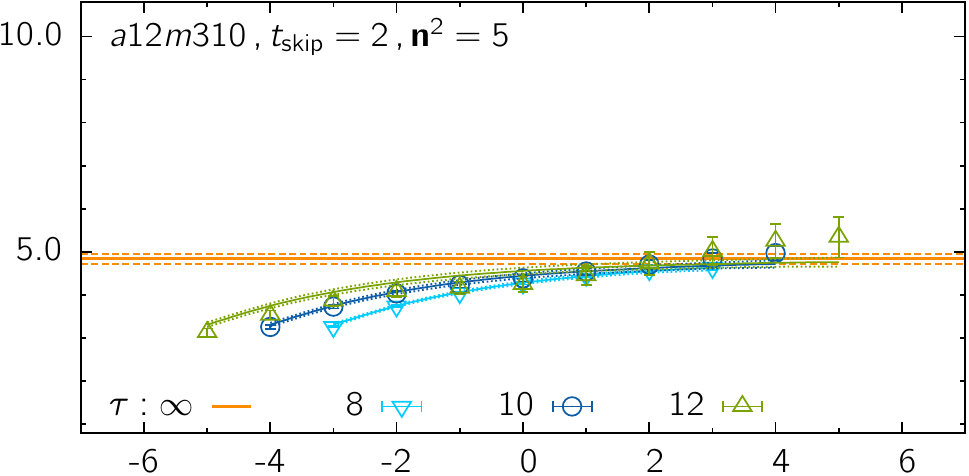}
}
\caption{Plots of the ratios that directly give the form factors:
  $G_A$ from ${\cal R}_{53}$ with $q_3=0$ (left), ${\tilde G}_P$ from
  ${\cal R}_{51}$ (middle), and the pseudoscalar $G_P$ (right) versus
  the operator insertion time $t-\tau/2$ for the $a12m310$
  ensemble. The top row shows data for ${\bf p}^2 = 2 (2\pi/La)^2$ and
  the bottom row for ${\bf p}^2 = 5 (2\pi/La)^2$. }
\label{fig:GP-GA-a12m310}
\end{figure*}

\begin{figure*}[tbp]
\centering
\subfigure{
\includegraphics[width=0.47\linewidth]{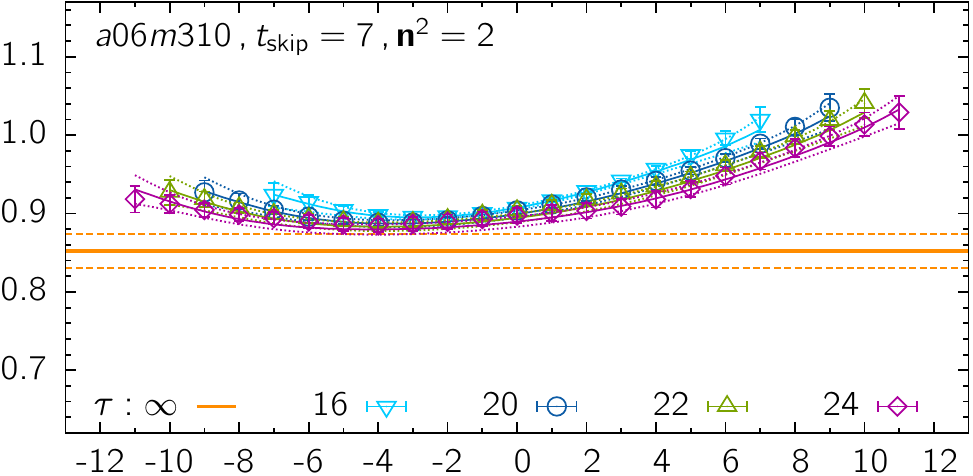}
\includegraphics[width=0.47\linewidth]{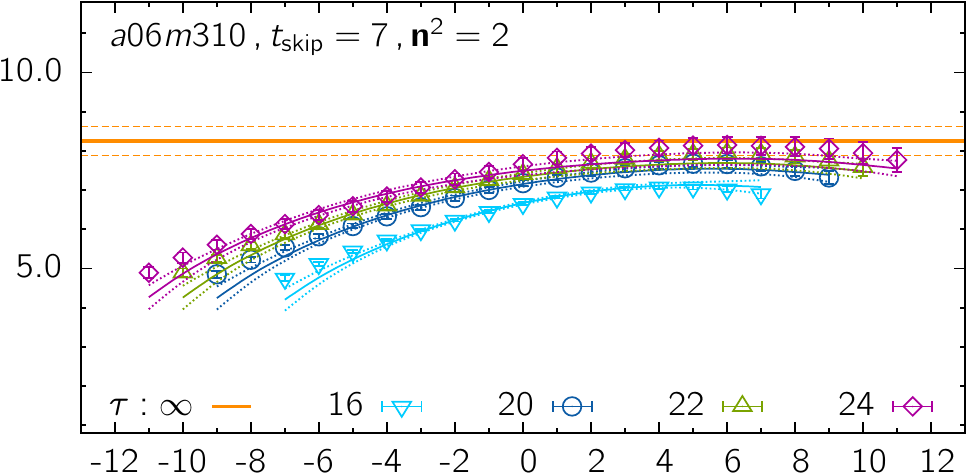}
}
\subfigure{
\includegraphics[width=0.47\linewidth]{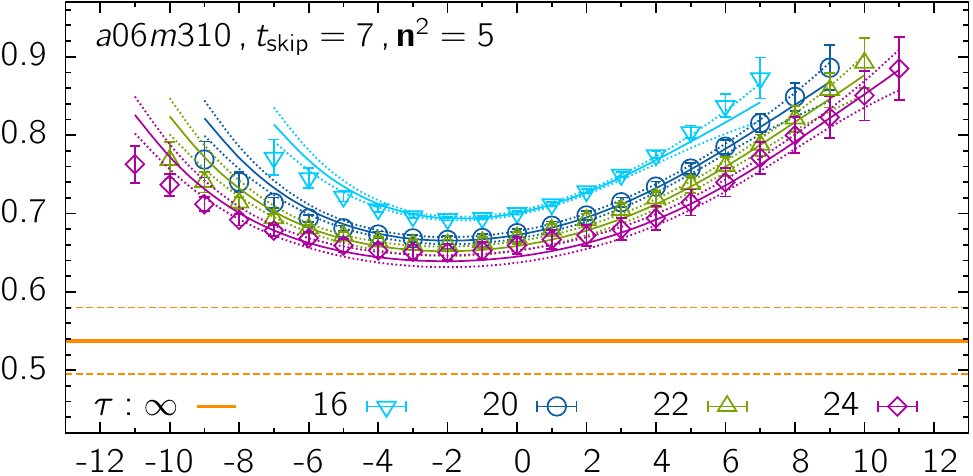}
\includegraphics[width=0.47\linewidth]{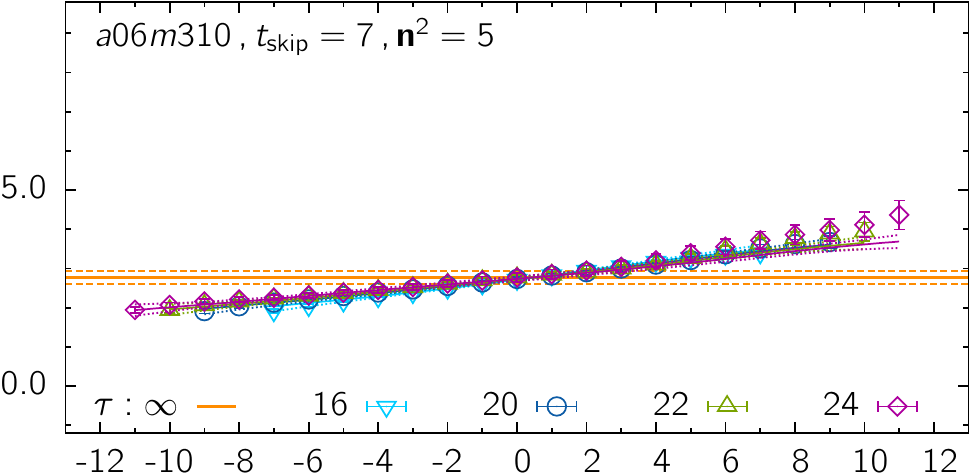}
}
\caption{Plots of the ratios that directly give the form factors:
  $G_A$ from ${\cal R}_{53}$ with $q_3=0$ (left) and ${\tilde G}_P$
  from ${\cal R}_{51}$ (right) versus the operator insertion time
  $t$ shifted by $\tau/2$ for the $a06m310$ ensemble. Data for $G_P(Q^2)$ has not
  been analyzed for this ensemble. The top row shows data for ${\bf p}^2 = 2
  (2\pi/La)^2$ and the bottom row for ${\bf p}^2 = 5 (2\pi/La)^2$.}
\label{fig:GP-GA-a06m310}
\end{figure*}

\begin{figure*}[tbp]
\centering
\subfigure{
\includegraphics[width=0.32\linewidth]{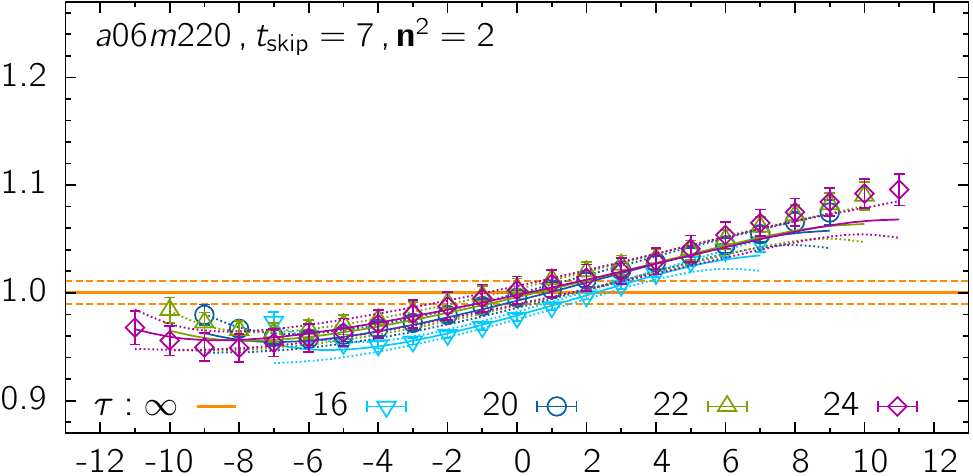}
\includegraphics[width=0.32\linewidth]{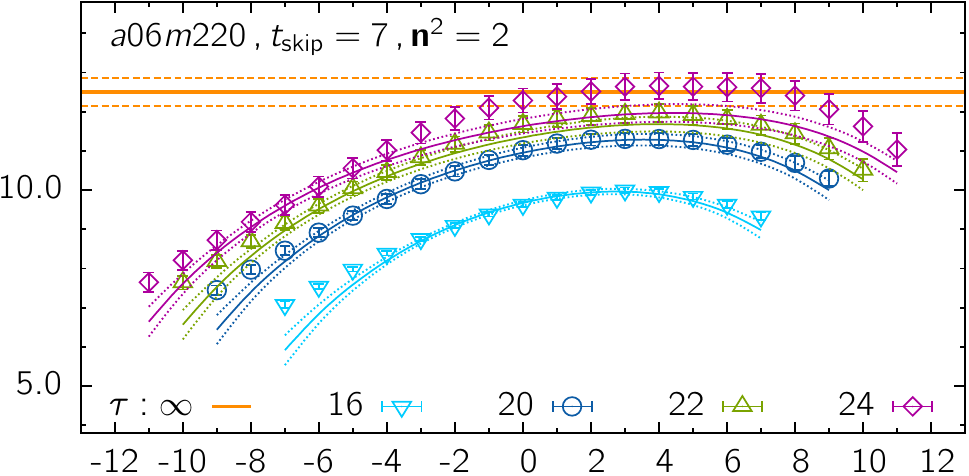}
\includegraphics[width=0.32\linewidth]{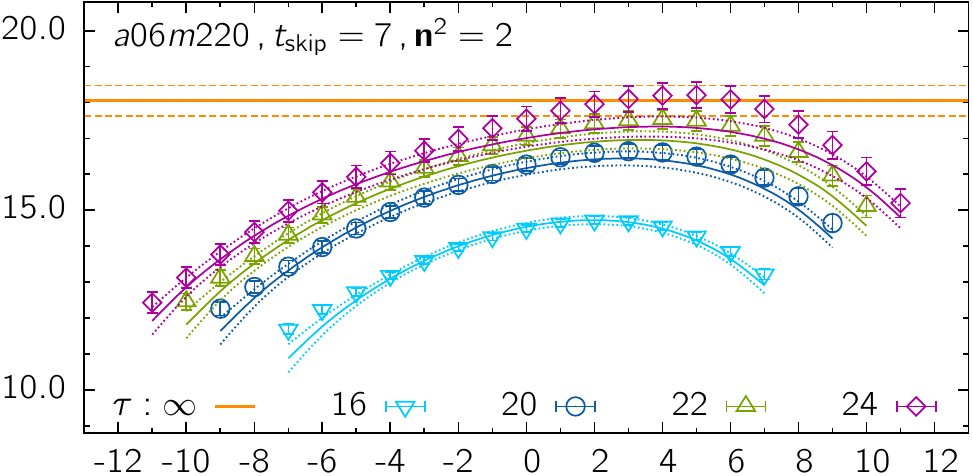}
}
\subfigure{
\includegraphics[width=0.32\linewidth]{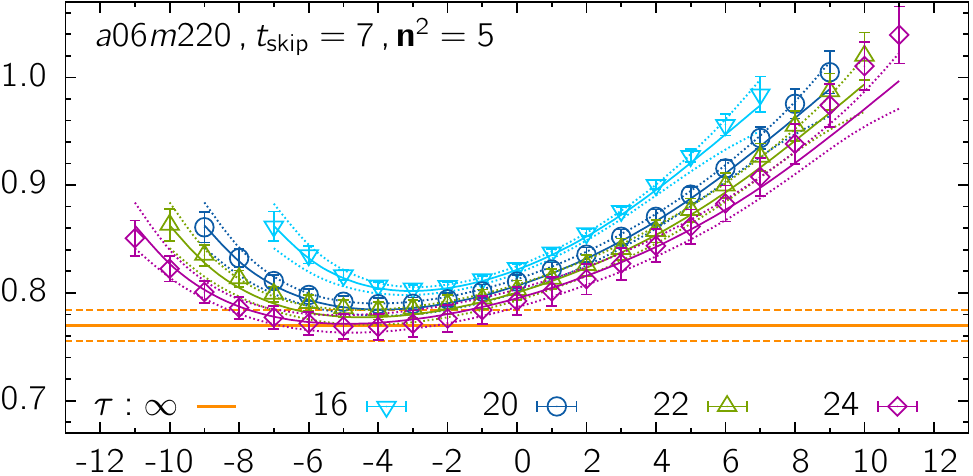}
\includegraphics[width=0.32\linewidth]{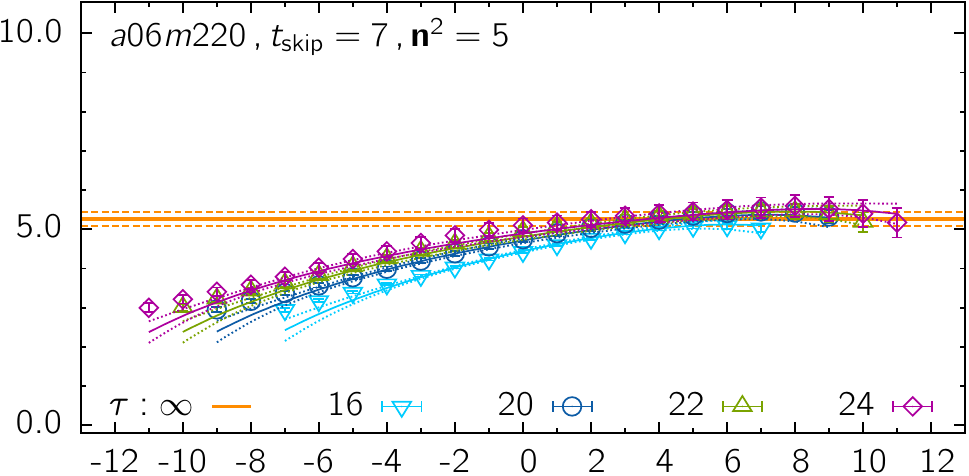}
\includegraphics[width=0.32\linewidth]{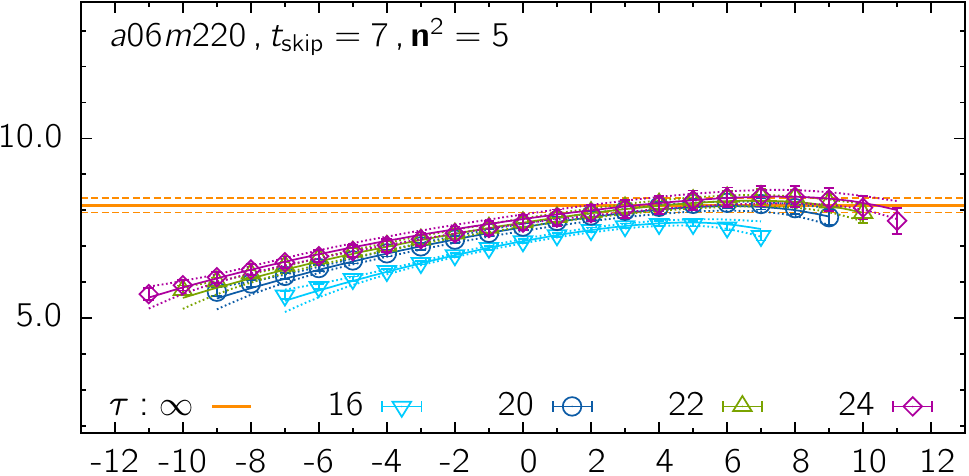}
}
\caption{Plots of the ratios that directly give the form factors:
  $G_A$ from ${\cal R}_{53}$ with $q_3=0$ (left), ${\tilde G}_P$ from
  ${\cal R}_{51}$ (middle), and the pseudoscalar $G_P$ (right) versus
  the operator insertion time $t$ shifted by $\tau/2$ for the
  $a06m220$ ensemble. The top row shows data for ${\bf p}^2 = 2
  (2\pi/La)^2$ and the bottom row for ${\bf p}^2 = 5 (2\pi/La)^2$.  }
\label{fig:GP-GA-a06m220}
\end{figure*}

In Figs.~\ref{fig:GP-GA-a12m310},~\ref{fig:GP-GA-a06m310},
and~\ref{fig:GP-GA-a06m220} we show the data for the ratio ${\cal R}_{53}$
defined in Eq.~\eqref{eq:ratio} versus the operator insertion time $t$
shifted by $\tau/2$.  The data for the ratio ${\cal R}_{53}$ with
$q_3=0$ gives the form factor $G_A$ while ${\cal R}_{51}$ gives 
${\tilde G}_P$.  In the right panels of Figs.~\ref{fig:GP-GA-a12m310}
and~\ref{fig:GP-GA-a06m220}, we also show the fits used to extract the
pseudoscalar form factor $G_P(Q^2)$ using Eqs.~\eqref{eq:ratio}
and~\eqref{eq:r2ff-GP}.  In each case, the horizontal band gives the
value in the limit $t \to \infty$, $\tau \to \infty$ and $\tau-t \to
\infty$ as defined in Eqs.~\eqref{eq:r2ff-GPGA3}
and~\eqref{eq:r2ff-GP}.

\cleardoublepage

%%%%%%%%%%%%%%%%%%%%%%%%%%%%%%%%%%%%%%%%%%%%%%%%%%%%%%%%%%%%%%%%%%%
%-----------
% reference
%-----------
%\bibliographystyle{abbrv} %%% physical review
%% \makeatletter
%% \ifx\@bibitemShut\undefined\let\@bibitemShut\relax\fi
%% \makeatother
\bibliography{ref} %%% ref.bib file

\end{document}

%% file: 2pt_Tables/twopt-mom-a12m310AMA-3_All.tex
\begin{table*}[!htb]
%% \begin{adjustwidth}{-8em}{-8em}
  \centering
  \renewcommand{\arraystretch}{0.9}
  \begin{tabular}{c|c|c|ccccccccc}
    \hline\hline
    \multicolumn{3}{r}{Priors} &
    & & 0.15(10) & 0.4(2) & 0.8(6) & 0.6(3) & 0.6(4) & 0.4(2) & 
    \\\hline
    $\bm{n^2}$ &
    $N_\text{2pt}$ & & 
    ${\cal A}_0 \times 10^{11}$ & $a E_0$ &
    $r_1$ & $a \Delta E_1$ &
    $r_2$ & $a \Delta E_2$ &
    $r_3$ & $a \Delta E_3$ &
    $\chi^2/\text{DOF}$
    \\\hline
\multirow{3}{*}{0}
& 2 & 03 - 15 & $6.86(11) $ & 0.671(002) & 1.011(186) & 0.837(098) &  &  &  &  & 0.916 \\
& 3 & 02 - 15 & $6.78(10) $ & 0.670(002) & 0.143(028) & 0.450(038) & 1.137(063) & 0.563(075) &  &  & 0.747 \\
& 4 & 02 - 15 & $6.75(10) $ & 0.669(002) & 0.137(030) & 0.420(037) & 0.732(038) & 0.500(066) & 0.518(066) & 0.396(023) & 0.738 \\
\hline
\multirow{3}{*}{1}
& 2 & 03 - 15 & $5.20(08) $ & 0.719(002) & 1.026(172) & 0.807(092) &  &  &  &  & 0.763 \\
& 3 & 02 - 15 & $5.15(08) $ & 0.718(002) & 0.152(029) & 0.438(039) & 1.204(066) & 0.566(076) &  &  & 0.652 \\
& 4 & 02 - 15 & $5.12(08) $ & 0.718(002) & 0.147(030) & 0.409(038) & 0.775(039) & 0.502(067) & 0.551(065) & 0.401(023) & 0.620 \\
\hline
\multirow{3}{*}{2}
& 2 & 03 - 15 & $3.93(08) $ & 0.763(003) & 1.008(151) & 0.751(090) &  &  &  &  & 0.802 \\
& 3 & 02 - 15 & $3.89(07) $ & 0.762(002) & 0.174(031) & 0.400(042) & 1.294(075) & 0.604(079) &  &  & 0.711 \\
& 4 & 02 - 15 & $3.86(08) $ & 0.762(002) & 0.169(033) & 0.368(040) & 0.826(043) & 0.537(071) & 0.609(065) & 0.409(021) & 0.636 \\
\hline
\multirow{3}{*}{3}
& 2 & 03 - 15 & $3.03(08) $ & 0.806(003) & 1.053(170) & 0.744(104) &  &  &  &  & 0.536 \\
& 3 & 02 - 15 & $3.00(06) $ & 0.806(003) & 0.176(029) & 0.394(039) & 1.376(077) & 0.608(070) &  &  & 0.522 \\
& 4 & 02 - 15 & $2.96(07) $ & 0.805(003) & 0.174(030) & 0.357(037) & 0.888(047) & 0.550(064) & 0.641(058) & 0.407(020) & 0.426 \\
\hline
\multirow{3}{*}{4}
& 2 & 03 - 15 & $2.30(09) $ & 0.843(005) & 1.085(173) & 0.703(118) &  &  &  &  & 1.652 \\
& 3 & 02 - 15 & $2.30(05) $ & 0.844(003) & 0.181(024) & 0.384(033) & 1.467(076) & 0.598(058) &  &  & 1.416 \\
& 4 & 02 - 15 & $2.27(06) $ & 0.843(003) & 0.181(025) & 0.342(029) & 0.958(050) & 0.545(056) & 0.675(051) & 0.406(020) & 1.294 \\
\hline
\multirow{3}{*}{5}
& 2 & 03 - 15 & $1.76(09) $ & 0.881(006) & 1.106(161) & 0.667(121) &  &  &  &  & 0.697 \\
& 3 & 02 - 15 & $1.78(04) $ & 0.883(003) & 0.185(022) & 0.377(031) & 1.541(078) & 0.592(056) &  &  & 0.758 \\
& 4 & 02 - 15 & $1.75(05) $ & 0.882(004) & 0.186(023) & 0.334(027) & 1.014(053) & 0.541(055) & 0.703(048) & 0.406(019) & 0.612 \\
 \hline\hline
  \end{tabular}
%% \end{adjustwidth}
  \caption{Results of the 2-, 3- and 4-state fits to the two-point
    nucleon correlator for the a12m310 ensemble.  The lattice momenta
    are $\bm{p}a$ = $2\pi\bm{n}/L$ with $\bm{n}^2$ listed in the first
    column. The third column gives the fit range $t_\text{min} -
    t_\text{max}$.  Priors, given in the first row, were used for the
    multistate fit when the number of states $N_{2\text{pt}}\geq 3$.
  }
\label{tab:multistates-twopt-mom-a12m310AMA-3}
\end{table*}

%% file: 2pt_Tables/twopt-mom-a12m220LAMA-3_All.tex
\begin{table*}[!htb]
%% \begin{adjustwidth}{-8em}{-8em}
  \centering
  \renewcommand{\arraystretch}{0.9}
  \begin{tabular}{c|c|c|ccccccccc}
    \hline\hline
    \multicolumn{3}{r}{Priors} &
    & & 0.4(3) & 0.3(2) & 1.0(8) & 0.8(4) & 0.8(6) & 0.4(2) & 
    \\\hline
    $\bm{n^2}$ &
    $N_\text{2pt}$ & & 
    ${\cal A}_0  \times 10^{11}$ & $a E_0$ &
    $r_1$ & $a \Delta E_1$ &
    $r_2$ & $a \Delta E_2$ &
    $r_3$ & $a \Delta E_3$ &
    $\chi^2/\text{DOF}$
    \\\hline
\multirow{3}{*}{0}
& 2 & 04 - 15 & $5.97(18) $ & 0.612(003) & 0.669(118) & 0.529(100) &  &  &  &  & 1.363 \\
& 3 & 02 - 15 & $5.75(22) $ & 0.609(003) & 0.400(067) & 0.350(071) & 1.461(171) & 0.878(102) &  &  & 0.885 \\
& 4 & 02 - 15 & $5.74(23) $ & 0.609(003) & 0.400(091) & 0.349(085) & 0.873(099) & 0.775(117) & 0.725(107) & 0.405(010) & 0.881 \\
\hline
\multirow{3}{*}{1}
& 2 & 04 - 15 & $5.30(19) $ & 0.630(003) & 0.638(086) & 0.475(092) &  &  &  &  & 1.497 \\
& 3 & 02 - 15 & $5.05(26) $ & 0.626(004) & 0.417(065) & 0.309(075) & 1.573(178) & 0.909(093) &  &  & 0.945 \\
& 4 & 02 - 15 & $5.04(29) $ & 0.626(004) & 0.417(077) & 0.306(089) & 0.944(109) & 0.806(104) & 0.773(101) & 0.402(010) & 0.931 \\
\hline
\multirow{3}{*}{2}
& 2 & 04 - 15 & $4.73(20) $ & 0.647(004) & 0.631(068) & 0.440(088) &  &  &  &  & 1.620 \\
& 3 & 02 - 15 & $4.50(24) $ & 0.644(004) & 0.441(067) & 0.292(066) & 1.661(188) & 0.930(090) &  &  & 1.047 \\
& 4 & 02 - 15 & $4.48(28) $ & 0.644(005) & 0.444(076) & 0.288(079) & 1.005(122) & 0.831(100) & 0.813(101) & 0.399(010) & 1.024 \\
\hline
\multirow{3}{*}{3}
& 2 & 04 - 15 & $4.19(23) $ & 0.664(005) & 0.638(053) & 0.404(087) &  &  &  &  & 1.740 \\
& 3 & 02 - 15 & $3.99(21) $ & 0.661(004) & 0.475(072) & 0.278(053) & 1.761(201) & 0.954(089) &  &  & 1.164 \\
& 4 & 02 - 15 & $3.95(27) $ & 0.660(005) & 0.484(083) & 0.272(066) & 1.078(138) & 0.860(098) & 0.861(104) & 0.397(010) & 1.125 \\
\hline
\multirow{3}{*}{4}
& 2 & 04 - 15 & $3.88(22) $ & 0.684(005) & 0.626(064) & 0.421(100) &  &  &  &  & 1.438 \\
& 3 & 02 - 15 & $3.71(20) $ & 0.681(004) & 0.445(074) & 0.287(059) & 1.777(209) & 0.946(091) &  &  & 0.973 \\
& 4 & 02 - 15 & $3.68(25) $ & 0.681(005) & 0.453(085) & 0.280(071) & 1.090(143) & 0.854(101) & 0.863(109) & 0.396(010) & 0.940 \\
\hline
\multirow{3}{*}{5}
& 2 & 04 - 15 & $3.46(22) $ & 0.700(006) & 0.637(058) & 0.396(095) &  &  &  &  & 1.406 \\
& 3 & 02 - 15 & $3.35(17) $ & 0.698(004) & 0.467(075) & 0.288(049) & 1.846(220) & 0.961(091) &  &  & 1.003 \\
& 4 & 02 - 15 & $3.31(21) $ & 0.698(005) & 0.479(087) & 0.282(060) & 1.143(154) & 0.874(102) & 0.897(112) & 0.394(009) & 0.960 \\

    \hline\hline
  \end{tabular}
%% \end{adjustwidth}
  \caption{Results of the 2-, 3- and 4-state fits to the two-point
    nucleon correlator for the a12m220L ensemble. 
    The rest is the same as in
    Table.~\protect\ref{tab:multistates-twopt-mom-a12m310AMA-3}.}
  \label{tab:multistates-twopt-mom-a12m220LAMA-3}
\end{table*}

%% file: 2pt_Tables/twopt-mom-a09m310-3_All.tex
\begin{table*}[!htb]
%%\begin{adjustwidth}{-8em}{-8em}
  \centering
  \renewcommand{\arraystretch}{0.9}
  \begin{tabular}{c|c|c|ccccccccc}
    \hline\hline
    \multicolumn{3}{r}{Priors} &
    & & 0.8(4) & 0.3(2) & 1.3(1.0) & 0.70(35) & 1.1(8) & 0.4(2) & 
    \\\hline
    $\bm{n^2}$ &
    $N_\text{2pt}$ & & 
    ${\cal A}_0  \times 10^{11}$ & $a E_0$ &
    $r_1$ & $a \Delta E_1$ &
    $r_2$ & $a \Delta E_2$ &
    $r_3$ & $a \Delta E_3$ &
    $\chi^2/\text{DOF}$
    \\\hline
\multirow{3}{*}{0}
& 2 & 05 - 20 & $13.3(1.2) $ & 0.493(007) & 0.943(097) & 0.331(078) &  &  &  &  & 1.268 \\
& 3 & 03 - 20 & $13.1(1.2) $ & 0.492(006) & 0.752(103) & 0.284(074) & 1.644(378) & 0.688(083) &  &  & 0.980 \\
& 4 & 03 - 20 & $13.1(1.4) $ & 0.492(007) & 0.796(105) & 0.287(084) & 1.187(261) & 0.686(093) & 1.018(187) & 0.404(012) & 0.976 \\
\hline
\multirow{3}{*}{1}
& 2 & 05 - 20 & $11.5(1.0) $ & 0.532(006) & 0.948(107) & 0.334(077) &  &  &  &  & 1.281 \\
& 3 & 03 - 20 & $11.4(9)   $ & 0.531(005) & 0.764(099) & 0.294(062) & 1.715(375) & 0.713(073) &  &  & 0.994 \\
& 4 & 03 - 20 & $11.3(9)   $ & 0.531(006) & 0.803(101) & 0.294(066) & 1.269(264) & 0.719(071) & 1.090(171) & 0.400(014) & 0.976 \\
\hline
\multirow{3}{*}{2}
& 2 & 05 - 20 & $10.3(9)   $ & 0.571(007) & 0.927(146) & 0.361(098) &  &  &  &  & 1.383 \\
& 3 & 03 - 20 & $10.2(7)   $ & 0.571(005) & 0.738(101) & 0.321(066) & 1.665(375) & 0.712(069) &  &  & 1.065 \\
& 4 & 03 - 20 & $10.1(8)   $ & 0.570(006) & 0.773(107) & 0.316(070) & 1.271(264) & 0.730(057) & 1.100(160) & 0.401(014) & 1.048 \\
\hline
\multirow{3}{*}{3}
& 2 & 05 - 20 & $6.75(2.67) $ & 0.586(023) & 1.210(662) & 0.229(096) &  &  &  &  & 0.738 \\
& 3 & 03 - 20 & $7.57(71)   $ & 0.594(006) & 0.820(161) & 0.226(036) & 1.894(326) & 0.691(071) &  &  & 0.618 \\
& 4 & 03 - 20 & $7.22(60)   $ & 0.591(006) & 0.900(152) & 0.213(034) & 1.413(243) & 0.672(085) & 1.135(178) & 0.393(014) & 0.586 \\
\hline
\multirow{3}{*}{4}
& 2 & 05 - 20 & $8.54(1.33) $ & 0.646(014) & 0.985(850) & 0.484(381) &  &  &  &  & 0.653 \\
& 3 & 03 - 20 & $8.18(74)   $ & 0.642(008) & 0.654(105) & 0.362(099) & 1.621(338) & 0.677(065) &  &  & 0.506 \\
& 4 & 03 - 20 & $7.93(1.15) $ & 0.640(011) & 0.667(131) & 0.326(149) & 1.305(273) & 0.707(045) & 1.100(157) & 0.399(015) & 0.493 \\
\hline
\multirow{3}{*}{5}
& 2 & 05 - 20 & $6.88(1.79) $ & 0.670(020) & 0.858(317) & 0.374(326) &  &  &  &  & 0.491 \\
& 3 & 03 - 20 & $6.86(75)   $ & 0.670(009) & 0.672(112) & 0.338(097) & 1.646(329) & 0.685(061) &  &  & 0.383 \\
& 4 & 03 - 20 & $6.54(1.16) $ & 0.667(013) & 0.694(148) & 0.293(138) & 1.341(281) & 0.710(046) & 1.121(160) & 0.397(015) & 0.365 \\
\hline\hline
  \end{tabular}
%%\end{adjustwidth}
  \caption{Results of the 2-, 3- and 4-state fits to the two-point
    nucleon correlator for the a09m310 ensemble. 
    The rest is the same as in
    Table.~\protect\ref{tab:multistates-twopt-mom-a12m310AMA-3}.} 
  \label{tab:multistates-twopt-mom-a09m310-3}
\end{table*}

%% file: 2pt_Tables/twopt-mom-a09m220-3_All.tex
\begin{table*}[!htb]
%% \begin{adjustwidth}{-8em}{-8em}
  \centering
  \renewcommand{\arraystretch}{0.9}
  \begin{tabular}{c|c|c|ccccccccc}
    \hline\hline
    \multicolumn{3}{r}{Priors} &
    & & 0.8(4) & 0.3(2) & 1.7(1.2) & 0.6(3) & 1.5(1.0) & 0.4(2) & 
    \\\hline
    $\bm{n^2}$ &
    $N_\text{2pt}$ & & 
    $A_0 \times 10^{11}$ & $E_0$ &
    $r_1$ & $\Delta E_1$ &
    $r_2$ & $\Delta E_2$ &
    $r_3$ & $\Delta E_3$ &
    $\chi^2/\text{DOF}$
    \\\hline
\multirow{3}{*}{0}
\multirow{3}{*}{0}
& 2 & 05 - 20 & $11.5(10)$   & 0.452(006) & 1.139(133) & 0.361(078) &  &  &  &  & 0.753 \\
& 3 & 03 - 20 & $11.0(10)$   & 0.450(006) & 0.805(116) & 0.270(054) & 2.247(423) & 0.668(077) &  &  & 0.541 \\
& 4 & 03 - 20 & $10.8(11)$   & 0.448(007) & 0.847(123) & 0.264(063) & 1.518(292) & 0.630(085) & 1.414(263) & 0.405(014) & 0.522 \\
\hline
\multirow{3}{*}{1}
& 2 & 05 - 20 & $10.6(11)$   & 0.470(007) & 1.124(122) & 0.340(083) &  &  &  &  & 0.783 \\
& 3 & 03 - 20 & $9.85(1.04)$ & 0.466(007) & 0.804(139) & 0.239(055) & 2.383(424) & 0.668(073) &  &  & 0.546 \\
& 4 & 03 - 20 & $9.47(1.27)$ & 0.464(008) & 0.854(153) & 0.225(067) & 1.623(293) & 0.619(088) & 1.455(258) & 0.400(016) & 0.516 \\
\hline
\multirow{3}{*}{2}
& 2 & 05 - 20 & $9.88(1.15)$ & 0.489(008) & 1.099(124) & 0.333(090) &  &  &  &  & 0.860 \\
& 3 & 03 - 20 & $9.21(1.02)$ & 0.485(007) & 0.787(146) & 0.233(055) & 2.416(425) & 0.670(071) &  &  & 0.607 \\
& 4 & 03 - 20 & $8.75(1.19)$ & 0.482(007) & 0.848(161) & 0.213(063) & 1.664(288) & 0.620(086) & 1.477(253) & 0.397(016) & 0.573 \\
\hline
\multirow{3}{*}{3}
& 2 & 05 - 20 & $9.18(1.28)$ & 0.506(010) & 1.068(130) & 0.319(099) &  &  &  &  & 0.935 \\
& 3 & 03 - 20 & $8.61(97)$   & 0.503(007) & 0.770(155) & 0.225(052) & 2.457(419) & 0.673(070) &  &  & 0.677 \\
& 4 & 03 - 20 & $8.07(97)$   & 0.500(007) & 0.847(161) & 0.202(053) & 1.714(271) & 0.624(084) & 1.505(242) & 0.394(016) & 0.639 \\
\hline
\multirow{3}{*}{4}
& 2 & 05 - 20 & $8.68(1.32)$ & 0.524(011) & 1.072(139) & 0.327(111) &  &  &  &  & 0.890 \\
& 3 & 03 - 20 & $8.38(95)$   & 0.522(007) & 0.755(145) & 0.248(060) & 2.375(430) & 0.670(072) &  &  & 0.661 \\
& 4 & 03 - 20 & $7.89(1.01)$ & 0.519(008) & 0.825(162) & 0.222(060) & 1.675(288) & 0.633(081) & 1.497(249) & 0.397(015) & 0.627 \\
\hline
\multirow{3}{*}{5}
& 2 & 05 - 20 & $7.65(1.52)$ & 0.536(013) & 1.115(210) & 0.292(103) &  &  &  &  & 1.027 \\
& 3 & 03 - 20 & $7.56(77)$   & 0.537(007) & 0.793(160) & 0.232(043) & 2.445(397) & 0.674(068) &  &  & 0.779 \\
& 4 & 03 - 20 & $7.11(69)$   & 0.533(006) & 0.876(160) & 0.211(041) & 1.724(267) & 0.635(080) & 1.529(235) & 0.395(015) & 0.738 \\
    \hline\hline
  \end{tabular}
%% \end{adjustwidth}
  \caption{Results of the 2-, 3- and 4-state fits to the two-point
    nucleon correlator for the a09m220 ensemble. 
    The rest is the same as in
    Table.~\protect\ref{tab:multistates-twopt-mom-a12m310AMA-3}.}
  \label{tab:multistates-twopt-mom-a09m220-3}
\end{table*}

%% file: 2pt_Tables/twopt-mom-a09m130LP1-3_All.tex
\begin{table*}[!htb]
%% \begin{adjustwidth}{-8em}{-8em}
  \centering
  \renewcommand{\arraystretch}{0.9}
  \begin{tabular}{c|c|c|ccccccccc}
    \hline\hline
    \multicolumn{3}{r}{Priors} &
    & & 1.0(5) & 0.20(15) & 2.0(1.5) & 0.6(3) & 1.7(1.2) & 0.4(2) & 
    \\\hline
    $\bm{n^2}$ &
    $N_\text{2pt}$ & & 
    ${\cal A}_0 \times 10^{11}$ & $a E_0$ &
    $r_1$ & $a \Delta E_1$ &
    $r_2$ & $a \Delta E_2$ &
    $r_3$ & $a \Delta E_3$ &
    $\chi^2/\text{DOF}$
    \\\hline
\multirow{3}{*}{0}
& 2 & 06 - 20 & $9.66(53) $ & 0.419(004) & 1.332(117) & 0.353(049) &  &  &  &  & 0.627 \\
& 3 & 04 - 20 & $8.79(67) $ & 0.414(005) & 1.032(090) & 0.253(039) & 2.736(519) & 0.703(059) &  &  & 0.684 \\
& 4 & 04 - 20 & $8.81(67) $ & 0.414(005) & 1.065(085) & 0.259(039) & 2.093(379) & 0.700(057) & 1.878(210) & 0.393(020) & 0.637 \\
\hline
\multirow{3}{*}{1}
& 2 & 06 - 20 & $8.74(57) $ & 0.427(004) & 1.326(078) & 0.312(041) &  &  &  &  & 0.364 \\
& 3 & 04 - 20 & $7.83(67) $ & 0.421(005) & 1.120(113) & 0.226(032) & 3.001(530) & 0.710(058) &  &  & 0.464 \\
& 4 & 04 - 20 & $7.84(70) $ & 0.421(005) & 1.148(111) & 0.231(033) & 2.263(404) & 0.699(063) & 1.953(224) & 0.385(023) & 0.399 \\
\hline
\multirow{3}{*}{2}
& 2 & 06 - 20 & $8.21(57) $ & 0.437(004) & 1.349(077) & 0.300(038) &  &  &  &  & 0.303 \\
& 3 & 04 - 20 & $7.38(61) $ & 0.431(005) & 1.168(119) & 0.223(027) & 3.117(528) & 0.720(054) &  &  & 0.451 \\
& 4 & 04 - 20 & $7.37(64) $ & 0.431(005) & 1.196(119) & 0.226(029) & 2.360(405) & 0.709(057) & 2.001(219) & 0.380(024) & 0.377 \\
\hline
\multirow{3}{*}{3}
& 2 & 06 - 20 & $7.61(63) $ & 0.445(005) & 1.390(090) & 0.284(037) &  &  &  &  & 0.472 \\
& 3 & 04 - 20 & $6.80(56) $ & 0.440(005) & 1.241(135) & 0.213(023) & 3.257(534) & 0.724(051) &  &  & 0.594 \\
& 4 & 04 - 20 & $6.77(60) $ & 0.440(005) & 1.273(138) & 0.215(024) & 2.465(417) & 0.710(056) & 2.050(224) & 0.374(025) & 0.510 \\
\hline
\multirow{3}{*}{4}
& 2 & 06 - 20 & $7.23(64) $ & 0.455(005) & 1.409(102) & 0.278(037) &  &  &  &  & 0.595 \\
& 3 & 04 - 20 & $6.47(52) $ & 0.450(004) & 1.271(140) & 0.211(020) & 3.325(529) & 0.731(048) &  &  & 0.718 \\
& 4 & 04 - 20 & $6.43(55) $ & 0.450(005) & 1.305(144) & 0.212(021) & 2.526(414) & 0.717(052) & 2.079(220) & 0.370(025) & 0.628 \\
\hline
\multirow{3}{*}{5}
& 2 & 06 - 20 & $6.68(67) $ & 0.463(006) & 1.474(132) & 0.265(035) &  &  &  &  & 0.493 \\
& 3 & 04 - 20 & $6.01(45) $ & 0.459(004) & 1.348(146) & 0.206(017) & 3.444(525) & 0.736(045) &  &  & 0.692 \\
& 4 & 04 - 20 & $5.96(48) $ & 0.458(004) & 1.385(151) & 0.207(017) & 2.623(416) & 0.722(049) & 2.127(219) & 0.365(026) & 0.593 \\
\hline
\multirow{3}{*}{6}
& 2 & 06 - 20 & $6.18(78) $ & 0.471(007) & 1.529(188) & 0.253(036) &  &  &  &  & 0.529 \\
& 3 & 04 - 20 & $5.60(40) $ & 0.468(004) & 1.401(153) & 0.198(014) & 3.514(524) & 0.730(046) &  &  & 0.711 \\
& 4 & 04 - 20 & $5.53(41) $ & 0.467(004) & 1.446(158) & 0.199(014) & 2.675(422) & 0.715(051) & 2.147(223) & 0.362(026) & 0.607 \\
\hline
\multirow{3}{*}{8}
& 2 & 06 - 20 & $5.18(90) $ & 0.485(010) & 1.754(331) & 0.235(034) &  &  &  &  & 1.092 \\
& 3 & 04 - 20 & $4.99(31) $ & 0.485(004) & 1.522(156) & 0.197(010) & 3.623(508) & 0.740(041) &  &  & 1.280 \\
& 4 & 04 - 20 & $4.92(31) $ & 0.484(004) & 1.573(159) & 0.198(011) & 2.786(409) & 0.728(044) & 2.206(214) & 0.355(026) & 1.165 \\
\hline
\multirow{3}{*}{9}
& 2 & 06 - 20 & $5.10(1.04) $ & 0.497(012) & 1.688(378) & 0.235(041) &  &  &  &  & 1.098 \\
& 3 & 04 - 20 & $4.90(30)   $ & 0.497(004) & 1.445(159) & 0.193(011) & 3.620(502) & 0.735(042) &  &  & 1.248 \\
& 4 & 04 - 20 & $4.82(30)   $ & 0.496(004) & 1.503(160) & 0.193(011) & 2.783(407) & 0.722(046) & 2.202(215) & 0.356(026) & 1.134 \\
\hline
\multirow{3}{*}{9$^\prime$}
& 2 & 06 - 20 & $5.91(87) $ & 0.505(009) & 1.456(185) & 0.272(053) &  &  &  &  & 0.619 \\
& 3 & 04 - 20 & $5.38(38) $ & 0.502(004) & 1.278(156) & 0.209(015) & 3.236(498) & 0.728(044) &  &  & 0.787 \\
& 4 & 04 - 20 & $5.30(39) $ & 0.501(004) & 1.330(158) & 0.208(016) & 2.506(394) & 0.718(045) & 2.067(207) & 0.371(024) & 0.703 \\
\hline
\multirow{3}{*}{10}
& 2 & 06 - 20 & $5.39(85) $ & 0.511(010) & 1.557(235) & 0.258(046) &  &  &  &  & 0.696 \\
& 3 & 04 - 20 & $5.02(34) $ & 0.509(004) & 1.360(158) & 0.206(014) & 3.333(493) & 0.737(041) &  &  & 0.931 \\
& 4 & 04 - 20 & $4.95(34) $ & 0.508(004) & 1.411(159) & 0.206(014) & 2.588(391) & 0.728(042) & 2.112(204) & 0.366(024) & 0.838 \\
 \hline\hline
  \end{tabular}
%% \end{adjustwidth}
  \caption{Results of the 2-, 3- and 4-state fits to the two-point
    nucleon correlator for the a09m130 ensemble. $\bm{n}^2=9$ has two
    combinations $(2,2,1)$ and $(3,0,0)$ 
    labeled $9$ and $9^\prime$, respectively.
    The rest is the same as in
    Table.~\protect\ref{tab:multistates-twopt-mom-a12m310AMA-3}.}
  \label{tab:multistates-twopt-mom-a09m130LP1-3}
\end{table*}

%% file: 2pt_Tables/twopt-mom-a06m310AMA-3_All.tex
\begin{table*}[!htb]
%\begin{adjustwidth}{-8em}{-8em}
  \centering
  \renewcommand{\arraystretch}{0.9}
  \begin{tabular}{c|c|c|ccccccccc}
    \hline\hline
    \multicolumn{3}{r}{Priors} &
    & & 1.0(5) & 0.16(10) & 2.4(1.5) & 0.3(2) & 2.2(1.5) & 0.3(2) & 
    \\\hline
    $\bm{n^2}$ &
    $N_\text{2pt}$ & & 
    ${\cal A}_0\times 10^{12}$ & $a E_0$ &
    $r_1$ & $a \Delta E_1$ &
    $r_2$ & $a \Delta E_2$ &
    $r_3$ & $a \Delta E_3$ &
    $\chi^2/\text{DOF}$
    \\\hline
\multirow{3}{*}{0}
& 2 & 10 - 30 & $5.56(35) $ & 0.326(003) & 1.362(097) & 0.199(026) &  &  &  &  & 1.371 \\
& 3 & 07 - 30 & $5.46(39) $ & 0.325(003) & 0.936(109) & 0.163(028) & 3.368(597) & 0.356(035) &  &  & 1.268 \\
& 4 & 07 - 30 & $5.40(43) $ & 0.325(003) & 0.964(116) & 0.161(031) & 2.554(366) & 0.338(037) & 2.323(334) & 0.276(042) & 1.238 \\
\hline
\multirow{3}{*}{1}
& 2 & 10 - 30 & $5.17(29) $ & 0.352(002) & 1.444(118) & 0.209(027) &  &  &  &  & 1.091 \\
& 3 & 07 - 30 & $5.09(29) $ & 0.352(002) & 1.022(110) & 0.175(024) & 3.096(598) & 0.348(038) &  &  & 1.010 \\
& 4 & 07 - 30 & $5.05(31) $ & 0.352(002) & 1.054(116) & 0.174(026) & 2.401(399) & 0.334(040) & 2.269(312) & 0.292(041) & 0.988 \\
\hline
\multirow{3}{*}{2}
& 2 & 10 - 30 & $4.74(31) $ & 0.376(003) & 1.507(135) & 0.211(030) &  &  &  &  & 0.819 \\
& 3 & 07 - 30 & $4.62(30) $ & 0.376(003) & 1.044(118) & 0.172(024) & 2.889(609) & 0.333(043) &  &  & 0.765 \\
& 4 & 07 - 30 & $4.57(32) $ & 0.375(003) & 1.076(125) & 0.171(027) & 2.263(419) & 0.319(044) & 2.180(316) & 0.307(042) & 0.753 \\
\hline
\multirow{3}{*}{3}
& 2 & 10 - 30 & $4.36(37) $ & 0.399(004) & 1.601(175) & 0.218(038) &  &  &  &  & 0.443 \\
& 3 & 07 - 30 & $4.12(34) $ & 0.397(003) & 1.046(131) & 0.164(024) & 2.730(592) & 0.315(046) &  &  & 0.416 \\
& 4 & 07 - 30 & $4.04(40) $ & 0.396(004) & 1.074(138) & 0.159(030) & 2.159(423) & 0.298(046) & 2.099(324) & 0.320(040) & 0.412 \\
\hline
\multirow{3}{*}{4}
& 2 & 10 - 30 & $4.18(45) $ & 0.423(005) & 1.504(234) & 0.224(055) &  &  &  &  & 0.884 \\
& 3 & 07 - 30 & $4.05(32) $ & 0.422(004) & 0.968(109) & 0.176(025) & 2.553(558) & 0.303(050) &  &  & 0.769 \\
& 4 & 07 - 30 & $3.94(37) $ & 0.420(004) & 1.009(111) & 0.169(030) & 2.107(416) & 0.296(045) & 2.022(318) & 0.324(038) & 0.775 \\
\hline
\multirow{3}{*}{5}
& 2 & 10 - 30 & $3.35(56) $ & 0.437(007) & 1.623(187) & 0.186(044) &  &  &  &  & 0.867 \\
& 3 & 07 - 30 & $3.27(31) $ & 0.437(004) & 1.107(170) & 0.149(017) & 2.703(520) & 0.307(046) &  &  & 0.756 \\
& 4 & 07 - 30 & $3.12(33) $ & 0.435(004) & 1.149(182) & 0.139(020) & 2.179(357) & 0.287(049) & 2.083(316) & 0.318(036) & 0.754 \\
\hline\hline
  \end{tabular}
%% \end{adjustwidth}
  \caption{Results of the 2-, 3- and 4-state fits to the two-point
    nucleon correlator for the a06m310 ensemble. 
    The rest is the same as in
    Table.~\protect\ref{tab:multistates-twopt-mom-a12m310AMA-3}.} 
  \label{tab:multistates-twopt-mom-a06m310AMA-3}
\end{table*}

%% file: 2pt_Tables/twopt-mom-a06m220AMA-3_All.tex
\begin{table*}[!htb]
%% \begin{adjustwidth}{-8em}{-8em}
  \centering
  \renewcommand{\arraystretch}{0.9}
  \begin{tabular}{c|c|c|ccccccccc}
    \hline\hline
    \multicolumn{3}{r}{Priors} &
    & & 2.0(1.0) & 0.25(20) & 3.0(1.5) & 0.3(2) & 2.8(1.8) & 0.3(2) & 
    \\\hline
    $\bm{n^2}$ &
    $N_\text{2pt}$ & & 
    ${\cal A}_0\times 10^{11}$ & $a E_0$ &
    $r_1$ & $a \Delta E_1$ &
    $r_2$ & $a \Delta E_2$ &
    $r_3$ & $a \Delta E_3$ &
    $\chi^2/\text{DOF}$
    \\\hline
\multirow{3}{*}{0}
& 2 & 10 - 30 & $10.8(4)  $ & 0.305(002) & 2.900(348) & 0.286(025) &  &  &  &  & 1.774 \\
& 3 & 07 - 30 & $10.6(4)  $ & 0.304(002) & 2.035(225) & 0.249(019) & 3.919(681) & 0.342(021) &  &  & 1.591 \\
& 4 & 07 - 30 & $10.5(4)  $ & 0.304(002) & 2.066(240) & 0.245(021) & 3.185(345) & 0.344(022) & 3.078(406) & 0.267(048) & 1.548 \\
\hline
\multirow{3}{*}{1}
& 2 & 10 - 30 & $10.0(4)  $ & 0.319(002) & 2.842(265) & 0.266(022) &  &  &  &  & 1.432 \\
& 3 & 07 - 30 & $9.74(42) $ & 0.318(002) & 1.998(208) & 0.228(019) & 4.119(680) & 0.341(022) &  &  & 1.342 \\
& 4 & 07 - 30 & $9.62(46) $ & 0.317(002) & 2.041(215) & 0.226(021) & 3.253(325) & 0.340(024) & 3.137(426) & 0.257(049) & 1.289 \\
\hline
\multirow{3}{*}{2}
& 2 & 10 - 30 & $9.38(46) $ & 0.333(002) & 2.869(238) & 0.256(021) &  &  &  &  & 1.356 \\
& 3 & 07 - 30 & $9.10(45) $ & 0.332(002) & 2.041(194) & 0.218(019) & 4.217(680) & 0.343(021) &  &  & 1.288 \\
& 4 & 07 - 30 & $8.95(50) $ & 0.331(002) & 2.083(200) & 0.214(021) & 3.302(321) & 0.340(024) & 3.203(439) & 0.250(049) & 1.227 \\
\hline
\multirow{3}{*}{3}
& 2 & 10 - 30 & $8.97(53) $ & 0.347(003) & 2.898(242) & 0.250(023) &  &  &  &  & 1.162 \\
& 3 & 07 - 30 & $8.60(52) $ & 0.346(003) & 2.019(189) & 0.208(020) & 4.249(701) & 0.336(022) &  &  & 1.130 \\
& 4 & 07 - 30 & $8.38(61) $ & 0.344(003) & 2.051(202) & 0.202(023) & 3.313(332) & 0.329(026) & 3.195(461) & 0.250(051) & 1.069 \\
\hline
\multirow{3}{*}{4}
& 2 & 10 - 30 & $8.69(66) $ & 0.361(004) & 2.843(242) & 0.244(026) &  &  &  &  & 0.824 \\
& 3 & 07 - 30 & $8.39(60) $ & 0.360(003) & 1.958(178) & 0.203(021) & 4.306(702) & 0.336(021) &  &  & 0.896 \\
& 4 & 07 - 30 & $8.02(77) $ & 0.358(004) & 1.979(188) & 0.192(026) & 3.378(335) & 0.327(025) & 3.268(495) & 0.240(054) & 0.828 \\
\hline
\multirow{3}{*}{5}
& 2 & 10 - 30 & $8.18(67) $ & 0.374(004) & 2.888(225) & 0.237(025) &  &  &  &  & 0.885 \\
& 3 & 07 - 30 & $7.95(58) $ & 0.373(003) & 2.049(182) & 0.200(019) & 4.250(693) & 0.337(021) &  &  & 0.951 \\
& 4 & 07 - 30 & $7.59(73) $ & 0.371(004) & 2.080(192) & 0.190(024) & 3.360(342) & 0.328(025) & 3.258(488) & 0.243(053) & 0.886 \\
 \hline\hline
  \end{tabular}
%% \end{adjustwidth}
  \caption{Results of the 2-, 3- and 4-state fits to the two-point
    nucleon correlator for the a06m220 ensemble. 
    The rest is the same as in
    Table.~\protect\ref{tab:multistates-twopt-mom-a12m310AMA-3}.} 
  \label{tab:multistates-twopt-mom-a06m220AMA-3}
\end{table*}

%% file: 2pt_Tables/twopt-mom-a06m130AMA-3_All.tex
\begin{table*}[!htb]
%% \begin{adjustwidth}{-8em}{-8em}
  \centering
  \renewcommand{\arraystretch}{0.9}
  \begin{tabular}{c|c|c|ccccccccc}
    \hline\hline
    \multicolumn{3}{r}{Priors} &
    & & 1.3(7) & 0.20(15) & 1.3(1.0) & 0.3(2) & 1.1(9) & 0.3(2) & 
    \\\hline
    $\bm{n^2}$ &
    $N_\text{2pt}$ & & 
    ${\cal{A}}_0 \times 10^{16}$ & $a E_0$ &
    $r_1$ & $a \Delta E_1$ &
    $r_2$ & $a \Delta E_2$ &
    $r_3$ & $a \Delta E_3$ &
    $\chi^2/\text{DOF}$
    \\\hline
\multirow{3}{*}{0}
& 2 & 08 - 30 & $3.03(21) $ & 0.276(004) & 1.867(235) & 0.283(042) &  &  &  &  & 1.089 \\
& 3 & 06 - 30 & $2.96(18) $ & 0.275(003) & 1.365(113) & 0.242(026) & 1.375(429) & 0.313(047) &  &  & 0.980 \\
& 4 & 06 - 30 & $2.92(19) $ & 0.274(003) & 1.383(169) & 0.237(031) & 1.161(338) & 0.323(027) & 1.098(181) & 0.311(027) & 0.976 \\
\hline
\multirow{3}{*}{1}
& 2 & 08 - 30 & $2.78(23) $ & 0.282(004) & 1.753(160) & 0.254(039) &  &  &  &  & 1.302 \\
& 3 & 06 - 30 & $2.73(23) $ & 0.282(004) & 1.275(142) & 0.218(036) & 1.680(506) & 0.327(036) &  &  & 1.158 \\
& 4 & 06 - 30 & $2.67(27) $ & 0.280(005) & 1.262(182) & 0.206(044) & 1.392(382) & 0.318(029) & 1.205(222) & 0.291(038) & 1.137 \\
\hline
\multirow{3}{*}{2}
& 2 & 08 - 30 & $2.58(24) $ & 0.288(004) & 1.737(130) & 0.238(036) &  &  &  &  & 1.123 \\
& 3 & 06 - 30 & $2.52(25) $ & 0.288(005) & 1.258(141) & 0.200(037) & 1.825(519) & 0.330(033) &  &  & 1.011 \\
& 4 & 06 - 30 & $2.43(32) $ & 0.286(006) & 1.243(166) & 0.185(048) & 1.499(387) & 0.314(032) & 1.256(238) & 0.281(042) & 0.982 \\
\hline
\multirow{3}{*}{3}
& 2 & 08 - 30 & $2.39(24) $ & 0.294(005) & 1.781(124) & 0.227(034) &  &  &  &  & 1.155 \\
& 3 & 06 - 30 & $2.31(26) $ & 0.293(005) & 1.292(141) & 0.186(036) & 1.905(522) & 0.329(033) &  &  & 1.043 \\
& 4 & 06 - 30 & $2.20(36) $ & 0.291(006) & 1.283(162) & 0.169(048) & 1.566(394) & 0.306(036) & 1.288(250) & 0.276(045) & 1.009 \\
\hline
\multirow{3}{*}{4}
& 2 & 08 - 30 & $2.44(23) $ & 0.305(005) & 1.745(144) & 0.245(039) &  &  &  &  & 1.186 \\
& 3 & 06 - 30 & $2.44(19) $ & 0.305(004) & 1.323(124) & 0.219(028) & 1.611(423) & 0.333(033) &  &  & 1.055 \\
& 4 & 06 - 30 & $2.38(21) $ & 0.304(004) & 1.319(151) & 0.209(032) & 1.355(323) & 0.329(025) & 1.195(190) & 0.292(032) & 1.037 \\
\hline
\multirow{3}{*}{5}
& 2 & 08 - 30 & $2.24(26) $ & 0.310(006) & 1.749(135) & 0.226(039) &  &  &  &  & 0.904 \\
& 3 & 06 - 30 & $2.22(24) $ & 0.310(005) & 1.288(142) & 0.194(034) & 1.836(474) & 0.337(030) &  &  & 0.847 \\
& 4 & 06 - 30 & $2.12(29) $ & 0.308(006) & 1.285(164) & 0.179(040) & 1.528(351) & 0.322(029) & 1.285(224) & 0.276(040) & 0.815 \\
\hline
\multirow{3}{*}{6}
& 2 & 08 - 30 & $2.07(30) $ & 0.315(007) & 1.800(178) & 0.215(040) &  &  &  &  & 1.143 \\
& 3 & 06 - 30 & $2.03(28) $ & 0.315(006) & 1.323(180) & 0.181(034) & 1.953(479) & 0.340(029) &  &  & 1.067 \\
& 4 & 06 - 30 & $1.91(32) $ & 0.313(007) & 1.342(217) & 0.164(036) & 1.614(337) & 0.320(032) & 1.339(228) & 0.267(041) & 1.027 \\
\hline
\multirow{3}{*}{8}
& 2 & 08 - 30 & $1.91(36) $ & 0.329(009) & 1.805(222) & 0.214(049) &  &  &  &  & 1.041 \\
& 3 & 06 - 30 & $1.89(28) $ & 0.329(007) & 1.313(189) & 0.180(036) & 1.894(466) & 0.334(030) &  &  & 0.968 \\
& 4 & 06 - 30 & $1.74(31) $ & 0.326(007) & 1.344(248) & 0.158(034) & 1.613(324) & 0.313(034) & 1.327(226) & 0.268(040) & 0.930 \\
\hline
\multirow{3}{*}{9}
& 2 & 08 - 30 & $1.73(44) $ & 0.333(012) & 1.869(411) & 0.199(049) &  &  &  &  & 1.386 \\
& 3 & 06 - 30 & $1.77(30) $ & 0.334(008) & 1.336(253) & 0.172(032) & 2.020(456) & 0.338(029) &  &  & 1.268 \\
& 4 & 06 - 30 & $1.61(28) $ & 0.331(007) & 1.394(311) & 0.151(024) & 1.699(282) & 0.316(034) & 1.386(217) & 0.257(038) & 1.222 \\
\hline
\multirow{3}{*}{9$^\prime$}
& 2 & 08 - 30 & $1.85(37) $ & 0.335(010) & 1.754(262) & 0.212(051) &  &  &  &  & 0.777 \\
& 3 & 06 - 30 & $1.94(27) $ & 0.338(007) & 1.264(176) & 0.197(038) & 1.848(406) & 0.341(028) &  &  & 0.760 \\
& 4 & 06 - 30 & $1.81(31) $ & 0.335(008) & 1.264(235) & 0.173(038) & 1.602(300) & 0.328(027) & 1.333(211) & 0.266(037) & 0.724 \\
\hline
\multirow{3}{*}{10}
& 2 & 08 - 30 & $1.76(38) $ & 0.342(010) & 1.757(286) & 0.207(052) &  &  &  &  & 1.018 \\
& 3 & 06 - 30 & $1.85(27) $ & 0.345(007) & 1.265(181) & 0.192(039) & 1.865(416) & 0.343(028) &  &  & 0.974 \\
& 4 & 06 - 30 & $1.72(30) $ & 0.342(008) & 1.274(245) & 0.168(037) & 1.615(301) & 0.329(028) & 1.344(210) & 0.265(037) & 0.936 \\
\hline\hline
  \end{tabular}
%% \end{adjustwidth}
\caption{Results of the 2-, 3- and 4-state fits to the two-point
  nucleon correlator for the a06m130 AMA ensemble.  $\bm{n}^2=9$ has
  two combinations $(2,2,1)$ and $(3,0,0)$ 
    labeled $9$ and $9^\prime$, respectively.
  The rest is the same as in
  Table.~\protect\ref{tab:multistates-twopt-mom-a12m310AMA-3}.}
  \label{tab:multistates-twopt-mom-a06m130AMA-3}
\end{table*}

%% file: paper.bbl
\begin{thebibliography}{39}%
\makeatletter
\providecommand \@ifxundefined [1]{%
 \@ifx{#1\undefined}
}%
\providecommand \@ifnum [1]{%
 \ifnum #1\expandafter \@firstoftwo
 \else \expandafter \@secondoftwo
 \fi
}%
\providecommand \@ifx [1]{%
 \ifx #1\expandafter \@firstoftwo
 \else \expandafter \@secondoftwo
 \fi
}%
\providecommand \natexlab [1]{#1}%
\providecommand \enquote  [1]{``#1''}%
\providecommand \bibnamefont  [1]{#1}%
\providecommand \bibfnamefont [1]{#1}%
\providecommand \citenamefont [1]{#1}%
\providecommand \href@noop [0]{\@secondoftwo}%
\providecommand \href [0]{\begingroup \@sanitize@url \@href}%
\providecommand \@href[1]{\@@startlink{#1}\@@href}%
\providecommand \@@href[1]{\endgroup#1\@@endlink}%
\providecommand \@sanitize@url [0]{\catcode `\\12\catcode `\$12\catcode
  `\&12\catcode `\#12\catcode `\^12\catcode `\_12\catcode `\%12\relax}%
\providecommand \@@startlink[1]{}%
\providecommand \@@endlink[0]{}%
\providecommand \url  [0]{\begingroup\@sanitize@url \@url }%
\providecommand \@url [1]{\endgroup\@href {#1}{\urlprefix }}%
\providecommand \urlprefix  [0]{URL }%
\providecommand \Eprint [0]{\href }%
\providecommand \doibase [0]{http://dx.doi.org/}%
\providecommand \selectlanguage [0]{\@gobble}%
\providecommand \bibinfo  [0]{\@secondoftwo}%
\providecommand \bibfield  [0]{\@secondoftwo}%
\providecommand \translation [1]{[#1]}%
\providecommand \BibitemOpen [0]{}%
\providecommand \bibitemStop [0]{}%
\providecommand \bibitemNoStop [0]{.\EOS\space}%
\providecommand \EOS [0]{\spacefactor3000\relax}%
\providecommand \BibitemShut  [1]{\csname bibitem#1\endcsname}%
\let\auto@bib@innerbib\@empty
%</preamble>
\bibitem [{\citenamefont {Fukuda}\ \emph {et~al.}(1998)\citenamefont {Fukuda}
  \emph {et~al.}}]{Fukuda:1998mi}%
  \BibitemOpen
  \bibfield  {author} {\bibinfo {author} {\bibfnamefont {Y.}~\bibnamefont
  {Fukuda}} \emph {et~al.} (\bibinfo {collaboration} {Super-Kamiokande}),\
  }\href {\doibase 10.1103/PhysRevLett.81.1562} {\bibfield  {journal} {\bibinfo
   {journal} {Phys. Rev. Lett.}\ }\textbf {\bibinfo {volume} {81}},\ \bibinfo
  {pages} {1562} (\bibinfo {year} {1998})},\ \Eprint
  {http://arxiv.org/abs/hep-ex/9807003} {arXiv:hep-ex/9807003 [hep-ex]}
  \BibitemShut {NoStop}%
%%CITATION = HEP-EX/9807003;%%
\bibitem [{\citenamefont {Ahmad}\ \emph {et~al.}(2001)\citenamefont {Ahmad}
  \emph {et~al.}}]{Ahmad:2001an}%
  \BibitemOpen
  \bibfield  {author} {\bibinfo {author} {\bibfnamefont {Q.~R.}\ \bibnamefont
  {Ahmad}} \emph {et~al.} (\bibinfo {collaboration} {SNO}),\ }\href {\doibase
  10.1103/PhysRevLett.87.071301} {\bibfield  {journal} {\bibinfo  {journal}
  {Phys. Rev. Lett.}\ }\textbf {\bibinfo {volume} {87}},\ \bibinfo {pages}
  {071301} (\bibinfo {year} {2001})},\ \Eprint
  {http://arxiv.org/abs/nucl-ex/0106015} {arXiv:nucl-ex/0106015 [nucl-ex]}
  \BibitemShut {NoStop}%
%%CITATION = NUCL-EX/0106015;%%
\bibitem [{\citenamefont {Ahmad}\ \emph {et~al.}(2002)\citenamefont {Ahmad}
  \emph {et~al.}}]{Ahmad:2002jz}%
  \BibitemOpen
  \bibfield  {author} {\bibinfo {author} {\bibfnamefont {Q.~R.}\ \bibnamefont
  {Ahmad}} \emph {et~al.} (\bibinfo {collaboration} {SNO}),\ }\href {\doibase
  10.1103/PhysRevLett.89.011301} {\bibfield  {journal} {\bibinfo  {journal}
  {Phys. Rev. Lett.}\ }\textbf {\bibinfo {volume} {89}},\ \bibinfo {pages}
  {011301} (\bibinfo {year} {2002})},\ \Eprint
  {http://arxiv.org/abs/nucl-ex/0204008} {arXiv:nucl-ex/0204008 [nucl-ex]}
  \BibitemShut {NoStop}%
%%CITATION = NUCL-EX/0204008;%%
\bibitem [{\citenamefont {{Philip Ball}}()}]{neutrinoOS}%
  \BibitemOpen
  \bibfield  {author} {\bibinfo {author} {\bibnamefont {{Philip Ball}}},\
  }\href@noop {} {\enquote {\bibinfo {title} {{Focus Nobel Prize--Neutrinos
  Oscillate}},}\ }\bibinfo {howpublished} {Physics Today,
  \url{https://physics.aps.org/articles/v8/97}}\BibitemShut {NoStop}%
\bibitem [{\citenamefont {Patrignani}\ \emph {et~al.}(2016)\citenamefont
  {Patrignani} \emph {et~al.}}]{Olive:2016xmw}%
  \BibitemOpen
  \bibfield  {author} {\bibinfo {author} {\bibfnamefont {C.}~\bibnamefont
  {Patrignani}} \emph {et~al.} (\bibinfo {collaboration} {Particle Data
  Group}),\ }\href {\doibase 10.1088/1674-1137/40/10/100001} {\bibfield
  {journal} {\bibinfo  {journal} {Chin. Phys.}\ }\textbf {\bibinfo {volume}
  {C40}},\ \bibinfo {pages} {100001} (\bibinfo {year} {2016})}\BibitemShut
  {NoStop}%
%%CITATION = CHPHD,C40,100001;%%
\bibitem [{\citenamefont {{Wikipedia List of Neutrino
  Experiments}}()}]{neutrino:expts}%
  \BibitemOpen
  \bibfield  {author} {\bibinfo {author} {\bibnamefont {{Wikipedia List of
  Neutrino Experiments}}},\ }\href@noop {} {\enquote {\bibinfo {title}
  {{Neutrino Experiments}},}\ }\bibinfo {howpublished}
  {\url{https://en.wikipedia.org/wiki/List\_of\_neutrino\_experiments}}\BibitemShut
  {NoStop}%
\bibitem [{\citenamefont {Meyer}\ \emph {et~al.}(2016)\citenamefont {Meyer}
  \emph {et~al.}}]{Meyer:2016oeg}%
  \BibitemOpen
  \bibfield  {author} {\bibinfo {author} {\bibfnamefont {A.~S.}\ \bibnamefont
  {Meyer}} \emph {et~al.},\ }\href {\doibase 10.1103/PhysRevD.93.113015}
  {\bibfield  {journal} {\bibinfo  {journal} {Phys. Rev.}\ }\textbf {\bibinfo
  {volume} {D93}},\ \bibinfo {pages} {113015} (\bibinfo {year} {2016})},\
  \Eprint {http://arxiv.org/abs/1603.03048} {arXiv:1603.03048 [hep-ph]}
  \BibitemShut {NoStop}%
%%CITATION = ARXIV:1603.03048;%%
\bibitem [{\citenamefont {Carlson}\ \emph {et~al.}(2015)\citenamefont
  {Carlson}, \citenamefont {Gandolfi}, \citenamefont {Pederiva}, \citenamefont
  {Pieper}, \citenamefont {Schiavilla}, \citenamefont {Schmidt},\ and\
  \citenamefont {Wiringa}}]{Carlson:2014vla}%
  \BibitemOpen
  \bibfield  {author} {\bibinfo {author} {\bibfnamefont {J.}~\bibnamefont
  {Carlson}}, \bibinfo {author} {\bibfnamefont {S.}~\bibnamefont {Gandolfi}},
  \bibinfo {author} {\bibfnamefont {F.}~\bibnamefont {Pederiva}}, \bibinfo
  {author} {\bibfnamefont {S.~C.}\ \bibnamefont {Pieper}}, \bibinfo {author}
  {\bibfnamefont {R.}~\bibnamefont {Schiavilla}}, \bibinfo {author}
  {\bibfnamefont {K.~E.}\ \bibnamefont {Schmidt}}, \ and\ \bibinfo {author}
  {\bibfnamefont {R.~B.}\ \bibnamefont {Wiringa}},\ }\href {\doibase
  10.1103/RevModPhys.87.1067} {\bibfield  {journal} {\bibinfo  {journal} {Rev.
  Mod. Phys.}\ }\textbf {\bibinfo {volume} {87}},\ \bibinfo {pages} {1067}
  (\bibinfo {year} {2015})},\ \Eprint {http://arxiv.org/abs/1412.3081}
  {arXiv:1412.3081 [nucl-th]} \BibitemShut {NoStop}%
%%CITATION = ARXIV:1412.3081;%%
\bibitem [{\citenamefont {Aguilar-Arevalo}\ \emph {et~al.}(2010)\citenamefont
  {Aguilar-Arevalo} \emph {et~al.}}]{AguilarArevalo:2010zc}%
  \BibitemOpen
  \bibfield  {author} {\bibinfo {author} {\bibfnamefont {A.~A.}\ \bibnamefont
  {Aguilar-Arevalo}} \emph {et~al.} (\bibinfo {collaboration} {MiniBooNE}),\
  }\href {\doibase 10.1103/PhysRevD.81.092005} {\bibfield  {journal} {\bibinfo
  {journal} {Phys. Rev.}\ }\textbf {\bibinfo {volume} {D81}},\ \bibinfo {pages}
  {092005} (\bibinfo {year} {2010})},\ \Eprint {http://arxiv.org/abs/1002.2680}
  {arXiv:1002.2680 [hep-ex]} \BibitemShut {NoStop}%
%%CITATION = ARXIV:1002.2680;%%
\bibitem [{\citenamefont {Bhattacharya}\ \emph {et~al.}(2012)\citenamefont
  {Bhattacharya}, \citenamefont {Cirigliano}, \citenamefont {Cohen},
  \citenamefont {Filipuzzi}, \citenamefont {Gonzalez-Alonso} \emph
  {et~al.}}]{Bhattacharya:2011qm}%
  \BibitemOpen
  \bibfield  {author} {\bibinfo {author} {\bibfnamefont {T.}~\bibnamefont
  {Bhattacharya}}, \bibinfo {author} {\bibfnamefont {V.}~\bibnamefont
  {Cirigliano}}, \bibinfo {author} {\bibfnamefont {S.~D.}\ \bibnamefont
  {Cohen}}, \bibinfo {author} {\bibfnamefont {A.}~\bibnamefont {Filipuzzi}},
  \bibinfo {author} {\bibfnamefont {M.}~\bibnamefont {Gonzalez-Alonso}},  \emph
  {et~al.},\ }\href {\doibase 10.1103/PhysRevD.85.054512} {\bibfield  {journal}
  {\bibinfo  {journal} {Phys.Rev.}\ }\textbf {\bibinfo {volume} {D85}},\
  \bibinfo {pages} {054512} (\bibinfo {year} {2012})},\ \Eprint
  {http://arxiv.org/abs/1110.6448} {arXiv:1110.6448 [hep-ph]} \BibitemShut
  {NoStop}%
%%CITATION = ARXIV:1110.6448;%%
\bibitem [{\citenamefont {Follana}\ \emph {et~al.}(2007)\citenamefont {Follana}
  \emph {et~al.}}]{Follana:2006rc}%
  \BibitemOpen
  \bibfield  {author} {\bibinfo {author} {\bibfnamefont {E.}~\bibnamefont
  {Follana}} \emph {et~al.} (\bibinfo {collaboration} {HPQCD Collaboration,
  UKQCD Collaboration}),\ }\href {\doibase 10.1103/PhysRevD.75.054502}
  {\bibfield  {journal} {\bibinfo  {journal} {Phys.Rev.}\ }\textbf {\bibinfo
  {volume} {D75}},\ \bibinfo {pages} {054502} (\bibinfo {year} {2007})},\
  \Eprint {http://arxiv.org/abs/hep-lat/0610092} {arXiv:hep-lat/0610092
  [hep-lat]} \BibitemShut {NoStop}%
%%CITATION = HEP-LAT/0610092;%%
\bibitem [{\citenamefont {Bazavov}\ \emph {et~al.}(2013)\citenamefont {Bazavov}
  \emph {et~al.}}]{Bazavov:2012xda}%
  \BibitemOpen
  \bibfield  {author} {\bibinfo {author} {\bibfnamefont {A.}~\bibnamefont
  {Bazavov}} \emph {et~al.} (\bibinfo {collaboration} {MILC Collaboration}),\
  }\href {\doibase 10.1103/PhysRevD.87.054505} {\bibfield  {journal} {\bibinfo
  {journal} {Phys.Rev.}\ }\textbf {\bibinfo {volume} {D87}},\ \bibinfo {pages}
  {054505} (\bibinfo {year} {2013})},\ \Eprint {http://arxiv.org/abs/1212.4768}
  {arXiv:1212.4768 [hep-lat]} \BibitemShut {NoStop}%
%%CITATION = ARXIV:1212.4768;%%
\bibitem [{\citenamefont {Bernard}\ \emph {et~al.}(2002)\citenamefont
  {Bernard}, \citenamefont {Elouadrhiri},\ and\ \citenamefont
  {Meissner}}]{Bernard:2001rs}%
  \BibitemOpen
  \bibfield  {author} {\bibinfo {author} {\bibfnamefont {V.}~\bibnamefont
  {Bernard}}, \bibinfo {author} {\bibfnamefont {L.}~\bibnamefont
  {Elouadrhiri}}, \ and\ \bibinfo {author} {\bibfnamefont {U.-G.}\ \bibnamefont
  {Meissner}},\ }\href {\doibase 10.1088/0954-3899/28/1/201} {\bibfield
  {journal} {\bibinfo  {journal} {J. Phys.}\ }\textbf {\bibinfo {volume}
  {G28}},\ \bibinfo {pages} {R1} (\bibinfo {year} {2002})},\ \Eprint
  {http://arxiv.org/abs/hep-ph/0107088} {arXiv:hep-ph/0107088 [hep-ph]}
  \BibitemShut {NoStop}%
%%CITATION = HEP-PH/0107088;%%
\bibitem [{\citenamefont {Smith}\ and\ \citenamefont
  {Moniz}(1972)}]{Smith:1972xh}%
  \BibitemOpen
  \bibfield  {author} {\bibinfo {author} {\bibfnamefont {R.~A.}\ \bibnamefont
  {Smith}}\ and\ \bibinfo {author} {\bibfnamefont {E.~J.}\ \bibnamefont
  {Moniz}},\ }\href {\doibase 10.1016/0550-3213(75)90612-4,
  10.1016/0550-3213(72)90040-5} {\bibfield  {journal} {\bibinfo  {journal}
  {Nucl. Phys.}\ }\textbf {\bibinfo {volume} {B43}},\ \bibinfo {pages} {605}
  (\bibinfo {year} {1972})},\ \bibinfo {note} {[Erratum: Nucl.
  Phys.B101,547(1975)]}\BibitemShut {NoStop}%
%%CITATION = NUPHA,B43,605;%%
\bibitem [{\citenamefont {Hill}\ and\ \citenamefont {Paz}(2010)}]{Hill:2010yb}%
  \BibitemOpen
  \bibfield  {author} {\bibinfo {author} {\bibfnamefont {R.~J.}\ \bibnamefont
  {Hill}}\ and\ \bibinfo {author} {\bibfnamefont {G.}~\bibnamefont {Paz}},\
  }\href {\doibase 10.1103/PhysRevD.82.113005} {\bibfield  {journal} {\bibinfo
  {journal} {Phys. Rev.}\ }\textbf {\bibinfo {volume} {D82}},\ \bibinfo {pages}
  {113005} (\bibinfo {year} {2010})},\ \Eprint {http://arxiv.org/abs/1008.4619}
  {arXiv:1008.4619 [hep-ph]} \BibitemShut {NoStop}%
%%CITATION = ARXIV:1008.4619;%%
\bibitem [{\citenamefont {Bhattacharya}\ \emph {et~al.}(2011)\citenamefont
  {Bhattacharya}, \citenamefont {Hill},\ and\ \citenamefont
  {Paz}}]{Bhattacharya:2011ah}%
  \BibitemOpen
  \bibfield  {author} {\bibinfo {author} {\bibfnamefont {B.}~\bibnamefont
  {Bhattacharya}}, \bibinfo {author} {\bibfnamefont {R.~J.}\ \bibnamefont
  {Hill}}, \ and\ \bibinfo {author} {\bibfnamefont {G.}~\bibnamefont {Paz}},\
  }\href {\doibase 10.1103/PhysRevD.84.073006} {\bibfield  {journal} {\bibinfo
  {journal} {Phys. Rev.}\ }\textbf {\bibinfo {volume} {D84}},\ \bibinfo {pages}
  {073006} (\bibinfo {year} {2011})},\ \Eprint {http://arxiv.org/abs/1108.0423}
  {arXiv:1108.0423 [hep-ph]} \BibitemShut {NoStop}%
%%CITATION = ARXIV:1108.0423;%%
\bibitem [{\citenamefont {Lee}\ \emph {et~al.}(2015)\citenamefont {Lee},
  \citenamefont {Arrington},\ and\ \citenamefont {Hill}}]{Lee:2015jqa}%
  \BibitemOpen
  \bibfield  {author} {\bibinfo {author} {\bibfnamefont {G.}~\bibnamefont
  {Lee}}, \bibinfo {author} {\bibfnamefont {J.~R.}\ \bibnamefont {Arrington}},
  \ and\ \bibinfo {author} {\bibfnamefont {R.~J.}\ \bibnamefont {Hill}},\
  }\bibfield  {booktitle} {\emph {\bibinfo {booktitle} {{Proceedings, Meeting
  of the APS Division of Particles and Fields (DPF 2015): Ann Arbor, Michigan,
  USA, 4-8 Aug 2015}}},\ }\href {\doibase 10.1103/PhysRevD.92.013013}
  {\bibfield  {journal} {\bibinfo  {journal} {Phys. Rev.}\ }\textbf {\bibinfo
  {volume} {D92}},\ \bibinfo {pages} {013013} (\bibinfo {year} {2015})},\
  \Eprint {http://arxiv.org/abs/1505.01489} {arXiv:1505.01489 [hep-ph]}
  \BibitemShut {NoStop}%
%%CITATION = ARXIV:1505.01489;%%
\bibitem [{\citenamefont {Goldberger}\ and\ \citenamefont
  {Treiman}(1958)}]{Goldberger:1958vp}%
  \BibitemOpen
  \bibfield  {author} {\bibinfo {author} {\bibfnamefont {M.~L.}\ \bibnamefont
  {Goldberger}}\ and\ \bibinfo {author} {\bibfnamefont {S.~B.}\ \bibnamefont
  {Treiman}},\ }\href {\doibase 10.1103/PhysRev.111.354} {\bibfield  {journal}
  {\bibinfo  {journal} {Phys. Rev.}\ }\textbf {\bibinfo {volume} {111}},\
  \bibinfo {pages} {354} (\bibinfo {year} {1958})}\BibitemShut {NoStop}%
%%CITATION = PHRVA,111,354;%%
\bibitem [{\citenamefont {Bhattacharya}\ \emph {et~al.}(2015)\citenamefont
  {Bhattacharya}, \citenamefont {Cirigliano}, \citenamefont {Cohen},
  \citenamefont {Gupta}, \citenamefont {Joseph}, \citenamefont {Lin},\ and\
  \citenamefont {Yoon}}]{Bhattacharya:2015wna}%
  \BibitemOpen
  \bibfield  {author} {\bibinfo {author} {\bibfnamefont {T.}~\bibnamefont
  {Bhattacharya}}, \bibinfo {author} {\bibfnamefont {V.}~\bibnamefont
  {Cirigliano}}, \bibinfo {author} {\bibfnamefont {S.}~\bibnamefont {Cohen}},
  \bibinfo {author} {\bibfnamefont {R.}~\bibnamefont {Gupta}}, \bibinfo
  {author} {\bibfnamefont {A.}~\bibnamefont {Joseph}}, \bibinfo {author}
  {\bibfnamefont {H.-W.}\ \bibnamefont {Lin}}, \ and\ \bibinfo {author}
  {\bibfnamefont {B.}~\bibnamefont {Yoon}} (\bibinfo {collaboration} {PNDME}),\
  }\href {\doibase 10.1103/PhysRevD.92.094511} {\bibfield  {journal} {\bibinfo
  {journal} {Phys. Rev.}\ }\textbf {\bibinfo {volume} {D92}},\ \bibinfo {pages}
  {094511} (\bibinfo {year} {2015})},\ \Eprint
  {http://arxiv.org/abs/1506.06411} {arXiv:1506.06411 [hep-lat]} \BibitemShut
  {NoStop}%
%%CITATION = ARXIV:1506.06411;%%
\bibitem [{\citenamefont {Andreev}\ \emph {et~al.}(2013)\citenamefont {Andreev}
  \emph {et~al.}}]{Andreev:2012fj}%
  \BibitemOpen
  \bibfield  {author} {\bibinfo {author} {\bibfnamefont {V.~A.}\ \bibnamefont
  {Andreev}} \emph {et~al.} (\bibinfo {collaboration} {MuCap}),\ }\href
  {\doibase 10.1103/PhysRevLett.110.012504} {\bibfield  {journal} {\bibinfo
  {journal} {Phys. Rev. Lett.}\ }\textbf {\bibinfo {volume} {110}},\ \bibinfo
  {pages} {012504} (\bibinfo {year} {2013})},\ \Eprint
  {http://arxiv.org/abs/1210.6545} {arXiv:1210.6545 [nucl-ex]} \BibitemShut
  {NoStop}%
%%CITATION = ARXIV:1210.6545;%%
\bibitem [{\citenamefont {Andreev}\ \emph {et~al.}(2015)\citenamefont {Andreev}
  \emph {et~al.}}]{Andreev:2015evt}%
  \BibitemOpen
  \bibfield  {author} {\bibinfo {author} {\bibfnamefont {V.~A.}\ \bibnamefont
  {Andreev}} \emph {et~al.} (\bibinfo {collaboration} {MuCap}),\ }\href
  {\doibase 10.1103/PhysRevC.91.055502} {\bibfield  {journal} {\bibinfo
  {journal} {Phys. Rev.}\ }\textbf {\bibinfo {volume} {C91}},\ \bibinfo {pages}
  {055502} (\bibinfo {year} {2015})},\ \Eprint
  {http://arxiv.org/abs/1502.00913} {arXiv:1502.00913 [nucl-ex]} \BibitemShut
  {NoStop}%
%%CITATION = ARXIV:1502.00913;%%
\bibitem [{\citenamefont {Schindler}\ \emph {et~al.}(2007)\citenamefont
  {Schindler}, \citenamefont {Fuchs}, \citenamefont {Gegelia},\ and\
  \citenamefont {Scherer}}]{Schindler:2006it}%
  \BibitemOpen
  \bibfield  {author} {\bibinfo {author} {\bibfnamefont {M.~R.}\ \bibnamefont
  {Schindler}}, \bibinfo {author} {\bibfnamefont {T.}~\bibnamefont {Fuchs}},
  \bibinfo {author} {\bibfnamefont {J.}~\bibnamefont {Gegelia}}, \ and\
  \bibinfo {author} {\bibfnamefont {S.}~\bibnamefont {Scherer}},\ }\href
  {\doibase 10.1103/PhysRevC.75.025202} {\bibfield  {journal} {\bibinfo
  {journal} {Phys. Rev.}\ }\textbf {\bibinfo {volume} {C75}},\ \bibinfo {pages}
  {025202} (\bibinfo {year} {2007})},\ \Eprint
  {http://arxiv.org/abs/nucl-th/0611083} {arXiv:nucl-th/0611083 [nucl-th]}
  \BibitemShut {NoStop}%
%%CITATION = NUCL-TH/0611083;%%
\bibitem [{\citenamefont {Baru}\ \emph {et~al.}(2011)\citenamefont {Baru},
  \citenamefont {Hanhart}, \citenamefont {Hoferichter}, \citenamefont {Kubis},
  \citenamefont {Nogga},\ and\ \citenamefont {Phillips}}]{Baru:2011bw}%
  \BibitemOpen
  \bibfield  {author} {\bibinfo {author} {\bibfnamefont {V.}~\bibnamefont
  {Baru}}, \bibinfo {author} {\bibfnamefont {C.}~\bibnamefont {Hanhart}},
  \bibinfo {author} {\bibfnamefont {M.}~\bibnamefont {Hoferichter}}, \bibinfo
  {author} {\bibfnamefont {B.}~\bibnamefont {Kubis}}, \bibinfo {author}
  {\bibfnamefont {A.}~\bibnamefont {Nogga}}, \ and\ \bibinfo {author}
  {\bibfnamefont {D.~R.}\ \bibnamefont {Phillips}},\ }\href {\doibase
  10.1016/j.nuclphysa.2011.09.015} {\bibfield  {journal} {\bibinfo  {journal}
  {Nucl. Phys.}\ }\textbf {\bibinfo {volume} {A872}},\ \bibinfo {pages} {69}
  (\bibinfo {year} {2011})},\ \Eprint {http://arxiv.org/abs/1107.5509}
  {arXiv:1107.5509 [nucl-th]} \BibitemShut {NoStop}%
%%CITATION = ARXIV:1107.5509;%%
\bibitem [{\citenamefont {Edwards}\ and\ \citenamefont
  {Joo}(2005)}]{Edwards:2004sx}%
  \BibitemOpen
  \bibfield  {author} {\bibinfo {author} {\bibfnamefont {R.~G.}\ \bibnamefont
  {Edwards}}\ and\ \bibinfo {author} {\bibfnamefont {B.}~\bibnamefont {Joo}}
  (\bibinfo {collaboration} {SciDAC Collaboration, LHPC Collaboration, UKQCD
  Collaboration}),\ }\href {\doibase 10.1016/j.nuclphysbps.2004.11.254}
  {\bibfield  {journal} {\bibinfo  {journal} {Nucl.Phys.Proc.Suppl.}\ }\textbf
  {\bibinfo {volume} {140}},\ \bibinfo {pages} {832} (\bibinfo {year}
  {2005})},\ \Eprint {http://arxiv.org/abs/hep-lat/0409003}
  {arXiv:hep-lat/0409003 [hep-lat]} \BibitemShut {NoStop}%
%%CITATION = HEP-LAT/0409003;%%
\bibitem [{\citenamefont {Bhattacharya}\ \emph {et~al.}(2016)\citenamefont
  {Bhattacharya}, \citenamefont {Cirigliano}, \citenamefont {Cohen},
  \citenamefont {Gupta}, \citenamefont {Lin},\ and\ \citenamefont
  {Yoon}}]{Bhattacharya:2016zcn}%
  \BibitemOpen
  \bibfield  {author} {\bibinfo {author} {\bibfnamefont {T.}~\bibnamefont
  {Bhattacharya}}, \bibinfo {author} {\bibfnamefont {V.}~\bibnamefont
  {Cirigliano}}, \bibinfo {author} {\bibfnamefont {S.}~\bibnamefont {Cohen}},
  \bibinfo {author} {\bibfnamefont {R.}~\bibnamefont {Gupta}}, \bibinfo
  {author} {\bibfnamefont {H.-W.}\ \bibnamefont {Lin}}, \ and\ \bibinfo
  {author} {\bibfnamefont {B.}~\bibnamefont {Yoon}},\ }\href {\doibase
  10.1103/PhysRevD.94.054508} {\bibfield  {journal} {\bibinfo  {journal} {Phys.
  Rev.}\ }\textbf {\bibinfo {volume} {D94}},\ \bibinfo {pages} {054508}
  (\bibinfo {year} {2016})},\ \Eprint {http://arxiv.org/abs/1606.07049}
  {arXiv:1606.07049 [hep-lat]} \BibitemShut {NoStop}%
%%CITATION = ARXIV:1606.07049;%%
\bibitem [{\citenamefont {Hasenfratz}\ and\ \citenamefont
  {Knechtli}(2001)}]{Hasenfratz:2001hp}%
  \BibitemOpen
  \bibfield  {author} {\bibinfo {author} {\bibfnamefont {A.}~\bibnamefont
  {Hasenfratz}}\ and\ \bibinfo {author} {\bibfnamefont {F.}~\bibnamefont
  {Knechtli}},\ }\href {\doibase 10.1103/PhysRevD.64.034504} {\bibfield
  {journal} {\bibinfo  {journal} {Phys.Rev.}\ }\textbf {\bibinfo {volume}
  {D64}},\ \bibinfo {pages} {034504} (\bibinfo {year} {2001})},\ \Eprint
  {http://arxiv.org/abs/hep-lat/0103029} {arXiv:hep-lat/0103029 [hep-lat]}
  \BibitemShut {NoStop}%
%%CITATION = HEP-LAT/0103029;%%
\bibitem [{\citenamefont {Sheikholeslami}\ and\ \citenamefont
  {Wohlert}(1985)}]{Sheikholeslami:1985ij}%
  \BibitemOpen
  \bibfield  {author} {\bibinfo {author} {\bibfnamefont {B.}~\bibnamefont
  {Sheikholeslami}}\ and\ \bibinfo {author} {\bibfnamefont {R.}~\bibnamefont
  {Wohlert}},\ }\href {\doibase 10.1016/0550-3213(85)90002-1} {\bibfield
  {journal} {\bibinfo  {journal} {Nucl. Phys.}\ }\textbf {\bibinfo {volume}
  {B259}},\ \bibinfo {pages} {572} (\bibinfo {year} {1985})}\BibitemShut
  {NoStop}%
%%CITATION = NUPHA,B259,572;%%
\bibitem [{\citenamefont {Bali}\ \emph {et~al.}(2010)\citenamefont {Bali},
  \citenamefont {Collins},\ and\ \citenamefont {Sch{\"a}fer}}]{Bali:2009hu}%
  \BibitemOpen
  \bibfield  {author} {\bibinfo {author} {\bibfnamefont {G.~S.}\ \bibnamefont
  {Bali}}, \bibinfo {author} {\bibfnamefont {S.}~\bibnamefont {Collins}}, \
  and\ \bibinfo {author} {\bibfnamefont {A.}~\bibnamefont {Sch{\"a}fer}},\
  }\href {\doibase 10.1016/j.cpc.2010.05.008} {\bibfield  {journal} {\bibinfo
  {journal} {Comput.Phys.Commun.}\ }\textbf {\bibinfo {volume} {181}},\
  \bibinfo {pages} {1570} (\bibinfo {year} {2010})},\ \Eprint
  {http://arxiv.org/abs/0910.3970} {arXiv:0910.3970 [hep-lat]} \BibitemShut
  {NoStop}%
%%CITATION = ARXIV:0910.3970;%%
\bibitem [{\citenamefont {Blum}\ \emph {et~al.}(2013)\citenamefont {Blum},
  \citenamefont {Izubuchi},\ and\ \citenamefont {Shintani}}]{Blum:2012uh}%
  \BibitemOpen
  \bibfield  {author} {\bibinfo {author} {\bibfnamefont {T.}~\bibnamefont
  {Blum}}, \bibinfo {author} {\bibfnamefont {T.}~\bibnamefont {Izubuchi}}, \
  and\ \bibinfo {author} {\bibfnamefont {E.}~\bibnamefont {Shintani}},\ }\href
  {\doibase 10.1103/PhysRevD.88.094503} {\bibfield  {journal} {\bibinfo
  {journal} {Phys.Rev.}\ }\textbf {\bibinfo {volume} {D88}},\ \bibinfo {pages}
  {094503} (\bibinfo {year} {2013})},\ \Eprint {http://arxiv.org/abs/1208.4349}
  {arXiv:1208.4349 [hep-lat]} \BibitemShut {NoStop}%
%%CITATION = ARXIV:1208.4349;%%
\bibitem [{\citenamefont {Yoon}\ \emph {et~al.}(2017)\citenamefont {Yoon} \emph
  {et~al.}}]{Yoon:2016jzj}%
  \BibitemOpen
  \bibfield  {author} {\bibinfo {author} {\bibfnamefont {B.}~\bibnamefont
  {Yoon}} \emph {et~al.},\ }\href {\doibase 10.1103/PhysRevD.95.074508}
  {\bibfield  {journal} {\bibinfo  {journal} {Phys. Rev.}\ }\textbf {\bibinfo
  {volume} {D95}},\ \bibinfo {pages} {074508} (\bibinfo {year} {2017})},\
  \Eprint {http://arxiv.org/abs/1611.07452} {arXiv:1611.07452 [hep-lat]}
  \BibitemShut {NoStop}%
%%CITATION = ARXIV:1611.07452;%%
\bibitem [{\citenamefont {Schmelling}(1995)}]{Schmelling:1994pz}%
  \BibitemOpen
  \bibfield  {author} {\bibinfo {author} {\bibfnamefont {M.}~\bibnamefont
  {Schmelling}},\ }\href {\doibase 10.1088/0031-8949/51/6/002} {\bibfield
  {journal} {\bibinfo  {journal} {Phys. Scripta}\ }\textbf {\bibinfo {volume}
  {51}},\ \bibinfo {pages} {676} (\bibinfo {year} {1995})}\BibitemShut
  {NoStop}%
%%CITATION = PHSTB,51,676;%%
\bibitem [{\citenamefont {Alexandrou}\ \emph {et~al.}(2017)\citenamefont
  {Alexandrou}, \citenamefont {Constantinou}, \citenamefont {Hadjiyiannakou},
  \citenamefont {Jansen}, \citenamefont {Kallidonis}, \citenamefont {Koutsou},\
  and\ \citenamefont {Vaquero Aviles-Casco}}]{Alexandrou:2017hac}%
  \BibitemOpen
  \bibfield  {author} {\bibinfo {author} {\bibfnamefont {C.}~\bibnamefont
  {Alexandrou}}, \bibinfo {author} {\bibfnamefont {M.}~\bibnamefont
  {Constantinou}}, \bibinfo {author} {\bibfnamefont {K.}~\bibnamefont
  {Hadjiyiannakou}}, \bibinfo {author} {\bibfnamefont {K.}~\bibnamefont
  {Jansen}}, \bibinfo {author} {\bibfnamefont {C.}~\bibnamefont {Kallidonis}},
  \bibinfo {author} {\bibfnamefont {G.}~\bibnamefont {Koutsou}}, \ and\
  \bibinfo {author} {\bibfnamefont {A.}~\bibnamefont {Vaquero Aviles-Casco}},\
  }\href@noop {} {\  (\bibinfo {year} {2017})},\ \Eprint
  {http://arxiv.org/abs/1705.03399} {arXiv:1705.03399 [hep-lat]} \BibitemShut
  {NoStop}%
%%CITATION = ARXIV:1705.03399;%%
\bibitem [{\citenamefont {Capitani}\ \emph {et~al.}(2017)\citenamefont
  {Capitani}, \citenamefont {Della~Morte}, \citenamefont {Djukanovic},
  \citenamefont {von Hippel}, \citenamefont {Hua}, \citenamefont {Jäger},
  \citenamefont {Junnarkar}, \citenamefont {Meyer}, \citenamefont {Rae},\ and\
  \citenamefont {Wittig}}]{Capitani:2017qpc}%
  \BibitemOpen
  \bibfield  {author} {\bibinfo {author} {\bibfnamefont {S.}~\bibnamefont
  {Capitani}}, \bibinfo {author} {\bibfnamefont {M.}~\bibnamefont
  {Della~Morte}}, \bibinfo {author} {\bibfnamefont {D.}~\bibnamefont
  {Djukanovic}}, \bibinfo {author} {\bibfnamefont {G.~M.}\ \bibnamefont {von
  Hippel}}, \bibinfo {author} {\bibfnamefont {J.}~\bibnamefont {Hua}}, \bibinfo
  {author} {\bibfnamefont {B.}~\bibnamefont {Jäger}}, \bibinfo {author}
  {\bibfnamefont {P.~M.}\ \bibnamefont {Junnarkar}}, \bibinfo {author}
  {\bibfnamefont {H.~B.}\ \bibnamefont {Meyer}}, \bibinfo {author}
  {\bibfnamefont {T.~D.}\ \bibnamefont {Rae}}, \ and\ \bibinfo {author}
  {\bibfnamefont {H.}~\bibnamefont {Wittig}},\ }\href@noop {} {\  (\bibinfo
  {year} {2017})},\ \Eprint {http://arxiv.org/abs/1705.06186} {arXiv:1705.06186
  [hep-lat]} \BibitemShut {NoStop}%
%%CITATION = ARXIV:1705.06186;%%
\bibitem [{\citenamefont {Bali}\ \emph {et~al.}(2015)\citenamefont {Bali},
  \citenamefont {Collins}, \citenamefont {Glässle}, \citenamefont {Göckeler},
  \citenamefont {Najjar} \emph {et~al.}}]{Bali:2014nma}%
  \BibitemOpen
  \bibfield  {author} {\bibinfo {author} {\bibfnamefont {G.~S.}\ \bibnamefont
  {Bali}}, \bibinfo {author} {\bibfnamefont {S.}~\bibnamefont {Collins}},
  \bibinfo {author} {\bibfnamefont {B.}~\bibnamefont {Glässle}}, \bibinfo
  {author} {\bibfnamefont {M.}~\bibnamefont {Göckeler}}, \bibinfo {author}
  {\bibfnamefont {J.}~\bibnamefont {Najjar}},  \emph {et~al.},\ }\href
  {\doibase 10.1103/PhysRevD.91.054501} {\bibfield  {journal} {\bibinfo
  {journal} {Phys.Rev.}\ }\textbf {\bibinfo {volume} {D91}},\ \bibinfo {pages}
  {054501} (\bibinfo {year} {2015})},\ \Eprint {http://arxiv.org/abs/1412.7336}
  {arXiv:1412.7336 [hep-lat]} \BibitemShut {NoStop}%
%%CITATION = ARXIV:1412.7336;%%
\bibitem [{\citenamefont {Bhattacharya}\ \emph {et~al.}(2006)\citenamefont
  {Bhattacharya}, \citenamefont {Gupta}, \citenamefont {Lee}, \citenamefont
  {Sharpe},\ and\ \citenamefont {Wu}}]{Bhattacharya:2005rb}%
  \BibitemOpen
  \bibfield  {author} {\bibinfo {author} {\bibfnamefont {T.}~\bibnamefont
  {Bhattacharya}}, \bibinfo {author} {\bibfnamefont {R.}~\bibnamefont {Gupta}},
  \bibinfo {author} {\bibfnamefont {W.}~\bibnamefont {Lee}}, \bibinfo {author}
  {\bibfnamefont {S.~R.}\ \bibnamefont {Sharpe}}, \ and\ \bibinfo {author}
  {\bibfnamefont {J.~M.~S.}\ \bibnamefont {Wu}},\ }\href {\doibase
  10.1103/PhysRevD.73.034504} {\bibfield  {journal} {\bibinfo  {journal} {Phys.
  Rev.}\ }\textbf {\bibinfo {volume} {D73}},\ \bibinfo {pages} {034504}
  (\bibinfo {year} {2006})},\ \Eprint {http://arxiv.org/abs/hep-lat/0511014}
  {arXiv:hep-lat/0511014 [hep-lat]} \BibitemShut {NoStop}%
%%CITATION = HEP-LAT/0511014;%%
\bibitem [{\citenamefont {Bulava}\ \emph {et~al.}(2015)\citenamefont {Bulava},
  \citenamefont {Della~Morte}, \citenamefont {Heitger},\ and\ \citenamefont
  {Wittemeier}}]{Bulava:2015bxa}%
  \BibitemOpen
  \bibfield  {author} {\bibinfo {author} {\bibfnamefont {J.}~\bibnamefont
  {Bulava}}, \bibinfo {author} {\bibfnamefont {M.}~\bibnamefont {Della~Morte}},
  \bibinfo {author} {\bibfnamefont {J.}~\bibnamefont {Heitger}}, \ and\
  \bibinfo {author} {\bibfnamefont {C.}~\bibnamefont {Wittemeier}} (\bibinfo
  {collaboration} {ALPHA}),\ }\href {\doibase 10.1016/j.nuclphysb.2015.05.003}
  {\bibfield  {journal} {\bibinfo  {journal} {Nucl. Phys.}\ }\textbf {\bibinfo
  {volume} {B896}},\ \bibinfo {pages} {555} (\bibinfo {year} {2015})},\ \Eprint
  {http://arxiv.org/abs/1502.04999} {arXiv:1502.04999 [hep-lat]} \BibitemShut
  {NoStop}%
%%CITATION = ARXIV:1502.04999;%%
\bibitem [{\citenamefont {Lin}\ \emph {et~al.}(2008)\citenamefont {Lin},
  \citenamefont {Blum}, \citenamefont {Ohta}, \citenamefont {Sasaki},\ and\
  \citenamefont {Yamazaki}}]{Lin:2008uz}%
  \BibitemOpen
  \bibfield  {author} {\bibinfo {author} {\bibfnamefont {H.-W.}\ \bibnamefont
  {Lin}}, \bibinfo {author} {\bibfnamefont {T.}~\bibnamefont {Blum}}, \bibinfo
  {author} {\bibfnamefont {S.}~\bibnamefont {Ohta}}, \bibinfo {author}
  {\bibfnamefont {S.}~\bibnamefont {Sasaki}}, \ and\ \bibinfo {author}
  {\bibfnamefont {T.}~\bibnamefont {Yamazaki}},\ }\href {\doibase
  10.1103/PhysRevD.78.014505} {\bibfield  {journal} {\bibinfo  {journal}
  {Phys.Rev.}\ }\textbf {\bibinfo {volume} {D78}},\ \bibinfo {pages} {014505}
  (\bibinfo {year} {2008})},\ \Eprint {http://arxiv.org/abs/0802.0863}
  {arXiv:0802.0863 [hep-lat]} \BibitemShut {NoStop}%
%%CITATION = ARXIV:0802.0863;%%
\bibitem [{\citenamefont {Yamazaki}\ \emph {et~al.}(2009)\citenamefont
  {Yamazaki}, \citenamefont {Aoki}, \citenamefont {Blum}, \citenamefont {Lin},
  \citenamefont {Ohta}, \citenamefont {Sasaki}, \citenamefont {Tweedie},\ and\
  \citenamefont {Zanotti}}]{Yamazaki:2009zq}%
  \BibitemOpen
  \bibfield  {author} {\bibinfo {author} {\bibfnamefont {T.}~\bibnamefont
  {Yamazaki}}, \bibinfo {author} {\bibfnamefont {Y.}~\bibnamefont {Aoki}},
  \bibinfo {author} {\bibfnamefont {T.}~\bibnamefont {Blum}}, \bibinfo {author}
  {\bibfnamefont {H.-W.}\ \bibnamefont {Lin}}, \bibinfo {author} {\bibfnamefont
  {S.}~\bibnamefont {Ohta}}, \bibinfo {author} {\bibfnamefont {S.}~\bibnamefont
  {Sasaki}}, \bibinfo {author} {\bibfnamefont {R.}~\bibnamefont {Tweedie}}, \
  and\ \bibinfo {author} {\bibfnamefont {J.}~\bibnamefont {Zanotti}},\ }\href
  {\doibase 10.1103/PhysRevD.79.114505} {\bibfield  {journal} {\bibinfo
  {journal} {Phys. Rev.}\ }\textbf {\bibinfo {volume} {D79}},\ \bibinfo {pages}
  {114505} (\bibinfo {year} {2009})},\ \Eprint {http://arxiv.org/abs/0904.2039}
  {arXiv:0904.2039 [hep-lat]} \BibitemShut {NoStop}%
%%CITATION = ARXIV:0904.2039;%%
\bibitem [{\citenamefont {Green}\ \emph {et~al.}(2017)\citenamefont {Green},
  \citenamefont {Hasan}, \citenamefont {Meinel}, \citenamefont {Engelhardt},
  \citenamefont {Krieg}, \citenamefont {Laeuchli}, \citenamefont {Negele},
  \citenamefont {Orginos}, \citenamefont {Pochinsky},\ and\ \citenamefont
  {Syritsyn}}]{Green:2017keo}%
  \BibitemOpen
  \bibfield  {author} {\bibinfo {author} {\bibfnamefont {J.}~\bibnamefont
  {Green}}, \bibinfo {author} {\bibfnamefont {N.}~\bibnamefont {Hasan}},
  \bibinfo {author} {\bibfnamefont {S.}~\bibnamefont {Meinel}}, \bibinfo
  {author} {\bibfnamefont {M.}~\bibnamefont {Engelhardt}}, \bibinfo {author}
  {\bibfnamefont {S.}~\bibnamefont {Krieg}}, \bibinfo {author} {\bibfnamefont
  {J.}~\bibnamefont {Laeuchli}}, \bibinfo {author} {\bibfnamefont
  {J.}~\bibnamefont {Negele}}, \bibinfo {author} {\bibfnamefont
  {K.}~\bibnamefont {Orginos}}, \bibinfo {author} {\bibfnamefont
  {A.}~\bibnamefont {Pochinsky}}, \ and\ \bibinfo {author} {\bibfnamefont
  {S.}~\bibnamefont {Syritsyn}},\ }\href@noop {} {\  (\bibinfo {year}
  {2017})},\ \Eprint {http://arxiv.org/abs/1703.06703} {arXiv:1703.06703
  [hep-lat]} \BibitemShut {NoStop}%
%%CITATION = ARXIV:1703.06703;%%
\end{thebibliography}%
